\input psfig.sty
\magnification=\magstep0
\hsize=13.5 cm               
\vsize=19.0 cm               
\baselineskip=12 pt plus 1 pt minus 1 pt  
\parindent=0.5 cm  
\hoffset=1.3 cm      
\voffset=2.5 cm      
\font\twelvebf=cmbx10 at 12truept 
\font\twelverm=cmr10 at 12truept 
\overfullrule=0pt
\nopagenumbers    
%
\newtoks\leftheadline \leftheadline={\hfill {\eightit Authors' name}
\hfill}
\newtoks\rightheadline \rightheadline={\hfill {\eightit the running title}
 \hfill}
\newtoks\firstheadline \firstheadline={{\eightrm Bull. Astron. Soc.
India (1999) {\eightbf 27}, 441-546 } \hfill}
\def\makeheadline{\vbox to 0pt{\vskip -22.5pt
\line{\vbox to 8.5 pt{}\ifnum\pageno=1\the\firstheadline\else%
\ifodd\pageno\the\rightheadline\else%
\the\leftheadline\fi\fi}\vss}\nointerlineskip}
%
\font\eightrm=cmr8  \font\eighti=cmmi8  \font\eightsy=cmsy8
\font\eightbf=cmbx8 \font\eighttt=cmtt8 \font\eightit=cmti8
\font\eightsl=cmsl8
\font\sixrm=cmr6    \font\sixi=cmmi6    \font\sixsy=cmsy6
\font\sixbf=cmbx6
%
\def\eightpoint{\def\rm{\fam0\eightrm}
\textfont0=\eightrm \scriptfont0=\sixrm \scriptscriptfont0=\fiverm
\textfont1=\eighti  \scriptfont1=\sixi  \scriptscriptfont1=\fivei
\textfont2=\eightsy \scriptfont2=\sixsy \scriptscriptfont2=\fivesy
\textfont3=\tenex   \scriptfont3=\tenex \scriptscriptfont3=\tenex
\textfont\itfam=\eightit  \def\it{\fam\itfam\eightit}%
\textfont\slfam=\eightsl  \def\sl{\fam\slfam\eightsl}%
\textfont\ttfam=\eighttt  \def\tt{\fam\ttfam\eighttt}%
\textfont\bffam=\eightbf  \scriptfont\bffam=\sixbf
\scriptscriptfont\bffam=\fivebf \def\bf{\fam\bffam\eightbf}%
\normalbaselineskip=10pt plus 0.1 pt minus 0.1 pt
\normalbaselines
\abovedisplayskip=10pt plus 2.4pt minus 7pt
\belowdisplayskip=10pt plus 2.4pt minus 7pt
\belowdisplayshortskip=5.6pt plus 2.4pt minus 3.2pt \rm}
%
%
\def\leftdisplay#1\eqno#2$${\line{\indent\indent\indent%
$\displaystyle{#1}$\hfil #2}$$}
\everydisplay{\leftdisplay}
%
\def\frac#1#2{{#1\over#2}}


%
%
\def\pmb#1{\setbox0=\hbox{$#1$}%
\kern-.025em\copy0\kern-\wd0
\kern.05em\copy0\kern-\wd0
\kern-.025em\raise.0433em\box0}
%
\pageno=1
\vglue 50 pt  
%
\leftline{\twelvebf Emerging trends of optical interferometry in astronomy}
%
\smallskip
\vskip 40 pt  
\leftline{\twelverm S. K. Saha} 
\vskip 4 pt
\leftline{\eightit Indian Institute of Astrophysics, Bangalore-560 034, India.}
\leftline{\eightit e-mail: sks@iiap.ernet.in
}
\bigskip
\noindent{\eightrm Received 4 August 1999; Accepted 15 September 1999}
%
%
\vskip 20 pt 
%
%
\leftheadline={\hfill {\eightit S. K. Saha} \hfill}
\rightheadline={\hfill {\eightit Emerging trends of optical interferometry in  
astronomy}  \hfill}

%
{\parindent=0cm\leftskip=1.5 cm

{\bf Abstract.}
\noindent
The current status of the high spatial resolution imaging interferometry in 
optical astronomy is reviewed in the light of theoretical explanation, as well
as of experimental constraints that exist in the present day technology. 
The basic mathematical interlude pertinent to the interferometric 
technique and its applications in astronomical observations using both single 
aperture, as well as diluted apertures are presented in detail. An elaborate 
account of the random refractive index fluctuations of the atmosphere producing 
random aberrations in the telescope pupil, elucidating the trade offs between 
long-exposure and short-exposure imaging is given. The formation of speckles 
and of ways to detect them in the case of astronomical objects are discussed. 
Further, the other methods viz., (i) speckle spectroscopy, (ii) speckle 
polarimetry, (iii) phase closure, (iv) aperture synthesis, (v) pupil plane 
interferometry, (vi) differential speckle interferometry etc., using single
moderate or large telescopes are described as well. The salient features of 
various detectors that are used for recording short-exposure images are 
summarized. The mathematical intricacies 
of the data processing techniques for both Fourier modulus and Fourier phase are 
analyzed; the various schemes of image restoration techniques are examined as 
well with emphasis set on their comparisons. A brief account of obtaining 
diffraction-limited high resolution features of a few extended objects is 
presented. The recent technological innovation to compensate the deleterious 
effects of the atmosphere on the telescope image in real-time is enumerated. The 
experimental descriptions of several working long baseline interferometers in 
the visible band using two or more telescopes are summarized. The astrophysical 
results obtained till date using both single aperture interferometry, as well as 
long baseline interferometry with diluted apertures are highlighted.  

}                                 
%
%
%
\vskip 20 pt
\centerline{\bf Table of contents}
\bigskip
\noindent
\item{1.} Prologue
\item{2.} Preamble

\indent{2.1.} Interference of two light waves

\indent{2.2.} Experiments to measure stellar diameter
\item{3.} Atmospheric turbulence

\indent{3.1.} Formation of eddies

\indent{3.2.} Inertial subrange and structure functions

\indent{3.3.} Wave propagation through the turbulent atmosphere

\indent{3.4.} Fried's parameter ($r_\circ$)

\indent\indent{3.4.1.} Benefit of short-exposure images

\indent\indent{3.4.2.} Measurement of $r_\circ$

\indent\indent{3.4.3.} Seeing at the telescope site
\item{4.} Single aperture interferometry  

\indent{4.1.} Speckle interferometry

\indent\indent{4.1.1.} Speckle interferometer

\indent\indent{4.1.2.} Estimation of Fourier modulus

\indent\indent{4.1.3.} Difficulties in data processing

\indent{4.2.} Speckle spectroscopy

\indent{4.3.} Speckle polarimetry

\indent{4.4.} Differential speckle interferometry

\indent{4.5.} Pupil plane interferometry

\indent{4.6.} Phase-closure imaging

\indent\indent{4.6.1.} Aperture synthesis
\item{5.} Detectors

\indent{5.1.} Frame transfer camera system

\indent{5.2.} Photon counting camera system

\indent\indent{5.2.1.} CCD based photon counting system

\indent\indent{5.2.2.} Precision analog photon address (PAPA)

\indent\indent{5.2.3.} MCP based photon counting detector

\indent\indent{5.2.4.} Multi anode micro-channel array (MAMA)
\item{6.} Image Processing

\indent{6.1.} Speckle holography method

\indent{6.2.} Shift-and-add algorithm

\indent{6.3.} Knox-Thomson technique (KT) 

\indent{6.4.} Triple correlation technique (TC)

\indent\indent{6.4.1.} Relationship between KT and TC

\indent{6.5.} Blind iterative deconvolution technique (BID)
\item{7.} Speckle imaging of extended objects

\indent{7.1.} Jupiter

\indent{7.2.} Sun

\indent\indent{7.2.1.} Solar speckle observation during eclipse
\item{8.} Adaptive optics 
\item{9.} Diluted aperture interferometry 

\indent{9.1.} Intensity interferometry 

\indent{9.2.} Amplitude and phase interferometry
 
\indent\indent{9.2.1.} Interferometers at Plateau de Calern 

\indent\indent{9.2.2.} Mount Wilson stellar interferometer

\indent\indent{9.2.3.} Sydney University stellar interferometer (SUSI)

\indent\indent{9.2.4.} Cambridge optical aperture synthesis telescope (COAST)

\indent\indent{9.2.5.} Infrared-optical telescope array (IOTA)
\item{10.} Astrophysical results

\indent{10.1} Results obtained with single aperture interferometry

\indent{10.2} Results obtained with adaptive optics

\indent{10.3} Results obtained with diluted aperture interferometry
\item{11.} Epilogue
\item{}Acknowledgment
\item{}References
\item{}Appendix
\vskip 20 pt
\centerline {\bf 1. Prologue}
\bigskip
\noindent
Diffraction-limited resolution achievable by any terrestrial large telescope far 
surpasses that imposed by atmospheric fluctuations above the telescope aperture
as against lone orbiting telescope. 
Turbulence with its associated random refractive index 
inhomogeneities in the atmosphere distorts the characteristics of light 
traveling through it. The limitation is due to warping of iso-phase surfaces and 
intensity variation across the wave-front, thereby, distorting the shapes of 
the wave-front (Fried, 1966). Due to the diffraction phenomenon, the image 
of a point source blurs at the focal point of the telescope. This phenomena 
is present in the sound waves, as well as in the electro-magnetic spectrum 
starting from gamma rays to radio waves. The blurring suffered by such images is 
modeled as convolution with the point spread function (PSF). Degradation in 
image quality can occur because of (i) opto-mechanical aberrations of the 
telescope and (ii) density inhomogeneities in the path of the optical wave-front.
Understanding of the effect of atmospheric turbulence on the structure of 
stellar images and of ways to overcome the hindrance would enable astronomers
to retrieve high spatial resolution of the object.  
\bigskip
Though interferometry at optical wavelengths in astronomy began more than a 
century and a quarter ago (Fizeau, 1868), the progress in achieving high angular 
resolution has been modest. The first successful 
measurement of the angular diameter of $\alpha$~Orionis was performed in 1920 
using stellar interferometer (Michelson and Pease, 1921), but the field lay 
dormant until it was revitalized by the development of
intensity interferometry (Brown and Twiss, 1958). On the other hand, real 
progress has been made at radio wavelengths in post war era. Development of long 
baseline and very long baseline interferometry (VLBI), as well as usage of 
sophisticated image processing techniques have brought high dynamic range
images with milliarc-second (mas) resolution. 
\bigskip
The implementation of imaging by
interferometry in optical astronomy is a challenging task. Over the last two
decades, a marked progress has been witnessed in the development of this field, 
offering to realize the potential of the interferometric technique.  
Single aperture speckle interferometry (Labeyrie, 1970) offers a new way of 
utilizing the large telescopes to obtain diffraction-limited spatial Fourier 
spectrum and image features of the object. A profound increase has been noticed 
in its contribution to measure fundamental
stellar parameters, viz., (i) diameter of stars, (ii) separation of close binary 
stars, (iii) imaging of emission line of the active galactic nuclei (AGN), 
(iv) the spatial distribution of circumstellar matter surrounding objects,  
(v) the gravitationally lensed QSO's etc. Further benefits
have been witnessed when the atmospherically degraded images of these objects
is applied to image restoration techniques (Liu and Lohmann, 1973, Rhodes and 
Goodman, 1973, Knox and Thomson, 1974, Lynds et al., 1976, Weigelt, 1977, 
Lohmann et al., 1983, Ayers and Dainty, 1988) for obtaining Fourier phase.  
\bigskip
Development of various interferometric techniques, namely, (i) speckle
spectroscopy (Grieger and Weigelt, 1992), (ii) speckle polarimetry (Falcke et
al., 1996), (iii) pupil plane interferometry (Roddier and Roddier, 1988),
(iv) Phase-closure method (Baldwin et al., 1986), (v) aperture synthesis using 
both partial redundant and non-redundant masking (Haniff et al., 1987, 1989, 
Nakajima et al., 1989, Busher et al., 1990, Bedding et al., 1992, 1994, 
Bedding, 1999), (vi) differential speckle interferometry (Petrov et al., 1986) 
using single aperture telescope have brought out a considerable amount of new 
results.
\bigskip 
Significant improvements in technological innovation over the past several years
have brought the hardware to compensate in real-time for telescope image 
degradation induced by the atmospheric turbulence. Wave-front sensing and 
adaptive optics (AO) are based on this hardware oriented correction (Babcock, 
1953, Rousset et al., 1990). Adaptive optical systems may become standard tool 
for the new generation large telescopes. 
\bigskip
Although large telescope helps in gathering more optical 
energy, the angular resolution is limited with its diameter. In the optical 
band, a large mirror of diameter more than 10~m class with high 
precision in figuring may not be possible to develop, thereby, 
restricting the resolution in the single aperture interferometry. Introduction
of long baseline interferometry using diluted apertures became necessary
(Labeyrie, 1975). Success in synthesizing 
images obtained from a pair of independent telescopes on a North-South baseline
configuration (Labeyrie, 1975, Labeyrie et al., 1986, Shao et al., 1988), 
impelled astronomers to venture towards ground-based very large
arrays (Davis et al., 1992). Potentials for progress in 
the direction of developing large interferometric 
arrays of telescopes (Labeyrie, 1996) are expected to provide 
images, spectra of quasar host galaxies, exo-planets that may be 
associated with stars outside the solar system (Labeyrie, 1995, 1998a, 1998b).
Plans are also on to put an interferometer of a 
similar kind on the surface of the moon by early next century. 
\bigskip
The surge of activities in this field over the past two decades reflect 
the motivation for obtaining true diffraction-limited images with the ground
based telescopes. Over the years optical interferometry has slowly gained in
importance and today it has become a powerful tool. 
This review focuses on the application of interferometry
to optical astronomy, as well as on the current activities of the various
groups across the globe. The problems encountered by the astronomers to 
develop various interferometers in the optical domain and possibilities offered 
to astronomy by high resolution imaging in the optical domain are addressed. 
The basic principle of interferometry and its applications, atmospheric 
turbulence and its behaviour in the case of wave propagation, speckle imaging 
and different interferometric techniques, adaptive optics system, the salient 
features of long baseline interferometers are enumerated. In addition, the 
various techniques applied to image restoration and their shortcomings are
illustrated. Some of the important results obtained so far using various
interferometers are also highlighted.
\vskip 20 pt
\centerline {\bf 2. Preamble}
\bigskip 
\noindent
Before venturing into the complicated equations pertinent to the 
interferometric systems, some of the fundamental equations describing a plane 
wave of monochromatic/quasi-monochromatic light impinging on an ideal instrument, 
say a telescope, are illustrated.
\bigskip
Let $V{\bf (r}, t)$ represent the monochromatic optical wave which at a point,
${\bf r}$, can be specified in the form,
${\it a}({\bf r})e^{-i[2\pi\nu_\circ t - \psi]}$, 
where, ${\it a}({\bf r})$ is the instantaneous complex amplitude of the wave,  

$$V{\bf (r}, t) = \Re\left\{{\it a}({\bf r})e^{-i[2\pi\nu_\circ t - \psi]}\right\}. \eqno(1)$$ 

\noindent
Here, ${\bf r} = (x, y, z)$ is the position vector of the point $(x, y, z)$, $t$ 
the time, and $\nu_\circ$ the frequency of the wave. 
\bigskip 
The complex representation of the analytic signal, 
${\cal U}{\bf (r}, t)$, associated with $V{\bf (r}, t)$ can be expressed as,

$${\cal U}{\bf (r}, t) = {\it a}({\bf r})e^{-i[2\pi\nu_\circ t - \psi]}  
= \Psi({\bf r})e^{-i2\pi\nu_\circ t}, \eqno(2)$$ 

\noindent
where, $\Psi({\bf r}) = {\it a}({\bf r})e^{i\psi}$, is a complex vector function
of position.  
\bigskip
From the equations (1) and (2), the
relationship can be translated in the form of following equation,  

$$V{\bf (r}, t) = \Re\left\{\Psi({\bf r})e^{-i\omega t}\right\} 
= \frac{1}{2}\left[\Psi({\bf r})e^{-i\omega t} + \Psi^\ast({\bf r})e^{i\omega t}
\right], \eqno(3)$$ 

\noindent
where, $\omega = 2\pi\nu_\circ$ is the angular frequency.
\bigskip
The intensity of light is defined as the time average of the amount of energy,
therefore, taking the latter over an interval much greater than the period, 
${\tt T} = 2\pi/\omega$, the intensity at the same point is calculated as,

$${\cal I} = <V^2> = \frac{1}{2}\Psi\Psi^\ast, \eqno(4)$$  

\noindent
where, $< \ >$ stands for the ensemble average of the quantity within the bracket.
and $\Psi^\ast$ represents for the complex conjugate of $\Psi$.
\bigskip
Since the complex amplitude is a constant phasor in the monochromatic case, the 
spectrum of the complex representation of the signal, ${\cal U}{\bf (r}, t)$, 
is given by,

$$\widehat{\cal U}{\bf (r}, \nu) = {\it a}({\bf r})e^{i\psi}\delta(\nu - \nu_\circ). \eqno(5)$$

It is equal to twice the positive part of the instantaneous spectrum,
$\widehat{V}{\bf (r}, \nu)$. In the polychromatic case, the expression for complex
amplitude can be formulated as,

$${\cal U}{\bf (r}, t) = 2\int_0^\infty\widehat{V}{\bf (r}, \nu) e^{-i2\pi\nu t}d\nu. \eqno(6)$$

The disturbance produced by a real physical source is calculated by the 
integration of the monochromatic signals over a optical band pass. 
In the case of quasi-monochromatic approximation (if the width of the 
spectrum, $\Delta\nu \ll \nu_\circ$), the expression can be read as,

$${\cal U}{\bf (r}, t) = \mid\Psi({\bf r}, t)\mid e^{i[\psi(t) - 2\pi\nu_\circ t]}.
\eqno(7)$$

The field is characterized by the complex amplitude, $\Psi{(\bf r}, t) = 
\mid\Psi({\bf r}, t)\mid e^{i\psi(t)}$; the phasor is time dependent,
although it is varying slowly with respect to $e^{-i2\pi\nu_\circ t}$.
\bigskip
In the case of 2-dimensional (2-d) distributions of the optical fields in the
object and image planes, the following steps are necessary to be incorporated.
The distribution of the complex amplitude, $\Psi({\bf x})$, in the image plane
for a monochromatic point source of wavelength $\lambda$, can be formulated as,

$$\Psi({\bf x}) = \int{\cal T}({\bf x, x_\circ})\Psi_\circ({\bf x_\circ}) 
d{\bf x_\circ}, \eqno(8)$$

\noindent
where, ${\bf x} = (x, y)$ is a 2-d space vector, 
$\Psi_\circ({\bf x_\circ})$ is complex amplitude in the telescope pupil plane, 
${\cal T}({\bf x, x_\circ})$ is the impulse response of the optical system. 
According to the diffraction theory, ${\cal T}({\bf x, x_\circ})$  
is proportional to $\int
{\cal P}({\bf x})e^{-i2\pi{\bf x\cdot x_\circ}/\lambda}d{\bf x}$. Therefore, the
equation (8) can be represented as,

$$\Psi({\bf x}) \propto \int{\cal P}({\bf x})\Psi_\circ({\bf x_\circ})
e^{-i2\pi{\bf x\cdot x_\circ}/\lambda}d{\bf x}, \eqno(9)$$

\noindent
where, ${\cal P}({\bf x})$ is the pupil transmission function.
\bigskip
The complex amplitudes exhibit a time-dependency in  
quasi-monochromatic case; therefore, the field in the image plane can be 
expressed as,

$$\Psi({\bf x}, t) = \int{\cal T}({\bf x, x_\circ})\Psi_\circ({\bf x_\circ}, t) 
d{\bf x_\circ}. \eqno(10)$$

Since the detectors in the visual band are not sensitive to the field but to
the intensity, ${\cal I}({\bf x})$, 

$${\cal I}({\bf x}) = <\Psi({\bf x}, t)\Psi^\ast({\bf x}, t)>. \eqno(11)$$

For an ideal telescope, ${\cal P}({\bf x})$ = 1, for inside the aperture and
0 for outside the aperture. In the space-invariant case,
the expression for the transfer function can be derived as,

$$\Psi({\bf x}) \propto \int{\cal P}({\bf u})\Psi({\bf u})
e^{-i2\pi{\bf x u}/\lambda}d{\bf u}, \eqno(12)$$

\noindent
where, $\int{\cal P}({\bf u})\Psi({\bf u})e^{-i2\pi{\bf x u}/\lambda}d{\bf u}
=\int\int_\infty^\infty{\cal P}(u, v)\Psi(u, v)e^{-i2\pi[u x + v y]/\lambda}
du dv$ and the dimensionless variable, ${\bf u}$, stands for ${\bf x}/\lambda$,
and hence, $\Psi({\bf u})$ can be written for $\Psi_\circ(\lambda{\bf u})$. 
\bigskip
Similarly, ${\cal P}({\bf u}) = {\cal P}_\circ(\lambda{\bf u})$. From  
equation (11), it is seen that the complex amplitude in the image plane  
is the convolution of the complex amplitude of the pupil 
plane and the pupil transmission function (Goodman, 1968). The 
mathematical description of the convolution of two functions is of the form:

$$\Psi({\bf x}) = {\cal P}({\bf x})\ast\Psi_\circ({\bf x}), \eqno(13)$$
 
\noindent
where, $\ast$ stands for the convolution operator. 
\bigskip
Convolution equations can be reduced to agreeable form using the Fourier 
convolution theorem. The Fourier transform of a convolution of two functions
is the product of the Fourier transform of the two functions. Therefore, in
the Fourier plane the effect becomes a multiplication, point by point, of
the transfer function of the pupil, $\widehat{\cal P}({\bf u})$, with the 
transform of the complex amplitude of the pupil plane, $\widehat{\Psi}_\circ
({\bf u})$. i. e.,
 
$$\widehat{\Psi}({\bf u}) = \widehat{\cal P}({\bf u})\cdot\widehat
{\Psi}_\circ({\bf u}). \eqno(14)$$
 
The illumination at the focal plane of the telescope observed as a function
of the image plane is,

$${\cal S}({\bf x}) = \mid\Psi({\bf x})\mid^2,  \eqno(15)$$

\noindent
where, ${\cal S}({\bf x})$, is the PSF produced
by both the telescope and the atmosphere and its Fourier transform, $\widehat
{\cal S}({\bf u})$, the optical transfer function (OTF), while its
modulus, $\mid\widehat{\cal S}({\bf u})\mid^2$, is the modulus transfer function
(MTF). 
\bigskip
In the ideal condition, the resolution that can be achieved in an imaging experiment,
${\cal R}$, is limited only by the imperfections in the optical system and
according to Strehl's criterion, is given by the integral of its transfer 
function. 

$${\cal R} = \int\widehat{\cal S}({\bf u})d{\bf u}. \eqno(16)$$

If ${\cal R} = {\cal S}({\bf 0})$, the central intensity in the image is
known as Strehl's intensity. 
\vskip 20 pt
\noindent
{\bf 2.1. Interference of two light waves}
\bigskip 
\noindent
The astronomical sources emit incoherent light consisting of the random
superposition of numerous successive short-lived waves sent out from many 
elementary radiators, and therefore, the optical coherence is related to the 
various forms of correlations of the random process. 
When two light beams from a single source are superposed, the intensity at
the place of superposition varies from point to point between maxima, which
exceed the sum of the intensities in the beams, and minima, which may be zero, 
known as interference. The correlated fluctuation can be partially or completely 
coherent (Born and Wolf, 1984). 
\bigskip
\noindent
\midinsert
{\eightpoint   
\noindent
\centerline{\psfig{figure=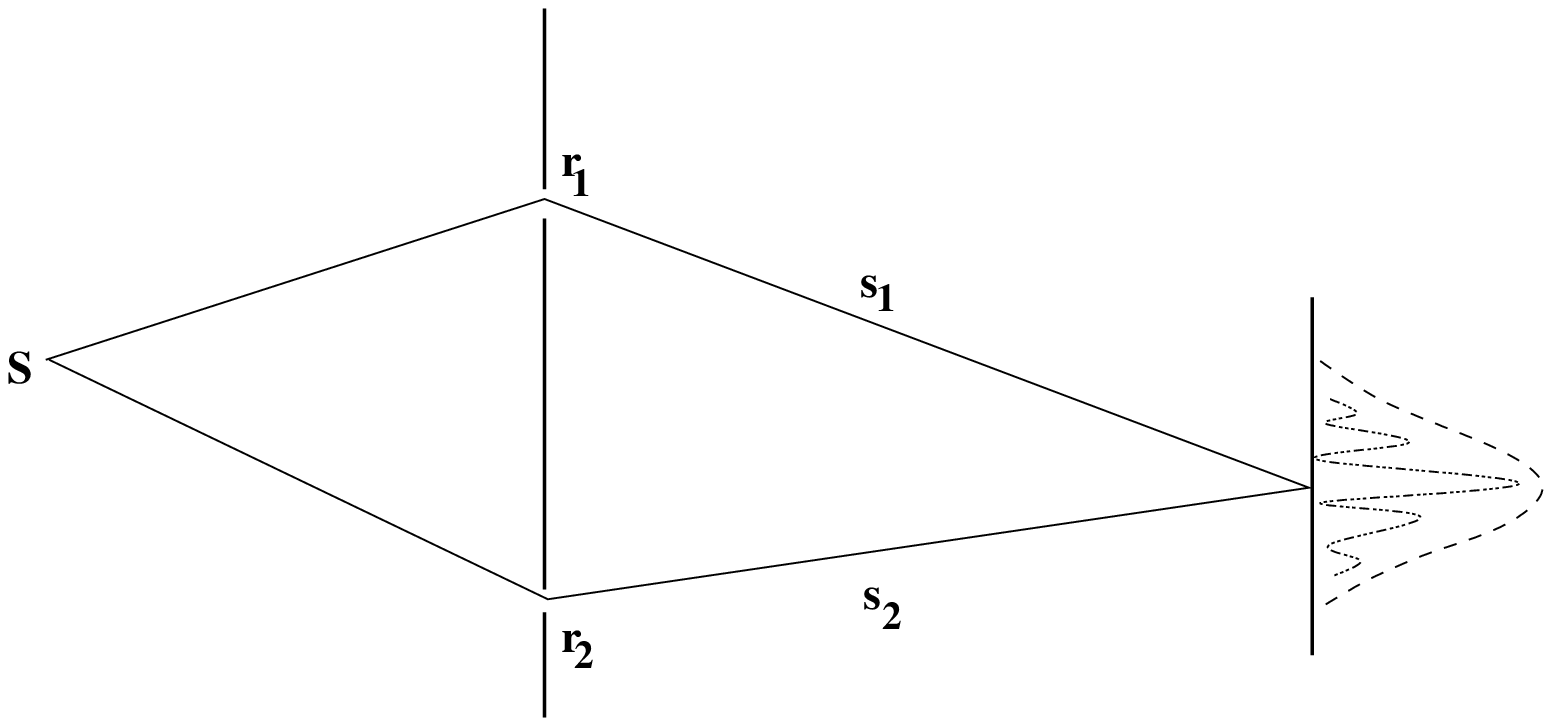,height=5.5cm,width=9cm}}
\bigskip
\noindent
{\bf Figure 1.} Concept of Young's interferometer.  
}     
\endinsert
\bigskip 
Figure 1 depicts the concept of Young's interferometer with 2 apertures.
Let the two monochromatic waves represented by the vector 
${\bf V}_1$ and ${\bf V}_2$ be superposed at the recombination 
point. The correlator sums the instantaneous amplitudes of the fields. 
The total field at the output is,

$${\bf V} = {\bf V}_1 + {\bf V}_2, \eqno(17)$$
$${\bf V}^2 = {\bf V}^2_1 + {\bf V}^2_2 + 2{\bf V}_1 {\bf V}_2. \eqno(18)$$ 

The total intensity, (see equation 4), at the same point is calculated as,

$${\cal I} = {\cal I}_1 + {\cal I}_2 + {\cal J}_{12}, \eqno(19)$$ 

\noindent
where, ${\cal I}_1 = <{\bf V}^2_1>$; ${\cal I}_2 = <{\bf V}_2^2>$ are the 
intensities of two waves; ${\cal J}_{12} = 2<{\bf V}_1 {\bf V}_2>$ 
is the interference term. 
\bigskip
Now, let $\Psi_1$ and $\Psi_2$ be the complex amplitudes of the 
two waves, with the corresponding phases $\psi_1$ and $\psi_2$.  

$${\cal J}_{12} = \frac{1}{2}\left(\Psi_1\Psi_2^\ast + 
\Psi_1^\ast\Psi_2\right) = 2\sqrt{{\cal I}_1 {\cal I}_2} 
cos\delta. \eqno(20)$$

This expression depicts that the interference term depends on the amplitude
components, as well as on the phase-difference between the two waves,
$\delta = 2\pi\Delta\varphi/\lambda_\circ$, where, $\Delta \varphi$, is the 
optical path difference (OPD) between the two waves 
from the common source to the intersecting point and $\lambda_\circ$ is the 
wavelength in vacuum. In general, two light beams are not correlated but the
correlation term, $\Psi_1\Psi_2^\ast$, takes significant values for
a short period of time and $<\Psi_1\Psi_2^\ast>$ = 0. 
The total intensity can be derived as,

$${\cal I} = {\cal I}_1 + {\cal I}_2 \pm 2 \sqrt{{\cal I}_1 {\cal I}_2} 
\ cos\delta. \eqno(21)$$ 

The maximum and minimum intensity can be obtained, when $\mid\delta\mid = 
0, 2\pi, 4\pi$ and $\mid\delta\mid = \pi, 3\pi, 5\pi$, respectively.
If ${\cal I}_1 = {\cal I}_2$,

$${\cal I} = 4{\cal I}_1cos^2\frac{\delta}{2}. \eqno(22)$$  

The intensity varies between 4${\cal I}_j$, and 0.
\bigskip 
Unlike the monochromatic wave field, where the amplitude of the vibration at any 
point is constant and the phase varies linearly with time, the amplitude and 
phase in the case of the quasi-monochromatic wave field, undergo irregular 
fluctuations (Born and Wolf, 1984). The rapidity of fluctuations depends on the 
effective width of the spectrum. The interferometers based on, viz., (i) 
wave-front division (Young's experiment), (ii) amplitude division (Michelson's 
interferometer) are generally used to measure spatial coherence and
temporal coherence respectively. On the other hand, Young's holes (see figure
1) set-up is sensitive to the bandwidth of the source. 
A mathematical description of the Young's experiment is as follows. 
\bigskip 
Let the analytical signal, ${\cal U}(t)$, obtained at the output be,

$${\cal U}(t) = {\bf K}_1{\cal U}(r_1, t-\tau_1) + {\bf K}_2{\cal U}(r_2, 
t-\tau_2), \eqno(23)$$

\noindent
where, ${\bf K}_j$'s are the constants, $r_j$'s the positions of two pin-holes 
in the wave field, j = 1, 2, $s_j$'s the distances of a meeting point of the 
two beams from the two pin-holes, $\tau_j = s_j/c$, the time needed to 
travel from the respective pin-holes to the meeting point, and $c$ the 
velocity of light. 
\bigskip
If the pin-holes are small and the diffracted fields are considered to be 
uniform, the intensity at the output can be expressed as,

$${\cal I} = \mid{\bf K}_1\mid^2<\mid{\cal U}(r_1, t - \tau_1)\mid^2> + 
<\mid{\bf K}_2\mid^2{\cal U}(r_2, t - \tau_2)\mid^2> + 
{\bf K}_1\cdot{\bf K}_2^\ast \; \times \eqno$$
$$<{\cal U}(r_1, t - \tau_1){\cal U}^\ast(r_2, t - \tau_2)>  
+ {\bf K}_1^\ast\cdot{\bf K}_2<{\cal U}^\ast(r_1, t - \tau_1){\cal U}(r_2, t - \tau_2)>,  
\eqno(24)$$

If the value of the constants can be put as, $\mid{\bf K}_j\mid = K_j$; 
it turns out to be,
${\bf K}_1^\ast\cdot{\bf K}_2 = {\bf K}_1\cdot{\bf K}^\ast_2 = K_1K_2$ and 
noting, ${\cal I}_j = \mid{\bf K}_j\mid^2<\mid{\cal U}(r_j, t - \tau_j)\mid^2>$,
and therefore, equation (24) leads to, 

$${\cal I} = {\cal I}_1 + {\cal I}_2 + 2K_1K_2 
\Re\left[\pmb{\gamma}_{12}\left(\frac{s_2 - s_1}{c}\right)\right]. \eqno(25)$$

The Van Cittert-Zernicke theorem states that the modulus of the complex degree 
of coherence (describes the correlation of vibrations at a fixed point and a 
variable point) in a plane illuminated by a incoherent quasi-monochromatic 
source is equal to the modulus of the normalized spatial Fourier transform of 
its brightness distribution (Born and Wolf, 1984). The complex degree of  
(mutual) coherence, $\pmb{\gamma}_{12}(\tau)$, 
of the observed source is defined as, 

$$\pmb{\gamma}_{12}(\tau) = \frac{\pmb{\Gamma}_{12}(\tau)}{\sqrt{\pmb{\Gamma}_
{11}(0)\pmb{\Gamma}_{22}(0)}} = \frac{\pmb{\Gamma}_{12}(\tau)}
{\sqrt{{\cal I}_1 {\cal I}_2}}. \eqno(26)$$

The function, $\pmb{\Gamma}_{12}(\tau) = 
<{\cal U}(r_1, t + \tau){\cal U}^\ast(r_2, t)>$, is measured at two points. At a
point where both the points coincide, the self coherence, $\pmb{\Gamma}_{11}(\tau) 
= <{\cal U}(r_1, t + \tau){\cal U}^\ast(r_1, t)>$, reduces to ordinary intensity.
When $\tau = 0$, $\pmb{\Gamma}_{11}(0) = {\cal I}_1; 
\pmb{\Gamma}_{22}(0) = {\cal I}_2$. 
The ensemble average can be replaced by a time average due to the 
assumed ergodicity of the fields. If both the fields are directed on a 
quadratic detector, it yields the desired cross-term (time average due to the
finite time response). The measured intensity at the detector would be,

$${\cal I} = {\cal I}_1 + {\cal I}_2 + 2\sqrt{{\cal I}_1. {\cal I}_2} 
\Re\left[\pmb{\gamma}_{12}\left(\frac{s_2 - s_1}{c}\right)\right]. \eqno(27)$$

In order to keep the time correlation close to unity, the delay, $\tau$, must
be limited to a small fraction of the temporal width or coherence time, $\tau_c
= 1/\Delta\nu$; here, $\Delta\nu$, is the spectral width.  
The relative coherence of the two beams diminishes with the increase of
path length difference, culminating in a drop in the visibility of the fringes.
A factor less than unity affects the degree of coherence. The corresponding 
limit for the OPD between two fields is the coherence length, defined by
$l_c = c\cdot\tau_c = (\lambda_\circ)^2 /\Delta\lambda$. If $\tau \ll \tau_c$; 
the function, $\pmb{\gamma}_{12}(\tau)$, can be expressed as, $\pmb{\gamma}_{12}(0)
e^{-2\pi i\nu_\circ\tau}$, The exponential term is nearly constant and
$\pmb{\gamma}_{12}(0)$, measures the spatial coherence. 
Let $\psi_{12}$, be the argument of $\pmb{\gamma}_{12}(\tau)$, then,

$${\cal I} = {\cal I}_1 + {\cal I}_2 + 2\sqrt{{\cal I}_1, {\cal I}_2}\Re\left
[\mid\pmb{\gamma}_{12}(0)\mid e^{i(\psi_{12}-2\pi\nu_\circ \tau)}\right]. \eqno(28)$$  

The measured intensity at a distance $x$ from the origin (point at zero OPD) on 
a screen at a distance, $z$, from the aperture (see figure 1) is

$${\cal I}(x) = {\cal I}_1 + {\cal I}_2 + 2\sqrt{{\cal I}_1,{\cal I}_2}
\mid\pmb{\gamma}_{12}(0)
\mid cos\left\{\frac{2\pi d(x)}{\lambda}-\psi_{12}\right\}, \eqno(29)$$ 

\noindent
where, $d(x) = b.x/(z,\lambda)$, is the OPD corresponding to $x$, 
and $b$ the distance between the two apertures.
\bigskip 
The modulus of the spatial coherence of collected fields at the aperture appears
through the contrast of the fringes, {\cal V}, and is measured by the 
equation, 

$${\cal V} = \frac{{\cal I}_{max}-{\cal I}_{min}}{{\cal I}_{max} + {\cal I}_{min}}  
= \mid\pmb{\gamma}_{12}(0)\mid\frac{2\sqrt{{\cal I}_1{\cal I}_2}}{{\cal I}_1 + 
{\cal I}_2}. \eqno(30)$$

\vskip 20 pt
\noindent
{\bf 2.2. Experiments to measure stellar diameter}
\bigskip 
\noindent
In order to produce Young's fringes at the focal plane of the telescope, Fizeau 
(1868) had installed a screen with two holes on top of the telescope
and found the relationship between the aspect of interference 
fringes and the angular size of the light source. 
According to him, these fringes remain visible in presence of seeing, therefore,
allow measurements of stellar diameters with diffraction-limited resolution.
Stefan attempted with 1 meter (m) telescope at Observatoire de Marseille and 
fringes appeared within the common Airy disk of the sub-apertures. But he could 
not notice any significant drop of fringe visibility. Since the maximum 
achievable resolution is limited by the diameter of the telescope, he concluded 
none of the observed stars approached 0.1 arc-second ($^{\prime\prime}$) in angular size. 
Later, Michelson measured the diameter of the satellites of Jupiter with Fizeau 
interferometer on top of the Yerkes refractor. Similar interferometer (Fizeau 
mask) was placed on top of the 100~inch telescope at Mt. Wilson Observatory 
(Anderson, 1920) 
and the angular separation of the spectroscopic binary star, Capella, was 
measured. A similar experiment has been conducted by Saha et al., (1987) at 
VBO, Kavalur using 1~m Carl-Zeiss telescope and recorded the fringes of several 
bright stars through a broadband filter in the blue region using a 16 millimeter 
(mm) movie camera giving an exposure of 16 milliseconds (ms) per frame;  
a Barlow lens was used to magnify the image scale. 
\bigskip 
\noindent
\midinsert
{\eightpoint   
\noindent
\centerline{\psfig{figure=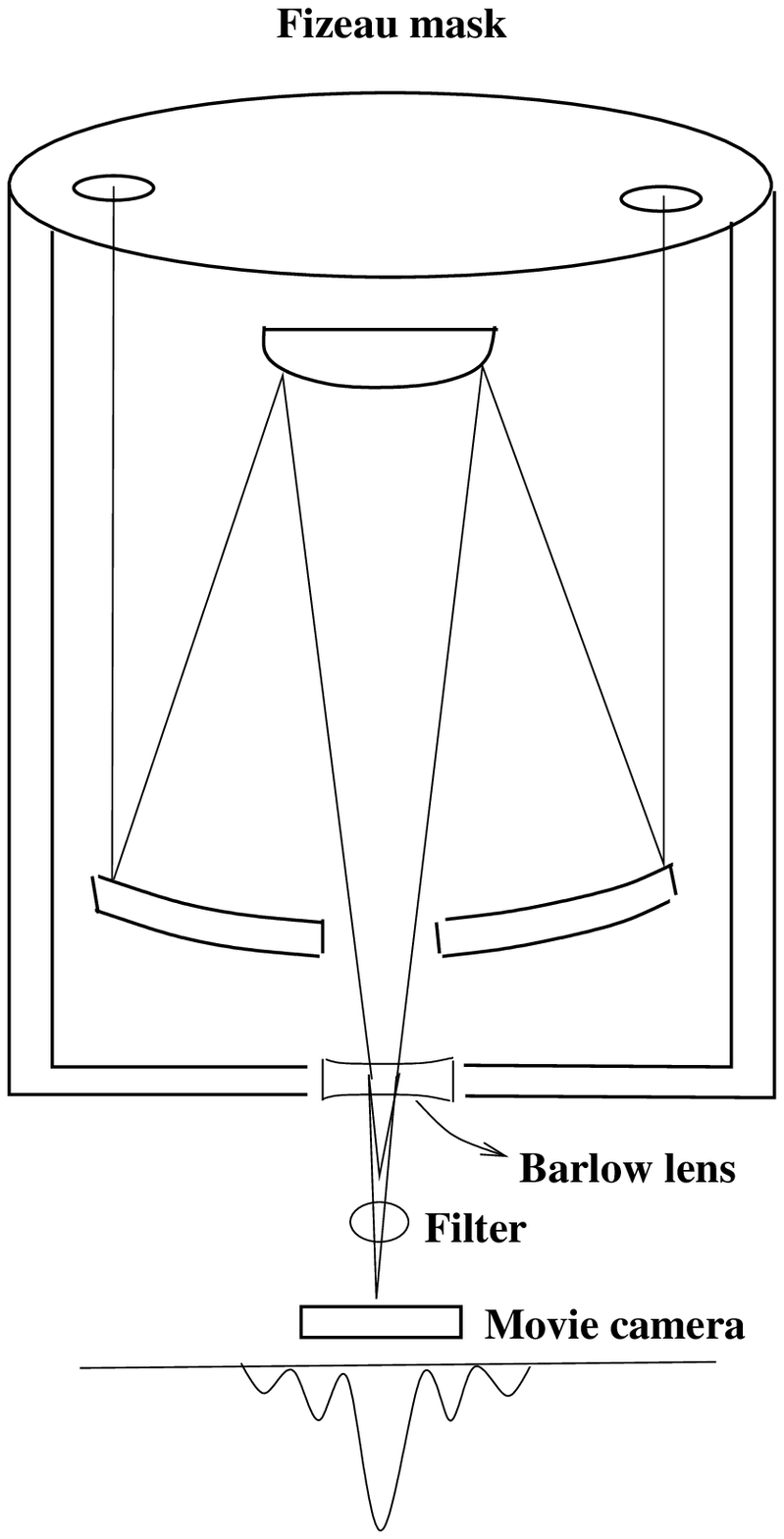,height=9.5cm,width=5cm}}
{\bf Figure 2.} Fizeau - Stefan interferometer using mask over the 1~m telescope
at Vainu Bappu Observatory, Kavalur, India (Saha et al., 1987).  
}     
\endinsert
\bigskip 
Figure 2 describes the concept of the Fizeau - Stefan interferometer.
In this, the beams are diffracted by the sub-apertures in front of a 
telescope which acts as both collector and correlator. A temporal
coherence is automatically obtained due to the built-in zero OPD. The spatial
modulation frequency, as well as the required sampling of the image change with
the separation of sub-apertures. The intensity in the focal plane can be 
written as follows:

$${\cal I}({\bf x}) = {\it a}(D.x)\left[{\cal I}_1 + {\cal I}_2 + 2\sqrt{{\cal I}_1
{\cal I}_2}\mid\pmb{\gamma}_B(0)\mid cos\left\{\frac{2\pi}{\lambda}\left
(\frac{B.x}{f} + d\right) - \psi_B\right\}\right], \eqno(31)$$ 

\noindent
where, $B$ is the baseline, $D$ the diameter of the sub-apertures, 
$f$ the focal length, the envelope, ${\it a}(D.x)$ the 
image of each sub-apertures (Airy disk), $\psi_B$, the phase of 
$\gamma_B(0)$, $2\pi d/\lambda$, the incidental non-zero OPD between the fields, 
and $d$ is the extra optical path in front of one aperture. 
\bigskip 
To overcome the restrictions of the baseline, Michelson (1920) constructed his 
stellar interferometer by installing a 7~m steel beam on top of the said
100~inch reflecting telescope at Mt. Wilson. It was equipped with 4 flat mirrors 
to fold the beams in periscopic fashion. Michelson and Pease (1921) were able to 
resolve the supergiant star, $\alpha$~Orionis, and a few other stars. 
In this design, the maximum resolution is limited by the length of the girder
bearing the collectors. The spatial modulation frequency in the focal plane
is independent of the distance between the collectors. This feature allows one to 
keep the same detection conditions when varying the baseline $B$. The 
telescope serves as correlator, thus, provides zero OPD.
This experiment had faced various difficulties in resolving stars. These are 
mainly due to the (i) effect of atmospheric turbulence, (ii) variations of 
refractive index above small sub-apertures of the interferometer causing the
interference pattern to move as a whole, (iii) 7 meter separation of outer 
mirrors is insufficient to measure the diameter of more stars, and 
(iv) mechanical instability prevents controlling the large interferometer.
\vskip 20 pt
\centerline {\bf 3. Atmospheric turbulence}
\bigskip 
\noindent
The turbulent phenomena associated with heat flow and winds in 
the atmosphere, causing the density of air fluctuating in space and time, 
occur predominantly due to the winds 
at various heights, convection in and around the building and the dome, 
any obstructed location near the ground, off the surface 
of the telescope structure, inside the primary mirror cell, etc. The cells 
of differing sizes and refractive indices produced by this phenomena, move 
rapidly across the path of the light, causing the distortion on the shape of 
the wave-front and variations on the intensity and phase. In what follows, the 
descriptions of the atmospheric turbulence theory,
metrology of seeing (Coulman, 1985), observational results, mathematical
details (Roddier, 1981, 1988a) etc., are enumerated.  
\vskip 20 pt
\noindent
{\bf 3.1. Formation of eddies}
\bigskip 
\noindent
Large scale temperature inhomogeneities caused by differential heating of
different portions of the Earth's surface produce random micro-structures
in the spatial distribution of temperature which, in turn,  
cause the random fluctuations in the refractive index of air. These large
scale refractive index inhomogeneities are broken up by turbulent wind and
convection, spreading the scale of the inhomogeneities to smaller sizes.
\bigskip 
When the average velocity, $v_a$, of a viscous fluid of characteristic size, $l$,
is gradually increased, two distinct states of fluid motion, viz., (i) laminar 
(regular and smooth in space and time), at very low, $v_a$, (ii) unstable and 
random fluid motion $-$ turbulence $-$ at $v_a >$ some critical value, are 
observed (Tatarski, 1967, Ishimaru, 1978). The dimensionless quantity, known 
as Reynolds number, $R_e = v_a l/\nu_v$, that characterizes a turbulent 
flow, is a function of the flow geometry, $v_a, l$, and kinematic viscosity of 
the fluid, $\nu_v$. When $R_e$ exceeds some value in a pipe (depending on 
its geometry), the transition of the flow from laminar to turbulent occurs. 
Between these two extreme conditions, the
flow undergoes a series of unstable states. The atmosphere is difficult to
study due to the high Reynolds number, of the order of $10^6$, which ensures 
that atmospheric air flow is nearly always turbulent (Ishimaru, 1978).
\bigskip 
According to the atmospheric turbulence model the 
energy enters the flow at low frequencies (Taylor, 1921, 1935, Richardson, 1922,
Kolmogorov, 1941b, 1941c) with scale length, $L_\circ$, (outer scale length) 
and spatial frequency, $k_{L_{\circ}} = 2\pi/L_\circ$, as a direct result of the 
non-linearity of the Navier-Stokes equation governing fluid motion. 
$L_\circ$ varies according to the local conditions, ranging from 2~m 
(Nightingale and Busher, 1991) to 100~m. A value for $L_\circ$, which is more 
than 2~km is also reported in the literature (Colavita et al., 1987).
\bigskip
The refractive index of the atmosphere, $n({\bf r, t}) [= n_\circ + n_1({\bf r, t})$, 
where, $n_\circ \approx 1$ is the mean refractive index of air, $n_1({\bf r, t})$, is 
the randomly fluctuating term, ${\bf r}$ is the 3-dimensional (3-d) position
vector and $t$ is the time], varies due to the temperature inhomogeneities. The 
dependence of the refractive index of air upon pressure, $P$ (millibar) and
temperature, $T$ (Kelvin), at optical wavelengths is given by 
$n_1 \cong n - 1 = 77.6~\times~10^{-6}P/T$ (Ishimaru, 1978). The packets of air 
that have a uniform refractive index, referred as turbulent eddies,  
affect the optical propagation. The statistical distribution of the size and
number of these eddies is characterized by the power spectral density,  
$\Phi_n({\bf k})$, of $n_1({\bf r})$. 
\vskip 20 pt
\noindent
{\bf 3.2. Inertial subrange and structure functions}
\bigskip 
\noindent
When the scale length associated with these eddies decreases the Reynolds number 
to a very low value, where the kinetic energy of the flow is lost as heat via 
viscous dissipation resulting in a rapid drop in power spectral density, 
$\Phi_n({\bf k})$, for $k > k_\circ$ ($k_\circ$ is 
critical wave number), imposes a highest possible spatial frequency on the 
flow; the break up of the turbulent eddies stops (Tatarski, 1967, 
1968). This is denoted as inner scale length, $l_\circ$, and spatial frequency,
$k_{l_{\circ}} = 2\pi/l_\circ$. $l_\circ$, varies 
from a few millimeter near the ground up to a centimeter (cm) high in the 
atmosphere. The value of inertial subrange, $l_\circ < r < L_\circ$ ({\bf r}, is the 
vector between the two points of interest) would be different at various 
locations at the site. The 3-d power spectrum, $\Phi_n$, 
for the wave number, $k > k_\circ$, (critical wave number), in the case of inertial 
subrange, can be equated as,

$$\Phi_n({\bf k}) = 0.033{\cal C}_n^2 k^{-11/3}, \eqno(32)$$

\noindent
where, ${\cal C}_n^2$ is known as structure constant of the refractive index 
fluctuations. 
\bigskip 
This spectrum for refractive index changes for a given structure constant and 
is valid within the inertial subrange, ($k_L{_\circ} < k < k_l{_\circ}$). 
This Kolmogorov-Obukhov model of turbulence, describing 
the power-law spectrum for the inertial intervals of wave numbers, is widely 
used for astronomical purposes (Tatarski 1993). 
\bigskip 
Owing to the non-integrable pole 
at $k = 0$, mathematical problems arise to use this equation for modeling the 
spectrum of the refractive index fluctuations when, $k \rightarrow 0$.
To overcome this defect, von K\'arm\'an model is often used (Goodman, 1985).
The structure functions of the refractive index and phase fluctuations are the 
main characteristics of the light propagation through the turbulent atmosphere, 
influencing the performance of the imaging system. The refractive index 
structure function, ${\cal D}_n({\bf r})$, can be defined as,

$${\cal D}_n({\bf r}) = <\mid n\left(\pmb{\rho} + {\bf r}\right) - n\left(\pmb{\rho}
\right)\mid^2>, \eqno(33)$$

\noindent
which expresses its variance at two points $r_1, r_2$. 
\bigskip 
The structure functions
are related to the covariance function, ${\cal B}_n({\bf r})$, through 

$${\cal D}_n({\bf r)} = 2[{\cal B}_n({\bf 0}) - {\cal B}_n({\bf r})], \eqno(34)$$

\noindent
where, ${\cal B}_n({\bf r}) = <n\left(\pmb{\rho}\right)n\left(\pmb{\rho} + {\bf r} \right)>$ 
and the covariance is the 3-d Fourier transform of the spectrum, 
$\Phi_n({\bf k)}$ (Roddier, 1981).
\bigskip
Kolmogorov (1941a) states that the structure function in the inertial range
(homogeneous and isotropic random field) depends on ${\bf r} = 
\mid{\bf r}\mid$, as well as on the values of the rate of production or 
dissipation of turbulent energy $\epsilon_\circ$ and the rate of production or
dissipation of temperature inhomogeneities $\eta_\circ$. From the Tatarski (1967)
approximation, the refractive index structure function can be written as,

$${\cal D}_n({\bf r}) = {\cal C}_n^2 r^{2/3}. \eqno(35)$$

Similarly, the velocity structure function, ${\cal D}_v ({\bf r}) = {\cal C}_v^2 r^{2/3}$, 
and temperature structure function, ${\cal D}_T({\bf r}) = {\cal C}_T^2 r^{2/3}$, can also be 
derived. The temperature structure constant, ${\cal C}_T^2$, is proportional to the
vertical temperature gradient but is not related to the velocity of the flow.
Several experiments confirm this two-thirds power law in the atmosphere 
(Wyngaard et al., 1971, Coulman, 1969, 1974, Hartley et al., 1981, Walters and 
Kunkel, 1981, Lopez, 1991). The rms fluctuation in the difference between the 
refractive index at any two points in earth's atmosphere is often approximated 
as a power law of the separation between the points. 
\bigskip
Seeing affects the measurements of fringe visibility with a long baseline 
optical amplitude interferometer (section 9.2), by introducing phase 
aberrations across the wave fronts incident on the interferometer, therefore, 
the relative phase of the wave fronts at the apertures changes with time, as 
well as varies the optical paths through the two arms. Venkatakrishnan and 
Chatterjee, (1987) pointed out that the afore-mentioned power law
is not valid for large separations. The rms fluctuation tends to a
constant for large separations which reduces the random excursions
of fringes in an interferometer from the value expected from the power
law. In turn, this pushes the observing limits to fainter magnitudes.
Venkatakrishnan, (1987) applied this idea to the question of astronomical 
seeing. It was shown that the correct statistical behaviour at large separations 
predicts a lower value for seeing than that expected from a power law behaviour. 
Davis and Tango, (1996) have developed a method for measuring the atmospheric 
coherence time that can be carried out in parallel with the determination of the 
fringe visibility in an amplitude interferometer. The measured coherence time
was found to be ranged between $\sim$~1 to $\sim~7$~ms.
\bigskip
Since the fluctuations in the air refractive index are proportional to 
the temperature fluctuations, the refractive index structure constant can be 
related by the following equation (Roddier, 1988a),

$${\cal C}_n = 75~\times~10^{-6}\frac{P}{T^2}{\cal C}_T. \eqno(36)$$

The value of the ${\cal C}_n^2$ depends on both local conditions, as well
as on the planetary boundary layer. The significant scale lengths, in the case
of former, depend on the local objects and the primary effect is to change the 
inertial subrange, as well as to introduce temperature differentials. While the 
latter can be generalized as: (i) surface boundary layer due to the ground 
convection, extending up to a few kilometer (km) height of 
the atmosphere, where shear is the 
dominant source of turbulence (scale lengths are roughly constant and 
${\cal C}_T^2 \propto z^{-2/3}$), (ii) the free convection layer associated with
orographic disturbances, where the scale lengths are height dependent, 
(${\cal C}_T^2 \propto z^{-4/3}$) and (iii) in the tropopause and above, where 
the turbulence is due to the wind shear and the temperature gradient vanishes 
slowly. The values of ${\cal C}_T^2$ have been measured by various techniques, 
viz., micro-thermal studies, radar and acoustic soundings, balloon and aircraft 
experiments (Coulman, 1969, 1974, Tsvang, 1969, Lawrence et al., 1970, 
Kallistratova and Timanovskiy, 1971, Ochs and Lawrence, 1972, Metcalk, 1975) 
\bigskip
The turbulence which reaches a minimum just after the sunrise and steeply increases 
until afternoon, is primarily due to the solar heating of the ground (Hess, 
1959). It decreases to a secondary minimum after sunset and slightly increases
during night. The typical values of ${\cal C}_n^2$, 12~m above the ground, during the 
daytime is found to be $10^{-13}~m^{-2/3}$ and $10^{-14}~m^{-2/3}$, during 
night time (Kallistratova and Timanovskiy, 1971). These values are 
height-dependent; (i) $h^{-4/3}$, under unstable daytime conditions,
(ii) $h^{-2/3}$, under neutral conditions, (iii) a slow decrease under stable 
condition, say during night time (Wyngaard et al., 1971). It reaches a minimum value 
of the order of $10^{-17}~m^{-2/3}$ around 6$-$9~km with slight increase to a
secondary maximum near the tropopause and decreases further in the stratosphere
(Roddier, 1981). Masciadri et al., (1999) have noticed the value of 
${\cal C}_n^2$ increases about 11~km over Paranal, Chile using Scidar and DIMM 
data. The turbulence concentrates into a thin layer of 100$-$200~m 
thickness where the value of ${\cal C}_n^2$ increases by more than an order of its 
background level. Orographic disturbances also play an important role, while the 
behaviour of the former is independent of the location (Barletti et al., 1976).
\vskip 20 pt
\noindent
{\bf 3.3. Wave propagation through the turbulent atmosphere}
\bigskip
\noindent
The random refractive index fluctuations deflect the rays of the light
propagating down the turbulence of the atmosphere; the longer the path, more it 
suffers the deflection. The spatial correlational properties of the 
turbulence-induced field perturbations are evaluated by combining the basic 
turbulence theory with the stratification and phase screen approximations. 
The variance of the ray can be translated into a variance of the phase 
fluctuations. For calculating the same, Roddier (1981) used
the correlation properties for propagation through a single (thin) turbulence 
layer and then extended procedure to account for many such
layers. It is assumed that the refractive index fluctuations 
between the individual layers are statistically independent (Tatarski, 1967). 
Several investigators (Goodman, 1985, Troxel et al., 1994) have argued that 
individual layers can be treated as independent provided the separation of the 
layer centres is chosen large enough so that the fluctuations of the log 
amplitude and phase introduced by different layers are uncorrelated. The 
method set out by Roddier (1981) for the wave propagation through the 
atmosphere runs as follows. 
\bigskip
Let a monochromatic plane wave of wavelength $\lambda$
from a distant star at zenith, propagate towards the ground based observer;
the complex amplitude at co-ordinate, $({\bf x}, h)$, is given by,

$$\Psi_h({\bf x}) = \mid\Psi_h({\bf x})\mid e^{i\psi_h({\bf x})}. \eqno(37)$$

\noindent
Here, for height h, the average value of the phase, $<\psi_h({\bf x})> = 0$, 
and the unperturbed complex amplitude outside the atmosphere is normalized
to unity $[\Psi_\infty({\bf x}) = 1]$. 
\bigskip
When this wave is allowed to pass through a thin layer of turbulent air 
of thickness $\delta h_j$, which is considered to be large compared to the
scale of turbulent eddies but small enough for the phase screen approximation
(diffraction effects is negligible over the distance, $\delta h_j$), 
the complex amplitude of the plane wave-front after passing through the layer 
can be expressed as,

$$\Psi_j({\bf x}) = e^{i\psi_j({\bf x})}. \eqno(38)$$

\noindent
Here, $\psi_j({\bf x})\left [=k\int_{h_j}^{h_j+\delta h_j}n({\bf x},z) dz\right ]$ is 
the phase-shift introduced by the refractive index fluctuations, n({\bf x},h) 
inside the layer, 
$k = 2\pi/\lambda$ is the wave number of the vibration. In this 
case, the rest of the atmosphere is thought to be calm and homogeneous.
\bigskip
The coherence function of the complex amplitude, $<\Psi_j\left({\bf x}\right)
\Psi_j^*\left({\bf x}+\pmb{\xi}\right)>$, at the layer output leads to

$${\cal B}_j\left(\pmb{\xi}\right) = <e^{i\left[\psi_j\left({\bf x}\right)-
\psi_j\left({\bf x}+\pmb{\xi}\right)\right]}>. \eqno(39)$$

The quantity $\psi_j({\bf x})$ can be considered to be the sum of a large 
number of independent variables, and therefore, have Gaussian statistics.
Roddier (1981) points out that the expression in square brackets is Gaussian
too with zero mean. He finds the similarity of this equation to Fourier
transform of the probability density function at unit frequency, therefore;

$${\cal B}_j\left(\pmb{\xi}\right) = e^{-\frac{1}{2}{\cal D}_{\psi_j}\left(\pmb{\xi}\right)}, 
\eqno(40)$$

\noindent
where, ${\cal D}_{\psi_j}\left(\pmb{\xi}\right)\left [=<\mid \psi_j\left({\bf x}\right)-
\psi_j\left({\bf x}+\pmb{\xi}\right)\mid^2>\right]$, is the 2-d structure
function of the phase, $\psi_j({\bf x})$. 
\bigskip
Let the covariance of the phase be defined as ${\cal B}_{\psi_j}\left(\pmb{\xi}\right) 
= <\psi_j\left({\bf x}\right)\psi_j\left({\bf x}+\pmb{\xi}\right)>$, and by
replacing $\psi_j({\bf x})$, 

$${\cal B}_{\psi_j}\left(\pmb{\xi}\right)=k^2\int_{h_j}^{h_j+\delta h_j} dz\int_{h_j-x}
^{h_j+\delta h_j-x}{\cal B}_n\left(\pmb{\xi}, \zeta\right)d\zeta, \eqno(41)$$ 

\noindent
where, ${\cal B}_n\left(\pmb{\xi}, \zeta\right) = <n({\bf x}, z)n\left({\bf x}+\pmb{\xi},
z^\prime\right)>$ is the 3-d refractive index covariance and $\zeta = z^\prime -z$.
\bigskip
Since the thickness of the layer, $\delta h_j$, is large compared to the 
correlation scale of the turbulence, the integration over $\zeta$  
from $-\infty \; to \; +\infty$, leads to

$${\cal B}_{\psi_j}\left(\pmb{\xi}\right)=k^2\delta h_j\int {\cal B}_n\left(\pmb{\xi},
\zeta\right)d\zeta. \eqno(42)$$ 

The phase structure function is related to its covariance (equation 34), and
therefore,

$${\cal D}_{\psi_j}\left(\pmb{\xi}\right) = 2k^2\delta h_j\int \left[{\cal B}_n\left({\bf 0},
\zeta \right)-{\cal B}_n\left(\pmb{\xi},\zeta\right)\right]d\zeta. \eqno(43)$$

The refractive index structure function defined in equation (35) can be 
evaluated as, ${\cal D}_n\left(\pmb{\xi}, \zeta\right) = 
{\cal C}_n^2\left(\xi^2 + \zeta^2\right)^{1/3}$, and recalling equation (34), 
equation (43) can be integrated and yields,

$${\cal D}_{\psi_j}\left(\pmb{\xi}\right) = 2.91k^2{\cal C}_n^2\xi^{5/3}\delta h_j. \eqno(44)$$

The covariance of the phase can be obtained by substituting equation (44), 
in equation (40),

$${\cal B}_{h_j}\left(\pmb{\xi}\right) = e^{-\frac{1}{2}\left(2.91k^2{\cal C}_n^2\xi^{5/3}
\delta h_j\right)}. \eqno(45)$$

Using Fresnel approximation, Roddier (1981) enunciates that the covariance of 
the phase at the ground level due to a thin layer of turbulence at some height
off the ground, ${\cal B}_\circ\left(\pmb{\xi}\right) = {\cal B}_{h_j}\left(\pmb{\xi}\right)$.  
For high altitude layers the complex field will fluctuate both in phase and in
amplitude, and therefore, the wave structure function, ${\cal D}_{\psi_j}\left(\pmb{\xi}
\right)$, is not strictly true as the ground level. The turbulent layer acts 
like a diffracting screen; however, correction in the condition of astronomical 
observation remains small (Roddier, 1981). 
\bigskip
The wave structure function after passing through N layers can be expressed as
the sum of the N wave structure functions associated with the individual layers,
For each layer, the coherence function is multiplied by the term,
$e^{-\frac{1}{2}\left[2.91k^2{\cal C}_n^2(h_j)\xi^{5/3} \delta h_j\right]}$; therefore,
the coherence function at ground level is given by

$${\cal B}_\circ\left(\pmb{\xi}\right) = \prod_{j=1}^N e^{-\frac{1}{2}
\left[2.91k^2{\cal C}_n^2(h_j)\xi^{5/3}\delta h_j\right]} = e^{-\frac{1}{2}
\left[2.91k^2\xi^{5/3} \sum_{j=1}^N {\cal C}_n^2(h_j)\delta h_j\right]}. \eqno(46)$$

This expression can be generalized for a star at an angular distance  
$\gamma$ away from the zenith viewed through all over the turbulent atmosphere, 

$${\cal B}_\circ\left(\pmb{\xi}\right) = e^{-\frac{1}{2}\left[2.91k^2\xi^{5/3}
/cos\gamma\int {\cal C}_n^2(z)dz\right]}. \eqno(47)$$
\vskip 20 pt
\noindent
{\bf 3.4. Fried's parameter ($r_\circ$)}
\bigskip
\noindent
When a plane wave-front passes down through refractive index inhomogeneities,
it suffers phase fluctuations and reaches the entrance pupil of a
telescope with patches of random excursions in phase (Fried, 1966).
Each patch of the wave-front with diameter, $r_\circ$, known as Fried's Parameter 
(atmospheric coherence diameter), would act independently of the rest of the 
wave-front. 
\bigskip
Let the modulation transfer function (MTF) of an optical system be composed of 
the atmosphere and a telescope. The long-exposure PSF is defined by the 
ensemble average, 
$<{\cal S}({\bf x})>$, which is independent of the direction. Figure 3 depicts 
the plane-wave propagation through multiple turbulent layers.
If the object emits incoherently, the average illumination, ${\cal I}({\bf x})$, 
of a resolved object, ${\cal O}({\bf x})$, obeys the convolution relationship, 

$$<{\cal I}({\bf x})> = {\cal O}({\bf x}) \ast <{\cal S}({\bf x})>. \eqno(48)$$

Using 2-d Fourier transform, the above equation can be written,

$$<\widehat{{\cal I}}({\bf u})> = \widehat{{\cal O}}({\bf u})\cdot 
<\widehat{{\cal S}}({\bf u)}>, \eqno(49)$$

\noindent
where, $<\widehat{{\cal S}}({\bf u})>$ denotes the transfer function for 
long-exposure images, ${\bf u}$, the spatial frequency vector with magnitude, 
$u$ and $\widehat{\cal O}({\bf u})$ the object spectrum. 
\bigskip
\noindent
\midinsert
{\eightpoint   
\noindent
\centerline{\psfig{figure=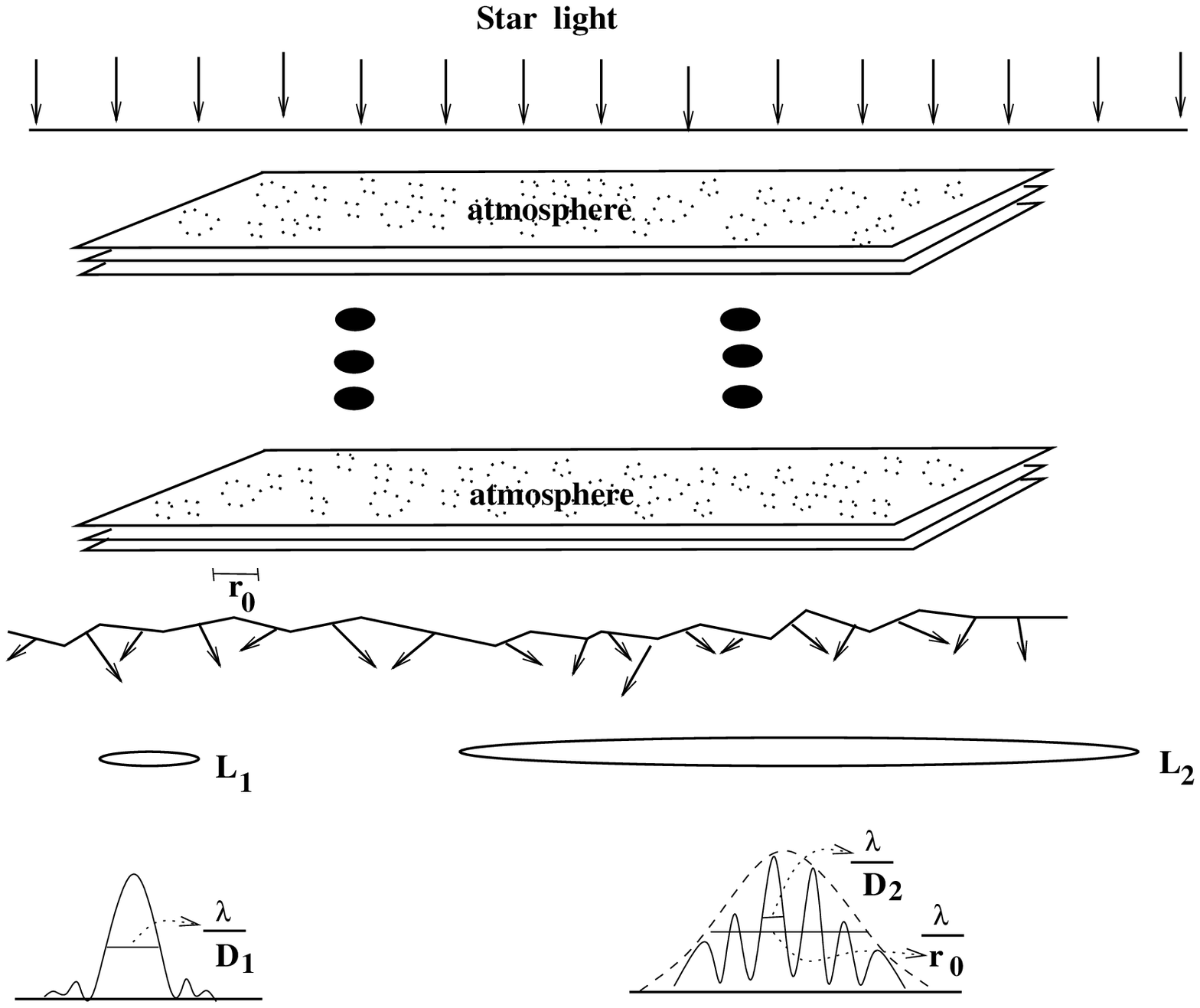,height=10.5cm,width=10.5cm}}
\bigskip
\noindent
{\bf Figure 3.} Plane-wave propagation through the multiple turbulent layers.
}     
\endinsert
\bigskip 
The transfer function is the product of 
the atmosphere transfer function (wave coherence function), {\cal B}({\bf u}),
and the telescope transfer function, {\cal T}({\bf u}),

$$<\widehat{{\cal S}}({\bf u})> = {\cal B}({\bf u})\cdot{\cal T}({\bf u}). 
\eqno(50)$$

For a long-exposure through the atmosphere, the resolving power, ${\cal R}$, of 
any optical telescope, is expressed as,

$${\cal R} = \int{\cal B}({\bf u})\cdot{\cal T}({\bf u})d{\bf u}. \eqno(51)$$

It is limited either by the telescope or by the atmosphere, depending on the
relative width of the two functions, ${\cal B}({\bf u})$ and ${\cal T}({\bf u})$.
The diffraction-limited resolving power, ${\cal R}_D$, of a small 
telescope with an unobscured circular aperture of diameter, $D \ll r_\circ$, depends 
on its optical transfer function (Roddier, 1981);

$${\cal R}_D = \int{\cal T}({\bf u})d{\bf u} = \frac{\pi}{4}\left(\frac{D}
{\lambda}\right)^2. \eqno(52)$$

The resolving power of a large telescope, $(D >> r_\circ)$ is dominated by
the turbulence effects;

$${\cal R}_\infty = \int{\cal B}({\bf u})d{\bf u}. \eqno(53)$$

According to equation (47), ${\cal B}({\bf u})$, can be expressed as,
 
$${\cal B}({\bf u}) = {\cal B}_\circ(\lambda{\bf u}) 
= e^{-\frac{1}{2}\left[2.91k^2(\lambda u)^{5/3}/cos\gamma\int
{\cal C}_n^2(z) dz\right]} \eqno(54)$$
$${\cal R}_\infty = (6\pi/5)\left[\frac{1}{2}\left(\frac{2.91k^2
\lambda^{5/3}}{cos\gamma}\int{\cal C}_n^2(z) dz\right)\right]^{-6/5}
\Gamma(6/5). \eqno(55)$$

Fried, (1966) had introduced the critical diameter $r_\circ$, for a telescope 
for which ${\cal R}_D = {\cal R}_\infty$. Therefore, placing $D = r_\circ$ in 
equation (52), the equation (44) takes the form as follows, 

$${\cal B}({\bf u}) = e^{-3.44\left(\lambda u/r_\circ\right)^{5/3}} \eqno(56)$$
$${\cal B}_\circ\left(\pmb{\xi}\right) = e^{-3.44\left(\xi/r_\circ\right)^{5/3}}. 
\eqno(57)$$

The expression for the phase structure function across the telescope aperture
can be written as (see equation 40),

$${\cal D}_{\psi}\left(\pmb{\xi}\right) = 6.88\left(\frac{\xi}{r_\circ}\right)^{5/3}.
\eqno(58)$$

Several observers have confirmed this relationship (O'Byrne, 1988,
Nightingale and Buscher, 1991). The resolving power of any large telescope 
(larger than $r_\circ$) is essentially limited by the size of the seeing disk, 
$1.22\lambda/r_\circ$. By replacing the value of 
${\cal B}_\circ\left(\pmb{\xi}\right)$, in equation (47), an expression for 
$r_\circ$ in terms of the distribution of the turbulence in the
atmosphere can be derived.
 
$$r_\circ = \left[0.423\frac{k^2}{cos\gamma}\int{\cal C}_n^2(z)dz\right]^{-3/5}. 
\eqno(59)$$

Computer simulations by Venkatakrishnan et al., (1989), demonstrated the
destructions of the finer details of an image of a star by the atmospheric
turbulence.
\vskip 20 pt
\noindent
{\bf 3.4.1. Benefit of short-exposure images}
\bigskip
\noindent
In the long-exposure images, the image is spread during the exposure
by its random variations of the tilt. The image sharpness and the
MTF are affected by the wave-front tilt, as well as by the more complex shapes, 
while in the case of a short-exposure image, the image sharpness and MTF are 
insensitive to the tilt. The random factor associated with the tilt is extracted
from the MTF before taking the average, where in the long-exposure case,
no such factor is removed (section 3.4).  
\bigskip
Let the two seeing cells be separated by a vector $\lambda {\bf u}$, in the 
telescope pupil. If a point source is imaged through the telescope by using 
pupil function consisting of two apertures ($\theta_1, \theta_2$), corresponding
to the two seeing cells, then a fringe pattern is produced with narrow spatial 
frequency bandwidth. Each sub-aperture is small enough for the field to be 
coherent over its extent. Any stellar object is too small to be resolved 
through a single sub-aperture. 
Atmospheric turbulence does not affect the amplitude of the fringes produced,
but introduces phase delays, which, in turn, shift the fringe pattern randomly
and blur the fringe pattern during long-exposure. 
\bigskip
The major component $\widehat{\cal I}({\bf u})$, at the frequency, 
${\bf u}$, is produced 
by contributions from all pairs of points with separations $\lambda {\bf u}$, 
with one point in each aperture. If the major component is averaged over many 
frames, the resultant for frequencies greater than $r_\circ/\lambda$, tends
to zero since the phase-difference, $\theta_1 - \theta_2$; mod $2\pi$,  
between the two apertures is distributed uniformly between $\pm\pi$, with two 
mean. The Fourier component performs a random walk in the complex 
plane and averages to zero: 

$$<\widehat{\cal I}{\bf (u)}> = 0, \ \ \ \  u > r_\circ / \lambda. \eqno(60)$$

The argument of equation (49) is expressed as,

$$arg\mid\widehat{\cal I}{\bf (u)}\mid = \psi{\bf (u)} + \theta_1 - \theta_2, \eqno(61)$$

\noindent
where, $\psi({\bf u})$ is the Fourier phase at ${\bf u}$, and
$arg\mid\mid$ stands for, `the phase of'.  
\bigskip
While in the case of autocorrelation technique, the autocorrelation of 
${\cal I}({\bf x})$ is the correlation of ${\cal I}({\bf x})$ and 
${\cal I}({\bf x})$ multiplied by the complex exponential factor with zero 
spatial frequency. The major Fourier component of 
the fringe pattern is averaged as a product with its complex conjugate and so 
the atmospheric phase contribution is eliminated and the averaged signal is 
non-zero. Therefore, the resulting representation in the Fourier space is,

$$<\widehat{\cal I}^A({\bf u})> = <\widehat{\cal I}({\bf u})\widehat{\cal I}^\ast
({\bf u})> = <\mid\widehat{\cal I}{\bf (u)}\mid^2> \neq 0.\eqno(62)$$ 

The argument of this equation is given by the expression,

$$arg\mid\widehat{\cal I}{\bf (u)}\widehat{\cal I}{\bf (-u)}\mid = \theta^A({\bf u}) = 
\psi({\bf u}) + \theta_1 - \theta_2 + \psi {\bf (-u)} - \theta_1 + \theta_2 = 0. 
\eqno(63)$$

For a large telescope, the aperture, ${\cal P}$, can be sub-divided into a set of
sub-apertures, $p_j$. Each pair of them, $p_n, p_m$, separated by a distance 
(baseline) $B$ would form fringes. The intensity in the focal plane 
of the telescope, ${\cal I}$, according to the diffraction theory (Born and 
Wolf, 1984), is given by the expression,

$${\cal I} = \sum_{n,m}<\Psi_n\Psi^\ast_m>. \eqno(64)$$ 

The term, $\Psi_n\Psi^\ast_m$, is multiplied by $e^{i\psi}$, where, $\psi$ is 
the random instantaneous shift in the fringe pattern.
With increasing distance between the two sub-apertures, 
the fringes move with an increasingly larger amplitude. On a long-exposure
image no shift is observed, which implies the loss of the high frequency 
components of the image. If the integration time is shorter
($<20$~ms) than the evolution time of the phase inhomogeneities, the 
interference fringes are preserved but their phases are randomly distorted. 
This produces the speckle pattern observed in short-exposure images. These
speckles can occur randomly along any direction within an angular patch of
diameter, $\lambda/r_\circ$. The sum of several statistically uncorrelated 
speckle patterns from a point source can result in an uniform patch of light
a few arcseconds wide (conventional image).
\vskip 20 pt
\noindent
{\bf 3.4.2. Measurement of $r_\circ$}
\bigskip
\noindent
Measurement of $r_\circ$ is of paramount importance to estimate the seeing at any 
astronomical site. Systematic studies of this parameter would help in 
understanding the various causes of the local seeing. 
Various methods of measuring $r_\circ$ have been discussed in the literature 
(Von der L\"uhe, 1984, Wood, 1985, Roddier, 1988a, Vernin et al., 1991). 
Using Fizeau mask interferometer with two apertures, the atmospheric transfer 
function can be estimated by measuring the coherence function, 
${\cal B}({\bf u})$. The intensity of the star light at telescope focus is,

$${\cal I} = \mid\Psi({\bf x})\mid^2 + \mid\Psi({\bf x}+\lambda{\bf u})\mid^2  
+ 2\Re[\Psi({\bf x})\Psi^*({\bf x} + \lambda{\bf u})]. \eqno(65)$$ 

The amplitude of the long-exposure fringes is proportional to  
$<\Psi({\bf x})\Psi^*({\bf x} + \lambda{\bf u})>$, and therefore,
provides the measurement of ${\cal B}({\bf u})$. Rotational shear 
interferometers (Roddier and Roddier, 1988) can be used to determine the 
coherence of the source by using a short integration time.
\bigskip
\noindent
\midinsert
{\eightpoint   
\noindent
\centerline{\psfig{figure=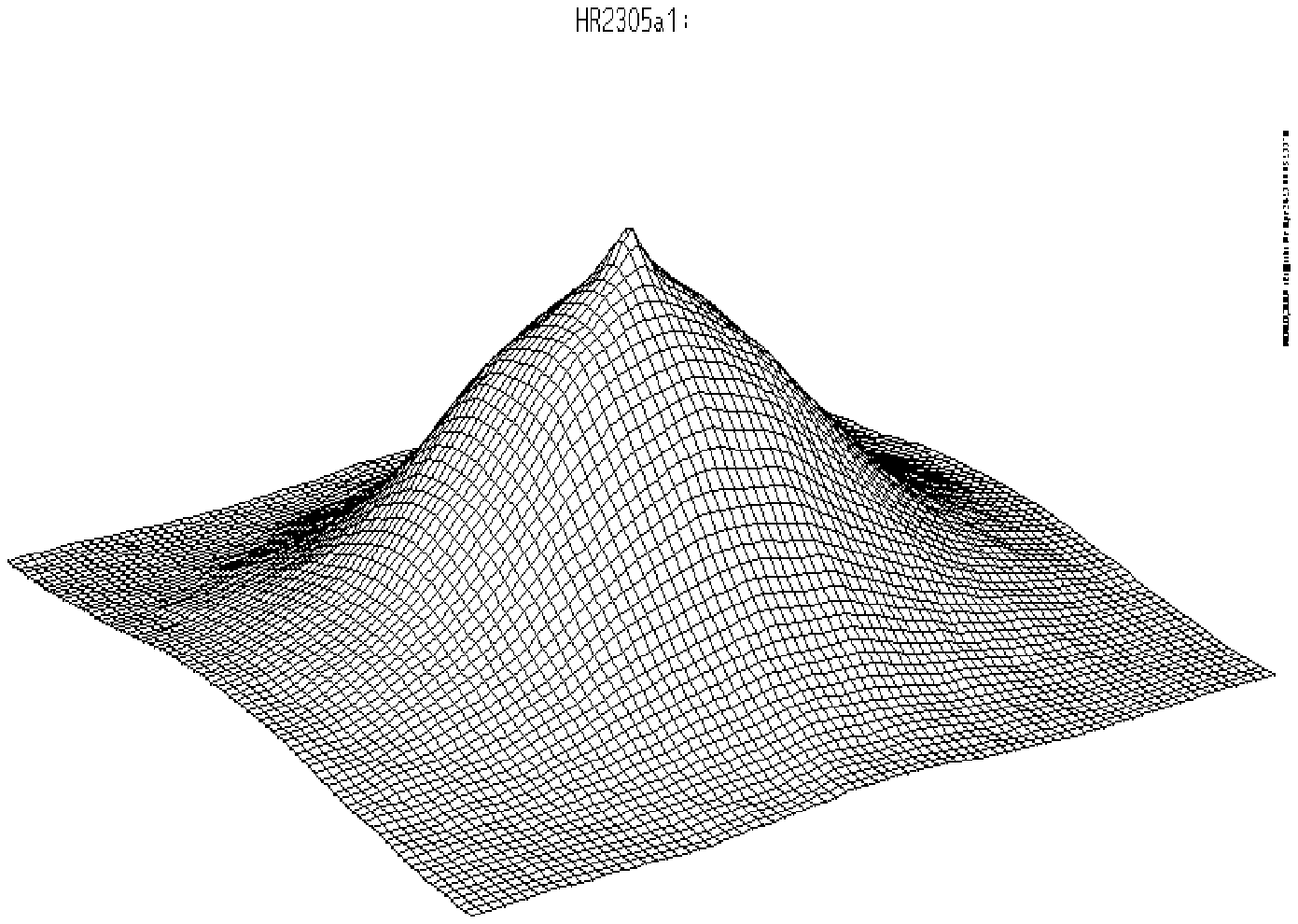,height=8.5cm,width=11cm,angle=270}}
\bigskip
\noindent
{\bf Figure 4.} 3-d picture of the autocorrelation of HR2305 observed at
at 2.34~m VBT, Kavalur, India, on 28$^{th}$ March, 1991, 1510 hr. UT (Saha and 
Chinnappan, 1999).  
}     
\endinsert
Another qualitative method is that based on from the short-exposure images using speckle 
interferometric technique (Labeyrie, 1970). In this, the area of the telescope 
aperture divided by the estimated number of speckles gives the wave-front 
coherence area $\sigma$, from which $r_\circ$ can be found by using relation, 

$${\sigma = {0.342 {\left( \frac {r_\circ} {\lambda}\right) }^2}}. \eqno(66)$$ 

The averaged autocorrelation of the these images contains both the 
autocorrelations of the seeing disk, as well as of the 
mean speckle cell. It is the width of the speckle component of the 
autocorrelation that provides the information on the size of the object being 
observed (Saha and Chinnappan, 1999, Saha et al., 1999a). The form of transfer
function, $<\mid\widehat{\cal S}({\bf u})\mid^2>$, can be obtained by calculating 
Wiener spectrum of the instantaneous intensity distribution from a point 
source (section 6). Figure 4 depicts the autocorrelation of HR2305 
observed at 1510 hrs UT, obtained at the Cassegrain focus of 2.34~m VBT, Kavalur,
India. 
\vskip 20 pt
\noindent
{\bf 3.4.3. Seeing at the telescope site}
\bigskip
\noindent
The term `seeing' is the total effect of distortion in the path of the star
light via different contributing layers of the atmosphere (see section 3.2)
to the detector placed at the focus of the telescope. Though the effect of the 
different layer turbulence has been receiving attention to identify the best 
site, the major sources of image degradation predominantly comes from the 
thermal and aero-dynamic disturbances in the atmosphere surrounding the 
telescope and its enclosure, viz., (i) thermal distortion of primary and 
secondary mirrors when they get heated up, (ii) dissipation of heat by the 
secondary mirror (Zago, 1995), (iii) rise in temperature at the primary cell, 
and (iv) at the focal point causing temperature gradient close to the 
detector etc. The aberrations in the design, manufacture and 
alignment of the optical train also add to this degradation. 
\bigskip
Various corrective measures, viz., (i) insulating the surface of the
floors and walls, (ii) introducing active cooling system to eliminate the
heat produced by electric equipments on the telescope and elsewhere in the
dome, (iii) installing ventilator to generate a sucking effect through the
slit to counteract the upward action of the bubbles (Racine, 1984,
Ryan and Wood, 1995), and (iv) maintaining a uniform temperature in and around 
the primary mirror of the telescope (Saha and Chinnappan, 1999) have been 
proposed to improve the seeing at the telescope site.
\bigskip
Mirror seeing is an important source of image spread and has the longest 
time-constant, of the order of several hours depending on the size and 
thickness of the mirror, for equilibrating with the ambient temperature (Woolf 
and Cheng, 1988). The spread amounts to 0.5$^{\prime\prime}$ for a 1$^\circ$ difference in
temperature. The production of mirror seeing takes place very close to its surface, 
$\sim$~2~cm above (Harding et al., 1979, Lowne, 1979, Zago, 1995 and references 
therein). The free convection above the mirror depends on the excess temperature 
of its surface above the ambient temperature with an exponent of 1.2. Iye et al., 
(1991) opined that a temperature difference of $<1^\circ~C$ should be maintained 
between the mirror and its ambient. The mirror seeing becomes weak and 
negligible if the mirror can be kept at a 1$^\circ$ lower temperature than the 
surrounding air (Iye et al., 1992). While, Racine et al., (1991) found that 
mirror seeing sets in as soon as the mirror is measurably warmer than the 
ambient air and is quite significant if it is warmer by $1^\circ$. Gillingham 
(1984a, 1984b) reported that the ventilation of the primary mirror of the 3.9~m
Anglo-Australian telescope (AAT) was found to improve the seeing when the mirror 
is warmer than the ambient dome air, and degrade the seeing when mirror is cooler 
than the latter (Barr et al., 1990).
\bigskip
On the other hand, a $4-5^\circ~C$ difference in temperature between the outside 
and inside of the dome causes a seeing degradation amounting to 0.5$^{\prime\prime}$ only (Racine et 
al., 1991). The reported improvement of seeing at the 3.6~m Canada-France-Hawaii 
telescope (CFHT), Mauna-Kea, Hawaii, is largely due to the implementation of 
the floor chilling system to damp the natural convection, which essentially 
keeps the temperature of the primary mirror closer to the air volume (Zago, 1995). 
Saha and Chinnappan (1999) have measured the night-time variation of Fried's
parameter, as obtained at the Cassegrain focus of the VBT, Kavalur, India, using
the speckle interferometer. It is found that average observed $r_\circ$ is 
higher during the later part of the night than the earlier part, implying that 
the PSF has a smaller FWHM during the former period. This might indicate that
the slowly cooling mirror creates thermal instabilities that decreases slowly 
over the night (Saha and Chinnappan, 1999).   
\vskip 20 pt
\centerline {\bf 4. Single aperture interferometry}
\bigskip 
\noindent
Labeyrie (1970) identified the short-exposure turbulence degraded 
images as speckles and suggested an interferometric technique $-$ speckle
interferometry $-$ to retrieve diffraction-limited informations of a stellar
object using a large telescope. Ever since the development of this technique, 
it is widely used both in the visible, as well as in the infrared (IR) bands.
With the photon counting detectors, it is able to record speckles of faint
stellar objects up to the visual magnitude $m_v \approx 16$. In what follows,
the stellar speckle interferometry and related techniques are elucidated.
\vskip 20 pt
\noindent
{\bf 4.1. Speckle interferometry}
\bigskip 
\noindent
The term, `Speckle', refers to the grainy structure observed when an uneven 
surface of an object is illuminated by a fairly coherent source. The formation
of speckles results from the summation of coherent vibrations having 
different random characteristics. The statistical properties of speckle pattern
$-$ the summation of many sine functions having different random characteristics
$-$ depend both on the coherence of the incident light and the random properties 
of medium. Since the positive and negative values cannot cancel out 
everywhere, adding an infinite number of such sine functions would result in a 
function with 100$\%$ constructed oscillations (Labeyrie, 1985). 
\bigskip 
Depending on the randomness of the source, spatial or temporal, speckles tend 
to appear (Labeyrie, 1985). Spatial speckles may be observed when all
parts of the source vibrate at same constant frequency but with different
amplitude and phase, while temporal speckles are produced if all parts of it
have uniform amplitude and phase. With a heterochromatic vibration spectrum, 
in the case of random sources of light, spatio-temporal speckles are produced. 
The speckle size is of the same order of magnitude as the Airy disc of
the telescope in the absence of turbulence. The number of correlation cells
is determined by the equation, $N = D/r_\circ$ and the number of photons
per speckle is independent of telescope diameter (Dainty, 1975).  
\bigskip
Speckle interferometry is a method to estimate the modulus of the Fourier 
transform from a set of short-exposure specklegrams of the object of interest. 
A specklegram represents the resultant of diffraction-limited incoherent 
imaging of the object irradiance, ${\cal O}({\bf x})$, 
convolved with the function representing the combined effects of the turbulent 
atmosphere and the image forming optical system, ${\cal S}({\bf x})$. 
This is averaged over the duration of the short, narrow bandpass exposure. An 
ensemble of such specklegrams, ${\cal I}_k({\bf x}), k = t_1, t_2, t_3, \ldots, 
t_M$, constitute an astronomical speckle observation.
The transfer function of ${\cal S}({\bf x})$, can be estimated by 
calculating Wiener spectrum of the instantaneous intensity from the unresolved 
star. The size of the data sets is constrained by the consideration of the 
signal-to-noise (S/N) ratio. Integration time of each exposure varies from a 
few milliseconds to twenty milliseconds, depending on the condition of seeing, 
to freeze the single realization of the turbulence. Usually specklegrams of the
brightest possible reference star are recorded to ensure that the S/N ratio of 
reference star is much higher than the S/N ratio of the programme star.
\vskip 20 pt
\noindent
{\bf 4.1.1. Speckle interferometer}
\bigskip 
\noindent
A speckle interferometer is a high quality diffraction-limited camera
where magnified ($\sim$~f/100) short-exposure images can be recorded. 
To compensate for the atmospherically induced dispersion at zenith angles larger 
than a few degrees, counter-rotating computer controlled dispersion correcting 
prisms are used in many conventional speckle cameras. The observation can also 
be carried out using a narrow bandwidth filter (Saha et al., 1997a) without 
using the correcting process. Instead of interchangeable filters, a concave 
grating was used by Labeyrie (1974) to provide the necessary filtering in a 
tunable way. In this system, the object spectrum is displayed by 
a video monitor when adjusting the wavelength and bandwidth selection 
decker. The spectral features of interest is isolated down to 1~nm
bandwidth. Subsequently, the system has been modified by replacing the concave
grating with a concave holographic grating providing a four-channel facility (Foy, 
1988). Labeyrie (1988) suggested for development of an 
interferometer with more spectral channels with narrow bandwidth 
down to 0.03~nm using a large spectrograph, facilitating
the tuning of the desired wavelength in real time simultaneously. 
\bigskip 
An exceptionally rigid camera system has been developed by Saha et al.,  
(1997a, 1999a) for the 2.34~m telescope, VBO, Kavalur, India. In this set up,
the wave-front falls on the focal plane of an optical flat made of low expansion 
glass with a high precision hole of aperture, ($\sim$~350~$\mu$m), at an angle of 
15$^{o}$, on its surface (Saha et al., 1997a). The image of the object passes on 
to a microscope objective through this aperture, which slows down the image 
scale of the telescope to f/130. A narrow band filter is set before 
the detector (cooled intensified-CCD) to avoid the chromatic blurring. The 
surrounding star field gets reflected from the optical flat on to a plane 
mirror and is re-imaged on an uncooled ICCD for guiding the object remotely.
Figure 5 shows an overview
of the speckle interferometer for VBT. In this set-up, the design
analysis has been carried out with the modern finite element method 
(Zienkiewicz, 1967) and computer controlled machines were used in manufacturing to 
get the required dimensional and geometrical accuracies. Another versatile
speckle interferometer has been developed by Prieur et al., (1998) recently
for use at the 2~m Bernard Lyot Telescope, Pic du Midi Observatory. It provides
the facilities of setting in various operating modes, such as, full pupil
imaging, masked-pupil speckle imaging and aperture synthesis, spectroscopy, 
wave front sensor, stellar coronography. The system is in operation 
at the said telescope (Carbillet et al., 1996, Aristidi et al., 1997, 1999). 
\bigskip
\noindent
\midinsert
{\eightpoint   
\noindent
\centerline{\psfig{figure=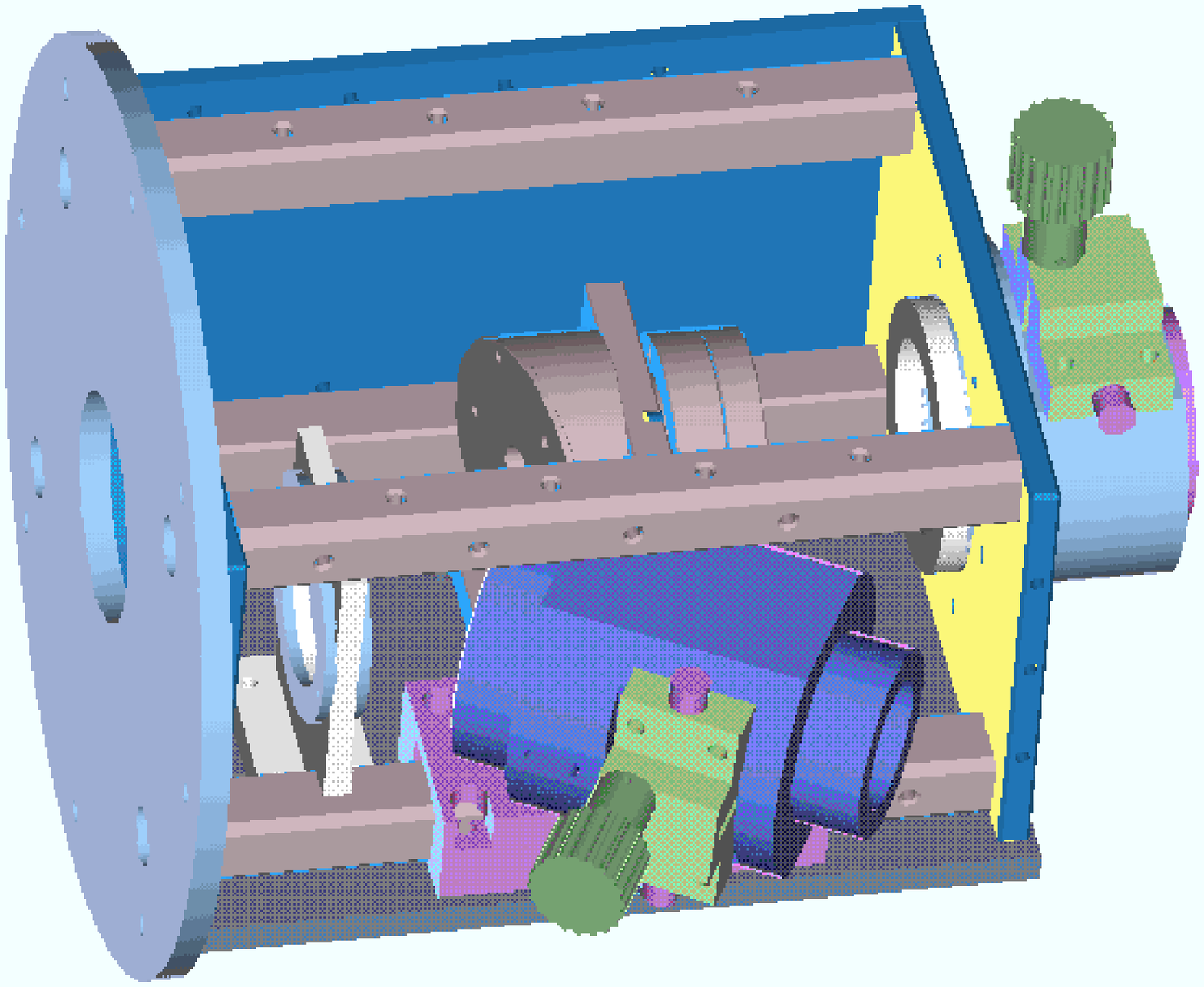,height=8.0cm,width=11cm,angle=270}}
\bigskip
\noindent
{\bf Figure 5.} Overview of the speckle interferometer (Saha et al., 1997a, 1999a).
}     
\endinsert
\vskip 20 pt
\noindent
{\bf 4.1.2. Estimation of Fourier modulus}
\bigskip 
\noindent
The post-imaging process of speckle imaging technique are to obtain the 
spatial autocorrelation of the objects at low light levels. 
The intensity distribution, ${\cal I}({\bf x)}$, of the specklegrams, 
in the case of quasi-monochromatic incoherent source, is described by the 
space-invariant imaging equations. Let the imaging system be composed of a 
simple lens based telescope in which the PSF is invariant to spatial shifts. An
object at a point, ${\cal O}({\bf x}^\prime)$, anywhere in the field of
view will, therefore, produce a pattern, ${\cal S}({\bf x} - {\bf x}^\prime)$, 
across the image, 

$${\cal I}({\bf x}) = \int{\cal O}({\bf x}^\prime){\cal S}({\bf x} - 
{\bf x}^\prime) d{\bf x}^\prime. \eqno(67)$$

This equation can be translated into a convolution product,

$${\cal I}{(\bf x)} = {\cal O}{(\bf x)}\ast{\cal S}{(\bf x)}. \eqno(68)$$

The Fourier space relationship between object and the image is

$$\widehat{\cal I}({\bf u}) = \widehat{\cal O}({\bf u})\cdot\widehat{\cal S}({\bf u}). 
\eqno(69)$$

Taking the modulus square of the expression and averaging over many frames,
the average image power spectrum is,

$$<\mid\widehat{\cal I}({\bf u})\mid^{2}> = \mid\widehat{\cal O}({\bf u})\mid^{2} 
\cdot<\mid\widehat{\cal S}({\bf u})\mid^{2}>. \eqno(70)$$ 

Since $\mid\widehat{\cal S}({\bf u})\mid^{2}$ is a random function in which the 
detail is continuously changing, its ensemble average becomes 
smoother. This form of transfer function can be obtained by calculating Wiener 
spectrum of the instantaneous intensity distribution from the reference star 
(unresolved star). The object is reconstructed from the power spectrum of the 
image knowing the PSF power spectrum and the true object
autocorrelation is then obtained by Fourier inversion.
\bigskip 
Reconstruction of object autocorrelation in case of the components in a 
group of stars retrieves the separation, position angle with a 180$^\circ$ 
ambiguity, and the relative magnitude difference. Saha and Venkatakrishnan, 
(1997), found the usefulness of the autocorrelation technique in obtaining the 
prior information on the object for certain applications of the image 
processing algorithms. 
\bigskip 
Most of the results obtained in the interferometric measurements of close binary 
stars are obtained from this technique (McAlister et al., 1987, 1989, 1990). 
Later Bagnuolo et al., (1992) have developed a PC-based `directed vector 
autocorrelation' (DVA) to eliminate the 180$^\circ$ ambiguity. It is found to be
effective in determining the duplicity of the star, as well as in measuring  
its Cartesian coordinates (McAlister et al., 1993, Hartkopf et al., 1997a, 
1997b). Saha et al., (1999d) developed an IRAF-based 
autocorrelation algorithm, where a Wiener parameter is added to PSF power 
spectrum in order to avoid zeros in the PSF power spectrum. This Wiener 
parameter is chosen according to the S/N ratio of image power spectrum. The 
notable advantage of this method is that the object can be reconstructed with a 
few frames. However, the phase ambiguities still remain. 
\bigskip
The group at Observatoire de la Cote d'Azur, France, has developed a digital 
correlator for on-line data reduction (Blazit et al., 1977a) which is used for 
double star measurements (Bonneau and Foy, 1980, Bonneau et al., 1980, 1986, 
Blazit et al., 1987). Singh (1999) has developed a system for on-line data 
processing by optical / hybrid means using the correlation techniques. This 
system is used, at present for the reduction of laboratory speckles.  
\vskip 20 pt
\noindent
{\bf 4.1.3. Difficulties in data processing} 
\bigskip 
\noindent
In general, the specklegrams have additive noise contamination, 
${\cal N}_j({\bf x})$, which includes all additive measurement of uncertainties.
This may be in the form of (i) photon statistics noise, and (ii) all distortions 
from the idealized iso-planatic model represented by the convolution of 
${\cal O}({\bf x})$ with ${\cal S}({\bf x})$ that includes non-linear geometrical 
distortions (Christou, 1988). 
\bigskip
The quality of the image degrades (Foy, 1988) due to the following reasons: (i) 
variations of air mass or of its time average between the 
object and the reference, (ii) seeing differences between the MTF for the object 
and its estimation from the reference, (iii) deformation of mirrors or to a 
slight misalignment while changing its pointing direction, (iv) bad focusing, 
(v) thermal effect from the telescope etc. These may lead to a dangerous 
artifact, yielding a wrong identification of the companion star.
\bigskip
To estimate the MTF, it is necessary to 
calibrate it on a unresolved star for which all observing conditions are 
required to be identical to those for the object. Such a comparison
is likely to introduce deviation in the statistics of speckles from the 
expected model based on the physics of the atmosphere. This, in turn,
would result either in the suppression or in the enhancement of intermediate
spatial frequencies which would also lead to artifacts (Foy, 1988). It is
essential to choose the point source calibrator as close as possible, preferably
within 1$^\circ$ of the programme star; the object and calibrator observations
should be interleaved to calibrate for changing seeing condition 
by shifting the telescope back and forth during the observing run to equalize
seeing distributions for both target and reference. 
\bigskip
Another difficulty arises from using the frame transfer CCD as detector. It 
is subjected to limitations in detecting fast photon-event pairs. A 
pair of photons closer than a minimum separation cannot be detected as a pair
by the afore-mentioned sensor. This yields a loss in HF information which, 
in turn, produces a hole in the centre of the autocorrelation, known as
Centreur hole (Foy, 1988), resulting the degradation of the power spectra or 
bispectra (section 6.4) of short exposures images. This requires to be
corrected (Thi\'ebaut, 1994, Berio et al., 1998).
\vskip 20 pt
\noindent
{\bf 4.2. Speckle spectroscopy}
\bigskip 
\noindent
The application of speckle interferometric technique to speckle spectroscopic
observations enables to obtain spectral resolution with high spatial resolution 
of astronomical objects simultaneously. Labeyrie (1980) enunciated the 
importance of combining angular and spectral resolution for long baseline 
optical interferometry. Information is concentrated in narrow spectral intervals 
in astrophysics and can be obtained from narrow band stellar observations.  
\bigskip 
There are various types of speckle spectrograph, viz., (i) objective speckle 
spectrograph (slitless) that yields objective prism
spectra with the bandwidth spanning from 400~nm~$-~$800~nm (Kuwamura et al., 1992),
(ii) wide-band projection speckle spectrograph that yields spectrally dispersed 
1-dimensional (1-d) projection of the object (Grieger et al., 1988),  
(iii) slit speckle spectrograph, where the width of the slit is comparable to 
the size of the speckle (Beckers et al., 1983) have been developed. A prism or 
a grism can be used to disperse 1-d specklegrams. In the case of projection 
spectrograph, the projection of 2-d specklegrams is carried out by a pair of 
cylindrical lenses and the spectral dispersion is done by a spectrograph 
(Grieger and Weigelt, 1992). Weigelt et al., (1992) suggested that an image 
slicer consisting of K slits can also be used for the slit spectrograph. 
\bigskip 
Baba et al., (1994a) have developed an imaging spectrometer where a reflection 
grating acts as disperser. Two synchronized detectors record the dispersed
speckle pattern and the specklegrams of the object. They have 
obtained stellar spectra of a few stars with the 
diffraction-limited spatial resolution of the 1.88~m telescope,
Okayama Astrophysical Observatory by referring to the latter that are
recorded simultaneously with dispersed speckle patterns. Another imaging 
spectrometer (see section 4.1.1) has been developed recently by Prieur et al., 
(1998) facilitating the conversion to a spectrograph by selecting the grism and 
a slit in the entrance image plane. It provides a low dispersion spectrographic
mode with spectral range of 350-500~nm and a spectral resolution of $\sim$0.7$^{\prime\prime}$.
Figure 6 depicts the concept of a speckle spectroscopic camera.
\bigskip
\noindent
\midinsert
{\eightpoint   
\noindent
\centerline{\psfig{figure=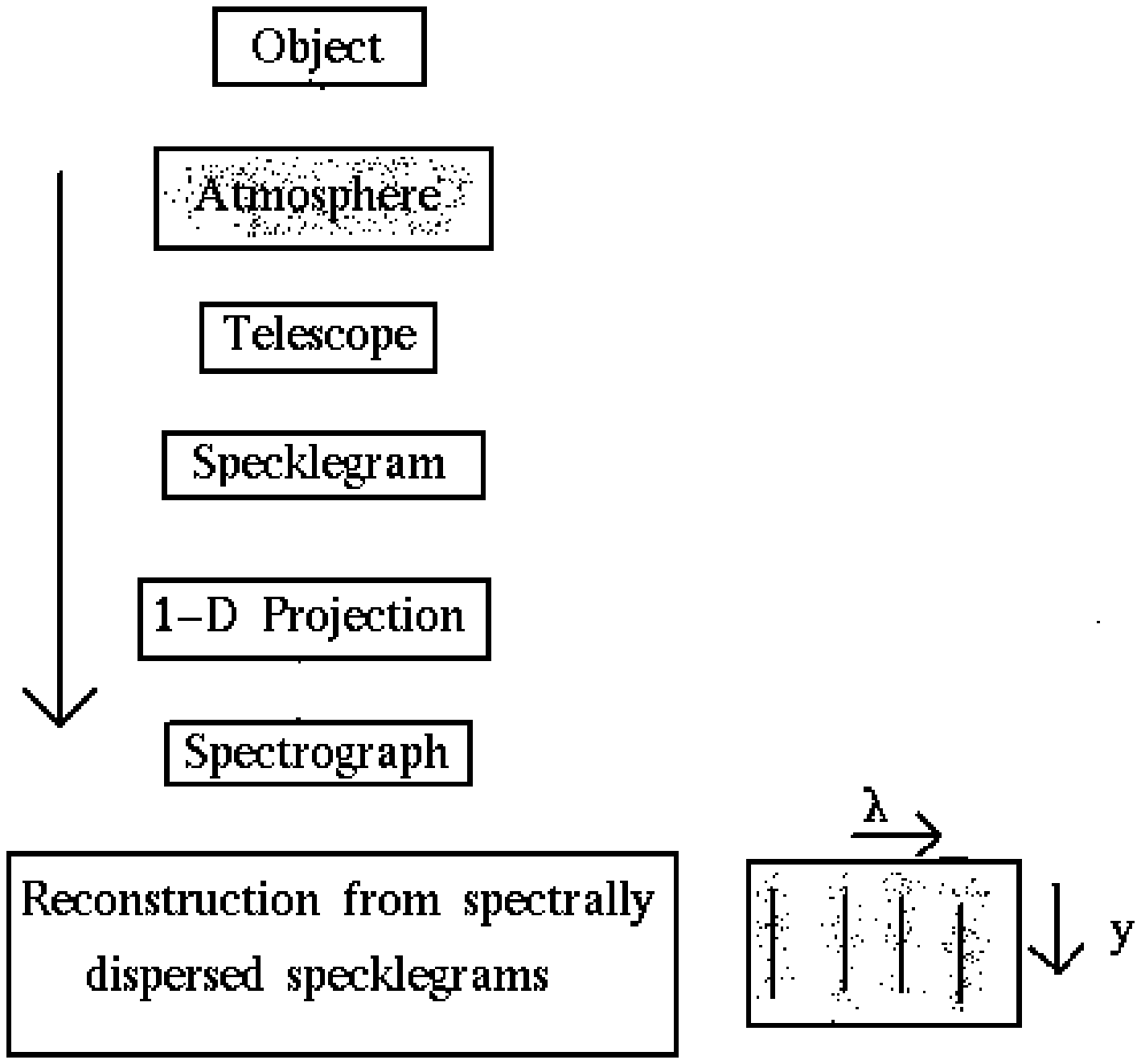,height=9.0cm,width=10cm,angle=270}}
{\bf Figure 6.} Concept of a speckle spectrograph.
}     
\endinsert
Mathematically, the intensity distribution, ${\cal W}({\bf x})$, of an 
instantaneous objective prism speckle spectrogram can be derived as,

$${\cal W}({\bf x}) = \sum_m{\cal O}_m({\bf x}-{\bf x}_m)\ast{\cal G}_m({\bf x})\ast
{\cal S}({\bf x}), \eqno(71)$$

\noindent
where, ${\cal O}_m({\bf x}-{\bf x}_m)$ denotes the m$^{th}$ object pixel and
${\cal G}_m({\bf x})$ is the spectrum of the object pixel. 
In the narrow wavelength bands ($<$30~nm), the PSF, ${\cal S}({\bf x})$, is 
wavelength independent. The objective prism spectrum,
$\sum_m{\cal O}_m({\bf x}-{\bf x}_m)\ast{\cal G}_m(\bf x)$, can be reconstructed
from the speckle spectrograms.
\vskip 20 pt
\noindent
{\bf 4.3. Speckle polarimetry}
\bigskip
\noindent
Polarized light carries valuable information about where the light is produced,
and the various physical parameters which have been responsible for its 
generation. Processes such as magnetic 
fields, chemical interactions, molecular structure and mechanical stress
cause changes in the polarization state of an optical beam. Applications
relying on the study of these changes cover a vast area, viz., astrophysics,
molecular biology, electric power generation etc. 
\bigskip
The importance of polarimetric observations in astronomy is to obtain 
information such as dust around stars, size and shape of the dust grains,
magnetic fields etc. Among other scientific objectives, 
(i) the wavelength dependence of the degree of polarization and the rotation 
of the position angle in stars with extended atmospheres, (ii) the wavelength 
dependence of degree of polarization and position angle of light emitted by 
stars present in very young ($\leq~2\times~10^6$ years) clusters and 
associations are the important ones to investigate. 
\bigskip 
The effect on the statistics of a speckle pattern is the degree of 
depolarization caused by the scattering at the surface. If the light is 
depolarized, the resulting speckle field is considered to be the sum of two
component speckle fields produced by scattered light polarized in two
orthogonal directions. The intensity at any point is the sum of the
intensities of these component speckle patterns (Goodman, 1975). These patterns
are partially correlated, therefore, a polarizer that transmits one of the
component speckle patterns, can be used in the speckle camera system.  
\bigskip 
The advantage of using speckle camera over conventional imaging
polarimeter is the capability of monitoring the short-time variability of
the atmospheric transmission. Another advantage of high resolution polarimetry
is that of using it as a tool to get insight on the binary star mechanism.
Recently, speckle polarimetry has been used to obtain polarimetric information 
with sub-arc-second resolution of astronomical objects (Falcke et al., 1996, 
Fischer et al., 1998). Falcke et al., (1996) developed a speckle imaging 
polarimeter consisting of a rotatable, achromatic $\lambda/2$~-~retardation mica 
plate in front of a fixed polarization filter. These elements were installed on 
a single mount and inserted into the optical axis in front of the telescope
focus of their speckle camera. 
\vskip 20 pt
\noindent
{\bf 4.4. Differential speckle interferometry}
\bigskip 
\noindent
Differential speckle interferometry is a method to observe the objects in
different wave modes simultaneously and computes the average cross-spectrum or
cross-correlation of pairs of speckle images (Beckers, 1982). This technique
can be used to resolve close binary system below the diffraction limit of
the telescope, as well as to determine (i) the relative orbit orientations in
multiple system, (ii) the masses of double-lined spectroscopic system, (iii) the 
masses of long period binary systems (Tokovinin (1992). Sub-diffraction limit of 
measurement of the displacement of the photo-centre of a unresolved object
as a function of wavelength has also been reported (Petrov et al., 1992).
\bigskip
Let ${\cal O}_1{(\bf x)}$ and ${\cal O}_2{(\bf x)}$, be respectively the object 
and reference brightness distributions, and ${\cal I}_1 {(\bf x)}$ and 
${\cal I}_2{(\bf x)}$ are their 
associated instantaneous image intensity distributions. The relation between 
the object and the image in the Fourier space becomes

$$\widehat{\cal I}_1({\bf u}) = \widehat{\cal O}_1({\bf u})\cdot\widehat{\cal S}_1
({\bf u}), \eqno(72)$$ 
$$\widehat{\cal I}_2({\bf u}) = \widehat{\cal O}_2({\bf u})\cdot\widehat{\cal S}_2
({\bf u}), \eqno(73)$$

\noindent
where, $\widehat{\cal S}_1{(\bf u)}$ and $\widehat{\cal S}_2{(\bf u)}$ are the 
related transfer functions. 
\bigskip
The average cross-spectrum between the object and the reference,

$$<\widehat{\cal I}_1{(\bf u)}\widehat{\cal I}_2^\ast{(\bf u)}> = \widehat
{\cal O}_1{(\bf u)}\widehat{\cal O}_2^\ast{(\bf u)}\cdot<\widehat{\cal S}_1
{(\bf u)}\widehat{\cal S}_2^\ast({\bf u})>. \eqno(74)$$

The transfer function for the cross-spectrum,
$<\widehat{\cal S}_1{(\bf u)}\widehat{\cal S}_2^\ast{(\bf u)}>$, can be 
calibrated on reference point source for which,  
$<\widehat{\cal O}_1{(\bf u)}> = <\widehat{\cal O}_2{(\bf u)}> = 1$. If 
the two spectral windows are close enough ($\Delta\lambda /\lambda \ll r_\circ/D$), 
the instantaneous transfer function can be assumed identical in both the
channels [${\cal S}_1 = {\cal S}_2 = {\cal S}$]. Therefore, equation (74) 
can be translated in to,

$$ \widehat{\cal O}_1({\bf u}) = \frac{<\widehat{\cal I}_1({\bf u})
\widehat{\cal I}_2^\ast({\bf u})>}{\widehat{\cal O}_2^\ast({\bf u})
<\mid\widehat{\cal S}({\bf u})\mid^2>} = \widehat{\cal O}_2({\bf u})
\frac{<\widehat{\cal I}_1({\bf u})\widehat{\cal I}_2^\ast({\bf u})>}
{<\mid\widehat{\cal I}_2({\bf u})\mid^2>}. \eqno(75)$$

The noise contributions from two different detectors are
uncorrelated, thereby, their contributions cancel out. Aime et al., (1985)
have used cross-spectral analysis to correlate the solar photospheric brightness
and velocity fields. 
\bigskip
Differential interferometry estimates the ratio,
$\widehat{\cal O}_1({\bf u})/\widehat{\cal O}_2({\bf u})$ and the differential 
image, ${\cal D}_{\cal I}({\bf x})$, can be obtained by performing inverse Fourier 
transform of this ratio,

$${\cal D}_{\cal I}({\bf x}) =  
{\cal F}^{-1}\left[\frac{<\widehat{\cal I}_1({\bf u})\widehat{\cal I}_2^\ast
({\bf u})>}{<\mid\widehat{\cal I}_2({\bf u})\mid^2>}\right]. \eqno(76)$$

${\cal D}_{\cal I}({\bf x})$ is self calibrating for seeing and 
represents an image of the object in the emission feature having the resolution 
of the object imaged in the continuum (Hebden et al., 1986). Petrov et al., 
(1986) found an increase in S/N ratio as the band-passes of the two 
components of the dual specklegram increase. 
\vskip 20 pt
\noindent
{\bf 4.5. Pupil plane interferometry}
\bigskip 
\noindent
In the shearing interferometer, interference fringes are produced by two
partially or totally superimposed pupil images created by introducing a beam
splitter. At each point, interference occurs from the combination of only
two points on the wave-fronts at a given baseline, and therefore, behaves as 
an array of Michelson-Fizeau interferometers.
The true object visibility can be recorded by employing the short-exposure.
The major advantage of this technique over speckle interferometry is to obtain
better S/N ratio on bright sources. The other noted advantages 
are the insensitivity to calibrate errors due to seeing fluctuations, as well
as to telescope aberrations (Roddier and Roddier, 1988). It also makes a better 
use of the detector dynamic range to detect faint extended structure like 
stellar envelopes. The following paragraphs enumerates 
the various types of shearing interferometers.
\bigskip 
(i) The lateral shearing interferometer (Cognet 1973, Ribak and Lipson, 1981)
shifts the wave-front laterally and mixing it with itself, thereby, obtains the 
interference patterns which correspond to the wave-front tilt in the shear 
direction. It is necessary to make two orthogonal measurements to assess the 
full wave-front tilt. To be precise, this typed interferometer splits the 
incident field into two optical beams (x and y direction beam). Before  
detection the beams are split again and
laterally shifted (sheared) with respect to each other. All the baselines
are identical and a single object Fourier component is measured at a time.
To achieve shear one can either make use of gratings (Wyant, 1974, Horwitz, 
1990) or use of beam separation by polarization (Hardy and MacGovern, 1987).
Depending on the specific type of interferometer, the optical field in each 
leg is split a number of times (Sandler et al., 1994). The sheared beams are
superimposed in a detector to form fringe.
\bigskip 
Ribak et al., (1985) enumerated a modified system in which fringes are detected
independently in several channels. At increasing baselines, there is a loss
of photons. Many sequential measurements are needed to explore the 
object Fourier spectrum (Roddier, 1988b). The concept of developing an 
interferometer based on using Babinet compensator (BC) has also been reported
(Saxena and Jayarajan, 1981, Saxena and Lancelot, 1982). The system is 
similar to that of a single beam lateral shearing polarizing interferometer. 
The cone of light passing through BC would produce fringe pattern due to the 
different phase change introduced between the extraordinary and the ordinary 
vibrations at different points during the oblique passage of the ray. 
\bigskip 
(ii) The wave-front folding interferometer produces 1-d Fourier 
transform, providing information on all the object Fourier components in one 
direction. The other direction may be obtained sequentially. In this 
interferometer, one pupil image is superimposed to its mirror image and is made 
of a beam splitter, a flat mirror in one arm and a roof prism in the other arm 
(Mertz, 1970). The shortcomings of this system comes from the loss of light as 
it is reflected back to the sky and produces low contrast fringes owing to mismatch
of orthogonal polarizations.
\bigskip 
(iii) The rotation shear interferometer produces 2-d Fourier transform
(Mertz, 1970). In this, one pupil image is rotated with respect to the other.
If the rotation axis coincides with the centre of the pupil, the two images
overlap. All the object Fourier components within telescope diffraction cutoff
frequency are measured simultaneously. Breckinridge (1978) suggested the use 
of $180^\circ$ rotational shear. In this system, light enters a beam 
splitter and is reflected back through a roof prism in each arm; the roof edges
are set at right angles to each other. Similar shortcomings can be noticed
(Roddier, 1988b) as in the case of wave-front folding interferometer described
above. Variable rotational shear was used to develop a similar interferometer by
Roddier and Roddier, (1983), where one roof prism is allowed to rotate around 
the optical axis. They have introduced a phase plate between the  beam splitter
and the roof prism to correct the mismatch of polarization. A similar
set-up has been developed by a group and was able to 
record fringes of a few bright objects (Saxena, 1993).
\vskip 20 pt
\noindent
{\bf 4.6. Phase-closure imaging}
\bigskip 
\noindent
The advantage of closure-phase technique (Jennison, 1958) in high resolution 
imaging is that its visibility remains uncorrupted in the presence of 
atmospheric turbulence. This technique has been applied in the field of radio 
astronomy. Its potential lies in exploiting fully the resolution attainable with 
large optical telescope (Baldwin et al., 1986). Equation (62) represents the 
simplest form of the phase closure technique but unfortunately phase 
information is not preserved. The phase-closure requirement can be redefined in 
terms of spatial frequency vectors forming a closed loop. The phase-closure 
is achieved (equation 62) in the power spectrum by taking pairs of vectors of 
opposite sign, ${\bf -u}$ and ${\bf +u}$, to form a closed loop. The closure 
phase, $\beta_{123} -$ the sum of observed phases around a triangle of baselines 
is the sum of phases of the source Fourier components and can be derived as,

$$\beta_{123} = \psi_{12} + \psi_{23} + \psi_{31}, \eqno(77)$$ 

\noindent
where, the subscripts refer to the antennae at each end of a particular 
baseline, $\psi_{ij}$ are the observed phases of the fringes produced by the
antennae i and j on a different baselines; contain the phases of the source 
Fourier components $\psi_{0,ij}$ and also the error terms.
\bigskip
The observed fringes, $\psi_{ij}$, represent the following equation,

$$\psi_{ij} = \psi_{0,ij} + \theta_i - \theta_j, \eqno(78)$$

\noindent
where, $\theta_i, \theta_j$, introduced by errors at the individual antenna,
as well as by the atmospheric variations at each antenna. Therefore,
by adding the corresponding equations the following relationship emerges,

$$\beta_{123} = \psi_{0,12} + \psi_{0,23} + \psi_{0,31}. \eqno(79)$$

Equation (79) implies cancelations of the antennae phase errors. Figure 7
describes the concept of a three element interferometer.
\bigskip
\noindent
\midinsert
{\eightpoint   
\noindent
\centerline{\psfig{figure=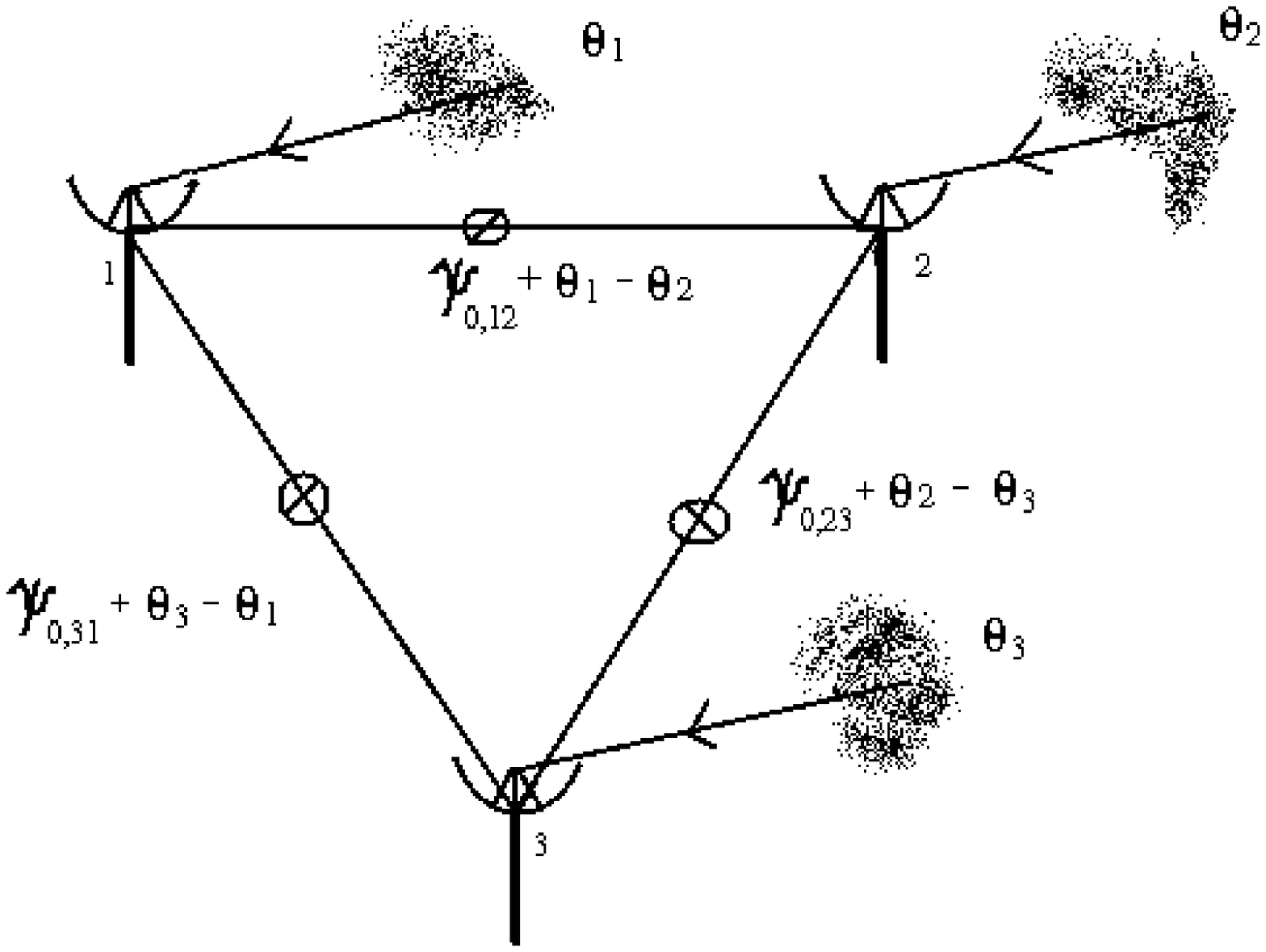,height=8cm,width=10cm,angle=270}}
{\bf Figure 7.} Concept of three element interferometer. 
}     
\endinsert
\bigskip 
Rogstad (1968) suggested the use of phase-closure technique in optics. Later,
several observers (Baldwin et al., 1986, Readhead et al., 1988) reported the 
measurements of the closure-phases obtained at
high light level with three hole aperture mask set in the pupil plane of 
the telescope and recorded interference pattern of the object using CCD as 
detector. Similar experiment was conducted by the author around the same time
using aperture mask of 3-hole, placed over the telescope, as well as in the
laboratory (Saha et al., 1987, 1988, Saha, 1999). 
\vskip 20 pt
\noindent
{\bf 4.6.1. Aperture synthesis}
\bigskip 
\noindent
The aperture synthesis imaging technique with telescope involves observing an 
object through a masked aperture of several holes and recording the resulting 
interference pattern in a series of short-exposures. Such a mask introduces
a series of overlapping two-hole interference patterns projected onto the 
detector allowing the Fourier amplitude and phase of each baseline can be 
recovered. By summing up the Fourier phases around each closed triangle, 
the closure-phase can be obtained (section 4.6).
\bigskip 
If a telescope aperture is made of elementary sub-apertures as enumerated in
section (3.4), the complex amplitude, $\Psi$, of the field in the telescope
plane is obtained by adding the contribution, $\Psi_n$, of each sub-aperture
(Born and Wolf, 1984). From the equation (64), the illumination 
$\mid\Psi\mid^2 = \mid\sum_{n,m}\Psi_n\mid^2$, can be written as,

$$\mid\sum_{n,m}\Psi_n\mid^2 = \sum_{n,m}\mid\Psi_n\mid^2 + 
\sum_{n\neq m}\sum_m\Psi_n\Psi^\ast_m. \eqno(80)$$

The second term of the right-hand side of this equation, describes the 
interference through the cross product, contains high resolution information.
An image can be reconstructed from sequential measurements of all the cross
products using pairs of sub-apertures. For $n$
apertures, there are $n(n~-~1)/2$ independent baselines, with, $n~-~1$ unknown
phase errors. This implies that by using many telescopes in an interferometric
array, most of the phase informations can be retrieved. 
\bigskip
Non-redundant masking method has been successfully developed and used at 
telescopes by several groups (Haniff et al., 1987, 1989, Nakajima et al., 1989,
Busher et al., 1990, Bedding et al., 1992, 1994, Vasisht et al., 1998,
Prieur et al., 1998, Monnier et al., 1999). Some of these 
experimental details can be found in the article (Saha, 1999 and references 
therein). In the laboratory, Saha et al., (1988) studied the shapes of fringe 
pattern of n-hole apertures by introducing different aperture masks of 
various sizes arranged in both redundantly and non-redundantly. 
\bigskip
The advantage of such a technique at telescopes is the built-in delay to observe 
objects at low declinations. It produces the optical aperture synthesis 
maps of high dynamic range but is restricted to only high light levels. The 
instantaneous coverage of spatial frequencies is sparse and most of the 
available light is discarded. Improvement can be foreseen in utilizing rotatable 
masks to several different position angles on the sky. Haniff and Busher (1992)
have used a doubly-redundant mask geometry for faint source. 
A new device using cylindrical lens in lieu of aperture mask has been 
developed recently by Bedding (1999) for fainter objects.  
\vskip 20 pt
\centerline {\bf 5. Detectors}
\bigskip 
\noindent
During the initial phases of the development in speckle interferometry, a few 
observers used photographic films with an intensifier attached to it for 
recording speckles of various stars (Breckinridge et al., 1979). Saha et al., 
(1987) used a bare movie camera to record the fringes and 
specklegrams of a few bright stars. Owing to low quantum efficiency of the 
photographic emulsion, usage of the modern ICCD camera system or a 
intensified photon counting system (Boksenburg, 1975) became necessary
to gather the speckles of faint objects. Though the water-cooled bare CCD was 
used for certain interferometric observations (Saha et al., 1997c), it is 
essential to obtain snap shots of very high time resolution of the order of (i) 
frame integration of 50~Hz (Blazit et al., 1977a, Blazit, 1986), (ii) photon 
recording rates of a few MHz (Papaliolios et al., 1985, Durand et al., 1987, 
Paresce et al., 1988, Timothy et al., 1989, Nightingale, 1991, Graves et al., 
1993). The following sub-sections enunciate the salient features of various 
modern recording devices. 
\vskip 20 pt
\noindent
{\bf 5.1. Frame transfer camera system}
\bigskip 
\noindent
The frame transfer intensified CCD (ICCD) detector consists of an image 
intensifier coupled to a CCD camera (Lemonier et al., 1988); the system employs 
micro-channel plate (MCP) as an intensifier. In the MCP image intensifier, the 
photo-electron is accelerated into a channel of the MCP releasing secondaries 
and producing an output charge cloud of about $10^3 - 10^4$ electrons with 5~-~10 
kilovolt (KV) potential. With further applied potential of $\sim$~5~-~7~KV, these 
electrons are accelerated to impact a phosphor, thus producing an output pulse of 
$\sim~10^5$ photons. These photons are directed to the CCD by fibre optic 
coupling and operate at commercial video rate with an exposure of 20~ms per 
frame (Chinnappan et al., 1991). The video frame grabber cards digitize and 
store the images in the memory buffer of the card. Depending on the buffer size, 
the number of interlaced frames stored in the personal computer (PC) can vary 
from 2 to 32 (Saha et al., 1997a). 
\bigskip 
McAlister et al., (1987) used an ICCD system which had 30~$\mu$m square pixels 
in a 244$\times$248 format for their interferometric binary star survey 
programmes. The frame rate varied from 1~-~15~ms by means of gating the 
voltage of the photo-cathode of the image intensifier. Subsequently, they
have obtained another system, ITT camera, with single-stage intensified CCD,
which allows detection to a magnitude limit of m$_v \sim$10 for 90~s integrations
(Mason et al., 1993). ICCDs with low duty cycles that are capable of recording 
a few frames per second (Weigelt et al., 1996) have also been in use. 
Aristidi et al., (1997) uses a system that has a single stage intensifier and is
thus unable to reach beyond m$_v \sim$10. The exposure time is tunable between
64~$\mu$s and 16~ms. The output is an analog video signal, recorded on SVHS 
video cassettes (Prieur et al., 1998).
\bigskip 
Another version of the ICCD (386$\times$576) camera with Peltier cooled system 
is currently available at IIA, Bangalore, India, which offers options of 
choosing the exposure time, viz., 1~ms, 5~ms, 10~ms, 20~ms etc. It can operate 
in full frame, frame transfer and kinetic modes. 
Since the CCD is cooled to $\sim$~40$^\circ$, the dark noise is low. Data are 
digitized to 12 bits and can be archived to a Pentium based PC.  
In full frame, as well as in frame transfer modes, the region of interests can 
be acquired at a faster speed. While in the kinetic mode, the image area
can be kept small to satisfy the requirements. Horch 
et al., (1998) came up with a new concept of using large format scientific 
grade CCD as the imager to collect speckle patterns. The system consists of an
optics module that contains a piezoelectric tip-tilt mirror capable of executing 
a timed sequence of movements to place many speckle patterns in a block pattern 
over the entire active area of the CCD chip. 
\vskip 20 pt
\noindent
{\bf 5.2. Photon counting camera system}
\bigskip
\noindent 
Ever since the successful development of a photon counting detector system
(Boksenburg, 1975), a number of different classes of photon counting systems 
are in use. The typical values for an object of m$_v$ = 12 over a field of 2.5$^{\prime\prime}$
are $<$~50~photons/ms within the narrow band filter. The marked 
advantage of such a scheme is that of reading the signal 
a posteriori to optimize the correlation time of short exposures in order to 
overcome the loss of fringe visibility due to the speckle lifetime. The other 
notable features are, (i) capability of determining the position of a detected
photon to 10~$\mu$m to 10~cm, (ii) ability to register individual photons with 
equal statistical weight and produces signal pulse (with dead time of ns), and 
(iii) low dark noise typically of the order of 0.2 counts $cm^{-2} s^{-1}$. 
\vskip 20 pt
\noindent
{\bf 5.2.1. CCD-based photon counting system}
\bigskip
\noindent 
Blazit et al., (1977a) had used a photon counting system which is coupled with a
micro-channel image intensifier to a commercial television camera. This camera
operates at the fixed scan line (312) with 20~ms exposure. A digital correlator 
discriminates the photon events and computes their positions in the digital 
window. It calculates the vector differences between the photon positions in 
the frame and integrates in memory a histogram of these difference vectors.  
\bigskip
Subsequently, Blazit (1986) has developed another version of the photon 
counting system (CP40), comprising a set of four Thomson 288$\times$384 CCDs 
image sensors with a common stack of a 40~mm diameter Varo image tube and a MCP. 
The readout of this system is standard, 20~ms. The amplified image
is split into four quadrants through a fibre optics reducer and four fibre
optics cylinders. Each of these quadrants is read out with a CCD device at the
video rate (50~Hz). This camera is associated with the CP40 processor $-$
a hardware photon centroiding processor to compute the photo-centre of each 
event with an accuracy of 0.25 pixel. 
\bigskip
The major shortcomings of such system arise
from the (i) calculations of the coordinates which are hardware-limited with
an accuracy of 0.25 pixel, and (ii) limited dynamic range of the detector.
This system is regularly used by the groups at Observatoire de la Cote d'Azur, 
France, for the speckle interferometric programmes as well as to record the 
fringes of stars (Mourard et al., 1989, 1992, 1994a) using Grand 
Interf\'erom\`etre \`a Deux T\'elescope (GI2T). Morel and Koechlin (1998) came 
out with a concept of using 3 linear CCDs with a 2.6~$\mu$s read-out at rates up 
to a million photons per second. A new photon counting camera that allows
a high photon rate and a direct numerisation of photon coordinates is under way
(Koechlin and Morel, 1998).
\vskip 20 pt
\noindent
{\bf 5.2.2. Precision analog photon address (PAPA)}
\bigskip
\noindent 
The Precision Analog Photon Address (PAPA) camera is a 2-d photon counting 
detector which allows recording of the address (position) and time of arrival of 
each detected photon (Papaliolios et al., 1985). The front-end of the
camera is a high gain image intensifier which produces a bright spot on its
output phosphor for events detected by the photo-cathode. The back face 
(phosphor) of the intensifier is then re-imaged by an optical system which is
made up of a large collimating lens and an array of smaller lenses. Each of the
small lenses produces a separate image of the phosphor on a mask
to provide position information of the detected photon. Behind
each mask is a field lens which relays the pupil of the small lens onto a small
photo-multiplier (PMT). A set of 19 PMTs used out of which 9 + 9 PMTs 
provides a format of 512$\times$512 pixels optical configuration. The 19th
tube acts as an event strobe, registering a digital pulse if the spot in the 
phosphor is detected by the instrument. 9 tubes are used to obtain positional 
information for an event in one direction, while the other 9 are used for 
that in the orthogonal direction. If the photon image falls on clear
area, an event is registered by the photo-tubes. The masks use grey code 
which ensures that mask stripes do not have edges located in the same place
in the field. 
\bigskip 
This first generation of the PAPA detector was used to record the 
specklegrams of the stellar objects by the group at Center
for Astrophysics, Harvard-Smithsonian Observatory, Harvard (Nisenson, 1988 and
references therein), as well as to record the fringes of stars using long
baseline interferometers by groups at Mt. Wilson (Kaplan et al., 1988) and at
Sydney (Lawson, 1994). A modified PAPA camera has been used to observe close
binary stars (Aristidi et al., 1999). This version has a new binary mask and
refurbished image intensifier. Another modified version of the system
with 2000$\times$2000 pixels has been developed by using a CeTe photo-cathode 
intensifier for the NASA UV applications (Nisenson, 1997). The 
photo-cathode is sensitive down to 100~nm and operates in vacuum. The 
system runs at a maximum count rate of 2 million per second. Figure 8 depicts
an overview of PAPA camera made for NASA UV applications. The new generation
PAPA would replace PMTs by a set of intensified photo-diodes (Nisenson, 1997).
\bigskip 
\noindent
\midinsert
{\eightpoint   
\noindent
\centerline{\psfig{figure=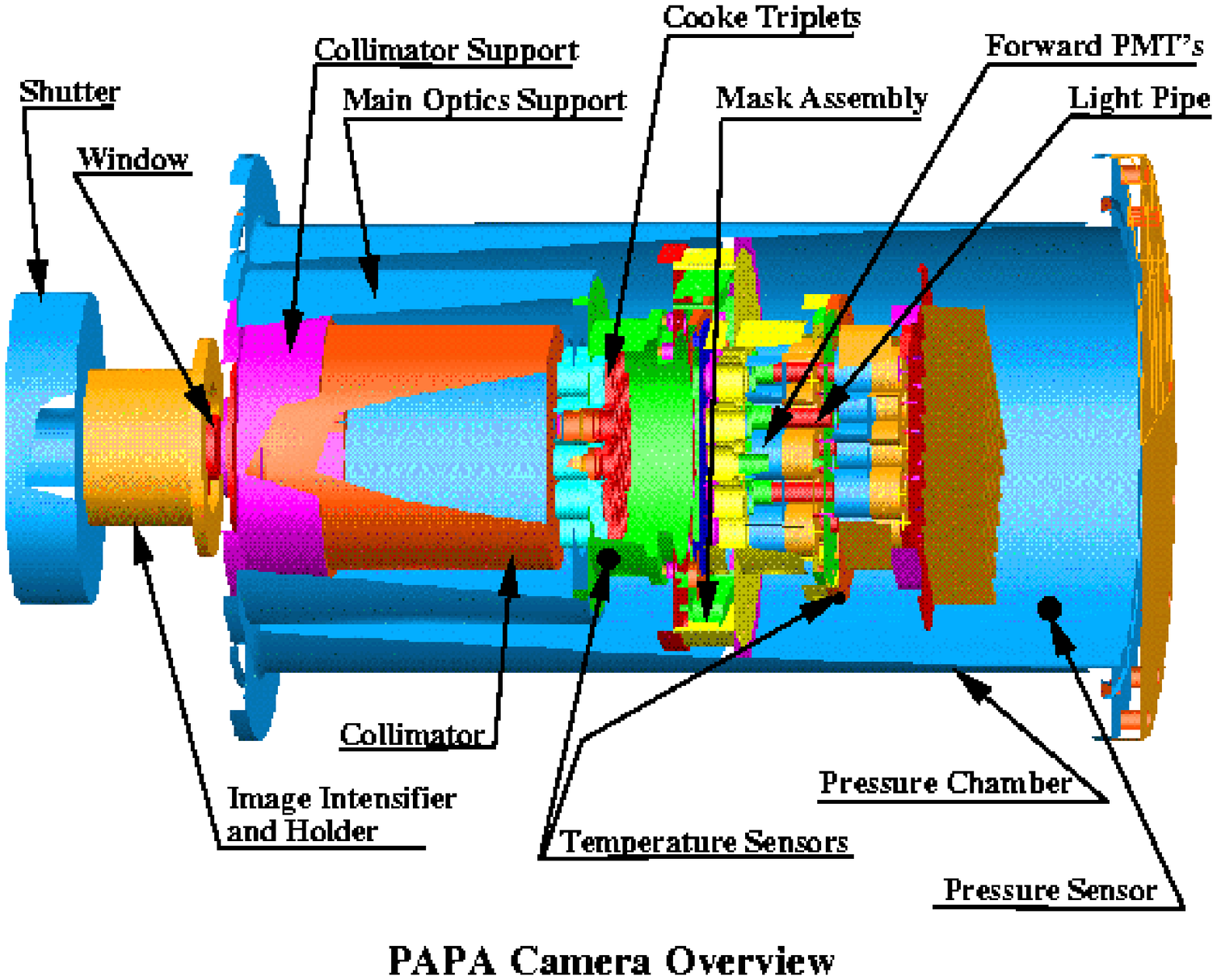,height=8.0cm,width=13cm,angle=270}}
\bigskip 
\noindent
{\bf Figure 8.} An overview of the PAPA camera (Courtesy: P. Nisenson)
}     
\endinsert
\vskip 20 pt
\noindent
{\bf 5.2.3. MCP based photon counting detector}
\bigskip
\noindent 
Several electronic readout techniques have been developed to detect the
charge cloud from a high gain MCP. Some of these systems provide 2-d
imaging capabilities (Lampton and Carlson, 1979, Martin et al., 1981, Timothy, 
1986, Siegmund et al., 1994). The main advantage of such systems is to obtain 
spatial event information by means of the position sensitive readout systems; 
the encoding systems identify each event's location. The short-comings of the
MCPs are notably due to its local dead-time which essentially restricts the
conditions for use of these detectors for high spatial resolution applications
(Rodriguez et al., 1992). These constraints are also related with the luminous
intensity and the pixel size. The following paragraphs elucidate a few of 
these systems.
\bigskip
(i) In the resistive anode position sensing detector system, 
a continuous uniform resistive sheet with appropriately shaped electrodes
provides the means for encoding the simultaneous location and arrival
time of each detected photon. The semi-transparent photo-cathode converts the
photons to electrons. This is coupled to a cascaded stack of MCPs
acting as the position sensitive signal amplifier. A net potential drop of
about 5~KV is maintained from the cathode to the anode. Each primary 
photo-electron results in an avalanche of $10^7~-~10^8$ secondary electrons
onto the resistive anode. The signals resulting from the charge redistribution 
on the plate are amplified and fed into a high speed signal processing 
electronics system that produces 12 bit x, y addresses for each event. A PC 
based data acquisition system builds up a 1024$\times$1024 image from this 
asynchronous stream of x, y values that can be stored in a PC (Clampin et al., 
1988, Paresce et al., 1988). Nakajima et al., (1989) used a similar one for
their programme. The drawback of this system is the large pixel response 
function; the nominal resolution of the system is about 60~$\mu$m. For certain
applications, this system is used regularly on the GI2T (Mourard et al., 
1994a). 
\bigskip
(ii) The photon counting system based on the wedge-and-strip anodes (Siegmund 
et al., 1986), uses conductive array structure, in which the geometrical
image distortions might be eliminated. It comprises multiple terminals
with the x, y co-ordinate of a charge cloud determined through ratios of
charge deposited onto the various terminals. The amplitudes of the signals
detected on the wedge, strip and electrodes are linearly proportional to
the x, y co-ordinates of the detected photon event. In this system, the 
spatial resolutions of the order of 40~-~70~$\mu$m FWHM and position sensitivities
of 10~$\mu$m are obtained at high MCP gains. High resolution wedge-and-strip 
detectors can operate at event rates up to about 5$\times~10^4$ photons per second.
\bigskip
(iii) Another photon counting system based on the delay line anodes has been
developed (Lampton et al., 1987, Siegmund et al., 1994) for micro-channel plate 
spectrometers. The delay line anodes offer high resolution pixels of the 
order of 15~$\mu$m FWHM, as well as high event rate of the order of 
more than $4\times~10^5$ photons per second (Siegmund et al., 1994). 
The delay line anode has a zigzag micro-strip transmission
line etched onto a low loss, high dielectric substrate. The position of the 
charge cloud event is encoded as the difference in arrival times of the charge 
pulse at both ends of the transmission line. 
\bigskip
(iv) A new photon counting system, silicon anode detector with integrated
electronics is underway (Chakrabarti, 1998). This system offers to integrate
the anode of the detector and all supporting electronics onto a single silicon
substrate. A charge cloud from the MCP strikes the conducting strips on an 
anode. The charge is gated into a serial CCD shift register through field 
effect transistor (FET) charge transfer gates. The captured charge in the CCD
can be clocked out through an amplifier using three phase shifting technique.
The maximum estimated possible count rate is of 1$\times~10^6$ photons per second.
\vskip 20 pt
\noindent
{\bf 5.2.4. Multi anode micro-channel array (MAMA)}
\bigskip
\noindent 
The multi anode micro-channel array (MAMA) detector allows high speed, discrete 
encoding of photon positions and makes use of numerous anode electrodes that 
identify each event's location (Timothy, 1993). In this system, the electron 
amplification is obtained by an MCP and the charge is collected on a crossed 
grid coincidence array (Timothy et al., 1989). The resulting electron cloud 
hits two sets of anode arrays beneath the MCP, where one set is perpendicular in 
orientation to the other; the charge collected on each anode is amplified. The 
position of the event is determined by coincidence discrimination. The C-plate 
type MCP, which is a single curved plate that prevents ion feedback, is used. The 
present generation MAMA detector has the pixel format of 224$\times$960 with 14 
or 25~$\mu$m square pixels and is equipped with the application specific integrated 
circuit pixel decoders (Kasle and Morgan, 1991). The pulse-pair resolution is 
reported to be better than 200~ns. MAMA detector is used for speckle 
interferometric observations by the group at Stanford University (Horch et al.,
1992, 1996).
\vskip 20 pt
\centerline {\bf 6. Image processing}
\bigskip 
\noindent
The autocorrelation technique enumerated in section 4.1.3. falls short of 
providing true image reconstructions. The diffraction-limited phase retrieval by 
various methods, viz., (i) speckle holography method, (ii) shift-and-add method, 
(iii) Knox-Thomson (KT) technique, (iv) triple-correlation (TC) technique, 
(v) non-redundant aperture masking technique (Rhodes and Goodman, 1973), 
(vi) maximum entropy (MEM) method (Jaynes, 1982), (vii) CLEAN algorithm (Hogbom, 
1974), (viii) blind iterative deconvolution (BID) technique, 
have, by and large, produced qualitative scientific results. 
The following sub-sections outline some of these techniques of estimating the 
Fourier phase.  
\vskip 20 pt
\noindent
{\bf 6.1. Speckle holography method}
\bigskip 
\noindent
If a reference point source is available within the iso-planatic patch ($\sim~$7$^{\prime\prime}$),
it can be used as a key to reconstruct the target source in the same way as a 
reference coherent beam is used in holographic reconstruction (Liu and Lohmann, 
1973, Weigelt, 1978). Let the point source be 
represented by a Dirac impulse, ${\cal A}\delta({\bf x})$, at the 
origin and ${\cal O}_1({\bf x})$ be the nearby object to be reconstructed. The 
intensity distribution in the field of view is 

$${\cal O}({\bf x}) = {\cal A}\delta({\bf x}) + {\cal O}_1({\bf x}). \eqno(81)$$

A regular speckle interferometric measurement will give the squared modulus of 
its Fourier transform, $\widehat{\cal O}({\bf u})$,

$$\mid\widehat{\cal O}({\bf u})\mid^2 = \mid{\cal A} + \widehat{\cal O}_1({\bf u}) 
\mid^2 = {\cal A}^2 + {\cal A}\widehat{\cal O}_1({\bf u}) + {\cal A}\widehat
{\cal O}_1^\ast({\bf u}) + \widehat{\cal O}_1({\bf u})\widehat{\cal O}_1^\ast({\bf u}). \eqno(82)$$ 

The inverse Fourier transform gives the autocorrelation, 
${\cal C}_{\cal O}({\bf x})$, of the field of view

$${\cal C}_{\cal O}{(\bf x)} = {\cal A}^2\delta{(\bf x)} + {\cal A}{\cal O}_1{(\bf x)} 
+ {\cal A}{\cal O}_1{(\bf - x)} + {\cal C}_{{\cal O}_1}{(\bf x)}, \eqno(83)$$ 

\noindent
where, ${\cal C}_{{\cal O}_1}{(\bf x)}$ is the autocorrelation of the object.  
\bigskip
The first and the last term in equation (83) are centred at the origin. If the 
object is far enough from the reference source, ${\cal O}({\bf x})$, its mirror 
image, ${\cal O}{(\bf - x)}$, is therefore recovered apart from a 180$^\circ$ 
rotation ambiguity. Weigelt (1978) had retrieved the deconvolved images of close 
multiple systems, $\zeta$~Cancri and ADS3358, using this algorithm.
\bigskip
Another method, known as cross-spectrum analysis, is to calculate the 
cross-spectrum between the object and the 
reference source (see section 4.4). The angular distance between the two sources
should be within the iso-planatic patch or the two spectral windows should be close 
enough. Equation (75) showed that the speckle holography transfer function, 
$<\mid\widehat{\cal S}{(\bf u)}\mid^2>$, is real. The method is insensitive to 
aberrations and the expected value of the phase of the cross-spectrum coincides with
the phase-difference between the object and the reference. However, when this
is not the case, one must use image processing methods by using the 
different light levels of the speckle clouds (reference and target) to 
extract the object.
\vskip 20 pt
\noindent
{\bf 6.2. Shift-and-add algorithm}
\bigskip 
\noindent
In this technique, the image data frame is shifted so that the pixel with
maximum S/N in each frame can be co-added linearly at the same location in
the resulting accumulated image (Lynds et al., 1976, Worden et al., 1976). In a 
blurred image, each speckle is considered as a distorted image of the object, 
the brightest one being the least distorted. In this processes, the position of the 
brightest pixel, ${\bf x}_k$, is necessary to be located in each specklegram,
${\cal I}_k({\bf x})$, [${\cal I}_k({\bf x}_k) > {\cal I}_k({\bf x})$ for
all ${\bf x} \neq {\bf x}_k$], followed by shifting the specklegram (without
any rotation) to place this pixel at the centre of the image space. The
shift-and-add image, ${\cal I}_{sa}({\bf x})$, is obtained by averaging over
the set of the shifted specklegrams,

$${\cal I}_{sa}(\bf x) = <{\cal I}_k({\bf x} + {\bf x}_k)>. \eqno(84)$$

The large variations in the brightness of the brightest pixels can be 
observed in a set of speckle images. The contamination level of a specklegram
may not be proportional to the brightness of its brightest pixel (Bates and
McDonnell, 1986). The adjusted shift-and-add image, ${\cal I}_{asa}({\bf x})$,
be defined as,

$${\cal I}_{asa}({\bf x}) = <{\it w}[{\cal I}_k({\bf x}_k)]{\cal I}_k({\bf x} +
{\bf x}_k)>, \eqno(85)$$

\noindent
where, ${\it w}[{\cal I}_k({\bf x}_k)]$ is the weighting in relation with
the brightness of the brightest pixel. The choice of the same quantity can be
made as ${\it w}\{{\cal I}_k({\bf x}_k)\} = {\cal I}_k({\bf x}_k)$.
\bigskip
An array of impulse is constructed by putting an impulse at each of the centre 
of gravity with a weight proportional to the speckle intensity. This impulse 
array is considered to be an approximation of the instantaneous PSF and
is cross-correlated with the speckle frame (see section 6.1). Disregarding the
peaks lower than the pre-set threshold as enumerated above, the m$^{th}$ speckle
mask, $mask_m({\bf x})$, is defined by,

$$mask_m({\bf x}) = \sum_{n=1}^M{\cal I}_m({\bf x}_{m,n})\delta({\bf x} - {\bf x}_{m,n}).
\eqno(86)$$

The m$^{th}$ masked speckled image, $m{\cal I}_m({\bf x})$, is expressed as,

$$ m{\cal I}_m({\bf x}) = {\cal I}_m({\bf x}) \otimes mask_m({\bf x}), \eqno(87)$$

\noindent
where, $\otimes$ stands for correlation.
\bigskip
The Lynds-Worden-Harvey image can be obtained by averaging $<m{\cal I}_m({\bf x})>$.
This technique contains more information in each ${\cal I}_m({\bf x})$ than
${\cal I}_{asa}({\bf x})$, and therefore, exhibits more S/N. For direct speckle
imaging, the shift-and-add image, ${\cal I}_{sa}({\bf x})$, is a 
contaminated one containing two complications - a convolution, 
${\cal S}_k({\bf x})$ and an additive residual, ${\cal C}({\bf x})$ - which
means,

$${\cal I}_{sa}({\bf x}) = {\cal O}({\bf x}) \ast {\cal S}({\bf x}) + 
{\cal C}({\bf x}), \eqno(88)$$

\noindent
where, ${\cal S}({\bf x}) = \sum_{k=1}^k \delta({\bf x} - {\bf x^\prime}_k)d_k, 
{\bf x^\prime}_k$ being the constant position vectors and $d_k$, the positive 
constant.
\bigskip
It is essential to calibrate ${\cal I}_{sa}({\bf x})$ with an unresolved point
source and reduce it in the same way to produce ${\cal S}({\bf x})$. The 
estimate for the object, ${\cal O}({\bf x})$, is evaluated from the inverse 
Fourier transform of the following equation,

$$\widehat{\cal O}({\bf u}) = \frac{\widehat{\cal I}_{sa}({\bf u})}{\widehat
{\cal I}_\circ({\bf u}) + \widehat{\cal N}({\bf u})}, \eqno(89)$$

\noindent
where, $\widehat{\cal N}({\bf u})$ stands for the noise spectrum. 
\bigskip
This is the first approximation of the object irradiance. If this technique
is applied to a point source, an image similar to Airy pattern yields. This 
method is found to be insensitive to the telescope aberrations. The limitation 
of this algorithm is not being applicable when photon noise dominates in 
addition to the accuracy with which speckle maxima are located. Ribak, (1986)
used an iterative matched filter approach to reconstruct the image.
\vskip 20 pt
\noindent
{\bf 6.3. Knox-Thomson technique (KT)}
\bigskip 
\noindent
The Knox-Thomson method (Knox and Thomson, 1974), which is a small modification
of the autocorrelation technique, involves the centering in each specklegram with
respect to its centroids and in finding the ensemble autocorrelation of the Fourier 
transform of the instantaneous image intensity. 
\bigskip 
In lieu of the image energy 
spectrum, $<\mid\widehat{\cal I}({\bf u})\mid^{2}>$, let the general 
second order moment be the cross spectrum, 
$<\widehat{\cal I}({\bf u}_1)\widehat{\cal I}^\ast({\bf u}_2)>$. The 
cross-spectrum takes significant values only 
if $\mid{\bf u}_1 - {\bf u}_2\mid < r_\circ/\lambda$. The typical value of 
$\mid{\bf \Delta u}\mid$ is $\sim~$0.2 - 0.5~$r_\circ/\lambda$. 
Invoking equation (67), a 2-d irradiance distribution, 
${\cal I}({\bf x})$ and its Fourier transform, $\widehat{\cal I}({\bf u})$, is 
defined by the equation,

$$\widehat{\cal I}{\bf (u)} = \int^{+\infty}_{-\infty} {\cal I}{\bf (x)}e^{-i2\pi 
{\bf u x}} d{\bf x}. \eqno(90)$$

Unlike the autocorrelation technique (see section 3.4.1.), KT technique 
defines the correlation of ${\cal I}({\bf x})$ and
${\cal I}({\bf x})$ multiplied by a complex exponential factor with a spatial
frequency larger than zero. In image space, the correlations of 
${\cal I}({\bf x})$, is derived as, 

$${\cal I}({\bf x}_1, {\bf \Delta u}) = \int^{+\infty}_{-\infty} {\cal I}^*({\bf x})
{\cal I}({\bf x} + {\bf x}_1) e^{i2\pi{\bf \Delta u x}} d{\bf x}, \eqno(91)$$

\noindent
where, ${\bf x}_1 = {\bf x}_{1x} + {\bf x}_{1y}$ are 2-d spatial 
co-ordinate vectors.
\bigskip 
In Fourier space, $\widehat{\cal I}({\bf u})$ gives the following relationship,

$$\widehat{\cal I}({\bf u}_1, {\bf \Delta u}) = \widehat{\cal I}({\bf u}_1) \widehat{\cal I}^* 
({\bf u}_1 + {\bf \Delta u}), \eqno(92)$$

\noindent
where, ${\bf u}_1 = {\bf u}_{1x} + {\bf u}_{1y}$, and ${\bf \Delta u}
= {\bf \Delta u}_x + {\bf \Delta u}_y$ are 2-d spatial frequency 
vectors. ${\bf \Delta u}$ is a small, constant offset spatial frequency.
\bigskip 
This technique consists of evaluating the three sub-planes in Fourier
space corresponding to 
${\bf \Delta u} = {\bf \Delta u}_x \ , \ {\bf \Delta u} = {\bf \Delta u}_y \ , 
\ {\bf \Delta u} = {\bf \Delta u}_y + {\bf \Delta u}_x$. If digitized 
images are used, ${\bf \Delta u}$ normally corresponds to the fundamental
sampling vector interval (Knox, 1976). A number of sub-planes can be used by
using different values of ${\bf \Delta u}$. Invoking equation (69), into
equation (92), the following relationship can be found as,

$$\widehat{\cal I}{(\bf u}_1) \widehat{\cal I}^\ast ({\bf u}_1 + {\bf \Delta u}) 
= \widehat{\cal O}({\bf u}_1)\widehat{\cal O}^\ast({\bf u}_1+{\bf \Delta u})\widehat{\cal S}
({\bf u}_1)\widehat{\cal S}^\ast({\bf u}_1+{\bf \Delta u}). \eqno(93)$$

The argument of the equation (92) provides the phase-difference between the two
spatial frequencies separated by ${\bf \Delta u}$ and can be expressed as, 

$$arg\mid\widehat{\cal I}^{KT}({\bf u}_1, {\bf \Delta u})
\mid = \theta^{KT}({\bf u}_1, {\bf \Delta u}) = \psi 
({\bf u}_1) - \psi ({\bf u}_1 + {\bf \Delta u}). \eqno(94)$$  

The equation (93) can be expressed as,   

$$\widehat{\cal I}{(\bf u}_1, {\bf \Delta u}) \eqno$$
$$= \mid\widehat{\cal O}{(\bf u}_1)\mid\mid\widehat{\cal O}({\bf u}_1 + {\bf \Delta u})\mid 
\mid\widehat{\cal S}{(\bf u}_1)\mid \mid \widehat{\cal S}({\bf u}_1 + 
{\bf \Delta u})\mid e^{i[\theta^{KT}_{\cal O}({\bf u}_1, {\bf \Delta u})
+ \theta^{KT}_{\cal S}({\bf u}_1, {\bf \Delta u})]}. \eqno(95)$$

The difference in phase between points in the object phase-spectrum is 
encoded in the term, $e^{i\theta^{KT}_{\cal O}({\bf u}_1, {\bf \Delta u})} 
= e^{i[\psi_{\cal O}({\bf u}_1) - \psi_{\cal O}({\bf u}_1 + {\bf \Delta u})]}$
of the equation (95). In a single image realization the
object phase-difference is corrupted by the random phase-differences due to the
atmosphere-telescope OTF, $e^{i\theta^{KT}_{\cal S}({\bf u}_1, {\bf \Delta u})} 
= e^{i[\psi_{\cal S}({\bf u}_1) - \psi_{\cal S}({\bf u}_1 
+ {\bf \Delta u})]}$. If equation (94) is averaged over a large number of 
frames, the feature $({\bf \Delta \psi_{\cal S}}) = 0$. When ${\bf \Delta u}$ 
is small, $\mid \widehat{\cal O}({\bf u}_1+{\bf \Delta u}) \mid \ \approx \ \mid 
\widehat{\cal O}({\bf u}_1) \mid$, etc. and so,

$$<\widehat{\cal I}{(\bf u}_1, {\bf \Delta u})> \eqno$$
$$= \mid\widehat{\cal O}({\bf u}_1)\mid\mid\widehat{\cal O}({\bf u}_1 + {\bf \Delta u})\mid 
e^{i[\theta^{KT}_{\cal O}({\bf u}_1, {\bf \Delta u})]} 
<\widehat{\cal S}({\bf u}_1)\widehat{\cal S}^\ast({\bf u}_1 + {\bf \Delta u})>, \eqno(96)$$

\noindent
from which, together with equation (70), the object phase-spectrum, 
$\theta^{KT}_{\cal O}({\bf u}_1, {\bf \Delta u})$, can be determined.
This technique is found to be sensitive to odd order aberrations, e.g., coma but
not defocusing, astigmatism, spherical etc., (Barakat and Nisenson, 1981).
\vskip 20 pt
\noindent
{\bf 6.4. Triple Correlation Technique (TC)}
\bigskip 
\noindent
The triple correlation technique or speckle masking method (Weigelt, 1977,
Lohmann et al., 1983) is 
based on the closure phase (section 4.6) that remains uncorrupted in the 
presence of atmospheric turbulence and is being widely applied to improve the 
resolution of night-time objects (Cruzal\'ebes et al., 1996, Falcke et al., 
1996, Osterbart et al., 1996, Weigelt et al., 1996, Saha et al., 1999b, 1999c), 
as well as to the extended objects (Beletic, 1988, Von der L\"uhe and Pehlemann, 
1988, Max, 1994, Denker, 1998, Sridharan and Venkatakrishnan, 1999). The 
advantages of this technique are: (i) it provides information about the object 
phases with better S/N ratio from a limited number of frames, and (ii) it serves 
as the means for recovery with diluted 
coherent arrays (Reinheimer and Weigelt, 1987) owing to its relationship with 
the phase closure technique (section 4.6) used in radio astronomy (Roddier, 
1986, Cornwell 1987). The disadvantage of this technique is that of demanding 
very severe constraints on the computing facilities with 2-d data since the 
calculations are 4-dimensional (4-d) (see equations 98, 100). It requires
extensive evaluation-time and data storage requirements,
if the correlations are performed by using digitized images on a computer.
\bigskip 
Unlike in shift-and-add where, a Dirac impulse at 
the centre of gravity of each speckle is put to estimate the same, Weigelt, 
(1977), suggested to multiply the object speckle pattern ${\cal I}({\bf x})$ by 
an appropriately shifted version of this ${\cal I}({\bf x} + {\bf x}_1)$. 
The result is correlated with ${\cal I}({\bf x})$. (For example, in the case 
of a close binary star, the shift is equal to the angular separation between 
the stars, masking one of the two components of each double speckle). 
\bigskip
The bispectrum planes are obtained as an intermediate step,
contain phase information encoded by correlating ${\cal I}({\bf x})$ with
${\cal I}({\bf x})$ multiplied by complex exponentials with higher spatial
frequencies. The calculation of the ensemble average TC is given by,

$${\cal I}{\bf (x}_1, {\bf x}_2) = <\int^{+\infty}_{-\infty}{\cal I}
{\bf (x)}{\cal I}({\bf
x} + {\bf x}_1){\cal I}({\bf x} + {\bf x}_2)d{\bf x}>, \eqno(97)$$

\noindent
where, ${\bf x}_j = {\bf x}_{jx} + {\bf x}_{jy}$ are 2-d spatial 
co-ordinate vectors.
\bigskip
The Fourier transform of the triple correlation is called bispectrum and its
ensemble average is given by,

$$\widehat{\cal I}({\bf u}_1, {\bf u}_2) = <\widehat{\cal I}({\bf u}_1) 
\widehat{\cal I}^\ast 
({\bf u}_1 + {\bf u}_2) \widehat{\cal I}({\bf u}_2)>, \eqno(98)$$

\noindent
where, ${\bf u}_j = {\bf u}_{jx} + {\bf u}_{jy}, 
\widehat{\cal I}({\bf u}_j) =\int{\cal I}({\bf x})e^{-i2\pi{\bf u}_j{\bf x}} 
d{\bf x}, \widehat{\cal I}^\ast({\bf u}_1 + {\bf u}_2) = \int{\cal I}({\bf x})
e^{i2\pi({\bf u}_1 + {\bf u}_2).{\bf x}}d{\bf x}$. 
\bigskip
In the second order moment or in the energy spectrum, phase of the object's
Fourier transform is lost, but in the third order moment or in the bispectrum it 
is preserved. The argument of equation (98) can be expressed as,

$$arg\mid\widehat{\cal I}^{TC}({\bf u}_1, {\bf u}_2)\mid = 
\theta^{TC}({\bf u}_1, {\bf u}_2) = \psi({\bf u}_1) - \psi({\bf u}_1 + {\bf u}_2) + 
\psi({\bf u}_2). \eqno(99)$$

Equation (99) gives the 
phase-difference. Invoking equation (69) into equation (97), it emerges as,

$$\widehat{\cal I}({\bf u}_1, {\bf u}_2) = \widehat{\cal O}({\bf u}_1)\widehat{\cal O}^
\ast({\bf u}_1+{\bf u}_2)\widehat{\cal O}({\bf u}_2) <\widehat{\cal S}({\bf u}_1)
\widehat{\cal S}^\ast({\bf u}_1+{\bf u}_2)\widehat{\cal S}({\bf u}_2)>. \eqno(100)$$

The relationship implies that the image bispectrum is equal to the object 
bispectrum times a bispectrum transfer function, 
$<\widehat{\cal S}({\bf u}_1)\widehat{\cal S}
^\ast({\bf u}_1 + {\bf u}_2)\widehat{\cal S}({\bf u}_2)>$.  
The object bispectrum is given by,

$$\widehat{\cal I}_{\cal O}({\bf u}_1, {\bf u}_2) = \widehat{\cal O}({\bf u}_1)
\widehat{\cal O}^\ast({\bf u}_1 + {\bf u}_2)\widehat{\cal O}({\bf u}_2)   
= \frac{<\widehat{\cal I}({\bf u}_1)\widehat{\cal I}^\ast({\bf u}_1 + {\bf u}_2) 
\widehat{\cal I}({\bf u}_2)>}{<\widehat{\cal S}({\bf u}_1)\widehat{\cal S}^\ast ({\bf u}_1 
+ {\bf u}_2)\widehat{\cal S}({\bf u}_2)>}. \eqno(101)$$ 

The modulus $\mid\widehat{\cal O}({\bf u})\mid$ and phase $\psi({\bf u})$ of the 
object Fourier transform $\widehat{\cal O}({\bf u})$ can be evaluated from the object 
bispectrum $\widehat{\cal I}_{\cal O}({\bf u}_1, {\bf u}_2)$ (Weigelt, 1988). 
The object phase-spectrum, is encoded in the term, 
$e^{i\theta^{TC}_{\cal O}({\bf u}_1, {\bf u}_2)} = e^{i[\psi_{\cal O}({\bf u}_1) 
- \psi_{\cal O}({\bf u}_1 + {\bf u}_2) + \psi_{\cal O}({\bf u}_2)]}$, of 
equation (98). 
\bigskip
The S/N ratio of the phase recovery contains S/N ratio for bispectrum, as well 
as a factor representing improvement due to redundancy of phase-information
stored in the bispectrum. Karbelkar and Nityananda, (1987) showed that
the calculation of Wirnitzer (1985), in the wave limit true for bright objects, 
the S/N ratio for the bispectrum overestimated by a factor of the order of the 
square root of the number of speckles. The bispectrum method has been tested 
with a computer simulated image by using a code developed by Saha et al., 
(1999b). Figure 9 depicts (a) simulated binary system, (b) 2-d
representation of a 4-d bispectrum, (c) triple correlation and (d)
reconstructed image of the same binary system. 
\bigskip
\noindent
\midinsert
{\eightpoint   
\noindent
\centerline{\psfig{figure=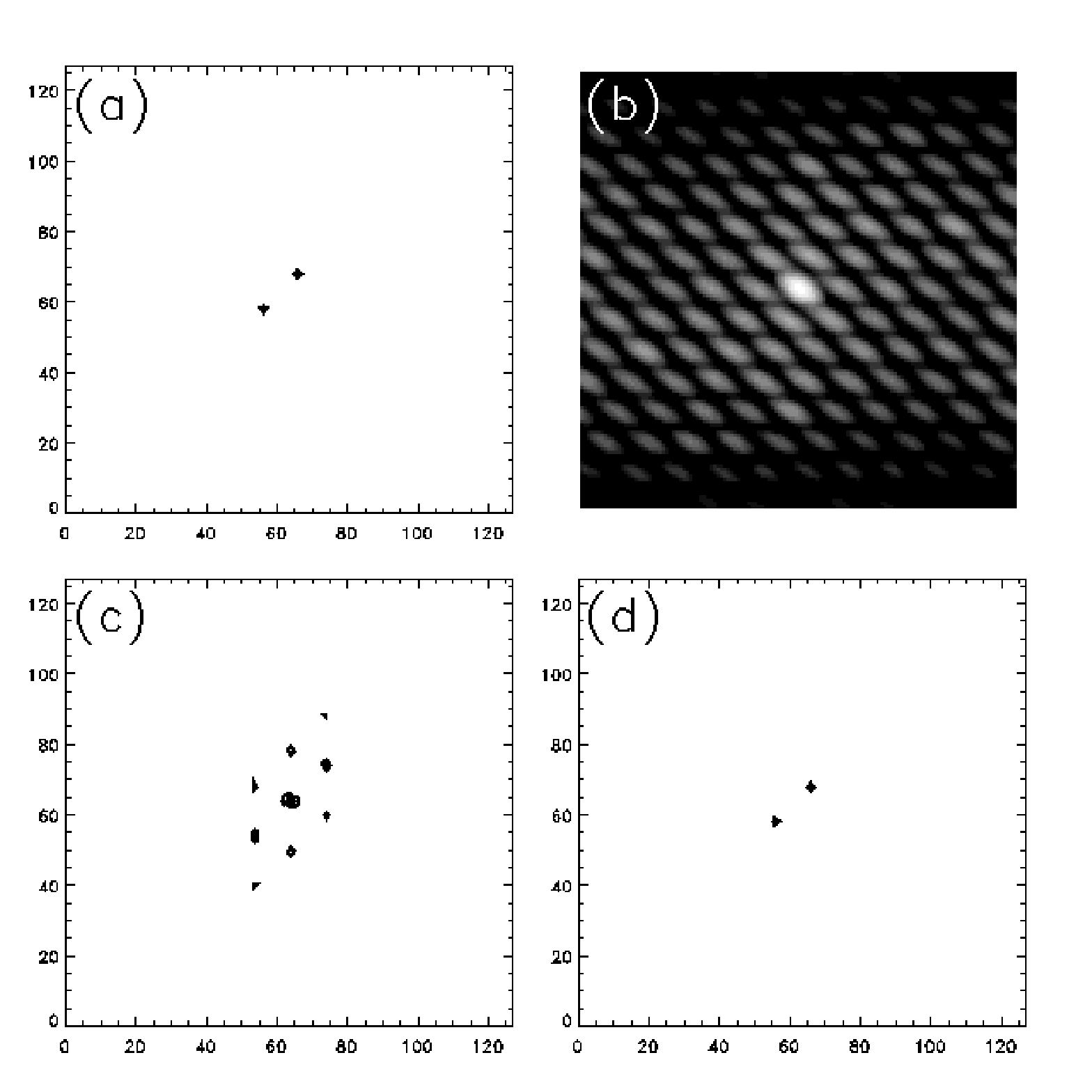,height=10cm,width=12cm}}
\bigskip
\noindent
{\bf Figure 9.} (a) 2-d maps of a simulated binary system, (b) 2-d
representation of its 4-d bispectrum, (c) its triple correlation
and (d) its reconstructed image. 
}     
\endinsert
The object phase-difference is corrupted by the random phase-differences due to 
the atmosphere-telescope OTF, $e^{i\theta^{TC}_{\cal S}({\bf u}_1, {\bf u}_2)} 
= e^{i[\psi_{\cal S}({\bf u}_1) - \psi_{\cal S} ({\bf u}_1 + {\bf u}_2)
+ \psi_{\cal S}({\bf u}_2)]}$, in a single image realization. If a sufficient 
number of specklegrams are averaged, one can overcome this shortcoming. 
Let $\theta^{TC}_{\cal O}({\bf u}_1, {\bf u}_2)$ be the phase of the object 
bispectrum; then,

$$\widehat{\cal O}({\bf u}) = \mid\widehat{\cal O}({\bf u})\mid e^{i\psi({\bf u})}, \eqno(102)$$ 
\noindent
and
$$\widehat{\cal I}_{\cal O}({\bf u}_1,{\bf u}_2) = \mid\widehat{\cal I}_{\cal O}
({\bf u}_1, {\bf u}_2)\mid e^{i\theta^{TC}_{\cal O}({\bf u}_1, {\bf u}_2)}. \eqno(103)$$

Equations (102) and (103) may be inserted into equation (101), yielding
the relations,

$$\widehat{\cal I}_{\cal O}({\bf u}_1, {\bf u}_2) \eqno$$
$$= \mid\widehat{\cal O}({\bf u}_1)\mid  
\mid\widehat{\cal O}({\bf u}_2 )\mid\mid\widehat {\cal O}({\bf u}_1+{\bf u}_2)\mid  
e^{i[\psi_{\cal O}({\bf u}_1) - \psi_{\cal O}({\bf u}_1 + {\bf u}_2)
+ \psi_{\cal O}({\bf u}_2)]}\rightarrow, \eqno(104)$$ 
$$\theta^{TC}_{\cal O}({\bf u}_1, {\bf u}_2) = \psi_{\cal O}({\bf u}_1) 
- \psi_{\cal O}({\bf u}_1 + {\bf u}_2) + \psi_{\cal O}({\bf u}_2). \eqno(105)$$

Equation (105) is a recursive equation for evaluating the phase of the object 
Fourier transform at coordinate ${\bf u} = {\bf u}_1 + {\bf u}_2$ (Weigelt, 1988). 
The reconstruction of the object phase-spectrum from the phase of the
bispectrum is recursive in nature. The object phase-spectrum at $({\bf u}_1
+ {\bf u}_2)$ can be expressed as,

$$\psi_{\cal O}({\bf u}_1 + {\bf u}_2) = \psi_{\cal O}({\bf u})  
= \psi_{\cal O}({\bf u}_1) + \psi_{\cal O}({\bf u}_2) - \theta^{TC}_{\cal O}
({\bf u}_1, {\bf u}_2). \eqno(106)$$  

If the object spectrum at ${\bf u}_1$ and ${\bf u}_2$ is known,
the object phase-spectrum at $({\bf u}_1 + {\bf u}_2)$ can be computed.
The bispectrum phases are mod $2\pi$, therefore, the recursive reconstruction
in equation (105) may lead to $\pi$ phase mismatches between the computed
phase-spectrum values along different paths to the same point in frequency 
space. Northcott et al., (1988) opined that phases from different paths to
the same cannot be averaged to reduce noise under this condition. A variation
of the nature of computing argument of the term, $e^{i\psi_{\cal O}({\bf u}_1 
+ {\bf u}_2)}$, is needed to obtain the object phase-spectrum and the equation 
(106) can be translated to,

$$e^{i\psi_{\cal O}({\bf u}_1 + {\bf u}_2)} = 
e^{i[\psi_{\cal O}({\bf u}_1) + \psi_{\cal O}({\bf u}_2) - 
\theta^{TC}_{\cal O}({\bf u}_1,{\bf u}_2)]}. \eqno(107)$$

The values obtained using the unit amplitude phasor recursive re-constructor
(see equation 106) are insensitive to the $\pi$ phase ambiguities and
has been used by several investigators in studies of the bispectrum
technique (Ayers et al., 1988, Northcott et al., 1988, Meng et al., 1990, 
Matson, 1991, Lawrence et al., 1992). Saha et al., (1999b, 1999c) have developed 
a code based on the unit amplitude phasor recursive re-constructor. The 
algorithm written in Interactive Data Language (IDL) takes about an hour for 
processing 10 frames of size 128$\times$128 using the SPARC ULTRA workstation. The 
memory needed for the calculation
exceeds 160 MB if the array size is more than the said number. Since the
bispectrum is a 4-dimensional function, it is difficult to represent it in
a 3-dimensional co-ordinate system. Therefore, they have stored the 
calculated values in 1-dimensional array and used later to calculate the 
phase by keeping track of the component frequencies (Saha et al., 1999b). 
For example, let $\psi(u_1, u_2)$ be the phase corresponding to the frequency, 
$(u_1, u_2)$ and $\theta^{TC}_{\cal O}$
be the phase of the bispectrum. Assuming $\psi(0, 0) = 0, \psi(0, \pm1) = 0$
and $\psi(\pm1, 0) = 0$, the phases are calculated by the unitary amplitude 
method. Saha et al., (1999b) have successfully determined
the phase from the average bispectrum of the specklegrams of a couple of stars. 
\bigskip
Like the phase-closure (see section 4.6),
the bispectrum is insensitive to the atmospherically induced random phase
errors, as well as to random motion of the image centroid;
therefore, images are not required to be shifted to common centroid prior to
computing the bispectrum. 
Further development of phase reconstruction from the bispectrum can be
seen in the form of (i) a least-square formulation of the phase reconstruction
(Meng et al., 1990, Glindemann et al., 1991, 1992), and (ii) a projection-slice 
theorem of tomography and the Radon transform (Northcott et al., 1988). 
Matson (1991) has developed two weighted least-squares
estimation formulations of the phase reconstructions problem. However, most
of the results, using the real data are limited to the recursive phase 
reconstruction (Weigelt, 1988 and references therein, Weigelt et al., 1996). 
\vskip 20 pt
\noindent
{\bf 6.4.1. Relationship between KT and TC}
\bigskip 
\noindent
Ayers et al., (1988) have illustrated the relationship of two of the widely used
algorithms, namely, KT and TC methods, which runs as follows. 
\bigskip
Substituting ${\bf u}_2 = {\bf \Delta u}$ in equation (98), the single plane
bispectrum can be expressed as,

$$\widehat{\cal I}^{TC}({\bf u}_1, {\bf \Delta u}) = \widehat{\cal I}({\bf u}_1) 
\widehat{\cal I}^\ast 
({\bf u}_1 + {\bf \Delta u}) \widehat{\cal I}({\bf \Delta u}). \eqno(108)$$

In the image space, the expression becomes,

$${\cal I}^{TC}{\bf (x}_1, {\bf \Delta u}) = \widehat{\cal I}({\bf \Delta u})
\int^{+\infty}_{-\infty}{\cal I}^\ast{\bf (x)}{\cal I}({\bf x} + {\bf x}_1)
e^{i2\pi{\bf \Delta u x}} d{\bf x} \eqno$$
$$ = \widehat{\cal I}({\bf \Delta u})I^{KT}({\bf x}_1, {\bf \Delta u}). \eqno(109)$$

Between the two equations (92) and (109), it is seen that for a spatial 
frequency difference, ${\bf \Delta u}$, the bispectrum plane corresponds to a 
weighted version, a signal depending factor, of the KT product in Fourier space.
On comparing equations (94) and (99), the shift invariant property of the
bispectrum can be expressed as,

$$\theta^{TC}({\bf u}_1, {\Delta u}) = \theta^{KT}({\bf u}_1, {\bf \Delta u})
+ \psi({\bf \Delta u}). \eqno(110)$$

In the KT method, the transform $\widehat{\cal I}({\bf u})$ interferes with 
itself after translation by a small shift vector ${\bf \Delta u}$. Figure 10
depicts the diagrammatic representation of pupil sub-apertures of diameter
$r_\circ$; (a) approximate phase-closure is achieved in KT method, and (b)
complete phase-closure in TC method.
\bigskip
\noindent
\midinsert
{\eightpoint   
\noindent
\centerline{\psfig{figure=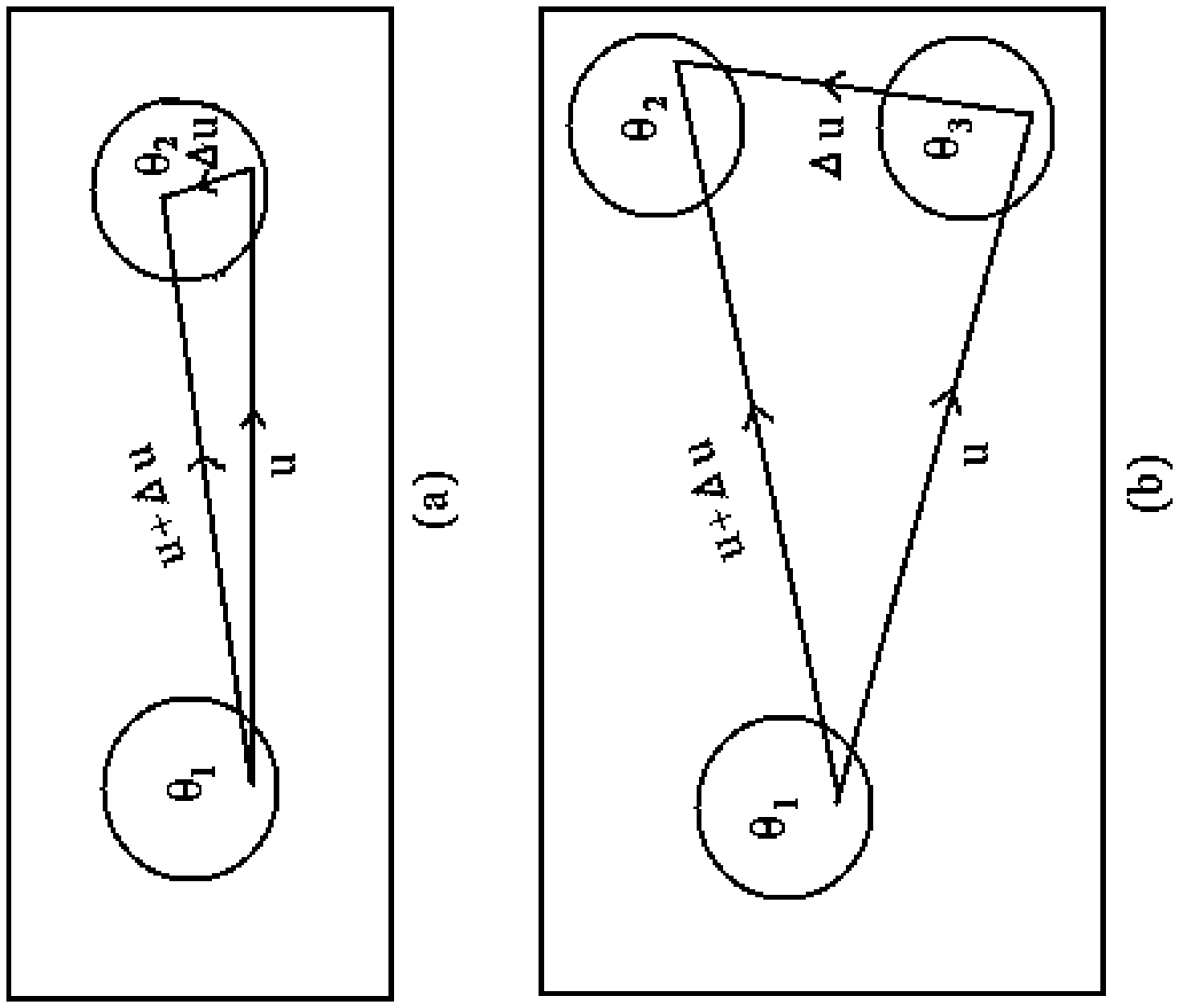,height=9.5cm,width=12cm,angle=270}}
{\bf Figure 10.} Pupil sub-apertures of diameter $r_\circ$, (a) approximate
phase-closure achieved in KT method, (b) while complete phase-closure is achieved 
in TC method.
}     
\endinsert
The approximate phase-closure in this case is achieved by two vectors, 
${\bf u}$ and ${\bf u} + {\bf \Delta u}$, assuming that the pupil phase is 
constant over ${\bf \Delta u}$. The major Fourier component of the fringe 
pattern is averaged with a component at a frequency displaced by 
${\bf \Delta u}$. If this vector does not force
the vector difference ${\bf -u - \Delta u}$ to be outside the spatial frequency
bandwidth of the fringe pattern, it preserves the Fourier 
phase-difference information in the averaged signal. The atmospheric phase 
effectively forms a closed loop. 

$$arg\mid\widehat{\cal I}{\bf (u)}\widehat{\cal I}{\bf (-u - \Delta u)}\mid  
= \psi{\bf (u)} + \theta_1
- \theta_2 + \psi {\bf (-u - \Delta u)} - \theta_1 + \theta_2, \eqno(111)$$
$$= \psi {\bf (u)} - \psi ({\bf u + \Delta u}), \eqno(112)$$ 
$$<\widehat{\cal I}{\bf (u)}\widehat{\cal I}({\bf -u - \Delta u})>  \neq 0, \ \  \Delta u < r_\circ/\lambda. \eqno(113)$$

Let this system of two apertures be extended to 3 and the 
$\Delta u > r_\circ/\lambda$, then,

$$arg\mid\widehat{\cal I}{\bf (u)}\widehat{\cal I}{\bf (-u - \Delta u)}\mid 
= \psi{\bf (u)} + 
\theta_1 - \theta_2 - \psi{\bf (-u - \Delta u)} - \theta_1 + \theta_3 \eqno$$ 
$$= \psi{\bf (u)} - \psi{\bf (u + \Delta u)} - \theta_2 + \theta_3, \eqno(114)$$  
$$<\widehat{\cal I}{\bf (u)}\widehat{\cal I}{\bf (-u -\Delta u)}> = 0, \ \  \Delta u < r_\circ/\lambda. \eqno(115)$$

The atmospheric phase contribution is not closed in this case. KT is limited 
to frequency differences $\Delta u < r_\circ/\lambda$. 
In the bispectrum method, a third vector, ${\bf \Delta u}>r_\circ/\lambda$, is 
added to form phase-closure. When $\lambda{\bf \Delta u}>r_\circ$, the 
third vector is essential; the KT method fails with this arrangement. 
If the bispectrum average is performed, the phase is closed and Fourier 
phase-difference information is preserved.

$$arg\mid\widehat{\cal I}{\bf (u)}\widehat{\cal I}{(\bf \Delta u)}\widehat{\cal I}
{\bf (-u - \Delta u)}\mid \eqno$$
$$= \psi{(u)} + \theta_1 - \theta_2 + \psi({\bf \Delta u}) 
+ \theta_2 - \theta_3 + \psi({\bf -u - \Delta  u}) - \theta_1 + \theta_3 \eqno$$
$$= \psi({\bf u}) - \psi{(\bf u + \Delta u)} + \psi(\Delta {\bf u}). \eqno(116)$$ 

Thus, the bispectrum method can give phase information for phase-differences 
$\Delta u > r_\circ/\lambda$. TC method is similar to the phase-closure technique.
Closure phases are insensitive to the atmospherically induced random phase
errors, as well as to the permanent phase errors introduced by telescope 
aberrations. Any linear phase term in the object phase cancels out in the 
closure phase. In the case of a TC method, the information resides within isolated 
patches in Fourier space, while the KT method is not applicable under these
conditions.
\bigskip
A comparative study was made by Weitzel et al., (1992) with the real data and
concluded that KT and TC (bispectrum) give the same results for a binary system
with a separation greater than the diffraction limit of the telescope. But
they found some improvement for a binary system at a separation about the
telescope diffraction limit in applying bispectrum method.
\vskip 20 pt
\noindent
{\bf 6.5. Blind iterative deconvolution technique (BID)}
\bigskip 
\noindent
The blind iterative deconvolution (BID) technique combines constrained 
iterative techniques (Gerchberg and Saxton, 1972, Fienup, 1978) with blind 
deconvolution (Lane and Bates, 1987).  
Essentially, it consists of using very limited information about the image, 
like positivity and image size, to iteratively arrive at a deconvolved image of 
the object, starting from a blind guess of either the object or both the 
convolving function. The iterative loop is repeated enforcing 
image-domain and Fourier-domain constraints until two images are found that 
produce the input image when convolved together (Bates and McDonnell, 1986,
Ayers and Dainty, 1988, Bates and Davey, 1988). The image-domain constraints 
of non-negativity is generally used in iterative algorithms associated with 
optical processing to find effective supports of the object and or PSF from a 
specklegram. The implementation of the algorithm of BID developed by Nisenson, 
(1992) runs as follows.
\bigskip
The algorithm has the degraded image, ${\cal I}({\bf x})$, as the operand. An initial
estimate of the PSF, ${\cal S}({\bf x})$, has to be provided. The degraded image is 
deconvolved from the guess PSF by Wiener filtering, which is an operation
of multiplying a suitable Wiener filter (constructed from the Fourier
transform, $\widehat{\cal S}({\bf u})$, of the PSF) with the Fourier transform, 
$\widehat{\cal I}({\bf u})$, of the degraded image. The technique of Wiener
filtering damps the high frequencies and minimizes the mean square error between
each estimate and the true spectrum. The Wiener filtering spectrum, 
$\widehat{\cal O}(\bf u)$, takes the form: 

$$\widehat{\cal O}({\bf u}) = \widehat{\cal I}({\bf u}){\frac{\widehat
{\cal S}^\ast({\bf u})}{\widehat{\cal S}({\bf u})\widehat{\cal S}^\ast({\bf u}) 
+ \widehat{\cal N}({\bf u})\widehat{\cal N}^\ast({\bf u})}}. \eqno(117)$$

The noise term, $\widehat{\cal N}({\bf u})$ can be replaced with a 
constant estimated as the rms fluctuation of the high frequency region in the 
spectrum where the object power is negligible. 
The result, $\widehat{\cal O}({\bf u})$, is transformed back to image space, 
the negatives in the image and the positives outside a prescribed domain 
(called object support) are set to zero. The average of 
negative intensities within the support are subtracted from all pixels. The 
process is repeated until the negative intensities decrease below the noise.
A new estimate of the PSF is next obtained by Wiener filtering the original
image, ${\cal I}({\bf x})$, with a filter constructed from the constrained object,
${\cal O}({\bf x})$; this completes one iteration. This entire process is repeated
until the derived values of ${\cal O}({\bf x})$ and ${\cal S}({\bf x})$ converge 
to sensible solutions. 
\bigskip
The comparative analysis of the recovery reveals that both the morphology and
the relative intensities are present in the retrieved diffraction-limited
image and PSF. The results are vulnerable to the choice of various parameters 
like the support radius, the level of high frequency suppression during the 
Wiener filtering, etc. The availability of prior knowledge on the object 
through autocorrelation of the degraded image is very useful for specifying 
the object support radius (Saha and Venkatakrishnan, 1997, Saha, 1999).  
Jefferies and Christou, (1993), have developed an algorithm which
requires more than a single speckle frame for improving the convergence. 
\bigskip
A major advantage of the BID
compared with the rest of the methods described above is the ability to 
retrieve the diffraction-limited image of an object from a single specklegram 
without the reference star data. Often, it may not be possible to gather a 
sufficient number of images within the time interval over which the statistics 
of the atmospheric turbulence remains stationary. 
\bigskip
Though BID requires high S/N ratios in the data, it 
appears to be a powerful tool for image restoration. The restoration for speckle 
data at the 12th magnitude level assuming 15\% photo-cathode sensitivity of the 
detector had been demonstrated (Nisenson et al., 1990).
Nisenson, (1992) opined that the technique can be improved by using multiple 
frames simultaneously as convergence constraints which may help in extending 
the magnitude level to fainter (m$_v >$~12). This technique is indeed an ideal 
one to process the degraded images of extended objects, viz., (i) Planets 
(Saha et al., 1997c), and (ii) Sun (Nisenson 1992). 
\vskip 20 pt
\centerline {\bf 7. Speckle imaging of extended objects}
\bigskip 
\noindent
Thus far, the discussions revolved around obtaining diffraction-limited 
informations of the stellar point sources. Reconstructions of high resolution 
features on the extended objects, viz., (i) Jupiter, (ii) Sun, were also made 
with interferometric techniques. The techniques for obtaining deconvolved
images with sub-arc-second resolution of these extended objects are illustrated
below. 
\vskip 20 pt
\noindent
{\bf 7.1. Jupiter}
\bigskip 
\noindent
Speckle interferometric technique was used to record the images of Jupiter 
during the collision of the fragments of comet Shoemaker-Levy 9 
(1993e) with the former during July 16-22, 1994. At Lick
observatory, the 3~m telescope equipped with speckle camera based on a bare CCD 
(roughly $1000\times1000$) with a pixel scale of 0.066$^{\prime\prime}$/pixel was used 
to image (each with 100~-~500~ms exposures) the said planet through 
several filters during the said period (Max, 1994). They used the bispectral 
reconstruction technique to retrieve the deconvolved image so as to  
achieve about 0.2~-~0.4$^{\prime\prime}$ resolution. 
\bigskip
\noindent
\midinsert
{\eightpoint   
\noindent
\centerline{\psfig{figure=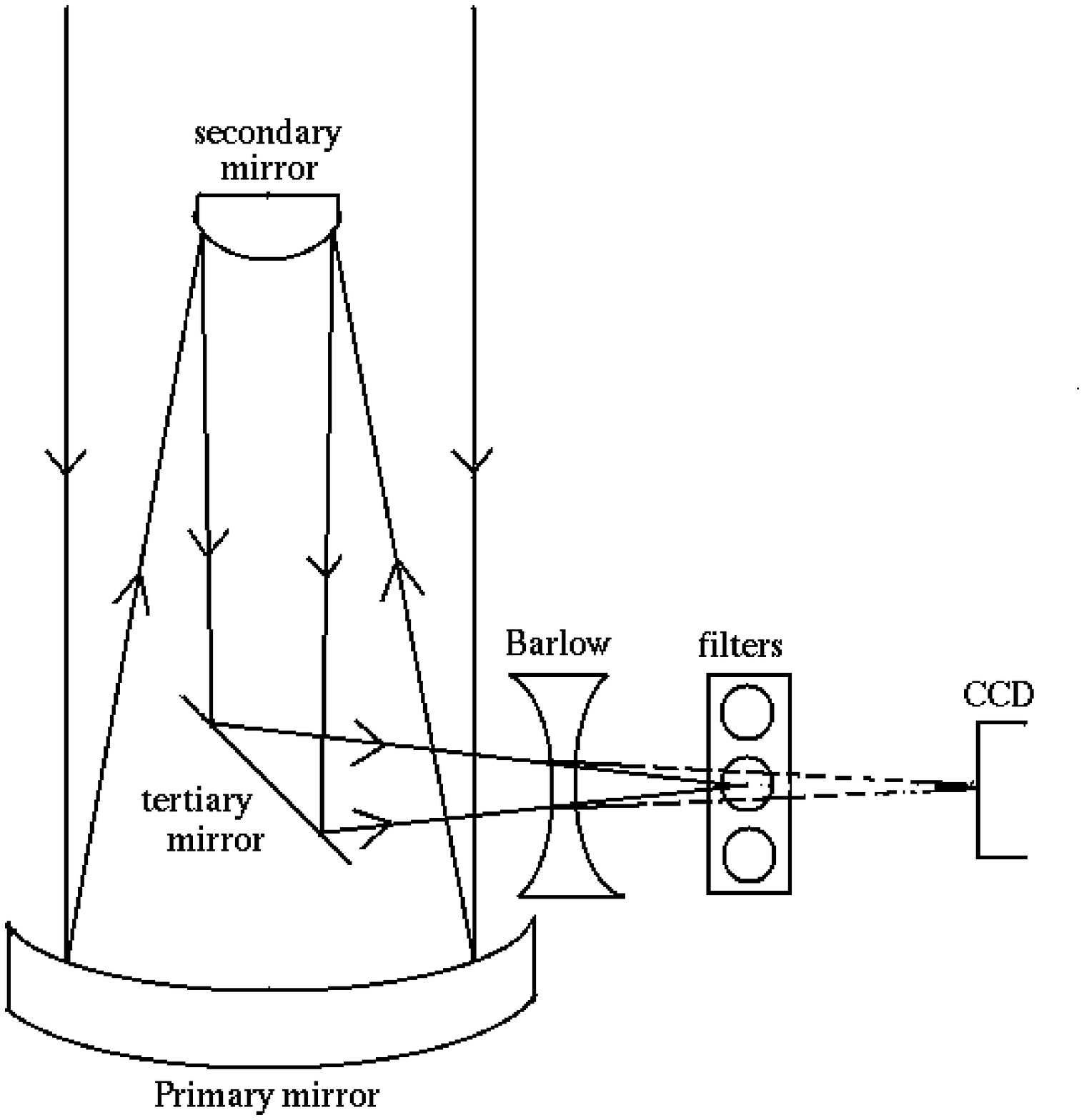,height=9cm,width=9cm}}
\bigskip
\noindent
{\bf Figure 11.} Experimental set-up of the interferometer to record the images
of Jupiter (Saha et al., 1997c).
}     
\endinsert
Large number of specklegrams of Jupiter in 
the visible band were recorded by Saha et al., (1997c), 
during the said period, at the Nasmyth focus of the 1.2~m 
telescope of Japal-Rangapur Observatory (JRO), Hyderabad. 
The interferometer developed for this purpose (Saha et al., 1997c) is depicted 
in figure 11. The image scale at the Nasmyth focus (f/13.7) of this telescope
was enlarged by a Barlow lens arrangement, with a sampling of 0.11$^{\prime\prime}$/pixel of the 
CCD. A set of 3 filters were used to image Jupiter, viz., (i) centred at 
550~nm, with FWHM of 30~nm, (ii) centred at 611~nm, with FWHM of 
9.9~nm and (iii) RG~9 with a lower wavelength cut-off at 800~nm (Saha et al.,
1997c). A water-cooled bare 1024$\times$1024 CCD was used as detector to record 
specklegrams of entire planetary disk of Jupiter with exposure times of 
100~-~200~ms. 
\bigskip
Saha et al., (1997c) have identified the complex spots due to impacts 
by the fragments using the technique of BID. The reconstructed image of Jupiter is 
shown in figure 12. The chief results of the construction is the enhancement in
the contrast of spots. The complex spots in the East are due to impacts by 
fragments Q, R, S, D, G and the spots close to the centre are due to K and L 
impacts.
\bigskip
\noindent
\midinsert
{\eightpoint   
\noindent
\centerline{\psfig{figure=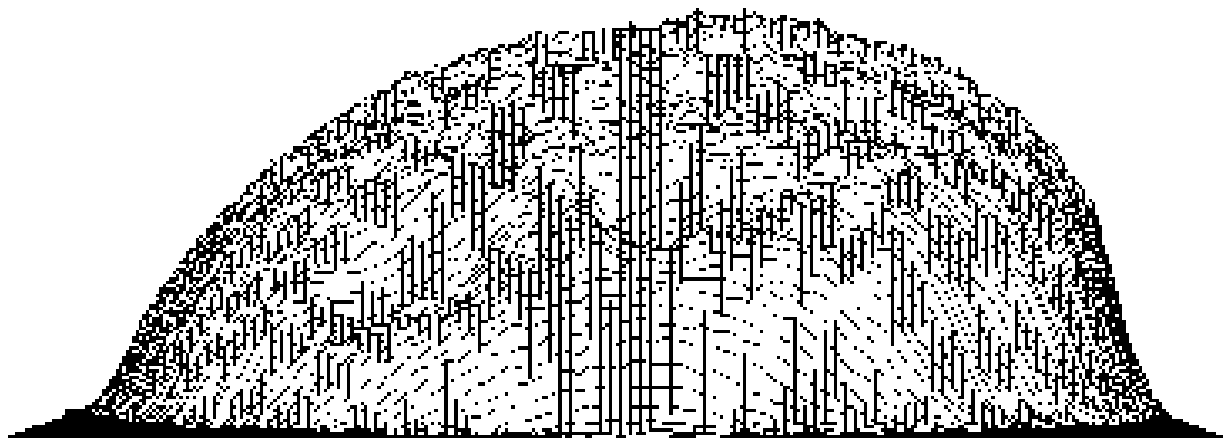,height=5.5cm,width=13cm,angle=270}}
\bigskip
\noindent
{\bf Figure 12.} The deconvolved image of Jupiter (Saha et al., 1997c).  
}     
\endinsert
\vskip 20 pt
\noindent
{\bf 7.2. Sun}
\bigskip 
\noindent
Interferometric techniques can bring out the high resolution informations of
the fundamental processes on the Sun that take place on sub-arc-second scales
concerning with convection and with magnetic fields (Von der L\"uhe and
Zirker, 1988). Several investigators (Harvey, 1972, Harvey and Breckinridge, 
1973, Harvey and Schwarzchild, 1975, Kinahan, 1975) found the existence of solar
features with sizes of the order of 100~km or smaller by applying the 
interferometric technique to solar observation. Subsequently, solar granulation
has been studied extensively with the said technique by many others (Aime et 
al., 1975, 1977, 1978, Aime, 1976, Ricort and Aime, 1979, Von der L\"uhe and 
Dunn, 1987). Solar small scale structure evolves fast and changes
its position by about 0.1$^{\prime\prime}$ within a fraction of a minute or so (Von
der L\"uhe, 1989). The rapid evolution of solar granulation
prevents the detection of the long sequence of specklegrams for reconstruction,
and therefore, the amount of time available for data accumulation is limited.
Another limitation comes from the lack of efficient detectors to record a 
large amount of frames within the stipulated time before the structure changes. 
\bigskip
From the observations of photospheric granulation from disc centre to limb at $\lambda
= 550\pm5~nm$, by means of speckle interferometric technique, at the Vacuum Tower 
Telescope (VTT), at the Observatorio del Teide (Tenerife), Wilken et al., (1997) 
found the decrease of the relative rms-contrast of the centre-to-limb  
of the granular intensity. Time series of high spatial resolution observations
with the same telescope reveal the highly dynamical evolution of 
sunspot fine structures, namely, umbral dots, penumbral filaments, facular 
points (Denker, 1998). The small-scale brightenings in the vicinity of 
sunspots, were also observed in the wings of strong 
chromospheric absorption lines (Denker et al., 1995, Denker, 1998). 
These structures which are associated with strong magnetic fields show
brightness variations close to the diffraction limit of the telescope 
($\sim~$0.16$^{\prime\prime}$ at 550~nm). 
\bigskip
To study the intensity enhancement in the inner line wings of H$\alpha$ (656.28~nm),
Denker et al (1995) used a speckle interferometer to obtain images of the solar
chromosphere. The set-up consists of a field stop at the prime focus of the
afore-mentioned telescope reducing stray-light, two achromats sampling 0.08$^{\prime\prime}$/pixel 
of the detector, an interference filter with FWHM of 3~nm and the 
Universal Birefringes filter with FWHM of 0.1~nm or 0.05~nm (transparency amounts
to 11\% at H$\alpha$). A beam splitter was inserted in the light path to feed
90\% light to the slow-scan CCD camera for speckle imaging and remaining light 
to the video CCD camera for guiding.
\bigskip
Magnetic fields in the solar photosphere are usually detected by measuring
the polarization in the wing of a Zeeman spectral line. Keller and Von der
L\"uhe (1992a, 1992b) applied differential speckle interferometric technique to 
solar polarimetry. This technique was used to image the quiet granulation,
as well as to make polarimetric measurements of a solar active region. The set-up 
used at the Swedish Vacuum Solar Telescope, La palma, consists of (i) an 
achromatic quarter-wave plate that transforms circularly polarized into 
linearly polarized light, (ii) two calcite plates rotated 
90$^\circ$ relative to each other to ensure that two beams have identical path
lengths, and (iii) a quarter-wave plate balancing the intensity in the two beams.
The two beams are split up by a non-polarizing beam-splitter cube; one passes 
through a 8.2~nm FWHM interference filter centred at 520~nm and the other passes
through a Zeiss tunable filter centred in the blue wing of FeI 525.02~nm with 
FWHM of about 0.015~nm. The former is used to determine the instantaneous
PSF; CCD video cameras were used to detect these channels.
\bigskip
Keller and Johannesson (1995) have developed another method to obtain
diffraction-limited spectrograms of sun consisting of
speckle polarimetry technique (Keller and Von der L\"uhe, 1992b) and a
rapid spectrograph (with a reflecting slit) scanning system (Johannesson et 
al., 1992). Two cameras record the spectrograms and 2-d slit-jaw 
images simultaneously. The slit of the spectrograph scans the solar 
surface during the observing run. 
\bigskip
To reconstruct solar images, various image processing algorithms, viz., 
(i) Knox-Thomson technique, (ii) speckle masking method, (iii) the technique of 
BID, have been applied (Stachnik et al., 1977, 1983, Beletic, 1988, Von der 
L\"uhe and Pehlemann, 1988, Nisenson, 1992, Max, 1994, Denker et al., 1995, 
Keller and Johannesson, 1995, Wilken et al., 1997, Denker, 1998). But the 
major problem comes from estimating PSF in the case of Sun due to the lack of
a reference point source, unlike stellar objects where this parameter 
can be determined from a nearby reference star. Von der L\"uhe (1984) 
proposed the spectral ratio technique based on a comparison between long and
short-exposure images and this was employed by the observers (Denker et al., 
1995, Wilken et al., 1997) to derive the Fried's parameter. The models of 
speckle transfer function (Korff, 1973) and of average short-exposure modulation 
transfer function (Fried, 1966) were applied to compare the observed spectral 
ratios with theoretical values. In this respect, the technique of BID, 
(see section 6.5), where a direct measurement of calibrating speckle transfer 
function is not required (Nisenson, 1992)  
has clear advantage over other techniques in retrieving the solar image. 
\vskip 20 pt
\noindent
{\bf 7.2.1. Solar speckle observation during eclipse}
\bigskip 
\noindent
The limb of the moon eclipsing the sun provides a sharp edge as a reference
object during solar eclipses, and therefore, helps in estimating the seeing effects
(Callados and V\`azquez, 1987). The intensity profile falls off sharply
at the limb; the departure of this fall off gives an indirect estimate
of the atmospheric PSF. Callados and V\`azquez, (1987) reported the measurement
of PSF during the observation of solar eclipse of 30 May, 1984 using the 
40~cm Newton Vacuum telescope at Observatorio dei Teide. Saha et al., (1997b) 
developed an experiment for the speckle reconstruction of solar features 
during the partial eclipse of Sun as viewed from Bangalore on 24 October,  
1995. The set-up is described as below.
\bigskip 
A Carl-Zeiss 15~cm Cassegrain-Schmidt reflector 
was used as telescope fitted with an aluminised glass plate in front that
transmits 20\% and reflects the rest back. A pair of polaroids were placed 
in the converging beam ahead of the focal plane. A 3~nm passband filter centred
at 600~nm was inserted between these polaroids. One of the polaroids was
mounted on a rotatable holder (see figure 13), so as to adjust the amount
of light falling onto the camera. A pin-hole of 1~mm
diameter was set at the focal plane for isolating a small field-of-view.
A microscope objective ($\times$5) re-imaged this pin-hole onto the EEV CCD
camera operated in the TV mode. The images were planned to be acquired with
exposure time of 20~ms using a Data Translation$^{TM}$ frame-grabber card
DT-2861. Unfortunately, unfavourable weather conditions at Bangalore prevented 
in recording any data.
\bigskip
\noindent
\midinsert
{\eightpoint   
\noindent
\centerline{\psfig{figure=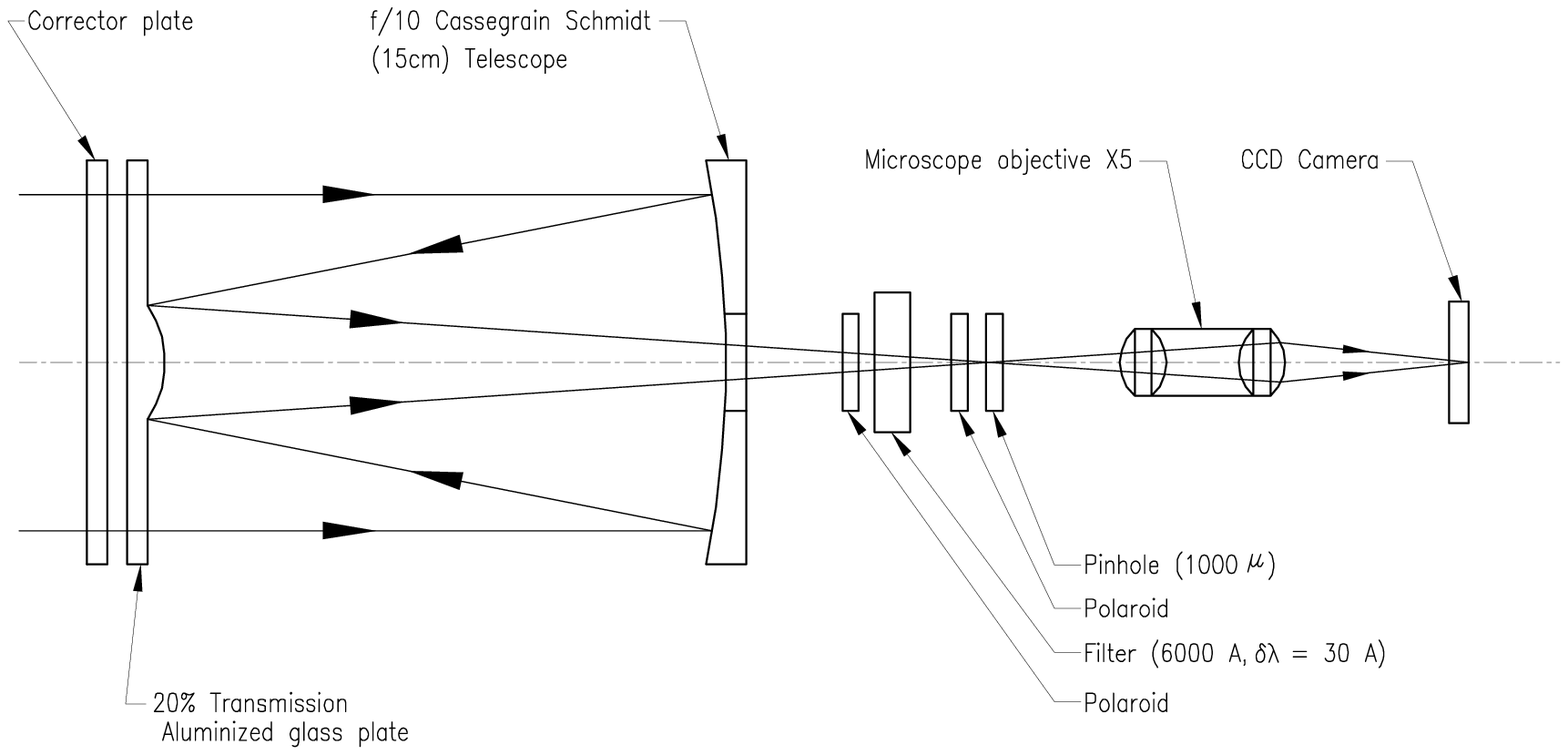,height=7cm,width=12cm}}
\bigskip
\noindent
{\bf Figure 13.} Schematic layout of the set-up 
for the speckle reconstruction of solar features 
during the partial eclipse of Sun as viewed from Bangalore on October 24$^{th}$, 
1995 (Saha et al., 1997b).
}     
\endinsert
\bigskip 
The image reconstruction involves the treatment of both amplitude errors,
as well as phase errors. The 20~ms exposure time is small enough to preserve
phase errors. Any of the schemes for phase reconstruction that satisfactorily
reproduces the lunar limb would be valid for solar features close to the
limb (within iso-planatic patch). Also, the limb reconstruction would be valid
only for phase distortions along one direction (in a direction normal to the
lunar limb). In spite of these shortcomings, the limb data would have provided
additional constraints for techniques, like BID. 
\bigskip
Another novel experiment was conducted by Koutchmy et al., 1994 using the 
modern 3.6~m CFHT telescope at Mauna-Kea during total solar eclipse of 
11 July, 1991 to probe the solar corona. In this venture, several cameras 
combining fast photographic 70~mm cameras and the video-CCD cameras were 
employed during the totality to acquire sub-arc-second spatial resolution
white light images. Two video-CCD cameras, viz., (i) one
fitted with a broad-band filter with $\times$2.5 magnification ratio 
to record video frames during the totality so as to enable them to perform
speckle interferometry image processing over the faintest coronal structures,
and (ii) a second one with a 
581$\times$756 pixels CCD (size of the pixel = 11~$\mu$m) fitted with a narrow-band
filter ($\lambda$ = 637~nm, FWHM = 7~nm) to detect coronal radiation, were used.
Fine-scale irregularities along coronal loops of very large aspect ratio were 
observed in a time series, confirming the 
presence of plasmoid-like activity in the inner corona.  
\bigskip
Intensive computations are generally required in post-detection image 
restoration techniques in solar astronomy. The introduction of adaptive optics 
systems (see section 8) may, therefore, will be useful for spectroscopic 
observations, as well as for photon-starved imaging with future very large solar 
telescope. A few higher order solar adaptive optics systems are in use or under
development (Beckers, 1999 and references therein). 
Images of sunspots on the solar surface were obtained with Lockheed adaptive
optics system (Acton and Smithson, 1992) at the Sacramento Peak Vacuum telescope.
\vskip 20 pt
\centerline {\bf 8. Adaptive optics }
\bigskip 
\noindent
Adaptive optics (AO) removes the turbulence induced wave-front distortions 
by introducing an optical imaging system in the 
light path that performs two main functions: (i) sense the wave-front 
perturbations, and (ii) compensate for them in real time (Beckers, 1993, Roggemann 
et al., 1997). In other words, this system can introduce controllable counter 
wave-front distortion which both spatially and temporally follows that of the 
atmosphere. Although this technique was first proposed by Babcock (1953), it 
could not be implemented until recently due to the lack of technological 
advancement. 
\bigskip 
The components required to perform these functions are wave-front sensing, 
wave-front phase error computation and a flexible mirror whose 
surface can be electronically controlled in real time to create a conjugate 
surface enabling to compensate the wave-front distortion. Figure 14 depicts a
schematic of the adaptive optics system. The principle of
the AO system runs as follows. The beam from the telescope is collimated and
fed to the tilt mirror to remove low frequency tilt errors. It travels further
and is reflected from the deformable mirror after eliminating high frequency
wave-front errors. This beam is split into two by 
introducing a beam-splitter; one is directed to the wave-front sensor and the other
is focused to form an image. The former measures the residual error in the 
wave-front and provides information to the actuator control computer to compute 
the deformable mirror actuator voltages. This process should be at speeds 
commensurate with the rate of change of the corrugated wave-front phase errors.
\bigskip
\noindent
\midinsert
{\eightpoint   
\noindent
\centerline{\psfig{figure=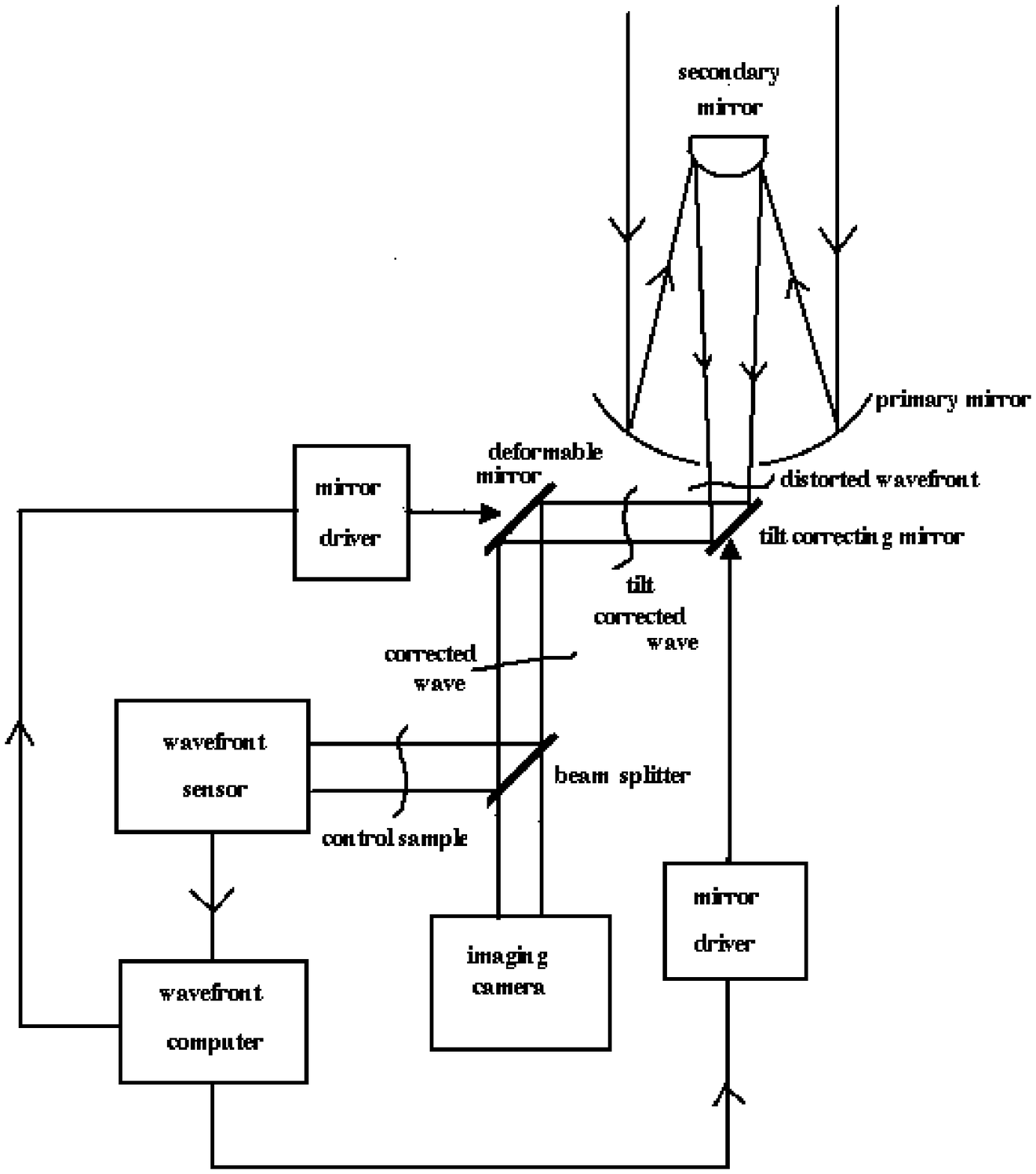,height=13cm,width=13cm}}
\bigskip
\noindent
{\bf Figure 14.} Schematic of the adaptive optics system.
}     
\endinsert
The utility of the AO system is greatly enhanced by use of the 
laser guide star source (Foy and Labeyrie, 1985) as a reference 
to measure the wave-front errors by means of a wave-front sensor, as well as
to map the phase on the entrance pupil. A pulsed laser is used to cause a bright 
compact glow in the upper atmosphere, which can serve as the source of measuring 
the turbulence of the atmosphere (Foy, 1996). Concerning the flux backscattered 
by a laser shot, Thompson and 
Gardner (1988) stressed the importance of investigating two basic problems: (i)
the cone effect which arises due to the parallax between the remote astronomical
source and artificial source (located 90~km high in the case Na~1D laser), 
and (ii) the angular anisoplanatic effects. This can be restored by
imaging the various turbulent layers of the atmosphere onto different adaptive
mirrors (Foy and Labeyrie, 1985, Beckers, 1988, Tallon et al., 1988). 
\bigskip 
In the low order AO systems, the fast tip-tilt correctors or image stabilization 
devices are able to correct turbulence-induced image motions; the residual 
telescope tracking error also needs to be corrected. These systems are 
limited to two Zernicke modes (x and y tilt), while higher order system 
compensating many Zernicke mode is required to remove high frequency errors. 
Glindemann, (1997) discussed the analytic formulae for the aberrations of the
tip-tilt corrected wave-front as a function of the tracking algorithm and of the
tracking frequency. A tip-tilt tertiary mirror system has been developed for the
Calar Alto 3.5~m telescope, Spain, that corrects the rapid image motion 
(Glindemann et al., 1997). The following paragraphs are dedicated to the various 
types of wave-front sensors.
\bigskip 
(i) The lateral shearing interferometer is used in AO system as wave-front 
sensor (Hardy et al., 1977, Sandler et al., 1994). The experimental details of
the system can be found in section 4.5. The 
shearing interferometer based on BC (Saxena and Jayarajan, 1981, Saxena and 
Lancelot, 1981) was reported to be very compact and rugged to use. It is planned 
to use this as the sensor for the AO programmes at IIA, Bangalore (Chinnappan et 
al., 1998). 
\bigskip
(ii) The Shack-Hartman sensor measures directly the angles of arrival, and
therefore, works well with incoherent white light extended sources (Rousset,
1999 and references therein); it divides the pupil into sub-pupils
(Wyant and Koliopoulos, 1981). This type of sensors have already been used in
AO systems (Fugate et al., 1991, Primmerman et al., 1991).
Tallon et al., (1988) have developed same type of set-up
to record wave-front surface, where the beam at the focal plane of the telescope 
is transmitted through a nearly field lens to a 
collimating doublet objective and imaged the exit pupil of the former on to a
lens-let mosaic which form an array of images on the detector. Each lens-let 
defines a sub-aperture in the telescope pupil. The position of the centre
of gravity of this spot is directly related to the phase mean slope over the
sub-aperture. Their positions are measured to give the full vectorial 
wave-front tilt in the areas of the pupil covered by each lens-let. 
\bigskip
The dimensions of the lens-lets are often taken to correspond approximately to 
$r_\circ$, though Tallon and Foy (1990) suggested that depending on the number
of turbulent layers, the size of the sub-pupils in this type of sensor can be
made significantly larger than the latter. The value of $r_\circ$ varies over 
the duration of observation, therefore, a minimal number of lens-let 
array for a given aperture size is required. The test consists of recording 
the ray impacts in a plane slightly before the focal plane. If optics were 
perfect, the recorded spots would be exactly distributed as the position of 
lens-lets but on a smaller scale. Due to aberrations, light rays are deviated 
from their ideal direction, producing spot displacements; the amount of 
displacement is a measure of the deviation of the ray which, in turn, the 
deviation of the local wave-front slope. 
\bigskip
(iii) The curvature sensor measures a signal proportional to the second
derivative of the wave-front phase (Roddier, 1988c, 1990) and works well with
incoherent white light (Rousset, 1999). This technique is a 
differential Hartman technique in which the spot displacement can 
be inverted. It is easy to see that if the ray impacts are recorded on the 
other side of the focal plane, the displacement occurs in the opposite 
direction. Hence, by comparing spot displacement on each side of the focal 
plane, one can double the test sensitivity. The difference between the two 
spot displacements is a measure of wave-front slope independent of the mask 
irregularities. Both the local wave-front slope and local wave-front curvature 
can be mapped with the same optical setup, doubling the number of 
reconstructed points on the wave front. The first astronomical images obtained
from a low-order adaptive optical imaging system using a curvature sensor was
reported by Roddier (1994). The CFHT adaptive optics bonnette (AOB), 
PUEO (Arsenault et al., 1994), is based on the variable curvature mirror 
(Roddier et al., 1991) and has a 19-zone bimorph mirror (Rigaut et al., 1998). 
\bigskip
The real-time computation of the wave-front error, as well as correction of 
wave-front distortion involves digital manipulation of wave-front sensor data
in the wave-front sensor processor, the re-constructor and the low-pass filter,
and converting to analog drive signals for the deformable-mirror actuators. The
functions are to (i) compute sub-aperture gradients, (ii) compute phases at the 
corners of each sub-aperture, (iii) compute low-pass filter phases, and (iv)
provide actuator offsets to compensate the fixed optical system errors and real-time
actuator commands for wave-front corrections.
\bigskip
The phase reconstruction method finds the relationship between the measured 
values and unknown wave-fronts and can be categorized as being either zonal or 
modal, depending on whether the estimate is either a phase value in a 
local zone or a coefficient of an aperture function. In the case of curvature 
sensing, the recorded two out-of-focus images are subtracted and the sensor 
signal is computed. Subsequently, the Poisson equation is solved numerically 
and the first estimate of the aberrations is obtained by least squares fitting 
Zernicke polynomials to the reconstructed wave-front. 
A conjugate shape is created using this data by controlling a deformable mirror,
which typically compose of many actuators in a square or hexagonal array. 
\bigskip
Deformable mirrors using discrete actuators are used in astronomical AO systems
at various observatories (Shelton and Baliunas, 1993, Wizinovitch et al., 1994).
In the operational deformable mirrors, the actuators use the ferroelectric
effect, in the piezoelectric or electrostrictive form (S\'echaud, 1999 and 
references therein). The primary parameters of deformable mirror 
based AO system are the number of actuators (usually more than 50), the control 
bandwidth and the maximum actuator stroke. The mirrors can be either segmented 
mirror or continuous faceplate mirrors. Actuators 
are normally push-pull type; though for curvature sensing 
actuators are used, these are difficult to fabricate and cost effective.
Since the number of actuators are large, there is a need for controlling all 
the actuators almost simultaneously; the frequency of control is about 1 KHz.
\bigskip 
The present generation piezoelectric actuators are no longer discrete, but 
ferroelectric wafers are bounded together and treated to isolate the different
actuators (Lillard and Schell, 1994).  
Flexible, mirrors micro-machined in silicon is being manufactured for use in 
AO system. An integrated electrostatically controlled adaptive mirror has the 
advantage of integrated circuit compatibility with high optical quality; then 
exhibit no hysteresis and hence are easy to control (Mali et al., 1997).
\bigskip 
A new concept of using adaptive secondary mirror to eliminate the optical 
components required to conjugate a deformable mirror at a reimaged pupil, as 
well as to minimize thermal emission has been proposed recently (Bruns et 
al., 1996). Among the new technologies developed for display components,
liquid crystal devices seem to be promising (Love et al., 1994) as an 
alternative to deformable mirrors.
\bigskip 
The performance of the wave-front sensing depends on the characteristics of the
detector, viz., (i) the spectral bandwidth, (ii) the quantum efficiency, (iii) 
the detector noise that includes dark current, read-out and amplifier noise,
(iv) the time lag due to the read-out of the detector, (v) the array size and 
the spatial resolution (Rousset, 1999). Among the various types of
detectors available till date, the cameras based on (i) photo-multiplier tubes
(Hardy et al., 1977), (ii) MAMA (section 5.2.4), (iii) ICCD (section 
5.1), as well as new generation electron bombarded CCD (Cuby et al., 1990),
(iv) avalanche photo-diodes (Zappa et al., 1996), (v) Back-illuminated bare CCD
(Twichell et al., 1990, Beletic, 1996), (vi) infra-red photo-diode arrays 
(Rigaut et al., 1992) may be implemented in the wave-front sensing (Rousset,
1999). For reducing the data obtained with AO systems may be processed with the
methods developed for high resolution imaging, speckle interferometry and
other image processing algorithms (L\'ena and Lai, 1999a).
\vskip 20 pt
\centerline {\bf 9. Diluted aperture interferometry}
\bigskip 
\noindent
The marked advantage 
of using independent telescopes is the increase in resolving capabilities, the
greater the distance between the two sub-apertures, the higher the angular
resolution. Modern technology has by and large solved the problems encountered 
by Michelson and Pease (see sub-section 2.2.). Several ground-based long 
baseline two- or multi-aperture interferometers are in operation, producing 
excessive scientific results both at optical, as well as at IR wavelengths. 
Building of different kinds of interferometers are in progress at several places. 
The following sub-sections dwell upon the salient features of the intensity 
interferometry, as well as descriptions of the direct interferometers in
use at various places.
\vskip 20 pt
\noindent 
{\bf 9.1. Intensity interferometry}
\bigskip 
\noindent
An amplitude interferometer, say Michelson's stellar interferometer, measures
the covariance $<\Psi_1\Psi_2>$ of the complex amplitudes
$\Psi_1, \Psi_2$, at two different points of the wave-fronts, while  
the intensity interferometer computes the fluctuations of the 
intensities ${\cal I}_1, {\cal I}_2$ at these points. The fluctuations
of the electrical signals from the two detectors are compared by a multiplier. 
The current output of each photo-electric detector
is proportional to the instantaneous intensity ${\cal I}$ of the incident light,
which is the squared modulus of the amplitude $\Psi$. The fluctuation of the
current output is proportional to $\Delta {\cal I} = \mid\Psi\mid^2 - <\mid\Psi
\mid^2>$. The covariance of the fluctuations, according to Goodman (1985), can 
be expressed as, 

$$<\Delta{\cal I}_1\Delta{\cal I}_2> = <\mid\Psi_1\Psi_2^\ast\mid^2>. \eqno(118)$$

This expression depicts that the covariance of the intensity fluctuations is
the squared modulus of the covariance of the complex amplitude.
\bigskip
Soon after the successful completion of the intensity
interferometer in radio wavelengths (Brown et al., 1952), in which the signals 
at the antennae are detected separately and the angular diameter of the source
is obtained by measuring the correlation of the intensity fluctuations of the 
signals as a function of antenna separation, Brown and Twiss (1956)  
demonstrated its potential at optical wavelengths by measuring the angular diameter of
$\alpha$~CMa (Brown and Twiss, 1958). The advantages of such a system over 
Michelson's interferometer are that of not requiring high mechanical stability
and of being unaffected by seeing. In this arrangement, light from a star
collected by two concave mirrors is focused on to two photo-electric cells
and the correlation of fluctuations in the photo-currents is measured as
a function of mirror separation. 
\bigskip
The same technique was applied later by Brown et al., (1967) in the 
construction of the intensity interferometer at Narrabri, Australia, 
which has a pair of 6.5~m light collectors on a circular railway
track spanning 188~m. The interferometer produced accurate measurement of the
diameters of 32 southern bright stars with an angular resolution limit of 
0.5~mas (Brown, 1974). 
The correlator and the control system are housed in a room at the centre. 
\bigskip
The degree of correlation depends on the detector spacing in comparison with
the size of speckle. In other words, the correlated fluctuations are obtained
if the two detectors are spaced by less than a speckle width. Therefore, it is
necessary to increase the speckle life time by reducing the spectral band
width (Labeyrie, 1985). In this experiment, narrow bandwidth filters were used. 
The sensitivity of this method is limited by the narrow spectral band 
used. Theoretical calculations (Roddier, 1988b) show that the limiting visual
magnitude of a star that can be observed with the intensity interferometry 
having a 6.5~m aperture is of the order of 2.
\vskip 20 pt
\noindent
{\bf 9.2. Amplitude and phase interferometry}
\bigskip 
\noindent
In contrast with the intensity interferometer, amplitude or phase 
interferometry provides a better sensitivity at short wavelengths. The light 
collected by an array of separated telescopes could be coherently combined. Each 
pair of telescopes in the array yields a measure of the amplitude of the spatial 
coherence function of the object at a spatial frequency ${\bf B}/\lambda$, where 
{\bf B} is the baseline vector and $\lambda$ the wavelength. According to the 
van Cittert-Zernicke theorem (Born and Wolf, 1984), if the number of samples of 
the coherent function can be made large, the spatial frequency spectrum of the 
object can be reconstructed. Figure 15 depicts the concept of the amplitude 
interferometer using two independent telescopes.
\bigskip
\noindent
\midinsert
{\eightpoint   
\noindent
\centerline{\psfig{figure=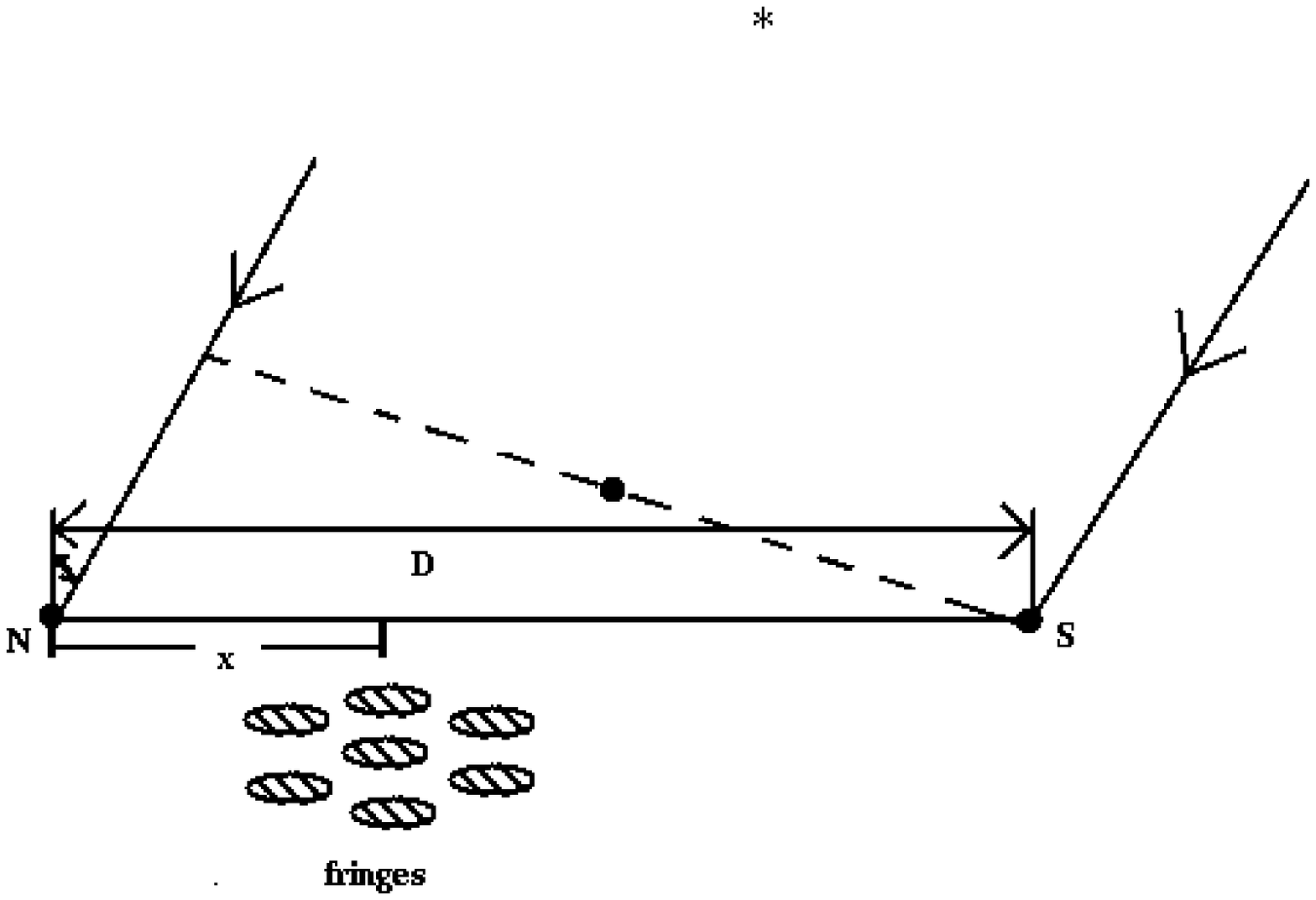,height=9cm,width=11cm,angle=270}}
{\bf Figure 15.} Concept of two telescopes interferometer
}     
\endinsert
\vskip 20 pt
\noindent
{\bf 9.2.1. Interferometers at Plateau de Calern} 
\bigskip 
\noindent
Labeyrie, (1975) had succeeded in obtaining interference fringes produced by a
pair of independent telescopes at Observatoire de Nice, France, in 1974. 
Subsequently, the interferometer $-$ Interf\'erom\`etre \`a Deux T\'elescope 
(I2T) $-$ was shifted to a relatively better site, Plateau de Calern, Observatoire 
de la Cote d'Azur (formerly CERGA). It combines the features
of the Michelson design and the radio interferometers and consists of
a pair of 26~cm telescopes on altitude-altitude mounts (Labeyrie, 1975)   
having a long coud\'e focus. These telescopes track the same source
(star) and send the collected light to the central laboratory where the star 
images are superposed at the foci in order to produce Young's fringes similar 
to those observed by Michelson and Pease (1921). Both the telescopes are run on 
tracks for variable North-South baseline (5~-~67~m). Each
telescope is a Cassegrain afocal system providing an angular magnification of 23
and the coud\'e beam is obtained by means of a single reflection on a rotating
flat mirror. 
\bigskip 
The beam-recombining optical device lies on a computer 
controlled motor driven carriage parallel to the baseline in order 
to correct the zero optical path length drift induced
by the diurnal rotation of the tracked star. The recombining optics are: (i) a
plane parallel plate for compensation of the chromatic phase effect induced
by the un-equal air travel, (ii) a collimator, (iii) removable optics for
control of the pupils' separation, (iv) an anamorphosor to lengthen and narrowing
the slit image so as to make full use of the camera target area, and (v) a dispersion
grating allowing to observe fringes simultaneously in several adjoining
spectral channels. A pair of photon counting cameras serve for guiding and
fringe monitoring (Blazit et al., 1977b). Apart from observing in visible
band, this interferometer has also been used for the near IR observation 
(DiBenedetto and Conti 1983). 
\bigskip
Following the success of the operation of I2T, Labeyrie (1978) undertook a 
project of building a large interferometer $-$ Grand Interf\'erom\`etre \`a Deux 
T\'elescope (GI2T) $-$ at the same site. The basic principle of this 
interferometer is similar to that of I2T; it comprises a pair of 1.52~m 
spherical telescopes (independent Cassegrain-Coud\'e telescopes) movable on 
North-South tracks (variable between 12 and 65~m), and an optical
recombining system feeding with 2 afocal beams housed in a central laboratory
(Labeyrie et al., 1986). Figure 16 shows the GI2T at Plateu de Calern.
\bigskip
\noindent
\midinsert
{\eightpoint   
\noindent
\centerline{\psfig{figure=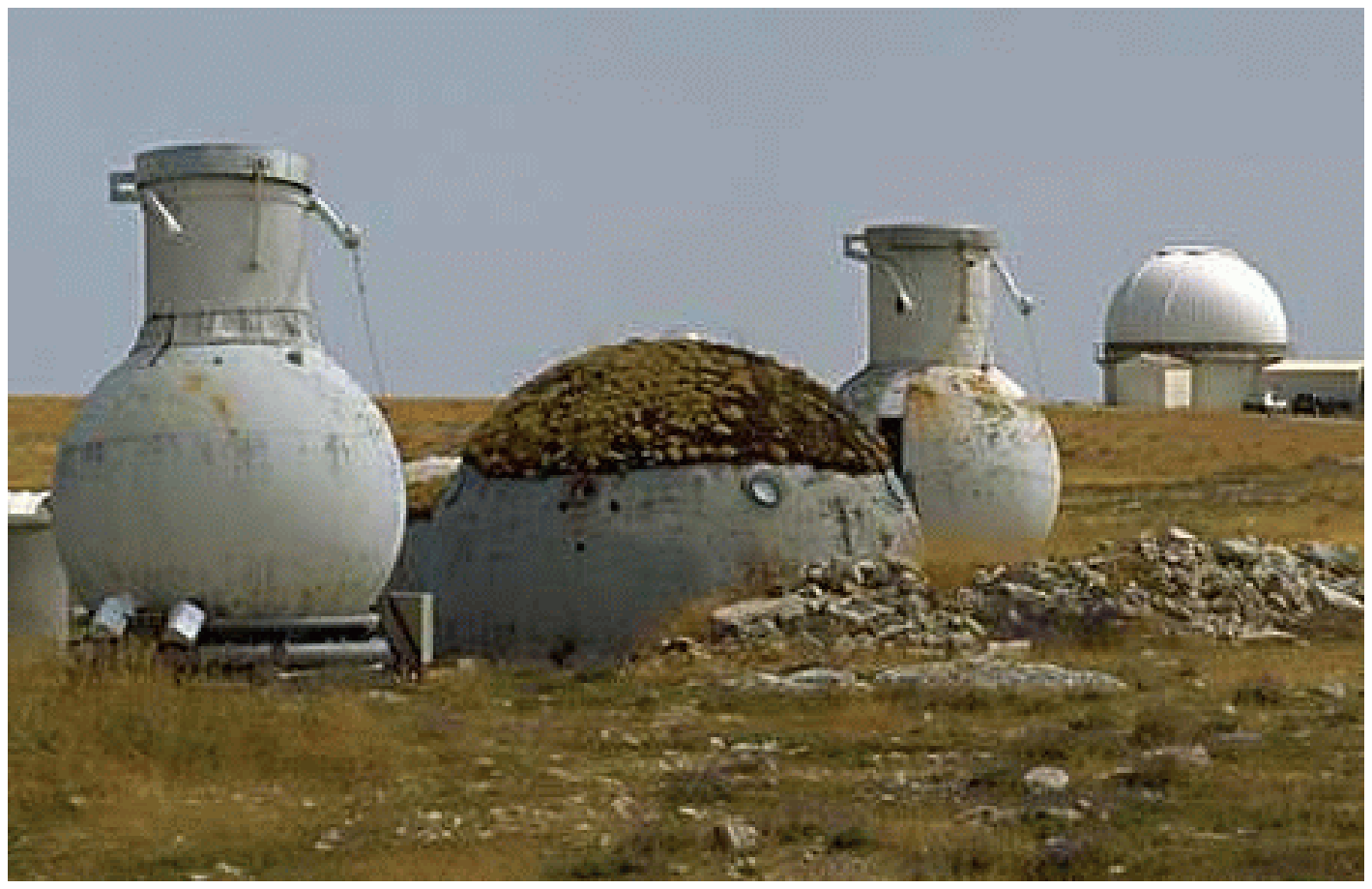,height=8cm,width=12.0cm}}
\bigskip
\noindent
{\bf Figure 16.} Grand interf\'erom\`etre \`a deux t\'elescope (GI2T) at Plateu
de Calern, France (Courtesy: P. Stee).
}     
\endinsert
As described in the preceding paragraph, here too,
the recombiner can be translated between these telescopes in order to compensate
for the optical delay. The domeless, 3.5~m diameter spherical telescope made of 
concrete with a surface accuracy of 1~mm, has 
three mirrors directing the horizontal afocal Coud\'e beam to the recombiner
optics. The driving system of this sphere consists of a pair of rings; each ring
is motorized by 3 actuators acting in 3 orthogonal directions within 2
different tangential planes (Mourard et al., 1994a). The two rings alternately
carry the sphere, which in turn, produce continuous motion. Each telescope
is equipped with 2 sets of 3 small DC-motors and 3 linear encoders. 
The resolution obtained here is
of the order of 1~$\mu$m - equivalent to 0.1$^{\prime\prime}$ on the sky. The clutching of 
the rings is performed by computer-controlled electro-valves and hydraulic 
pistons, which carry each of the 60~mm balls that raise the rings by 2~mm to carry 
the sphere. The main drawback comes from the slow pointing of the telescopes; 
4 or 5 stars can be tracked during the night (Mourard et al., 1994a). 
\bigskip
\noindent
\midinsert
{\eightpoint   
\noindent
\centerline{\psfig{figure=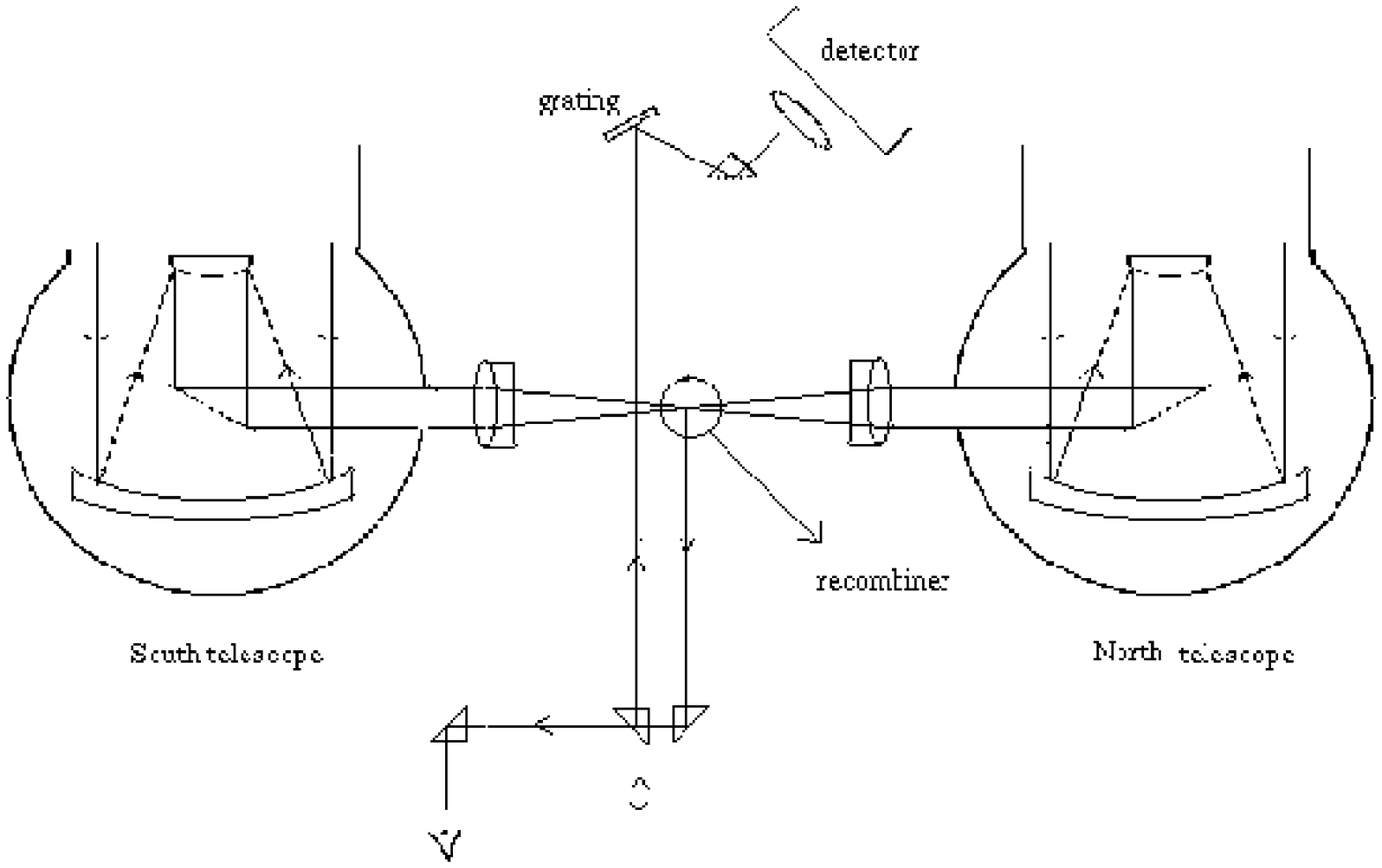,height=8.0cm,width=13cm}}
\bigskip
\noindent
{\bf Figure 17.} Concept of acquiring fringes of an object using GI2T. 
}     
\endinsert
Figure 17 depicts the concept of acquiring fringes of an object using GI2T. 
The beams from these telescopes are recombined in an image plane after 
reconfiguring the pupils; the fringed speckles are dispersed and the spectra
are recorded at short-exposure using a photon counting detector (Labeyrie et
al 1986). [Fringed speckle can be visualized when a speckle from one telescope 
is merged with the speckle from other telescope]. The optical table carrying
the focal point instruments moves along the North-South direction to maintain
constant zero optical path different (OPD) within the coherence length between 
the two beams (OPD changes due to the diurnal motion). The instruments consist
of (i) a recombining element for reconfiguring the pupil and fixing the fringe
spacing, (ii) an imager slicer (a series of 10 wedges of different angles stuck
on a field lens slices the image), (iii) the compensating system of the 
atmospheric dispersion, (iv) gratings (maximum spectral resolution 0.15~nm), 
(v) detectors to record the fringes, etc. A new addition is an implementation of
a fringe tracker based on photon counting detector and real-time image 
processing unit on the said interferometer (Koechlin et al., 1996).
\bigskip
The slicing of the image at the entrance of the spectrograph is required to obtain
better visibility since the large aperture (1.5~m telescope) produces more than 
100 speckles in the image and the fringe pattern within each speckle is randomly 
phased (Bosc, 1988). The advantages of the dispersion mode are the capabilities
to (i) allow continuous observation of fringes across the spectral bandwidth,
(ii) record the fringes with longer integration time, and (iii) select different 
spectral channels for differential visibility measurements (continuum and 
spectral line). The continuum channel is supposed to originate from the 
unresolved region of a star, say photosphere and the other one centred on
a part of the spectra created in an extended region, say circumstellar medium.
The modulus of the fringe visibility is estimated as the ratio of high frequency
to low frequency energy in the average spectral density of the short exposure.
The calibration of the resulting visibility is performed with a reference star,
for example, ${\cal V}_{cal}^2 = {\cal V}^2/{\cal V}^2_{ref}$ (Mourard et al., 
1994b, Thureau et al., 1998). Two different types of detectors, (i) CP40 
(Blazit, 1986), and (ii) RENICON (Clampin et al., 1988) are being used for data 
acquisition.
\vskip 20 pt
\noindent
{\bf 9.2.2. Mount Wilson stellar interferometer}
\bigskip
\noindent
Since 1978, an interferometer with two independent small apertures was in
operation (Shao and Staelin, 1980) at Mt. Wilson, USA. Improvements of this
interferometer have taken place in two stages, viz., mark II (Shao et al., 
1986a, 1986b, Shao and Colavita, 1987), and mark III (Shao et al., 1988). The 
fringes of the objects are obtained with automated operation. The basic 
programme of this interferometer, among others, is wide-angle astrometry (Shao
et al., 1990).
\bigskip
Unlike the interferometers described above (see section 9.2.1), this 
interferometer uses (i) siderostats and star trackers to point the instrument
at a star, (ii) an optical delay line to control the internal optical
path lengths, and (iii) a laser metrology system to monitor the position of the
siderostat mirrors. The 25~mm optical flat (siderostat mirror), is attached
to a polished hemisphere. A set of 4 laser interferometers measure the position
of the sphere relative to a set of 4 corner cubes embedded in a reference plate
(Shao, 1988). The delay line has a peak to peak range of 20~m with a small
signal bandwidth. For the astrometric observations, the measurement of the 
central fringe is performed in two colours to correct for the error caused by 
atmospheric turbulence, while for stellar diameter programmes, it is done through
narrow band filters. 
\vskip 20 pt
\noindent
{\bf 9.2.3. Sydney University stellar interferometer (SUSI)}
\bigskip
\noindent
After the successful venture of Narrabri intensity interferometer (Brown, 1974),
the Sydney University had built a prototype diluted aperture amplitude 
interferometer with a 11~m baseline using two coelostats as light collector 
(Davis and Tango, 1985a, 1985b) and determined the angular diameter of 
$\alpha$~CMa (Davis and Tango, 1986). Subsequently, they have built an optical 
interferometer with a baseline ranging from 5~m to 640~m on North-South direction 
(Davis, 1994). The required baselines are achieved with an array of 11 input stations, 
each equipped with a siderostat and relay optics, located to give a minimal 
baseline redundancy (Davis et al., 1998, 1999a); the intermediate baseline forms 
a geometric progression increasing in steps of $\sim$~40\%. The 
size of each aperture is 14~cm (single $r_\circ$). The wave-front tilt correction
and dynamic path length equalization are the two main features of this system.
\bigskip
Starlight is steered by two siderostats of 20~cm diameter using an Altitude/Azimuth 
mount placed upon large concrete piers, via relay mirrors into the evacuated 
pipe system that carries the light to the central laboratory (Davis et al., 
1992). At the central laboratory, the light enters a beam reducing telescope 
followed by the atmospheric refraction correctors systems consisting of the pairs of 
counter-rotating Risley prisms. Then the beam either enters the optical path 
length compensator (OPLC) or is diverted towards the acquisition camera. On 
leaving OPLC, the beams from the two arms of this interferometer is switched to 
one of the optical tables (blue or red) for recombination. 
\vskip 20 pt
\noindent
{\bf 9.2.4. Cambridge optical aperture synthesis telescope (COAST)}
\bigskip
\noindent
As discussed earlier in section 6.4, the corrupted phase of the spatial 
coherence function due to the instrumental and atmospheric phase errors can be 
corrected by using data processing algorithms. But the direct measurements of 
closure phase (see section 4.6) of any object using three or more independent 
telescopes in visible band have been made feasible at Cambridge by Baldwin et 
al., (1998). 
\bigskip
Four telescopes, each comprises a 50~cm siderostat flat feeding a fixed 
horizontal 40~cm Cassegrain telescope (f/5.5) with a magnification of 16 times, 
are arranged in a {\tt Y}-layout with one telescope on each arm, movable to a 
number of fixed stations and one telescope at the centre 
of the {\tt Y}. Light from each siderostat passes through pipes
containing air at ambient pressure into the beam combining laboratory. The 
laboratory accommodates the optical path compensation delay lines, the beam
combining optics, the detectors, the fringe acquisition systems etc. (Baldwin, 
1992, Baldwin et al., 1994). The
laboratory is a tunnel (32~m length $\times$ 6~m width $\times$ 2.4~m height),
covered sufficiently with earth having thick insulating end walls that
provides a stable thermal environment internally for the path compensator
delay lines, combining optics and detectors. 
\bigskip 
Each of the four telescope beams is provided with a variable delay 
line (maximum delay is 37~m); the equalization of the path delays for each
of them is carried out by a movable trolley, carrying a roof mirror running
on a rail track (Baldwin et al., 1994). 
The four beams emerging from the path compensator are each split at a dichroic;
the longer wavelength ($\lambda >$650~nm) of the visible band passes into the 
beam combining optics and the shorter ones are used for acquisition and
autoguiding. A cooled CCD detector system is used for both acquisition and 
guiding. The 4-way beam combiner accepts four input beams, one from each
telescope provides four output beams, each having equal contributions from the
light from the four telescopes. Each output beam passes through an iris
diaphragm and is focused by a long focus lens on to a fibre fed single-element 
avalanche photo-diode detector (Nightingale, 1991) for fringe detection (Baldwin 
et al., 1994). A similar beam combiner
is added recently for the near IR capability (Young et al., 1998).
\vskip 20 pt
\noindent
{\bf 9.2.5. Infrared-optical telescope array (IOTA)}
\bigskip
\noindent
The IR-optical telescope array at Mt. Hopkins, Arizona, USA, 
consists of two 45~cm collector assembly; each assembly comprises 
a siderostat, an afocal Cassegrain telescope and an active relay mirror. 
The collectors can be located at various stations on the L-shaped baseline,
to achieve baselines between 5 to 38~m. The tip-tilt corrected afocal star
light beams (4.5~cm diameter) from the telescopes are brought to the beam 
combining table. Optical differences are compensated by fixed and 
variable delays and the beams are recombined onto a beam splitter, producing 
two complementary interference signals (Carleton et al., 1994). These are 
focused onto a pair of InSb photovoltaic detectors through a
K-band filter. Fast autoguiding system is used to correct the atmospheric
wave-front tilt errors and a precise optical delay line is employed to 
compensate the effect of earth's rotation. Two active delay lines for three
telescopes are provided. A scanning piezo mirror is used to modulate the 
optical path difference between the two telescopes. A thinned back-illuminated
SITs 512$\times$512 CCD with a quantum efficiency 90\%, is used as sensor for 
IOTA's visible light detector. 
\bigskip
In addition to the afore-mentioned interferometers, several long baseline
interferometers and imaging interferometric array at optical/IR bands, 
viz., (i) European Southern Observatory's (ESO) very large telescope 
interferometer (VLTI) at Cerro Paranal, 
Chile (Mariotti et al., 1998), (ii) Palomar testbed interferometer (Malbet et 
al., 1998), (iii) Keck interferometer with a pair of 10~m telescopes on a 85~m
baseline at Mauna Kea, Hawaii (Colavita et al., 1998, van Belle et al., 1998), 
(iv) US Navy prototype optical interferometer (Armstrong, 1994, Hutter, 
1994), (v) CHARA array (McAlister et al., 1998) etc. are at various stages of 
development. The VLTI comprises of four 8.2~m telescopes and three movable
1.8~m auxiliary telescopes, situated at Cerro Paranal peak, Chile. Beams are
supposed to be received from these movable telescopes in the central laboratory 
for recombination and are to be made to interfere after introducing suitable 
optical delay lines (L\'ena and Lai, 1999a); it is planned to use AO systems at
the telescopes for efficient interferences.
\vskip 20 pt
\centerline {\bf 10. Astrophysical results}
\bigskip
\noindent
The single aperture interferometry has made impacts 
in several important fields in astrophysics, viz., 
(a) in studying the separation and orientation of close binary stars,
(b) in measuring diameter of giant stars, (c) in resolving the heavenly dance 
of Pluto-Charon system (Bonneau and Foy, 1980, Baier and Weigelt, 1987, Beletic 
et al., 1989), (d) in determining shapes of asteroids (Drummond et al., 1988), 
(e) in mapping the finer features of extended objects (section 7),
(f) in estimating sizes and mapping certain types of circumstellar 
envelopes, (g) in revealing structures of active galactic nuclei, and of 
compact clusters of a few stars like R136a complex, (h) in resolving the 
gravitationally lensed QSO's, etc. 
\bigskip
The results obtained with diluted aperture interferometry in the visible 
wavelength are from the area of stellar angular diameters with implications for 
emergent fluxes, effective temperatures, luminosities and structure of the 
stellar atmosphere, dust and gas envelopes, binary star orbits with impact on 
cluster distances and stellar masses, relative sizes of emission-line stars and 
emission region, stellar rotation, limb darkening, astrometry. 
Some of the results mentioned above can be found in various articles (Labeyrie, 
1985, McAlister, 1988, Ridgway, 1988, 1992, Roddier, 1988b, Foy, 1992, 
and references therein). In what follows, the important 
astrophysical results barring extended objects (noted already in section 7), 
obtained in the visible band in the recent past are addressed. 
\vskip 20 pt
\noindent
{\bf 10.1. Results obtained with single aperture interferometry}
\bigskip
\noindent
I. Studies of close binary stars are of paramount importance in stellar 
astrophysics, and play a fundamental role in measuring   
stellar masses, providing a benchmark for stellar evolution calculations; a 
long term benefit of interferometric imaging is a better calibration of the 
main-sequence mass-luminosity relationship (McAlister, 1988). An important 
parameter in obtaining masses of stars involves combining the spectroscopic 
orbit with the astrometric orbit as projected on the sky from the 
interferometric data (Torres et al., 1997). The atmosphere of the component
of a close binary system is distorted mainly by physical effects, viz.,
(i) rotation of the component, (ii) the tidal effect due to the presence 
of its component. The modelling of the radiative transfer concerning the effects 
of irradiation on the line formation in the expanding atmospheres of the 
components of close binary systems (Peraiah and Rao, 1998) can be studied.
\bigskip
Speckle interferometric observations of close binary stars have been 
carried out using telescopes of various sizes by several groups.  
Major contributions in resolving close binary systems came from the Center
for High Angular Resolution Astronomy (CHARA) at Georgia State University, USA.
In a span of a little more than 20~yr, this group had observed more than
8000 objects; 75\% of all published interferometric observations are of binary 
stars (Hartkopf et al., 1996, 1997a, 1997b, McAlister et al., 1996,
Mason et al., 1998). The separation of most of the new components 
discovered by interferometric observations are found to be less than 0.25$^{\prime\prime}$ 
(McAlister et al., 1993). During the period of 1989 - 1994, Hartkopf et
al., (1996) reported observations of 694 binary star systems using said
technique with the 4~m Cerro Tololo telescope. From an inspection of the 
interferometric data obtained with the cameras at CHARA and US naval 
observatories (USNO),
Mason et al., (1999) have confirmed the binary nature of 848 objects,  
discovered by the Hipparcos satellites. 
\bigskip 
In order to derive accurate orbital elements and 
masses, luminosities and distances, emphasis laid down in their
programmes was the continuous observation of spectroscopic binaries. A survey of
chromospheric emission (H and K lines of Ca II) in more than 800 southern 
stars (solar type) reveals that about 70\% of them are inactive (Henry et al., 
1996). In a programme of bright Galactic O stars for duplicity, Mason et al.,  
(1998) could resolve 15 new components. They opined that at least one-third
of the O stars, especially those among the members of clusters and associations, 
have close companions; a number of them, may even have a third companion. Among 
a speckle survey of several Be stars, Mason et al., (1997) were able to 
resolve 5 binaries including a new discovery. From a survey for
duplicity among white dwarf stars, McAlister et al., (1996) reported faint
red companions to GD~319 and HZ~43. 
\bigskip
Survey of visual and interferometric binary stars with orbital motions  
have also been reported by other groups 
(Balega et al., 1984, 1994, Balega and Balega, 1985, Bonneau et al., 
1986, Blazit et al., 1987, Horch et al., 1996, 1999, Douglass et al., 1997). 
Leinert et al., (1997) have resolved 11 binaries by means of near IR speckle
interferometry, out of 31 Herbig Ae/Be stars, of which 5 constitute 
sub-arc-second binaries. In a recent article, Germain et al., (1999) have 
reported position angles and separations of 547 binaries using 66~cm refractor
at USNO in Washington.
\bigskip 
Reconstructing the phase of binary stars using various image processing 
algorithms have been made (Karovska et al., 1986a, Miura et al., 1992, Horch et 
al., 1997, 1999, Saha and Venkatakrishnan, 1997, Saha et al., 1999b, 1999c). 
Speckle interferometric observations in the IR band have detected the close 
companions of $\Theta^1$~Ori~A and $\Theta^1$~Ori~B (Petr et al., (1998).
Subsequently, Simon et al., (1999) have found an additional faint companion of
the latter. Recently, Weigelt et al., (1999) have detected a close
companion of $\Theta^1$~Ori~C with a separation of $\sim$~33~mas. These 
Trapezium system, $\Theta^1$~Ori~ABCD, are massive O-type and early B-type stars
and are located at the centre of the Orion nebula. The speckle masking 
reconstruction of a close binary star, 41~Dra was carried out (Balega et al.,
1997a, 1997b); the separation of the binary components was found to be about 
25~mas, the resolution at the wavelength of 656~nm was 23~mas ($\lambda$/D). 
Figure 18 demonstrates the reconstructed image of said binary star 41~Dra. 
\bigskip
\noindent
\midinsert
{\eightpoint   
\noindent
\centerline{\psfig{figure=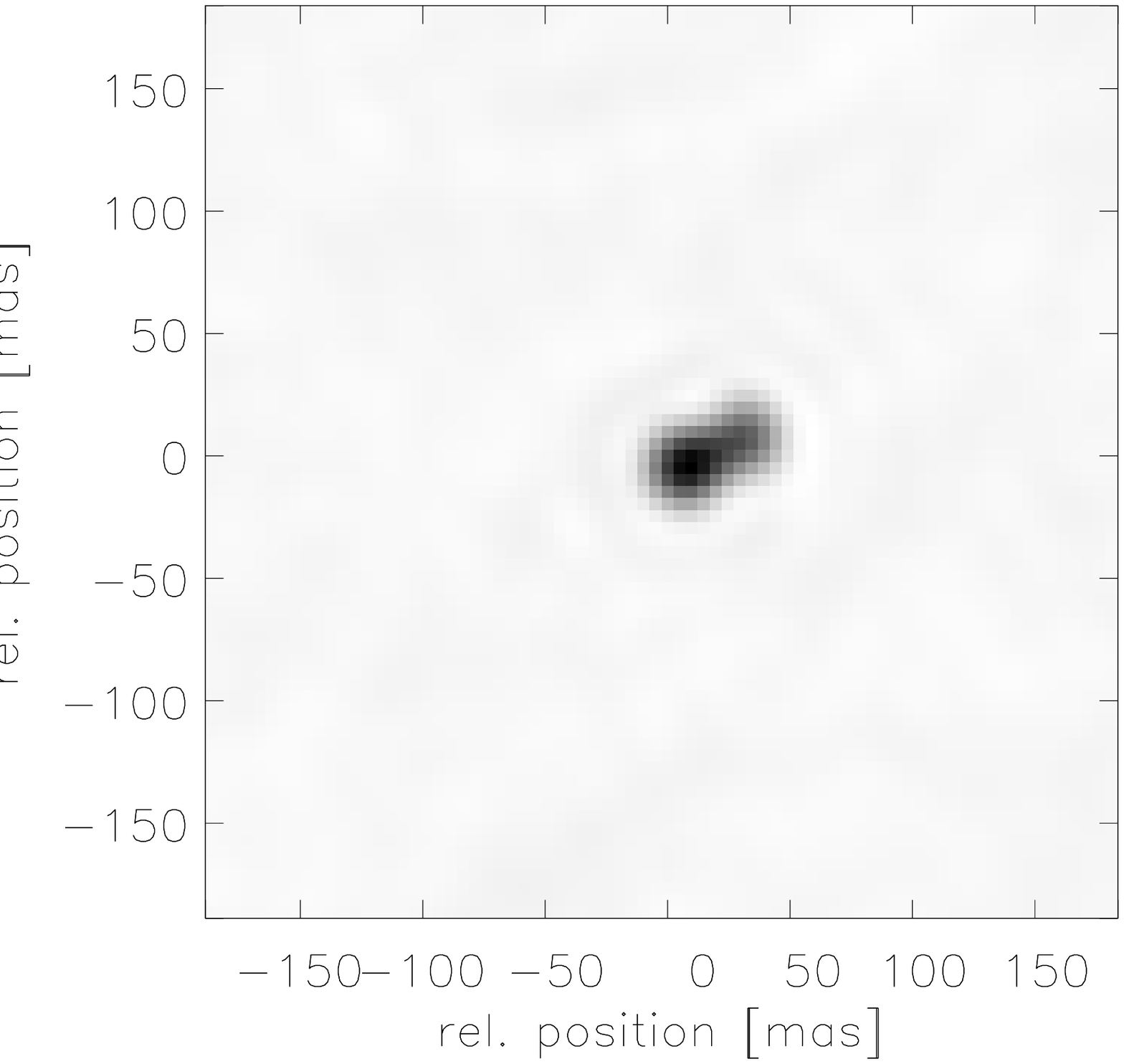,height=7cm,width=7cm}}
\bigskip
\noindent
{\bf Figure 18.} The speckle masking reconstruction 41~Dra (Balega et al., 1997a,
1997b). The separation of the binary was about 25~mas (Courtesy: R. Osterbart).
}     
\endinsert
\bigskip 
Various investigators have also calculated the orbital characteristics
of many close binary stars (Duquennoy et al., 1996, Gies et al., 1997, Torres et 
al., 1997, Mason et al., 1998, 1999, Aristidi et al., 1999). Torres et al., 
(1997) derived individual masses for $\theta^1$~Tau using the distance 
information from $\theta^2$~Tau. They found the empirical mass-luminosity 
relation from the data in good agreement with the theoretical models. Orbital
motions of DF~Tauri from the speckle imaging have also been reported
(Thi\'ebaut et al., 1995). Gies et 
al., (1997) measured the radial velocity for the massive binary 15 Monocerotis.  
Discovery of two binary stars, viz., (a) MOAI~1 (Carbillet et al., 1996), 
(b) $\nu$~Cyg (Hipparcos catalogue, 1997) have been reported recently. Nisenson 
et al., (1985) have resolved a faint optical source close to T~Tauri with a 
separation of 0.27$^{\prime\prime}$. The visual magnitude difference was found to be 4.33.
\bigskip
Kuwamura et al., (1992) obtained a series of spectra using objective speckle 
spectrograph with the bandwidth spanning from 400~nm~-~800~nm  
and applied shift-and-add algorithm for retrieving the diffraction-limited 
object prism spectra of $\zeta$~Tauri and ADS16836. They have resolved
spatially two objective prism spectra corresponding to the primary and the 
secondary stars of ADS16836 with an angular separation of $\approx$~0.5$^{\prime\prime}$ using 
speckle spectroscopy. Reconstructing emission and absorption lines of several
single stars have been carried out by Kuwamura et al., (1993a, 1993b). 
Using imaging spectroscopic (Baba et al., 1994a) method, Baba et al., (1994b) 
have observed a binary
star, $\phi$~And (separation 0.53$^{\prime\prime}$) at the Cassegrain focus of the 1.88~m 
telescope of the Okayama Astrophysical Observatory; the reconstructed spectra 
using algorithm based on cross-correlation method revealed that the primary star 
(Be star) has an H$\alpha$ emission line while the secondary star has an 
H$\alpha$ absorption line. Grieger and Weigelt, (1992) used a wide-band speckle 
spectrograph to record interferograms of a binary star (Grieger and Weigelt, 
1992) with the ESO 2.2~m telescope, La Silla, Chile. These authors have 
reconstructed the diffraction-limited image and spectrum of a binary star with 
speckle masking technique (section 6.4). 
\bigskip
The high angular polarization measurements of the pre-main sequence binary 
system Z~CMa at 2.2~$\mu$m have been reported recently by Fischer et al., (1998).
They have found that both the components are polarized; the secondary showed an
unexpected large polarization degree. Measurements of the close binary systems 
have also been carried out by means of aperture masking technique (Nakajima et al., 
1989, Busher et al., 1990, Bedding et al., 1992). Recently, Robertson et al., 
(1999) reported from the measurements with aperture masking technique at 3.9~m 
AAO telescope that $\beta$~Cen, a $\beta$~Cephei star is a binary systems with
components separated by 0.015$^{\prime\prime}$.
\bigskip
Another interesting programme, occultation binary star survey, is being carried
out by CHARA group (Mason, 1995). When any planetary body of notable size passes 
in front of a star, the light coming from the latter is occulted. The 
method is sensitive enough to detect the presence of an atmosphere on 
Ganymede $-$ a Jovian satellite, from its occultation of SAO 186800 (Carlson et 
al., 1973), and will remain useful because of the extraordinary geometric precision
it provides. More recently, Tej et al., (1999) have reported the measurements of 
the angular diameter of the Mira variables R~Leonis in near IR bands at the 
1.2~m telescope, Mt. Abu, India.
\bigskip
The notable advantage of occultation of binary stars is that of
determining relative intensities and measure the separations comparable
to those measured by long baseline interferometers. The contribution of speckle
survey of occultation binaries till date, at the smallest separation region, is of 
the order of $<$~0.025$^{\prime\prime}$. Further, this method provides a means of determining 
the limiting magnitude difference of speckle interferometry. The shortcomings of 
this technique can be noted as its singular nature; the object may not occult 
again until one Saros cycle later (18.6 yr), and limited to a belt of the sky (10\% 
of the celestial sphere). The direct speckle interferometric measurement of more 
than 2 dozens new occultation binaries have been reported (Mason, 1995, 1996). 
\bigskip
II. Several multiple stars (Tokovinin, 1997) were observed by means of 
speckle interferometric method (Cole et al., 1992, Hartkopf et al., 1992, 
Goecking et al., 1994, Aristidi et al., 1997, 1999). The star-like object, Luminous 
Blue Variable (LBV), $\eta$~Carina, was found to be a multiple object (Weigelt 
and Ebersberger, 1986, Hofmann and Weigelt, 1988). Image reconstruction with 
speckle masking method of the same object showed 4 components with separations 
0.11$^{\prime\prime}$, 0.18$^{\prime\prime}$ and 0.21$^{\prime\prime}$ 
(Weigelt, 1988, Hofmann and Weigelt, 1993). Falcke et 
al., (1996) recorded speckle polarimetric images of the same object with the ESO 
2.2~m telescope. The polarimetric reconstructed images with 0.11$^{\prime\prime}$ 
resolution in the H$\alpha$ line exhibit a compact structure 
elongated in consistent with the presence of a 
circumstellar equatorial disk. Karovska et al., (1986b) detected two close
optical companions to the supergiant $\alpha$~Orionis using photon counting
PAPA detector. The reconstructed image depicted that the separations
of the closest and the furthest companions from the said star are 0.06$^{\prime\prime}$ 
and 0.51$^{\prime\prime}$ 
respectively. The respective magnitude differences with respect to the 
primary at H$\alpha$ were found to be 3.4 and 4.6. 
\bigskip
Ground-based conventional observations of another important luminous central 
object, R136 (HD38268), of the 30 Doradus nebula in the LMC depict three 
components R136; a, b, and c, of which R136a was thought to be the most massive 
star with a solar mass of $\sim$~2500M$_\odot$ (Cassinelli et al., 1981). 
Later, it was found to be a dense cluster of stars with speckle interferometric 
observations (Weigelt and Baier, 1985, Neri and Grewing, 1988). The 
reconstructed image with speckle masking technique depicts more than 
40 components in the 4.9$^{\prime\prime}$ $\times$ 4.9$^{\prime\prime}$ 
field; the closest binaries were found to be 0.03$^{\prime\prime}$ and 
0.05$^{\prime\prime}$ (Pehlemann et al., 1992). Observations of 
R64, HD32228, the dense stellar core of the OB association LH9 in the LMC, 
revealed 25 stellar component within a 6.4$^{\prime\prime}$ $\times$ 6.4$^{\prime\prime}$ 
field of view (Scherti et 
al., 1996). Specklegrams of this object were recorded through the Johnson V
spectral band, as well as in the strong Wolf-Rayet emission lines between 
450 and 490~nm. Several sets of speckle data through different filters, viz., 
(a) RG~695~nm, (b) 658~nm, (c) 545~nm, and (d) 471~nm of the central object HD97950 in 
the giant HII region starburst cluster NGC3603 at the 2.2~m ESO telescope,
were also recorded (Hofmann et al., 1995). The speckle masking reconstructed 
images depict 28 stars within the field of view of
6.3$^{\prime\prime}$ $\times$ 6.3$^{\prime\prime}$,
down to the diffraction-limited resolution of $\sim$~0.07$^{\prime\prime}$
with m$_v$ in the range from 11.40~-~15.6. 
\bigskip
III. Speckle interferometric observations carried out at large telescopes
depict that the diameters of red supergiants $\alpha$~Orionis and Mira are 
wavelength dependent (Bonneau and Labeyrie, 1973, Labeyrie et al., 1977, 
Balega et al., 1982, Bonneau et al., 1982, Karovska et al., 1991, Weigelt et 
al., 1996). A recent study by means of bispectrum image reconstructions of the 
former star, based on the data obtained at the Nasmyth focus of William Herschel 
Telescope, Kl\"uckers et al., (1997) established the evidence of asymmetry on 
its surface. Prior to this, Karovska and Nisenson (1992) also found in the 
reconstructed image of the same object, the evidence for the presence of a 
large bright feature 
on the surface. The reconstructed images of Mira variable R~Cas (Weigelt et al., 
1996) showed that the disk of the star a non-uniform and elongated
along the position angle 52$^\circ~\pm~7^\circ$ and 57$^\circ~\pm~7^\circ$ in the
700~nm (moderate TiO band absorption), as well as in the 714~nm (strong
TiO band absorption) respectively.
\bigskip
Many supergiants have extended gaseous atmosphere which can be imaged in
their chromospheric lines. By acquiring specklegrams in the continuum and in
the chromospheric emission lines simultaneously, differential image can be 
constructed. Hebden et al., (1987) used differential speckle interferometric 
technique to study the chromospheric envelope of $\alpha$~Orionis in the 
H$\alpha$ wavelength. The rotation shear interferometer (Roddier and Roddier, 
1988) had also been applied in the visible band to map the visibility of fringes 
produced by the star, and the reconstruction of the image revealed the presence
of light scattered by a highly asymmetric dust envelope. 
\bigskip
Apart from measuring the angular diameters of cool stars (Busher et al., 1990, 
Bedding et al., 1992, 1997a, Haniff et al., 1995), surface imaging of long
period variables stars (Tuthill et al., 1999a), aperture synthesis using 
non-redundant masking technique at various large telescopes too depicted
the presence of hotspots and other asymmetries on the surface of red supergiants
and Mira variables (Busher et al., 1990, Haniff et al., 1992, Wilson et al., 
1992, Bedding et al., 1997a, Tuthill et al., 1997). 
\bigskip 
Observations carried out at 4.2~m William Herschel telescope, La Palma, 
Wilson et al., (1997) have detected a complex bright structure in the surface 
intensity distribution of Betelgeuse that changes in the relative flux
and positions of the spots over a period of eight weeks; a new circularly
symmetric structure around the star with a diameter of $\geq$~0.3$^{\prime\prime}$ 
is found.  
As far as Mira variables are concerned, Haniff et al., (1995) have reported
that the derived linear diameters of the same are not compatible with 
fundamental mode of pulsation. Bedding et al., (1997a) have found that the 
diameter of R Doradus (57$\pm$5~mas) exceeded that of Betelgeuse; an asymmetric 
brightness distribution has also been detected from non-zero closure phases
measurement. The diameter of a small amplitude Mira, W~Hya, are reported to be
44$\pm$4~mas (Bedding et al., 1997a). 
\bigskip
The reconstructed images in the optical, as well as near IR speckle 
interferometric observations of another interesting object, the red super giant 
VY~CMa, an unusual star that displays large amplitude variability in the visible 
and strong dust emission and high polarization respectively in the mid and near 
IR, Wittkowski et al., (1998a) found to have non-spherical circumstellar 
envelope. They opined that the star was an immediate progenitor of IRC+10420,
a post red supergiant during its transformation into a Wolf-Royet (WR) star.
The visibility function in the 2.11~$\mu$m band image reconstruction depicted the
contribution to the total flux from the dust shell to be $\sim$~40\%, and the 
rest from the unresolved central object (Bl\"ocker et al., 1999).
From the reconstructed images of the non-redundant masking of 21-hole aperture 
observations of VY~CMa carried out at the 10~m Keck telescope  
in the IR wave bands, Monnier et al., (1999) have found emission to be 
one-sided, inhomogeneous and asymmetric in
the near IR. They were able to derive the line-of-sight optical depths of 
circumstellar dust shell; the results allow the bolometric luminosity of VY~CMa 
to be estimated independent of dust shell geometry. Haas et al., (1997) have
detected a halo of Elias I with near IR speckle interferometry.
\bigskip
IV. High resolution interferometric imagery may depict the spatial distribution of 
circumstellar matter surrounding objects which eject mass, particularly young 
compact planetary nebulae (YPN) or newly formed stars in addition to T~Tau 
stars, late type giants or supergiants. The large, older and evolved planetary 
nebulae (PN) show a great variety of structure (Balick, 1987) that are (a) 
spherically symmetric (A39), (b) filamentary (NGC6543), (c) bipolar (NGC6302), 
and (d) peculiar (A35). The structure may form in the very early phases of the 
formation of the nebulae itself which is very compact and unresolved. Due to
the poor spatial resolution of the conventional imaging (section 3.4),
the first $\sim~$10$^3$ years of a PN are spent in a phase that remains obscured
for structural details. In order to understand the processes that determine the
structure and dynamics of the nebular matter in the PN, one needs to resolve and 
map the same when they are young and compact. By making maps at many epochs, as
well as by following the motion of specific structural features, it
would enable one to understand the dynamical processes at work. The structures
could be different in different spectral lines e.g., ionisation stratification 
in NGC6720 (Hawley and Miller, 1977), and hence maps can be made in various
atomic and ionic emission lines too. 
\bigskip  
The angular diameters of several young PNs in the Magellanic Clouds were 
determined with speckle interferometric technique (Barlow et al., 1986,
Wood et al., 1986, 1987). The high spatial resolution images 
of Red Rectangle (AFGL 915), a reflection nebula around the star HD44179, 
were recorded by several groups (Leinert and Haas, 1989,  
Cruzal\'ebes et al., 1996, Osterbart et al., 1996, 1997). These observations 
exhibit two lobes with the separation of $\sim$~0.15$^{\prime\prime}$.
Osterbart et al., (1996) 
argued that the dark lane between the lobes is due to an obscuring dust disk 
and the central star is a close binary system. Figure 19 depicts
the speckle masking reconstruction of the evolved object Red Rectangle; 
observations were carried out with NICMOS3 NIR-camera at the 6~m telescope,
Special Astrophysical Observatory (SAO), Russia (Osterbart et al., 1996). The 
resolution of this object is 75~mas for the H~band.
\bigskip
\noindent
\midinsert
{\eightpoint   
\noindent
\centerline{\psfig{figure=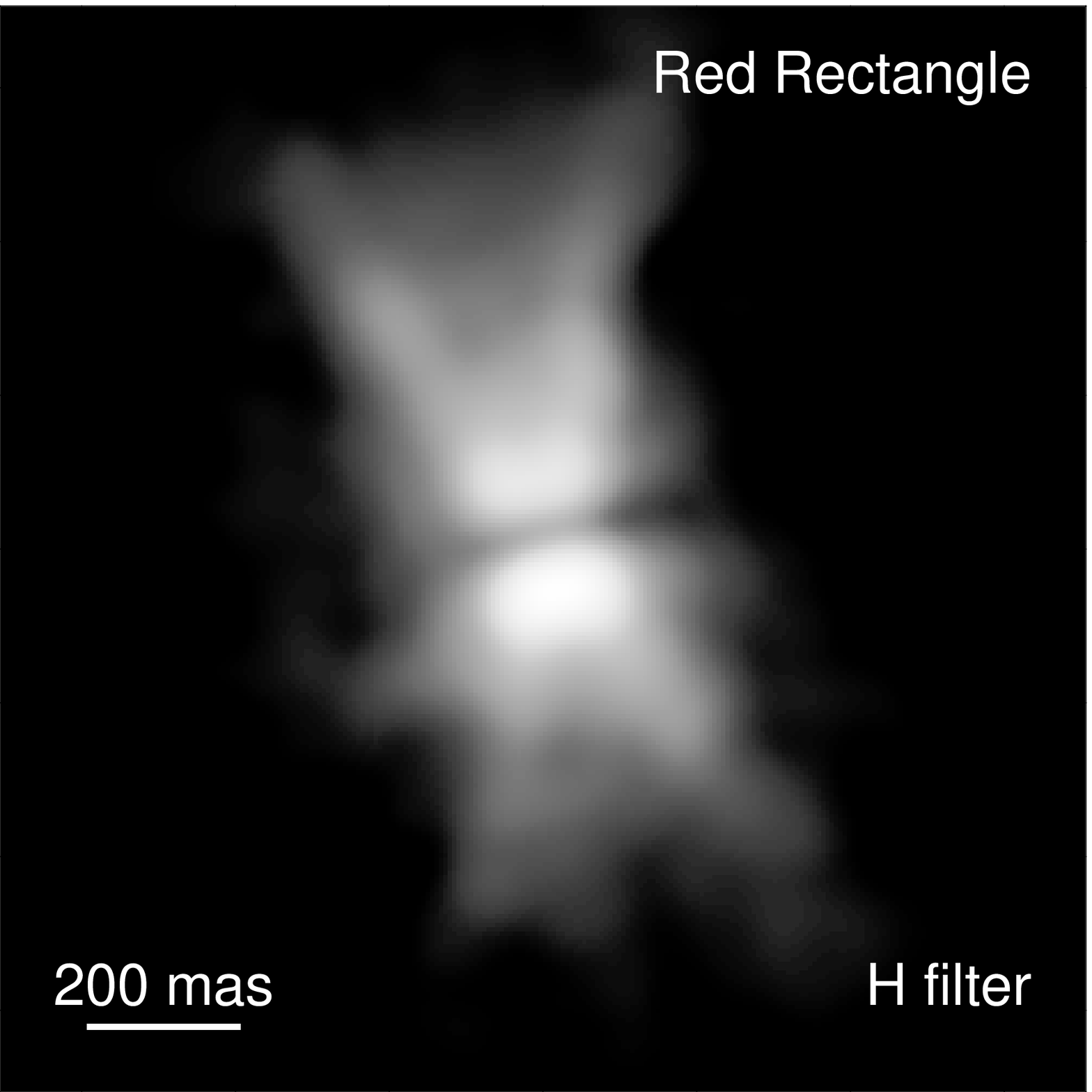,height=8cm,width=8cm}}
\bigskip
\noindent
{\bf Figure 19.} Speckle masking reconstruction of the evolved object Red 
Rectangle (Courtesy: R. Osterbart);
observations were carried out with NICMOS3 NIR-camera at the Russian SAO 6~m
telescope (Osterbart et al., 1996).  
}     
\endinsert
Osterbart et al., (1996) reported from the reconstructed speckle images
of the carbon star IRC+10216 (CW Leo) obtained at the SAO 6~m telescope, 
about a resolved central peak surrounded by patchy circumstellar matter. They 
found that the separation between bright clouds was 0.13$^{\prime\prime}$~-~0.21$^{\prime\prime}$, 
implying a stochastic behaviour
of the mass outflow in pulsating carbon stars. Weigelt et al., (1998)
found that five individual clouds were resolved within a 0.21$^{\prime\prime}$ 
radius of the 
central object in their high resolution K$^\prime$~band observation at the afore-mentioned 
telescope. They argued that the structures were produced by circumstellar dust 
formation. Figure 20 depicts the speckle masking reconstruction of IRC+10216; 
The resolution of this object is 76~mas for the K$^\prime$~band.
Recently, Tuthill et al., (1999b) recorded high resolution IR 
(1.65~$\mu$m and 2.27~$\mu$m) images of WR104 by means of aperture masking 
technique at the 10~m Keck telescope, Hawaii. The reconstructed images of the same 
at two epochs depict a spiral pinwheel in the dust around the star with a 
rotation period of 220~$\pm$~30 days. They opined that the circumstellar dust
and its rotation are the consequence of a binary companion. The aspherical dust
shell of the oxygen-rich AGB star AFGL~2290 (Gauger et al., 1999) has also been
reported. 
\bigskip
\noindent
\midinsert
{\eightpoint   
\noindent
\centerline{\psfig{figure=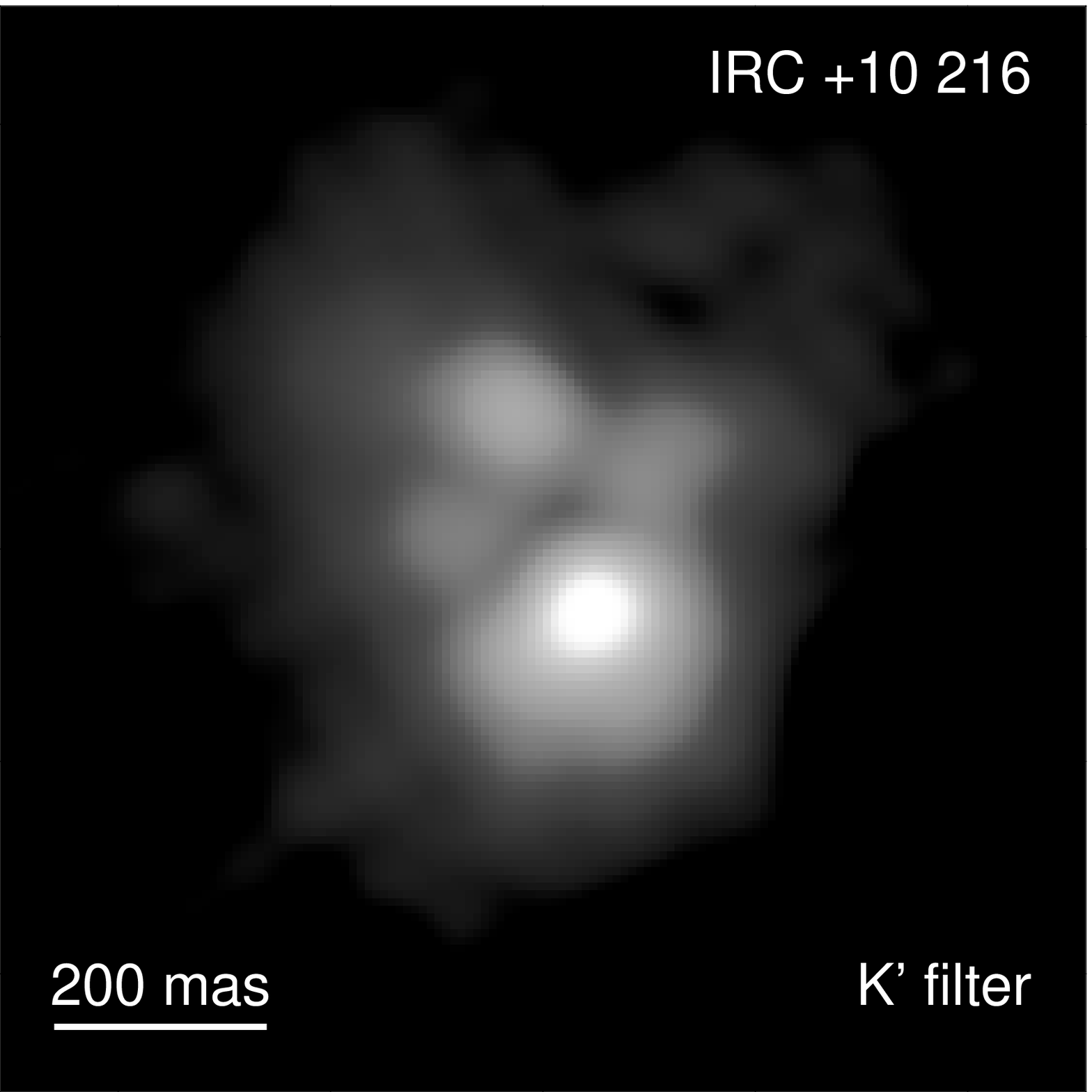,height=8cm,width=8cm}}
\bigskip
\noindent
{\bf Figure 20.} Speckle masking reconstruction of IRC+10216 (Courtesy: R. 
Osterbart); observations were carried out with NICMOS3 NIR-camera at the Russian 
SAO 6~m telescope (Osterbart et al., 1996).  
}     
\endinsert
The detailed information that is needed for the modelling of the 2-d radiative transfer 
concerning the symmetry $-$ spherical, axial or lack of clouds, plumes etc. $-$ 
can also be determined (Peraiah, 1999). Recently, 1-d, as well as 2-d radiative 
transfer modelling of AFGL~2290 (Gauger et al., 1999), Red Rectangle 
(Men'shchikov and Henning, 1997) have been 
carried out and used for the interpretation of the observation. The radiative
transfer calculations in the case of the latter depicted that the dust properties
are spatially inhomogeneous (Men'shchikov et al., 1998). In the radiative
transfer calculation, using the code DUSTY developed by Ivezi\'c et al., (1997), 
Bl\"ocker et al., (1999) noticed the ring like shell's 
intensity distribution of the rapidly evolved hypergiant object IRC+10420.
\bigskip  
V. Both nova and supernova (SN) have complex nature of shells viz., multiple, 
secondary and asymmetric; high resolution mapping may depict the events near the 
star and the interaction zones between gas clouds with different velocities. 
Soon after the explosion of the supernova SN1987A, various groups of observers 
monitored routinely the expansion of the shell in different wavelengths by means 
of speckle imaging (Nisenson et al., 1987, Papaliolios et al., 1989, Wood et al., 
1989). Karovska et al., (1989, 1991) measured the increasing diameter of the 
SN in several spectral lines and the continuum and Nulsen et 
al., (1990) have derived the velocity of the expansion. These observers found
that the size of this object was strongly wavelength dependent at the early
epoch $-$ pre-nebular phase indicating stratification in its envelope. 
A bright source at 0.06$^{\prime\prime}$ 
away from the said SN with a magnitude difference of 
2.7 at H$\alpha$ had been detected 30 and 38 days after the explosion by 
Nisenson et al., (1987) and 50 days after by Meikle et al., (1987). Chalabaev
et al., (1989) detected asymmetries in IR speckle observations that were in
the same direction as the bright object. Papaliolios et al., (1989) reported the 
asymmetry of the shell too. Based on the KT algorithm, that preserves the phase 
of the object Fourier transform, Karovska and Nisenson (1992) reported the
presence of knot-like structures. They opined that the knot-like structure
might be due to a light echo from material
located behind the supernova. Recent studies by Nisenson and Papaliolios 
(1999) with a image reconstruction algorithm based on iterative transfer
algorithm (Fienup, 1984) reveal a second spot, a fainter one (4.2 
magnitude difference) on the opposite side of the SN with 160~mas separation. 
Figure 21 depicts the recent image reconstruction based on modified iterative
transfer algorithm of SN1987A (Nisenson and Papaliolios, 1999).
\bigskip
\noindent
\midinsert
{\eightpoint   
\noindent
\centerline{\psfig{figure=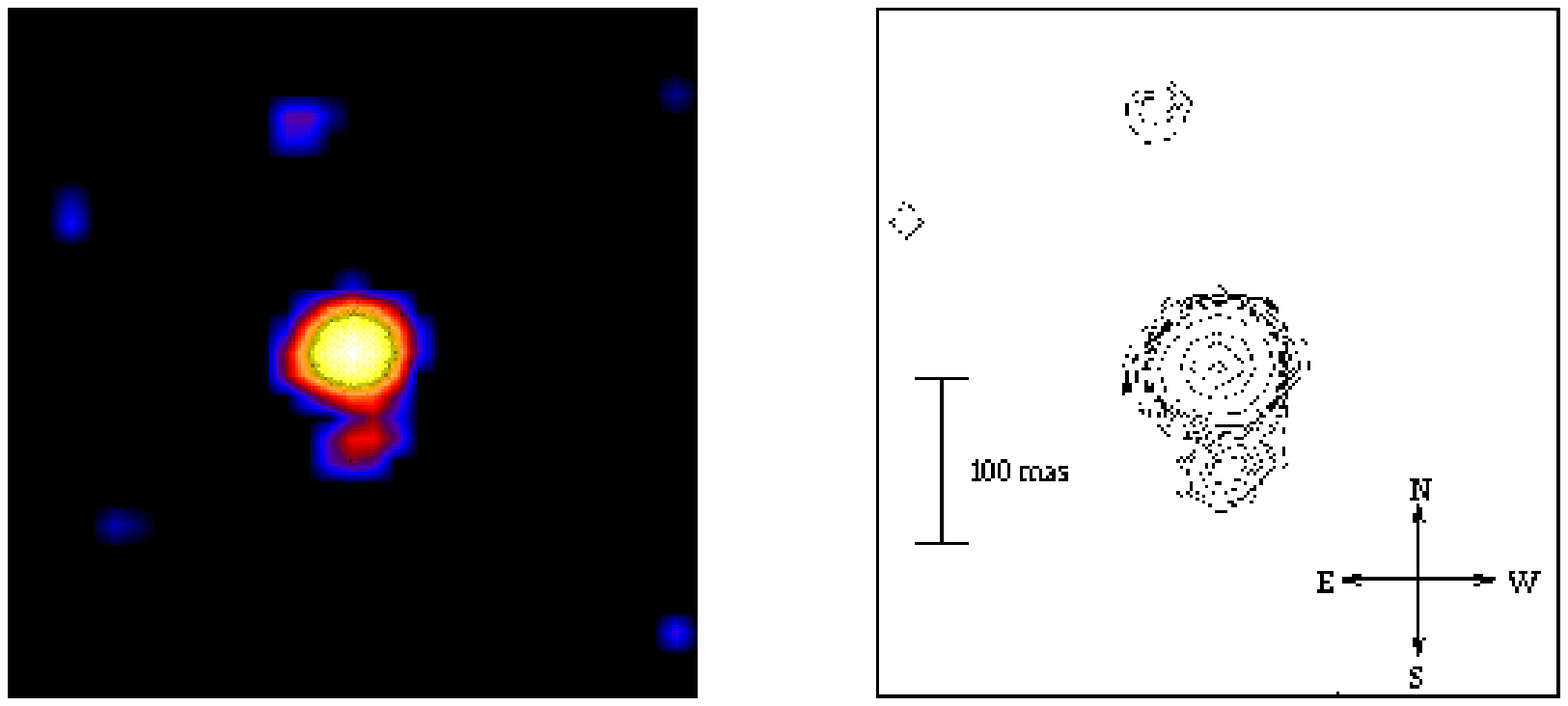,height=6.5cm,width=13cm}}
\bigskip
\noindent
{\bf Figure 21.} The reconstructed image and contour plot of SN1987A (Nisenson
and Papaliolios, 1999, Courtesy: P. Nisenson).
}     
\endinsert
VI. Another important field of observational astronomy is the study of 
the physical processes, viz., temperature, density and velocity of gas in the
active regions of the active galactic nuclei (AGN); optical imaging in the light 
of emission lines on sub-arc-second scales can reveal the structure of the 
narrow-line region. The scale of narrow-line regions is well resolved by 
the diffraction limit of a moderate-sized telescope (Afanas'jev et al., 1988, 
Ulrich, 1988). The time variability of AGNs ranging from minutes to decades 
(Krishan and Witta, 1994) is an important phenomena, which can also be studied with 
high resolution interferometric technique. 
\bigskip
Two of the brightest Seyfert galaxies, (i) NGC1068, and (ii) NGC4151, have strong
emission lines. In the case of latter, Shields (1999) points out the relative 
prominence of narrow forbidden lines in comparison with the former. The high resolution IR 
observation of NGC1068 depicted a compact core (Chelli et al., 1987,
Weinberger et al., 1996, Young et al., 1996). Wittkowski et al., (1998b) have 
resolved this compact source with the diameter of 0.03$^{\prime\prime}$ 
in the K-band. 
\bigskip
\noindent
\midinsert
{\eightpoint   
\noindent
\centerline{\psfig{figure=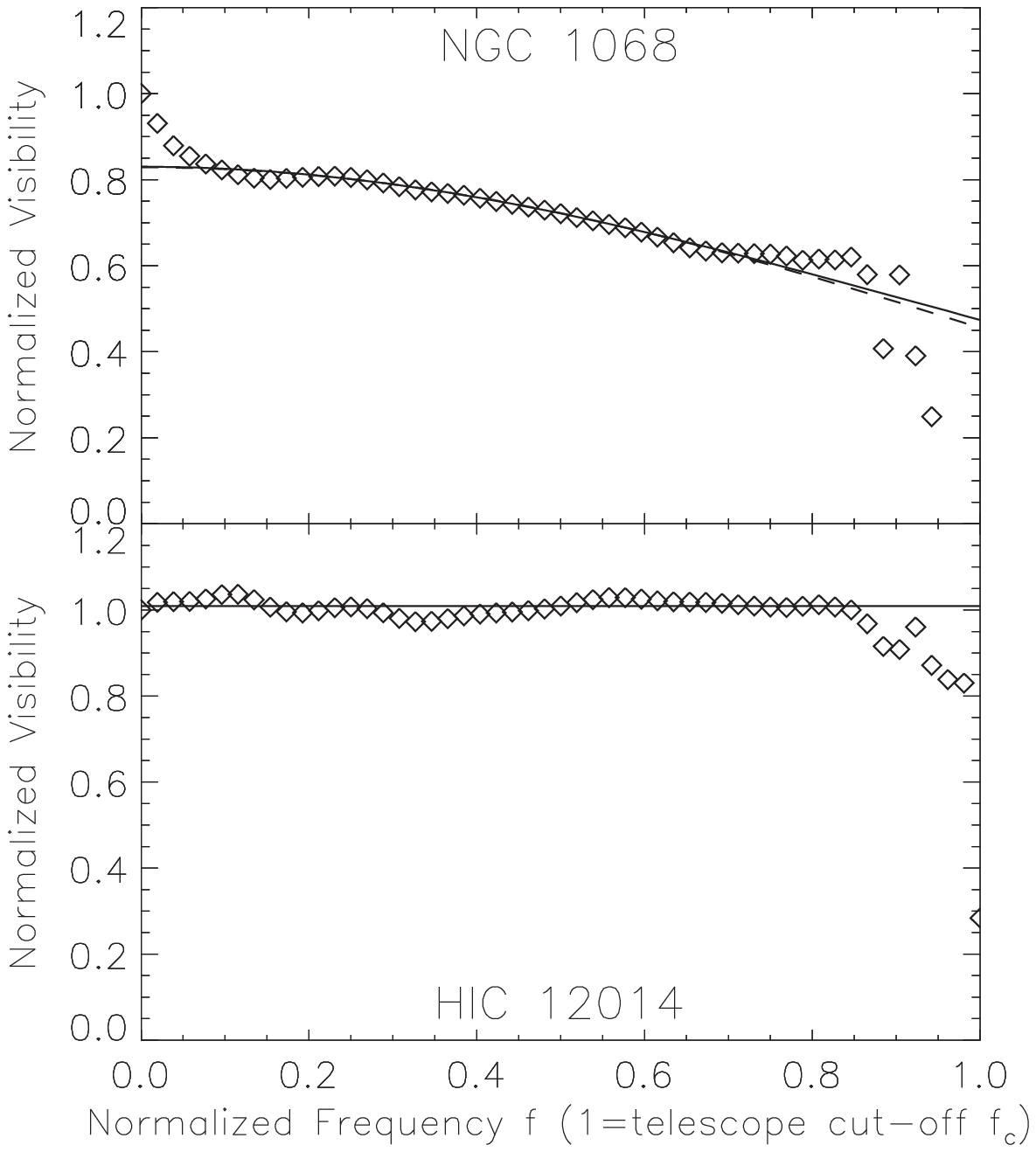,height=9cm,width=7cm}}
\bigskip
\noindent
{\bf Figure 22.} Azimuthally averaged visibilities of NGC1068 (top) and  
of unresolved reference star HIC12014 (bottom). The diamonds
indicate the observed visibilities, the solid lines the Gaussian fits, and the
dashed line is UD fit (Courtesy: M. Wittkowski).
}     
\endinsert
Figures 22 and 23 depict respectively the azimuthally averaged visibilities  
and the diffraction-limited speckle masking reconstruction of NGC1068. From the 
high resolution mapping of the  
object in the optical band, O~III emission line, Ebstein et al., (1989), found a
bipolar structure, extending over 0.4$^{\prime\prime}$.
Hofmann et al., (1992) resolved the
extension into 5 components. With the speckle spectroscopic method, Afanasiev 
et al., (1992) determined the radial velocity of both NE and
SW clouds located at 0.34$^{\prime\prime}$ and 0.33$^{\prime\prime}$
respectively from the centre. Ebstein et al.,
(1989) have resolved NGC4151 in O~III lines and from the reconstructed image 
they found the diameter to be 0.4$^{\prime\prime}$.
Zeidler et al., (1992) have resolved the 2.3$^{\prime\prime}$
central region of another AGN, NGC1346, into 4 clouds, distributed 
along the position angle of 20$^\circ$.
\bigskip 
\noindent
\midinsert
{\eightpoint   
\noindent
\centerline{\psfig{figure=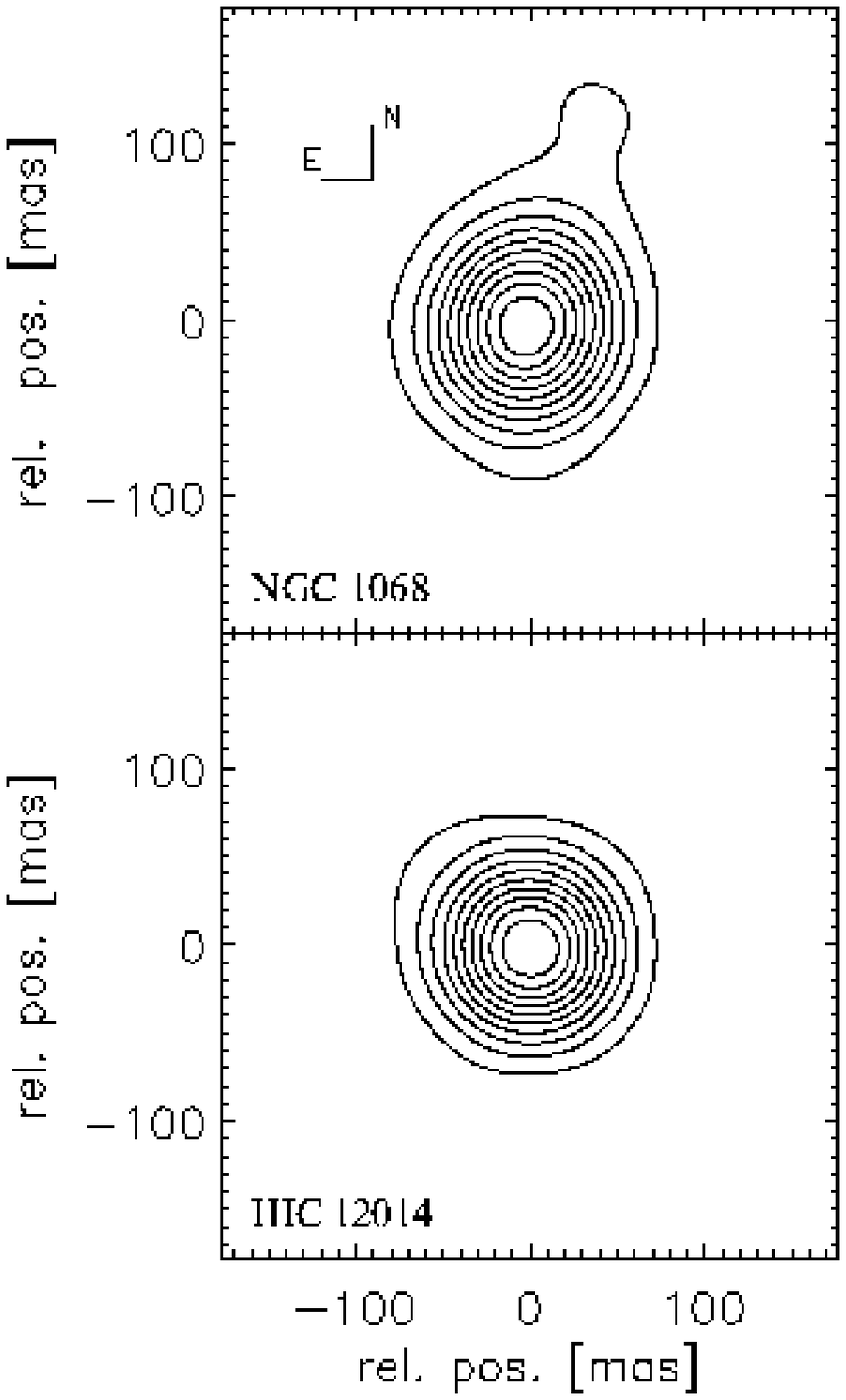,height=9cm,width=6cm}}
\bigskip
\noindent
{\bf Figure 23.} The speckle masking reconstruction of NGC1068 (top) and the
unresolved star HIC12014 (bottom). The contours are from 6\% to 100\% of peak
intensity (Courtesy: M. Wittkowski).
}     
\endinsert
VII. Quasars (QSO) may be gravitationally lensed by stellar objects, viz., 
stars, galaxies, clusters of galaxies etc., located along the line of sight.
The aim of the high angular imagery of these QSO's 
is to find their structure and components; their number and
structure as a probe of the distribution of the mass in the Universe (Ulrich,
1988). Capability of resolving these objects in the range of 0.2$^{\prime\prime}$
to 0.6$^{\prime\prime}$ would
allow the discovery of many more lensing events (Foy, 1992). With the speckle
interferometric technique, the gravitational image of the multiple QSO~PG1115+08 
was resolved (Hege et al., 1981, Foy et al., 1985). Foy et al., (1985) have
found one of the bright components, discovered to be double by Hege et al., 
(1981), was elongated. The possible causes of the structure, according to them, 
was due to a fifth component of the gravitational image of the QSO.  
\vskip 20 pt
\noindent
{\bf 10.2. Results obtained with adaptive optics}
\bigskip
\noindent
The noted advantages of the adaptive optics (AO) system over the conventional
techniques are the ability to recover near diffraction-limited images 
and to improve the point source sensitivity; but these need excellent seeing
conditions; an exact knowledge of point spread function is necessary.
Amplitude fluctuations are generally small and their effect on image degradation
remains limited, and therefore, their correction is not needed, except for
detection of exo-solar planets (Stahl and Sandler, 1995, Love and Gourlay, 1996).
\bigskip
AO system lacks capability of restoring wave-front and of retrieving fully 
diffraction-limited images of the objects (L\'ena and Lai, 1999a), nevertheless, 
the system began to produce a wide variety of astrophysical results, viz., 
(i) studying of the planetary meteorology (Poulet and Sicardy, 1996, Marco et al., 
1997, Roddier et al., 1997); images of Neptune's ring arcs are obtained (Sicardy
et al., 1999) that are interpreted as gravitational effects by one or more 
moons, (ii) imaging of the nucleus of M31 (Davidge et 
al., 1997a), (iii) surveying of young stars and multiple star systems (Bouvier 
et al., 1997), (iv) resolving the galactic centre (Davidge et al., 1997b), (v) 
imaging of Seyfert galaxies, QSO host galaxies (Hutchings et al., 1998a, 1998b),
and (vi) mapping of the circumstellar environment (Roddier et al., 1996) etc. 
Most of the results obtained so far are from the arena of 
near IR band (L\'ena, 1997, L\'ena and Lai, 1999b); the results from the area of 
visible wave length are continued to be sparse.
\bigskip
In a recent article, Rouan (1996, and references therein) addressed many 
interesting results obtained with the AO systems at CFHT, as 
well as with the systems at 3.6~m ESO telescope; the results obtained in the
near IR images of (a) NGC3690, one component of the Ultra 
Luminous system of merging galaxies Arp 299 (allowing to separate the 
nucleus and several compact sources), (b) the starburst/AGN galaxy NGC863, 
NGC7469, NGC1365, NGC1068, and (c) R136 in the 30 Doradus complex of the LMC 
depict the spectacular resolutions. From the diffraction-limited near IR images of the 
nucleus of NGC1068 obtained with the AO system, PUEO, at CFHT, Rouan et al.,
(1998), found several components that include: (i) an unresolved conspicuous 
core, (ii) an elongated structure, and (iii) large and small scale spiral 
structures. 
\bigskip
By means of AO imaging technique, Roddier et al., (1996) have detected a binary 
system consisting of a K7-MO star with an M4 companion that rotates 
clockwise. According to them, the system might be surrounded with a warm
unresolved disk. High resolution AO imagery of the massive star 
Sand-66$^\circ$41 in the LMC enabled Heydari and Beuzit (1994) to resolve it 
into 12 components. In order to 
perform differential photometry on close binary stars, Brummelaar et al., 
(1996) used a medium size 1.5~m telescope equipped with a laser guide star and AO 
system; the effective temperatures and the spectral types were determined for 
the secondaries of seven systems. Brandner et al., (1995) have resolved the 
close companion (sep. 0.128$^{\prime\prime}$) to NX~Pup using the AO system at ESO 3.6~m telescope 
and opined that the NX~Pup~B might be a pre-main sequence star, which is 
exhibiting strong IR excess. 
\bigskip
Brandl et al., (1996) have reported 0.15$^{\prime\prime}$ 
resolution near IR imaging of the R136 star 
cluster in 30 Doradus (LMC), an unusual high concentration of massive and bright
O, B, Wolf-Rayet stars, with ESO 3.6~m telescope equipped with the AO system,
ADONIS. Over 500 stars are detected within the field of view 12.8$^{\prime\prime}$ 
$\times$ 12.8$^{\prime\prime}$ 
covering a magnitude range of 11.2; $\sim$~110 are reported to be red stars. 
\bigskip
The improved resolution of crowded fields like globular clusters using
AO systems would enable to derive luminosity functions and spectral types,
to analyse proper motions in their central area etc. Simon et al., (1999)
have detected 292 stars in the dense Trapezium star cluster of the Orion nebula 
and resolved pairs to the diffraction limit of the 2.2~m University of Hawaii 
telescope using University of Hawaii (UH) AO system. Using both the
speckle masking technique and AO system, respectively at
ESO 2.2~m and ESO 3.6~m telescopes, quasi-simultaneous observations of the close 
Herbig Ae/Be binary star NX~Pup, associated with the cometary globular cluster I,
in the optical and near IR bands, Sch\"oller et al., (1996) estimated the mass
and age of both the components, NX~Pup~A and B. The circumstellar matter around
the former, according to them, could be described by a viscous accretion disk.   
\bigskip
Bedding et al., (1997b) have applied AO systems to study the resolved 
stellar populations in galaxies with the ESO 3.6~m telescope using ADONIS systems;
observations of several systems including the Sgr window in the bulge of the
milky way were made. In spite of small area covered, they were able to produce 
a infrared luminosity function and 
colour-magnitude diagram for 70 stars down to $\simeq$~19.5. These are the 
deepest yet measured for the galactic bulge, reaching beyond the turn-off. The
marked advantage over traditional approach is the usages of near IR region, 
where the peak of the spectral energy distribution for old populations is found
by them. Figure 24 depicts the ADONIS K$^\prime$ image of the Sgr window.
\bigskip
\noindent
\midinsert
{\eightpoint   
\noindent
\centerline{\psfig{figure=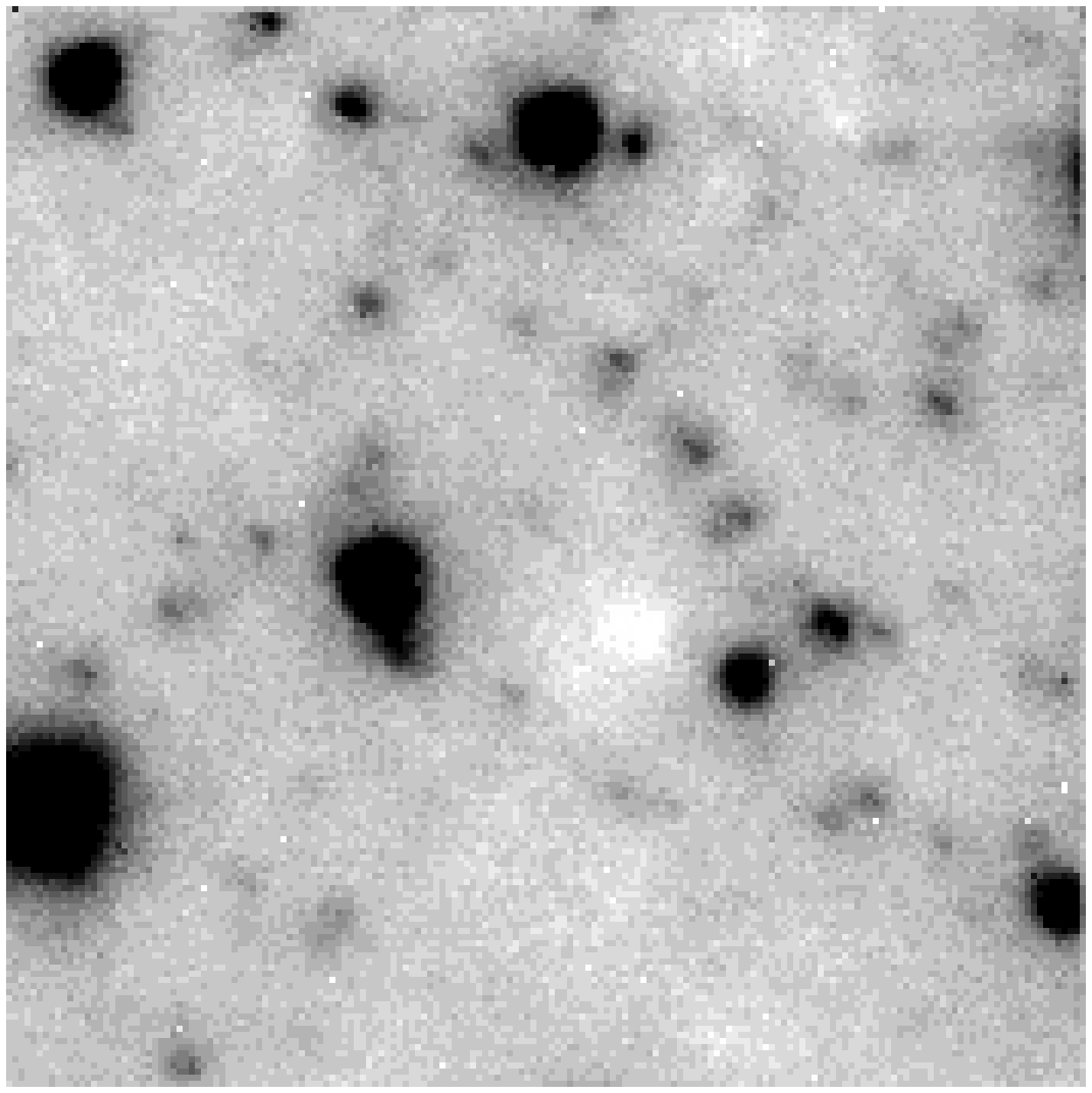,height=8cm,width=8cm}}
\bigskip
\noindent
{\bf Figure 24.} The ADONIS K$^\prime$ image of the Sgr window in the bulge of the
milky way (Bedding et al., 1997b). The image is 8$^{\prime\prime}$ $\times$ 
8$^{\prime\prime}$ (Courtesy: T. Bedding).
}     
\endinsert
AO systems have been used to observe extragalactic objects; imaging of central
area of active galaxies where cold molecular gas and star formation occur is an
important programme. High spatial resolution images of star forming region 
Messier 16 (Currie et al., 1996), the reflection nebula NGC2023 in Orion 
(Rouan et al., 1997) have been obtained with the ESO 3.6~m telescope equipped with 
ADONIS AO system. Observations of the latter revealed small scale structure
in the associated molecular cloud, close to the exciting star. Close et al., 
(1997) have mapped near IR polarimetric observations of the reflection nebula 
R~Monocerotis using UH AO system of CFHT. They have detected a faint source, 
0.69$^{\prime\prime}$ 
away from R~Mon and identified it as a T~Tauri star. Lai et al., (1998) have 
recorded images of Markarian 231, a galaxy 160~Mpc away with PUEO AO system at 
the said telescope, demonstrating the limits of achieving in terms morphological
structures on distant objects. The same system has also been used to image the 
nuclear region of NGC3690 in the interacting galaxy Arp~299 (Lai et al., 1999).
\bigskip
Aretxaga et al., (1998) reported the unambiguous detection of the host galaxy 
of a normal radio-quiet QSO at high-redshift in K-band using the AO system at the 
ESO 3.6~m telescope. Observations of the z = 3.87, broad absorption line quasar 
APM08279+5255 (Irwin et al., 1998) with the AOB of the CFHT, Ledoux et al., 
(1998) found the object to be a double source (separation = 0.35$^{\prime\prime}$ 
$\pm$~0.02$^{\prime\prime}$;
intensity ratio = 1.21~$\pm$~0.25 in H band). They opined in favour of 
gravitational lensing hypothesis which came from the uniformity of the quasar 
spectrum as a function of the spatial position in the image obtained with the
spectrograph, OASIS, at the same telescope. Search for molecular
gas in high red-shift normal galaxies in the foreground of the gravitationally
lensed quasar Q1208+1011 has also been made (Sams et al., 1996). 
\bigskip
From the observations with ADONIS adaptive optics system with infrared cameras on the 
3.6~m ESO telescope, Monnier et al., (1999) found a variety of dust 
condensations that include a large scattering plume, a bow shaped dust feature
around the red supergiant VY~CMa. A bright knot of emission 1$^{\prime\prime}$
away from the star is also reported. In view of the afore-mentioned findings, 
the authors argued in favour of the presence of chaotic and violent dust
formation processes around the star. The AO system at CFHT has also been 
used for searching brown dwarfs and giant planets (Walker et al., 1998).
AO imaging of proto-planetary nebulae (PPN), Frosty Leo and the Red Rectangle 
were also reported by Roddier et al., (1995). In both cases, they found a binary 
star at the origin of these PPNs and argued in favour of the
mechanism proposed by Morris (1987) for the formation of bipolar PPN. 
\bigskip
Another novel method of using coronographic mode has been tried recently with the AO 
system at the 3.6~m ESO telescope to observe $\alpha$~CMa (Malbet, 1995). The 
noted advantages of the on-line occulting mask system (Labeyrie, 1975, Bhattacharyya 
and Rajamohan, 1990, Roddier and Roddier, 1997, Prieur et al., 1998) are the 
reduction of the light 
coming from the central star, and filtering out of the light at low 
spatial frequency; the remaining light at the edge of the pupil corresponds to 
high frequencies. The high resolution imagery of a few interesting objects by 
means of the AO coronographic method detected a very low mass companion to the 
astrometric binary Gliese~105~A (Golimowski et al., 1995). 
\bigskip
Mouillet et al., (1997) have applied the same technique to record the images of 
proto-planetary disk around the star $\beta$~Pictoria at the 3.6~m ESO
telescope; a warp of the disc is detected. Images of the nebula 
around LBV AG Carina, with the John Hopkins AO coronograph at the 1~m telescope, in the 
H$\alpha$ and N~II revealed the presence of highly asymmetric features in AG 
Carina's circumstellar environment (Nota et al., 1992). Imaging of the bipolar 
nebula around the LBV~R127 has also been reported by Clampin et al., (1993). 
Nakajima and Golimowski, (1995) have obtained very high contrast images of several
pre-main sequence stars and noticed the remnant envelope of star formation. 
\bigskip
A new technique has been proposed by Labeyrie, (1995), called dark speckles 
method exploiting the light cancelation effect in random field (highly 
destructive interferences may occur occasionally depicting near black spots in 
the speckle pattern). The technique, which features the combination of both speckle 
interferometry and AO system, may improve the possibility of detecting faint 
companions of stars.  Recently, Boccaletti et al., (1998a) have 
found from the laboratory simulations the capability of detecting a stellar
companion of relative intensity 10$^6$ at 5 Airy radii from the star using an
avalanche photo-diode as detector. They also have recorded dark speckle data at 
the 1.52~m telescope of Haute-Provence using an AO system and detected a faint 
component of the spectroscopic binary star HD144217 ($\Delta$m = 4.8, 
separation = 0.45$^{\prime\prime}$). Subsequently, Boccaletti et al., (1998b) 
have applied the same technique at the said telescope to observe 
the relatively faint companions of $\delta$~Per and $\eta$~Psc and were able
to estimate of their position and magnitude difference.
\vskip 20 pt
\noindent
{\bf 10.3. Results obtained with diluted aperture interferometry}
\bigskip
\noindent
Considerable amount of astrophysical results, viz., (i) diameters (Labeyrie, 
1985) (ii) effective temperatures of giant stars (Faucherre et al., 1983), (iii) 
resolving the gas envelope of the Be star $\gamma$~Cassiopeiae 
in the H$\alpha$ line (Thom et al., 1986), (iv) diameters of cool bright giants
and their effective temperature at 2.2~$\mu$m (DiBenedetto and Rabbia, 1987) etc., 
have been obtained from interference fringes recorded using I2T interferometer
at Observatoire de Calern (section 9.2.1).
The angular diameter for more than 50 stars have been measured
(DiBenedetto and Rabbia, 1987, Mozurkewich et al., 1991, Dyck et al., 1993) with
accuracy better than 1\% in some cases with the long baseline amplitude 
interferometers at optical and IR wavelengths. 
\bigskip
The scientific programmes using GI2T are restricted by the low limiting 
magnitude down to 5 (seeing and visibility dependent) that
include the Be stars, Luminous Blue variables, spectroscopic and eclipsing 
binaries, wavelength dependent objects, bright stars (measuring diameters), 
circumstellar envelope etc., (Mourard et al., 1992, Thureau et al., 1998). 
The reason for concentrating on low magnitude object is mainly due to the visual 
detection and tracking of the fringes. The noted results obtained so far in 
recent time with the said interferometer include the diameter of $\delta$~Cephei 
(Mourard et al., 1997), subtle structures in circumstellar
environment such as, jets in the binary system $\beta$~Lyrae (Harmanec et al.,
(1996), clumpiness in the wind of P~Cygni (Vakili et al., 1997, 1998a); a 
rotating arm in $\zeta$~Tau (Vakili et al., 1998b).
\bigskip
\noindent
\midinsert
{\eightpoint   
\noindent
\centerline{\psfig{figure=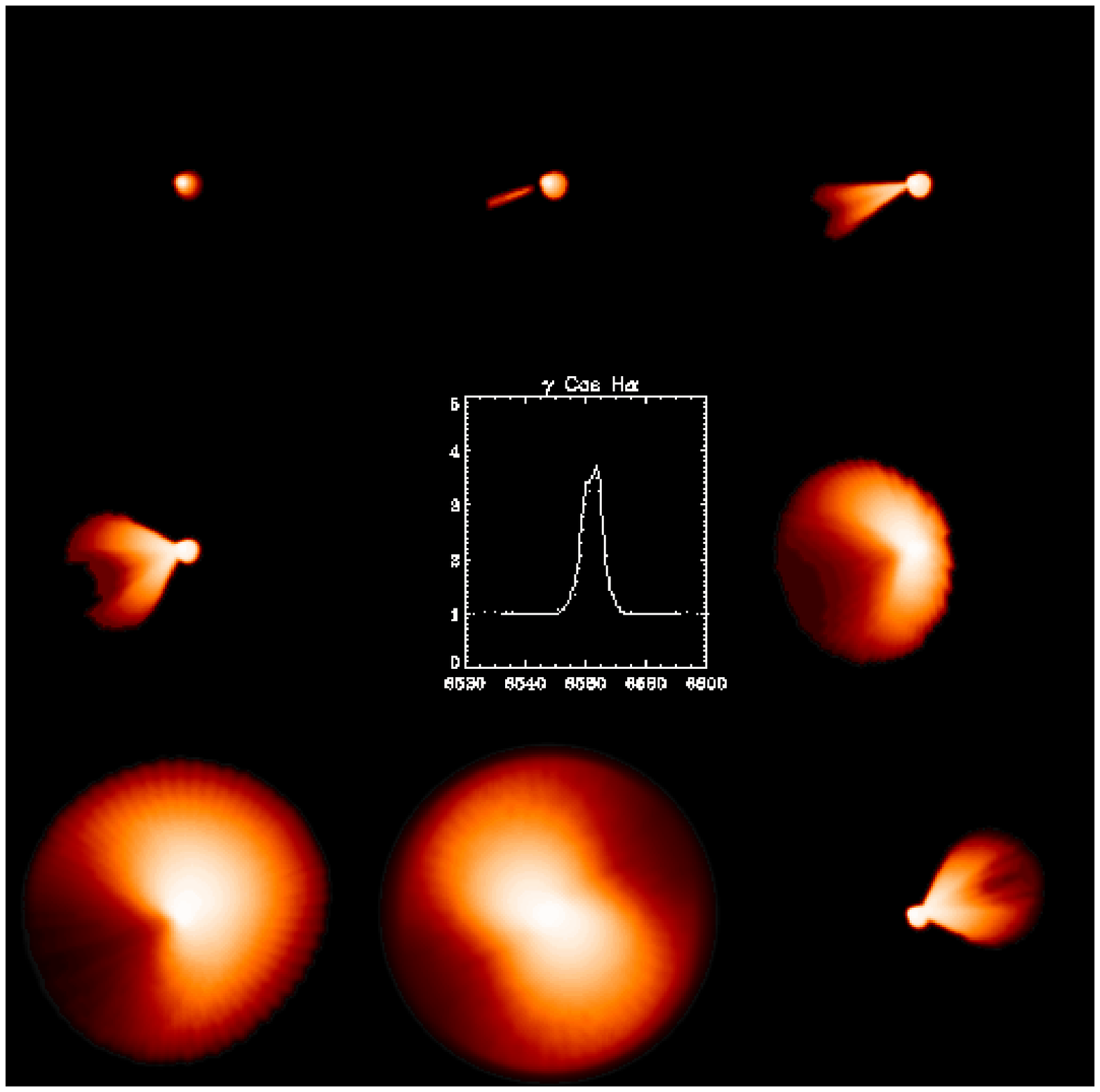,height=9cm,width=9cm}}
\bigskip
\noindent
{\bf Figure 25.} Intensity maps from the model of $\gamma$~Cassiopeiae of 
different Doppler shifts across the H$\alpha$ emission line (Courtesy: P. Stee).
}     
\endinsert
The measurements of spectrally resolved
visibilities of the emission-line star, $\gamma$~Cassiopeiae (HD5394), with GI2T 
by Mourard et al., (1989) distinguish the hydrogen emission in the envelope from 
the continuum photospheric emission. With the central star as a reference 
(unresolved), they determined the relative phase of the shell visibility and
showed clearly the rotation of the envelope. From the subsequent observations on 
later dates with the interferometer, Stee et al., (1995, 1998) derived 
the radiative transfer model for the star based on the spectroscopic and 
interferometric data. Figure 25 depicts the intensity maps from the model of the Be star 
$\gamma$~Cassiopeiae of different Doppler shifts across the H$\alpha$ emission
line (Stee et al., 1995). Each map is computed within a spectral band of 0.4~nm;
theoretical visibilities are computed from these maps and compared to GI2T data.
Figure 26 depicts the global picture of circumstellar environment of the said 
star.
\bigskip 
\noindent
\midinsert
{\eightpoint   
\noindent
\centerline{\psfig{figure=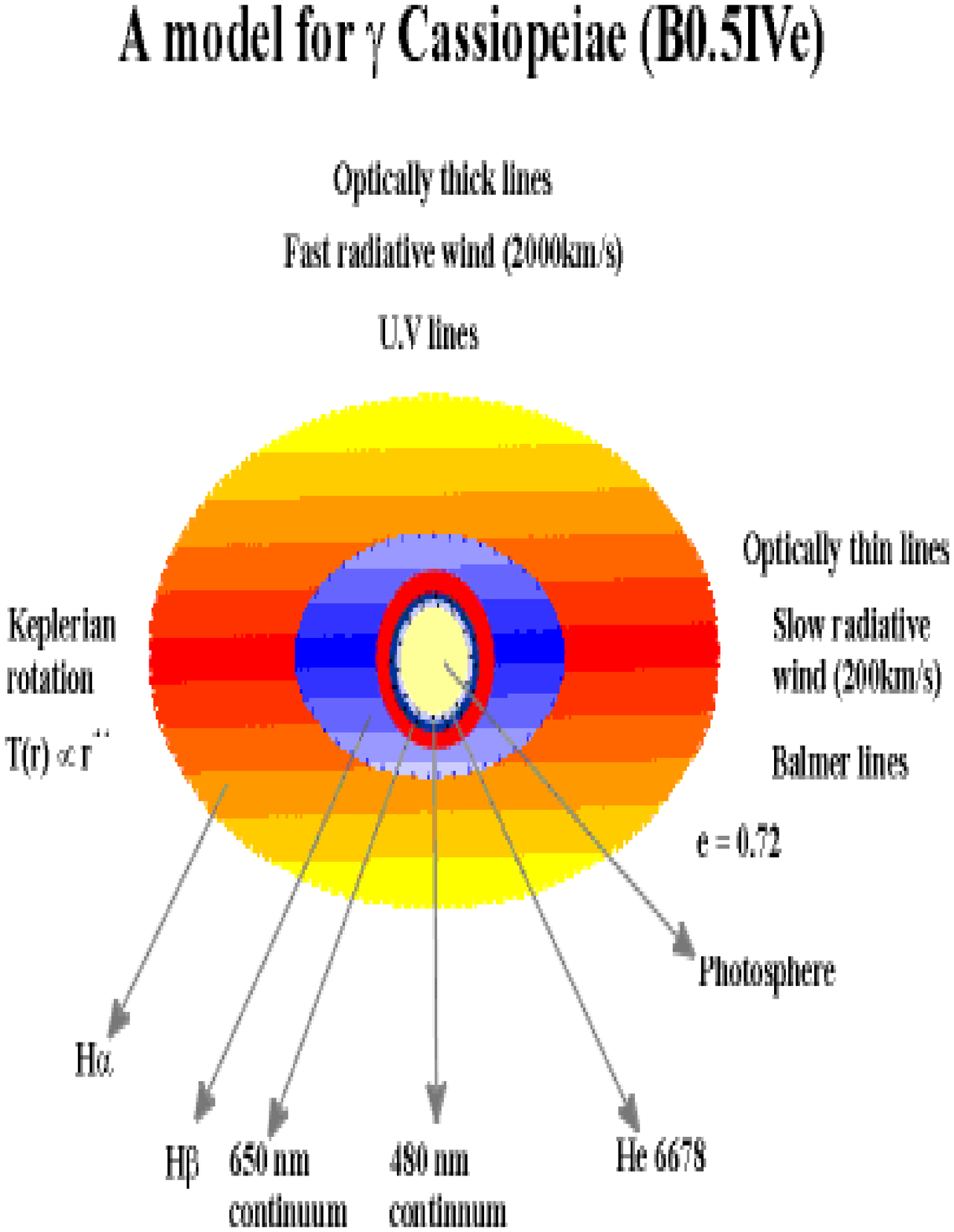,height=9cm,width=9cm}}
\bigskip
\noindent
{\bf Figure 26.} The global picture of circumstellar environment of 
$\gamma$~Cassiopeiae (Courtesy: P. Stee).
}     
\endinsert
Using high spatial 
resolution data at H$\alpha$ and He~I (667.8~nm) emission lines, Vakili et al., 
(1997) resolved the extended envelope of the LBV, P~Cygni, with the GI2T. 
Detection of a localized asymmetry at 0.8~mas to the south of the 
same star's photosphere has also been reported (Vakili et al., 1998a).
Harmanec et al., (1996) reported the results on the interacting binary star
$\beta$~Lyr. Figure 27 depicts the jet-like structure of the said star (Stee,
1999).
\bigskip
\noindent
\midinsert
{\eightpoint   
\noindent
\centerline{\psfig{figure=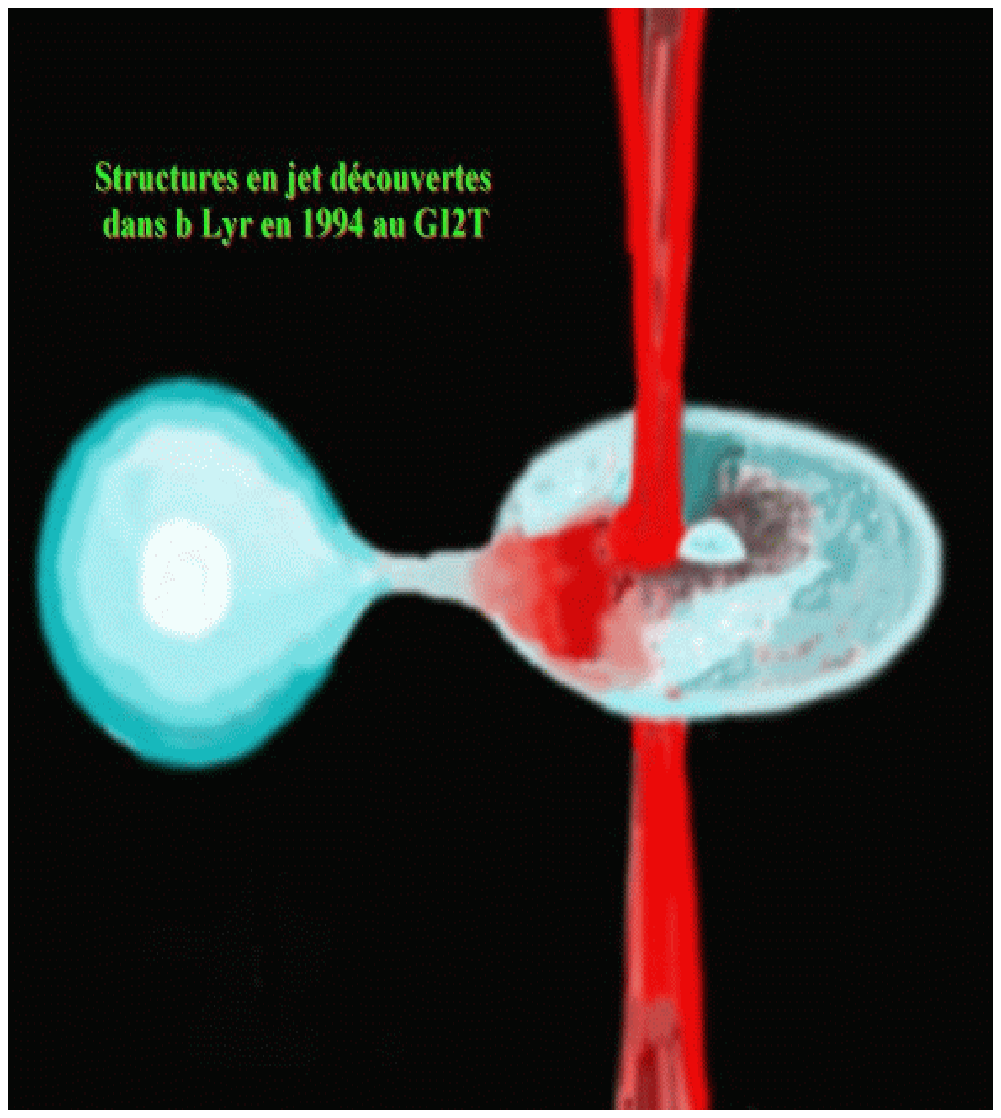,height=9cm,width=9cm}}
\bigskip
\noindent
{\bf Figure 27.} Jet-like structure of $\beta$~Lyr (Stee, 1999, Courtesy: 
P. Stee).
}     
\endinsert
Measurements of precise stellar positions and motions of the stars have been
carried out with the mark III interferometer at Mt. Wilson.  
It was designed with the internal metrology to support its 
operation to obtain astrometric observations of several stars. Shao et al., 
(1990) have determined astrometric positions of 12 stars with 1$\sigma$ errors 
6~mas in declination and 10~mas in right ascension. 
Hummel (1994) reported the measurement of the position of
11 FK5 stars. This set up has also been used
to derive the fundamental stellar parameters, like the orbits for 
spectroscopic binaries (Armstrong et al., 1992a, Pan et al., 1992), 
structure of circumstellar shells (Bester et al., 1991) etc. Armstrong et al., 
(1992a, 1992b) have determined the orbit of the double-lined spectroscopic 
binary stars, $\alpha$~Equulei, $\phi$~Cygni, as well as derived their masses 
and absolute magnitudes for its components. Observations of 5 eclipsing binary 
stars, $\beta$~Aur, $\alpha$~CrB, $\beta$~Per, $\zeta$~Aur, $\epsilon$~Aur, have been
reported (Shao and Colavita, 1994).
\bigskip
The results obtained recently from the other long baseline amplitude 
interferometers can also be recorded here. 
With SUSI interferometer Davis et al., (1998, 1999b) have determined the 
diameter of $\delta$~CMa with an accuracy of $\pm$1.8\%. Malbet et al., (1998) 
have resolved the young stellar object FU~Orionis using the long baseline Palomar 
testbed interferometer in the near infrared with a projected resolution
better than 2~AU. Observations of the young binary stars, $\iota$~Peg have
been conducted by Pan et al., (1996) with the testbed interferometer. They
have determined its visual orbit with separation of 1~mas in R. A., having a
circular orbit with a radii of 9.4~mas. 
\bigskip
The measurements of closure phase together with the measurements of
visibility amplitude allow one to reconstruct an image of the object. 
With the Cambridge optical aperture synthesis telescope (COAST), using a 
coherent array of 4 telescopes with 6 baselines (maximum baseline was 10~m), 
Baldwin et al., (1998) have resolved $\alpha$~Tau. Aperture synthesis maps
of the double-lined spectroscopic binary $\alpha$~Aurigae  
obtained by Baldwin et al., (1996) with three elements of the same 
interferometer showed the milliarc-second orbital motion of the system over a
15 day interval. Images of the resolved stellar disk of $\alpha$~Orionis 
depicting the presence of a circularly symmetric data with an unusual 
flat-topped and limb darkening profile have been
reported (Burns et al., 1997). The variations of the cycle of pulsation of the 
angular size of the Mira variable R~Leonis have also been measured with this 
interferometer directly (Burns et al., 1998). Pauls et al., (1998) have measured 
the limb darkened angular diameters of late-type giant stars using the Navy prototype
interferometer with three optical elements; measurement of non-zero closure 
phase has been performed on a single star.
\bigskip
Though the visible light measurements from the IOTA long baseline interferometer
have been reported (Coldwell et al., 1998), most of the results obtained so 
far are from the IR wave bands, particularly at 
near IR bands. They are in the form of measuring the angular diameters and
effective temperatures of carbon stars (Dyck et al., 1996a), carbon Miras and S
types (van Belle et al., 1997), K and M giants and supergiants (Dyck et al., 
1996b, 1998, Perrin et al., 1998), Mira variables (van Belle et al., 1996), and
cool giant stars (Dyck et al., 1995). Evidence of circumstellar dust and gas 
shell around two stars, RS~Cnc, RX~Boo, were also reported (Dyck et al., 1995).
\vskip 20 pt
\centerline{\bf 11. Epilogue}
\bigskip 
\noindent
Turbulence in the atmosphere restricts the resolution of optical telescopes to
about 1$^{\prime\prime}$, irrespective of their sizes (D~$>$~10~cm). The deployment
of space-bound telescopes may improve the resolving power to its diffraction
limit but the size and the cost of such a venture is its shortcoming.
The understanding of the basic random interference phenomenon $-$ speckle $-$
is of paramount importance to observational astronomy. In recent years the
uses of speckle pattern and a wide variety of applications have been found 
in many other branches of physics and engineering. Though the
statistical properties of speckle pattern is complicated, detailed analysis  
of this is useful in information processing. 
\bigskip
The single aperture speckle interferometric technique has made a major 
break-through in observational astronomy by counteracting the deleterious effect 
of the atmosphere on the structure of stellar images. A host of basic problems 
need high angular resolution for their solution. This may be in the form of 
measuring the separation of the components of close binaries, mapping the finer features of 
certain types of circumstellar envelopes, resolving the central core of 
AGNs and gravitationally lensed QSOs etc. Studies of the morphology of stellar 
atmospheres, the circumstellar environment of nova or supernova, YPN, long 
period variables (LPV), rapid variability of AGNs etc. are of paramount 
importance in astrophysics. Details of the structure of a wide range of stellar 
objects at scales of 0.015$^{\prime\prime}$~-~0.03$^{\prime\prime}$ are 
routinely observed. The physical properties of red dwarfs 
in the vicinity of sun can also be looked into; some dwarfs may often be close 
binaries. Speckle interferometric technique has been extended to IR domain too. 
With the photon counting detector system which is an essential tool in the 
application of optical interferometric imaging that allows the accurate photon 
centroiding, as well as provides dynamic range needed for measurements of  
source characteristics, one can record the specklegrams of the object of
faintest limiting magnitude. 
\bigskip
With the image reconstruction algorithms, which preserve the phase in the object 
Fourier transform permits true image restoration. Recently, phase-diversity 
technique (Gonsalves, 1982, Paxman et al., 1992) that uses two simultaneous 
short-exposure images taken at different focus/defocus positions near the
focal plane has found to be promising method in astronomy (Baba et al., 1994c, 
Seldin and Paxman, 1994). The single aperture interferometry by means
of various other techniques too brought considerable amount of results.
\bigskip
Adaptive optics technology has become an affordable tool at all new large
astronomical telescopes. This technique has been able to bring many new
results not only in the field of observational astronomy, but in other branches
of physics as well. Liang et al., (1997) have constructed a fundus camera equipped 
with adaptive optics allowing the imaging of microscopic size of single
cell in the living human retina. They have exhibited that eyes with adaptive
optics correction can resolve fine gratings that were invisible under normal 
viewing conditions. Observations using AO system on large telescopes of 10~m
class could surpass the resolution achievable with the present day
orbital telescope. 
\bigskip
Though the stellar speckle interferometry is capable of detecting relatively 
faint objects ($\sim$~16th. magnitude), the angular resolution is limited by 
the diameter of the telescope. On the other hand, the angular resolution of any
stellar object in the visible wavelength can vastly be improved by using long 
baseline interferometry. Among others, this technique can detect the 
morphological details, viz., (i) granulations, (ii) oblateness etc. of giant 
stars. Eclipsing binaries (Algol type) which show evidence of detached gas rings 
around the primary are also good candidates for long baseline interferometry.
The potential of this type of interferometer can be envisaged in determining the
fundamental astrophysical informations of circumstellar envelopes such as, the 
diameter of inner envelope, colour, symmetry, radial profile etc. 
\bigskip
Several long baseline interferometers are either in operation or under 
development at various stages. Rapid increase in the scientific output at optical, 
as well as at infrared wave bands using these interferometers can be foreseen at 
the fall of the next millennium. With improved technology, the long
baseline interferometric arrays of large telescopes fitted with 
high level adaptive optics system that applies dark speckle coronograph 
(Boccaletti et al., 1998b) may provide snap-shot images at their
recombined focus using the concept of densified-pupil imaging (Pedretti and
Labeyrie, 1999), and yield improved images and spectra of objects. One of the
key areas where the new technology would make significant contributions is the 
astrometric detection and characterisation of exo-planets.
\vskip 20 pt

\noindent
{\bf Acknowledgment}: The author expresses his gratitude to Dr A. Labeyrie 
and Prof. J. C. Bhattacharyya for comments on the article and indebtedness to
Dr P. Nisenson, Dr T. R. Bedding, Dr R. Osterbart, Dr P. Stee, and Dr M.
Wittkowski for providing the images, figures etc., and granting permission for
reproduction. Thanks are also due to Prof. R. K.
Kochhar and Dr A. V. Raveendran for reading the manuscript and to Prof. V. Krishan
for constant inspiration. The help 
rendered by Messrs V. Chinnappan, B. A. Varghese, B. S. Nagabhushana, K. 
Sankarsubramaian, and R. Sridharan for valuable communications are gratefully 
acknowledged.
 
\bigskip
\centerline{\bf References}
\bigskip
{\eightpoint\parindent=0pt\everypar={\hangindent=0.5 cm}

Acton D. S., Smithson R. C., 1992, Appl. Opt., 31, 3161.

Afanas'jev V. S., Balega I. I., Balega Y. Y., Vasyuk V. A., Orlov V. G.,
1988, Proc. ESO-NOAO conf. `High Resolution Imaging Interferometry', 
ed., F. Merkle, Garching bei M\"unchen, FRG, 127. 

Afanasiev V. L., Balega Y. Y., Orlov V. G., Vasyuk V. A., 1992,
Proc. ESO-NOAO conf. `High Resolution Imaging Interferometry', 
ed., J. M. Beckers \& F. Merkle, Garching bei M\"unchen, FRG, 53.

Aime C., 1976, A \& A, 47, 5.

Aime C., Petrov R., Martin F., Ricort G., Borgnino J., 1985, Proc. SPIE., 
556, 297.
 
Aime C., Ricort G., Grec G., 1975, A \& A, 43, 313.

Aime C., Ricort G., Grec G., 1977, A \& A, 54, 505.

Aime C., Ricort G., Harvey J. W., 1978, Ap J, 221, 362.

Anderson J. A., 1920, Ap J, 51, 263.

Aretxaga I., Mignant D. L., Melnick J., Terlevich R. J., Boyle B. J., 1998,
astro-ph/9804322, MNRAS.

Aristidi \'E., Carbillet M., Prieur J. -L., Lopez B., Bresson Y., 1997, A \& AS, 
126, 555.

Aristidi \'E., Prieur J. -L., Scardin M., Koechlin L., Avila R., Carbillet M.,
Lopez B., Rabbia Y., Nisenson P., Gezari D., 1999, A \& AS, 134, 545.

Armstrong J. T., 1994, Proc. SPIE., 2200, ed. J. B. Breckinridge, 62.

Armstrong J. T., Hummel C. A., Mozurkewich D., 1992a, Proc. ESO-NOAO conf. `High 
Resolution Imaging Interferometry', ed., J. M. Beckers \& F. Merkle, Garching 
bei M\"unchen, FRG, 673.

Armstrong J. T., Mozurkewich D., Vivekanand M., Simon R. S., Denison C. S.,
Johnston K. J., Pan X. -P., Shao M., Colavita M. M., 1992b, A J, 104, 241. 

Arsenault R., Salmon D. A., Kerr J., Rigaut F., Crampton D., Grundmann W. A.,
1994, SPIE conf., 2201, 883.

Ayers G. R., Dainty J. C., 1988, Opt. Lett., 13, 457.

Ayers G. R., Northcott M. J., Dainty J. C., 1988, J. Opt. Soc. Am. A., 
5, 963.

Baba N., Kuwamura S., Miura N., Norimoto Y., 1994b, Ap J, 431, L111.

Baba N., Kuwamura S., Norimoto Y., 1994a, App. Opt., 33, 6662.

Baba N., Tomita H., Miura N., 1994c, App. Opt., 33, 4428.

Babcock H. W., 1953, PASP, 65, 229.

Bagnuolo Jr. W. G., Mason B. D., Barry D. J., Hartkopf W. I., McAlister H. A., 
1992, A J, 103, 1399.

Baier G., Weigelt G., 1987, A \& A, 174, 295.

Baldwin J. E., 1992, Proc. ESO-NOAO conf. `High Resolution Imaging 
Interferometry', ed., J. M. Beckers \& F. Merkle, Garching bei M\"unchen, FRG,
747. 

Baldwin J. E., Beckett R. C., Boysen R. C., Burns D., Buscher D. F., Cox G. C.,
Haniff C. A., Mackay C. D., Nightingale N. S., Rogers J., Scheuer P. A. G.,
Scott T. R., Tuthill P. G., Warner P. J., Wilson D. M. A., Wilson R. W., 1996, 
A \& A, 306, L13.
 
Baldwin J. E., Boysen R. C., Cox G. C., Haniff C. A., Rogers J., Warner P. J.,
Wilson D. M. A., Mackay C. D., 1994, Proc. SPIE., on `Amplitude and Intensity
Spatial Interferometry II', 2200, 112.

Baldwin J. E., Boysen R. C., Haniff C. A., Lawson P. R., Mackay C. D., Rogers J., 
St-Jacques D., Warner P. J., Wilson D. M. A., Young J. S., 1998, Proc. SPIE., 
on `Astronomical Interferometry', 3350, 736.

Baldwin J. E., Haniff C. A., Mackay C. D., Warner P. J., 1986, Nature, 320, 
595.

Balega I. I., Balega Y. Y., Belkin I. N., Maximov A. E., Orlov V. G.,
Pluzhnik E. A, Shkhagosheva Z. U., Vasyuk V. A., 1994, A \& AS, 105, 503.

Balega I. I., Balega Y. Y., Falcke, H., Osterbart R., Reinheimer T., Sch\"oeller
M., Weigelt G., 1997a, Astron. Letters, 23, 172.

Balega I. I., Balega Y. Y., Falcke, H., Osterbart R., Sch\"oeller
M., Weigelt G., 1997b, in Visual Double Stars: Formation, dynamics and
evolutionary tracks, eds. J. A. Docobo, A. Elipe, H. McAlister, Astrophysics and 
Space Science Library, Kluwer Acad. Publ. Dordrecht, Boston, London,
223, 73.

Balega Y. Y., Balega I. I., 1985, Sov. Astron., 11, 47.

Balega Y., Blazit A., Bonneau D., Koechlin L., Foy R., Labeyrie A., 1982,
A \& A, 115, 253.

Balega Y., Bonneau D., Foy R., 1984, A \& AS, 57, 31.

Balick B., 1987, A J, 94, 671.

Barakat R., Nisenson P., 1981, J. Opt. Soc. Am., 71, 1390.

Barletti R., Ceppatelli G., Paterno L., Righini A., Speroni N., 1976, J. Opt.
Soc. Am., 66, 1380.

Barlow M. J., Morgan B. L., Standley C., Vine H., 1986, MNRAS,
223, 151.

Barr L. D., Fox J., Poczulp G. A., Roddier C. A., 1990, Proc. SPIE, 1236,
492.

Bates R. H. T., Davey B. L. K., 1988, Proc. NATO-ASI, `Diffraction Limited 
Imaging with Very Large Telescopes', ed. D. M. Alloin \& J. -M. Mariotti, 
Carg\'ese, France, 293.
 
Bates R.H.T., McDonnell M.J., 1986, `Image Restoration and Reconstruction',
Oxford Engineering Science, Clarendon Press, Oxford.
 
Beckers J. M., 1982, Opt. Acta., 29, 361.

Beckers J. M., 1988, ESO Conf. on `Very Large Telescopes and
their Instrumentation', ed. M.-H. Ulrich, 693.

Beckers J. M., 1993, Annual Rev. A \& A, 31, 13.

Beckers J. M., 1999, `Adaptive Optics in Astronomy', ed. F. Roddier, Cambridge
Univ. Press, 235.

Beckers J. M., Hege E. K., Murphy H. P., 1983, Proc. SPIE, 444, 85.

Bedding T. R., 1999, astro-ph/9901225, PASP (to appear).

Bedding T. R., Minniti D., Courbin F., Sams B., 1997b, A \& A, 326, 936.

Bedding T. R., Robertson J. G., Marson R. G., 1994, A \& A, 290, 340.

Bedding T. R., Robertson J. G., Marson R. G., Gillingham P. R., Frater R. H.,
O'Sullivan J. D., 1992, Proc. ESO-NOAO conf. `High Resolution Imaging 
Interferometry', ed., J. M. Beckers \& F. Merkle, Garching bei M\"unchen, FRG, 
391.

Bedding T. R., Zijlstra A. A., Von der L\"uhe O., Robertson J. G., Marson R. G., 
Barton J. R., Carter B. S., 1997a, MNRAS, 286, 957.

Beletic J. W., 1988, Proc. ESO-NOAO conf. `High Resolution Imaging 
Interferometry', ed., F. Merkle, Garching bei M\"unchen, FRG, 357.

Beletic J. W., 1996, `Adaptive Optics' OSA Technical Digest series, 13, 216.
 
Beletic J. W., Goody R. M., Tholen D. J., 1989, Icarus, 79, 38.
 
Berio P., Vakili F., Mourard D., Bonneau D., 1998, A \& AS, 129, 609.

Bester M., Danchi W. C., Degiacomi C. G., Townes C. H., 1991, Ap J, 367,
L27.

Bhattacharyya J. C., Rajamohan R., 1990, VBT News, No. 4., 1.

Blazit A., 1986, Proc., `Image Detection and Quality' - SFO, ed., SPIE,
702, 259.

Blazit A., Bonneau D., Foy R., 1987, A \& AS, 71, 57.
 
Blazit A., Bonneau D., Koechlin L., Labeyrie A., 1977a, Ap J, 214, L79. 

Blazit A., Bonneau D., Jose M., Koechlin L., Labeyrie A., On\'eto J. L., 1977b, 
Ap J, 217, L55.

Bl\"ocker T. Balega A., Hofmann K. -H., Lichtenth\"aler J., Osterbart R.,
Weigelt G., 1999, astro-ph/9906473 (to appear in A \& A)

Boccaletti A., Labeyrie A., Ragazzoni R., 1998a, astro-ph/9806144 (to appear in
A \& A).
 
Boccaletti A., Moutou C., Labeyrie A., Kohler D., Vakili F., 1998b, A \& A,
340, 629.

Boksenburg A., 1975, Proc., `Image Processing Techniques in Astronomy', ed.
C. de Jager \& H. Nieuwenhuizen.

Bonneau D., Balega Y., Blazit A., Foy R., Vakili F., Vidal J. L., 1986, A \& AS,
65, 27.

Bonneau D., Blazit A., Foy R., Labeyrie A., 1980, A \& AS, 42, 185.

Bonneau D., Foy R., 1980, A \& A, 86, 295.

Bonneau D., Foy R., 1980, A \& A, 92, L1.

Bonneau D., Foy R., Blazit, A., Labeyrie A., 1982, A \& A, 106, 235.

Bonneau D., Labeyrie A., 1973, Ap J, 181, L1.

Born M., Wolf E., 1984, Principles of Optics, Pergamon Press.

Bosc I., 1988, Proc. ESO-NOAO conf. `High Resolution Imaging 
Interferometry', ed. F. Merkle, Garching bei M\"unchen, FRG, 735.

Bouvier J., Rigaut F., Nadeau D., 1997, A \& A, 323, 139.

Brandl B., Sams B. J., Bertoldi F., Eckart A., Genzel R., Drapatz S., Hofmann R.,
Lowe M., Quirrenbach A., 1996, Ap J, 466, 254.

Brandner W., Bouvier J., Grebel E., Tessier E., de Winter D., Beuzit J. L., 
1995, A \& A, 298, 816.
 
Breckinridge J. B., 1978, Opt. Eng., 17, 156.

Breckinridge J. B., McAlister H. A., Robinson W. A., 1979, App. Opt., 18,
1034.

Brown R. H., 1974, `The Intensity Interferometry, its Applications
to Astronomy', Taylor \& Francis, London.

Brown R. H., Davis J., Allen L. R., 1967, MNRAS, 137, 375.

Brown R. H., Twiss R. Q., 1956, Nature, 178, 1046.

Brown R. H., Twiss R. Q., 1958, Proc. Roy. Soc. A, 248, 222.

Brown R. H., Jennison R. C., Das Gupta M. K., 1952, Nature, 170, 1061.

Brummelaar t. T. A., Mason B. D., Bagnuolo W. G., Hartkopf W. I., McAlister
H. A., Turner N. H., 1996, A J, 112, 1180.

Bruns D., Barrett T., Brusa G., Biasi R., Gallieni D., 1996, OSA conf. on
`Adaptive Optics' Hawaii, Tech. Digest Series 13, 302.

Burns D., Baldwin J. E., Boysen R. C., Haniff C. A., Lawson P. R., Mackay C. D., 
Rogers J., Scott T. R., Warner P. J., Wilson D. M. A., Young J. S., 1997, 
MNRAS, 290, L11.
 
Burns D., Baldwin J. E., Boysen R. C., Haniff C. A., Lawson P. R., Mackay C. D., 
Rogers J., Scott T. R., St-Jacques D., Warner P. J., Wilson D. M. A., Young 
J. S., 1998, MNRAS, 297, 467.

Busher D. F., Haniff C. A., Baldwin J. E., Warner P. J., 1990, MNRAS., 
245, 7.

Callados M., V\`azquez M., 1987, A \& A, 180, 223.

Carbillet M., Lopez B., Aristidi \'E., Bresson Y., Aime C., Ricort G., Prieur 
J. -L., Koechlin L., Helmer G., Lef\`evre J., Cruzal\`ebes., 1996, A \& A, 
314, 112.

Carleton N. P., Traub W. A., Lacasse M. G., Nisenson P., Pearlman M. R.,
Reasenberg R. D., Xu X., Coldwell C. M., Panasyuk A. V., Benson J. A.,
Papaliolios C. D., Predmore R., Schloerb F. P., Dyck H. M., Gibson D., 1994,
SPIE, 2200, 152.

Carlson R. W., Bhattacharyya J. C., Smith B. A., Johnson T. V., Hidayat B.,
Smith S. A., Taylor G. E., O'Leary B., Brinkmann R. T., 1973, Science, 182,
52.

Cassinelli J. P., Mathis J. C., Savage B. D., 1981, Science, 212, 1497.

Chakrabarti S., 1998, Private communication.

Chalabaev A. A., Perrier C., Mariotti J. -M., 1989, A \& A, 210, L1.

Chelli A., Perrier C., Cruz-Gonzalez I., Carrasco L., 1987, A \& A, 177,
51.

Chinnappan V., Saha S. K., Faseehana, 1991, Kod. Obs. Bull., 11, 87.

Chinnappan V., Saxena A. K., Sreenivasan A., 1998, BASI., 26, 371. 

Christou J. C., 1988, Proc. ESO-NOAO conf. `High Resolution Imaging 
Interferometry', ed., F. Merkle, Garching bei M\"unchen, FRG, 97.

Clampin M., Croker J., Paresce F., Rafal M., 1988, Rev. Sci. Instru., 59,
1269.

Clampin M., Nota A., Golimowski D. A., Leitherer C., Ferrari A., 1992,
Ap J, 410, L35.
 
Close L. M., Roddier F., Hora J. L., Graves J. E., Northcott M. J., 
Roddier C., Hoffman W. F., Doyal A., Fazio G. G., Deutsch L. K., 1997, 
Ap J, 489, 210.

Cognet M., 1973, Opt. Communication, 8, 430.

Colavita M., Shao M., Staelin D. H., 1987, App. Opt., 26, 4106.

Colavita M., Boden A. F., Crawford S. L., Meinel A. B., Shao M., Swanson P. N.,
van Belle G. T., Vasisht G., Walker J. M., Wallace J. K., Wizinowich P. L.,
1998, Proc. SPIE conf. on `Astronomical Interferometry', 3350, 776.

Coldwell C. M., Papaliolios C. D., Traub W. A., 1998, Proc. SPIE conf. on
`Astronomical Interferometry', 3350, 424.

Cole W. A., Fekel F. C., Hartkopf W. I., McAlister H. A., Tomkin J., 1992,
A J, 103, 1357.

Cornwell T. J., 1987, A \& A, 180, 269.

Coulman C. E., 1969, Solar Phys., 7, 122.

Coulman C. E., 1974, Solar Phys., 34, 491.

Coulman C. E., 1985, Ann. Rev. A \& A, 23, 19.

Cruzal\'ebes P., Tessier E., Lopez B., Eckart A., Tiph\'ene D., 1996,
A \& AS, 116, 597.

Cuby J. -G., Richard J. -C., Lemonier M., 1990, Proc. SPIE., 1235, 294.

Currie D. Kissel K., Shaya E., Avizonis P., Dowling D., Bonnacini D., 1996,
The Messenger, no. 86, 31.

Dainty J. C., 1975, `Laser Speckle and Related Phenomena', ed., J. C. Dainty,
Springer-Verlag, N Y., 255.

Davidge T. J., Rigaut F., Doyon R., Crampton D., 1997a, A J, 113, 2094.

Davidge T. J., Simons D. A., Rigaut F., Doyon R., Becklin E. E., Crampton D., 
1997b, A J, 114, 2586.

Davis J., 1994, Proc. IAU Symposium No., 158, 135.

Davis J., Tango W. J., 1985a, Proc. Astr. Soc. Australia, 6(1), 34.
 
Davis J., Tango W. J., 1985b, Proc. Astr. Soc. Australia, 6(1), 38.

Davis J., Tango W. J., 1986, Nature, 323, 234.
 
Davis J., Tango W. J., 1996, PASP., 108, 456.

Davis J., Tango W. J., Booth A. J., Minard R. A., Brummelaar t. T. A., 
Shobbrook R. R., 1992, Proc. ESO-NOAO conf. `High Resolution Imaging 
Interferometry', ed., J. M. Beckers \& F. Merkle, Garching bei 
M\"unchen, FRG, 741. 

Davis J., Tango W. J., Booth A. J., O'Byrne J. W., 1998, Proc. SPIE conf. on
`Astronomical Interferometry', 3350, 726.

Davis J., Tango W. J., Booth A. J., Brummelaar t. T. A., Minard R. A., Owens
S. M., 1999a, MNRAS, 303, 773.

Davis J., Tango W. J., Booth A. J., Thorvaldson E. D., Giovannis J.,
1999b, MNRAS, 303, 783.

Denker C., 1998, Solar Phys., 81, 108.

Denker C., de Boer C. R., Volkmer R., Kneer F., 1995, A \& A, 296, 567.

DiBenedetto G. P.,  Conti G., 1983, Ap J, 268, 309.

DiBenedetto G. P., Rabbia Y., 1987, A \& A, 188, 114.

Douglass G. G., Hindsley R. B., Worley C. E., 1997, Ap JS, 111, 289.

Drummond J., Eckart A., Hege E. K., 1988, Icarus, 73, 1.

Duquennoy A., Tokovinin A. A., Leinert CH., Glindemann A., Halbwachs J. L.,
Mayor M., 1996, A \& A, 314, 846.

Durand D., Hardy E., Couture J., 1987, Astron. Soc. Pacific, 99, 686.

Dyck H. M., Benson J. A., Carleton N. P., Coldwell C. M., Lacasse M. G., 
Nisenson P., Panasyuk A. V., Papaliolios C. D., Pearlman M. R., Reasenberg R. 
D., Traub W. A., Xu X., Predmore R., Schloerb F. P., Gibson D., 1995, A J,
109, 378.

Dyck H. M., Benson J. A., Ridgway S. T., 1993, PASP, 105, 610.

Dyck H. M., Benson J. A., van Belle G. T., Ridgway S. T., 1996b, A J, 111, 
1705.

Dyck H. M., van Belle G. T., Benson J. A., 1996a, A J, 112, 294.
 
Dyck H. M., van Belle G. T., Thomson R. R., 1998, A J, (to appear).
 
Ebstein S., Carleton N. P., Papaliolios C., 1989, Ap J, 336, 103.

Falcke H., Davidson K., Hofmann K. -H., Weigelt G., 1996, A \& A, 306, L17.

Faucherre M., Bonneau D., Koechlin L., Vakili F., 1983, A \& A, 120, 263.

Fienup J. R., 1978, Opt. Lett., 3, 27.

Fienup J. R., 1984, Proc. SPIE., 373, 147.

Fischer O., Stecklum B., Leinert Ch., 1998, A \& A, 334, 969.

Fizeau H., 1868, C. R. Acad. Sci. Paris, 66, 934.

Foy R., 1988, Proc., `Instrumentation for Ground Based Optical Astronomy - 
Present and Future', ed., L. Robinson, Springer Verlag, New York, 345.

Foy R, 1992, Proc. ESO-NOAO conf. `High Resolution Imaging Interferometry', 
ed., J. M. Beckers \& F. Merkle, Garching bei M\"unchen, FRG, 5. 

Foy R, 1996, Proc. NATO-ASI, `High Resolution in Astrophysics', Les Houches,
France, ed. A. M. Lagrange, D. Mourard \& P. L\'ena, 193.
 
Foy R., Bonneau D., Blazit A., 1985, A \& A, 149, L13.

Foy R., Labeyrie A., 1985, A \& A, 152, L29.

Fried D. C., 1966, J. Opt. Soc. Am., 56, 1972.

Fugate R. Q., Fried D. L., Ameer G. A., Boeke B. R., Browne S. L., Roberts P. H.,
Roberti P. H., Ruane R. E., Tyler G. A., Wopat L. M., 1991, Nature, 353, 
144.

Gauger A., Balega Y. Y., Irrgang P., Osterbart R., Weigelt G., 1999,
A \& A, 346, 505.

Gerchberg R. W., Saxton W. O., 1972, Optik, 35, 237.

Germain M. E., Douglass G. G., Worley C. E., 1999, A J, 117, 1905.

Gies R. D., Mason B. D., Bagnuolo W. G. (Jr.), Haula M. E., Hartkopf W. I.,
McAlister H. A., Thaller M. L., McKibben W. P., Penny L. R., 1997, Ap. J.,
475, L49.

Gillingham P. R., 1984a, in `Very Large Telescopes, their Instrumentation and
Programs', IAU Colloq. no. 79, ESO, ed., M. -H. Ulrich \& Kj\"ar, Garching 
bei M\"unchen, FRG, 415.

Gillingham P. R., 1984b, `Advanced Technology Optical Telescopes II', ed.,
L. D. Barr \& B. Mark, Proc. SPIE, 444, 165.

Glindemann A., 1997, PASP, 109, 682.

Glindemann A., Lane R. G., Dainty J. C., 1991, Proc. `Digital Signal Processing',
ed., V. Cappellini \& A. G. Constantinides, 59.
 
Glindemann A., Lane R. G., Dainty J. C., 1992, Proc. ESO-NOAO conf. `High 
Resolution Imaging Interferometry', ed., J. M. Beckers \& F. Merkle, Garching 
bei M\"unchen, FRG, 243.

Glindemann A., McCaughrean M. J., Hippler S., Birk C., Wagner K., Rohloff R. -R.,
1997, PASP, 109, 688.

Goecking K. D., Duerbeck H. W., Plewa T., Kaluzny J., Scherti D., Weigelt G.,
Flin P., 1994, A \& A, 289, 827.

Golimowski D. A., Nakajima T., Kulkarni S. R., Oppenheimer B. R., 1995,
Ap J, 444, L101.

Gonsalves S. A., 1982, Opt. Engr., 21, 829.
 
Goodman J. W., 1968, `Introduction to Fourier Optics', McGraw Hill, N. Y.

Goodman J. W., 1975, in `Laser Speckle and Related Phenomena', ed., J. C. Dainty,
Springer-Verlag, Berlin, 9.

Goodman J. W., 1985, `Statistical Optics', Wiley, N. Y.

Graves J. E., Northcott M. J., Roddier C., Roddier F., Anuskiewicz J., 
Monet G., Rigaut F., Madec P. Y., 1993, Proc. ICO-16 Satellite Conf. on 
"Active and Adaptive Optics" ed. F. Merkle, Garching bei M\"unchen, 
Germany, 47.   

Grieger F., Fleischman F., Weigelt G. P., 1988, Proc. ESO-NOAO conf. `High 
Resolution Imaging Interferometry', ed. F. Merkle, Garching bei M\"unchen, 
FRG, 225.
 
Grieger F., Weigelt G., 1992, Proc. ESO-NOAO conf. `High Resolution Imaging 
Interferometry', ed., J. M. Beckers \& F. Merkle, Garching bei M\"unchen, 
FRG, 481.

Haniff C. A., Busher D. F., 1992, J. Opt. Soc. Am. A., 9, 203.

Haniff C. A., Busher D. F., Christou J. C., Ridgway S. T., 1989, MNRAS,
241, 694.

Haniff C. A., Ghez A. M., Gorham P. W., Kulkarni S. R., Mathews K., Neugebauer
G., 1992, A J, 103, 1662.

Haniff C. A., Mackay C. D., Titterington D. J., Sivia D., Baldwin J. E.,
Warner P. J., 1987, Nature, 328, 694.

Haniff C. A., Scholz M., Tuthill P. G., 1995, MNRAS, 276, 640.

Harding G. A., Mack B., Smith F. G., Stokoe J. R., 1979, MNRAS, 188, 241.

Hardy J. W., Lefebvre J. E., Koliopoulos C. L., 1977, J. Opt. Soc. Am., 67, 
360.

Hardy J. W., MacGovern A. J., 1987, Proc. SPIE, 816, 180.
 
Harmanec P., Morand F., Bonneau D., Jiang Y., Yang S., Guinan E. P., Hall D. S.,
Mourard D., Hadrava P., Bozic H., Sterken C., Tallon-Bosc I., Walker G. A. B.,
McCook P. M., Vakili F., Stee P., 1996, A \& A, 312, 879.

Hartkopf W. I., Mason B. D., McAlister H. A., Turner N. H., Barry D. J.,
Franz O. G., Prieto, C.M. 1996, A J, 111, 936.
 
Hartkopf W. I., McAlister H. A., Mason B. D., 1997a, CHARA Contrib. No. 4,
`Third Catalog of Interferometric Measurements of Binary Stars', W.I.

Hartkopf W. I., McAlister H. A., Mason B. D., Brummelaar t. T., Roberts 
Jr., L. C., Turner N. H., Wilson J. W., 1997b, A J, 114, 1639.

Hartkopf W. I., McAlister H. A., Yang X., Fekel F. C., 1992, A J, 103, 
1976.

Hartley M., Mcinnes B., Graham Smith F., 1981, Q. J. Astr. Soc., 22, 272.

Harvey J. W., 1972, Nature, 235, 90.

Harvey J. W., Breckinridge J. B., 1973, Ap J, 182, L137.

Harvey J. W., Schwarzchild M., 1975, Ap J, 196, 221.

Haas M., Leinert C., Richichi A., 1997, A \& A, 326, 1076.

Hawley S. A., Miller J. S., 1977, Ap J, 212, 94.

Hebden J. C., Eckart A., Hege E. K., 1987, Ap J, 314, 690.

Hebden J. C., Hege E. K., Beckers J. M., 1986, Opt. Eng., 25, 712.

Hege E. K., Hubbard E. N., Strittmatter P. A., Worden S. P., 1981, Ap J,
248, 1.

Henry T. J., Soderblom D. R., Donahue R. A., Baliunas S. L., 1996, A J, 
111, 439.

Hess S. L., 1959, Introduction to Theoretical Meteorology (Holt, New York).

Heydari M., Beuzit J. L., 1994, A \& A, 287, L17.

The Hipparcos catalogue, 1997, ESA, SP-1200.

Hofmann K. -H., Mauder W., Weigelt G., 1992, Proc. ESO-NOAO conf. `High 
Resolution Imaging Interferometry', ed., J. M. Beckers \& F. Merkle, Garching 
bei M\"unchen, FRG, 61.
 
Hofmann K. -H., Seggewiss W., Weigelt G., 1995, A \& A, 300, 403.

Hofmann K. -H., Weigelt G., 1988, A \& A, 203, L21.

Hofmann K. -H., Weigelt G., 1993, A \& A, 278, 328.

Hogbom J., 1974, Ap. JS, 15, 417.

Horch E. P., Dinescu D. L., Girard T. M., van Altena W. F., L\'opez C. E.,
Franz O. G., 1996, A J, 111, 1681.

Horch E. P., Morgan J. S., Giaretta G., Kasle D. B., 1992, PASP, 104, 939.

Horch E. P., Ninkov Z., Slawson R. W., 1997, A J, 114, 2117.

Horch E. P., Ninkov Z., van Altena W. F., 1998, Proc. SPIE on `Optical
Astronomical Instrumentation', 3355, 777.
 
Horch E. P., Ninkov Z., van Altena W. F., Meyer R. D., Girard T. M., Timothy J. 
G., 1999, A J, 117, 548.

Horwitz B. A., 1990, Opt. Eng., 29, 1223.

Hummel C. A., 1994, IAU Symp. 158, `Very high resolution imaging' ed., J. G.
Robertson and W. J. Tango, 448.

Hutchings J. B., Crampton D., Morris S. L., Durand D., Steinbring E., 1998a, 
astro-ph/9812159 (to appear A J).

Hutchings J. B., Crampton D., Morris S. L., Steinbring E., 1998b, PASP, 110,
374.

Hutter D. J., 1994, Proc. SPIE., 2200, ed. J. B. Breckinridge, 81.

Irwin M. J., Ibata R. A., Lewis G. F., Totten E. J., 1998, Ap J, 505, 529.

Ishimaru A., 1978, `Wave Propagation and Scattering in Random Media', Academic,
N. Y.

Ivezi\'c Z., Nenkova M., Elitzur M., 1997, User Manual for DUSTY, http://www.pa.uky.edu/\-moshe/dusty.

Iye M., Nishihara E., Hayano Y., Okada T., Takato N., 1992, PASP, 104, 760.

Iye M., Noguchi T., Torti Y., Mikami Y., Ando H., 1991, PASP, 103, 712.

Jaynes E. T., 1982, Proc. IEEE, 70, 939. 

Jefferies S. M., Christou J. C., 1993, Ap J, 415, 862.

Jennison R. C., 1958, MNRAS, 118, 276.

Johanneson A., Bida T., Lites B. W., Scharmer G. B., 1992, A \& A, 258,
572.

Kallistratova M. A., Timanovskiy D. F., 1971, Izv. Akad. Nauk. S S S R.,
Atmos. Ocean Phys., 7, 46.

Kaplan G. H., Hershey J. L., Hughes J. A., Hutter D. J., Johnston K. J.,
Mozurkewich D., Simon R. S., Colavita M. M., Shao M., Hines B. E., Staelin D. 
H., 1988, Proc. ESO-NOAO conf. `High Resolution Imaging Interferometry', 
ed., F. Merkle, Garching bei M\"unchen, FRG, 841. 

Karbelkar S. N., Nityananda R., 1987, J A A, 8, 271.
 
Karovska M., Koechlin L., Nisenson P., Papaliolios C., Standley C., 1989, Ap 
J, 340, 435.

Karovska M., Nisenson P., 1992, Proc. ESO-NOAO conf. `High Resolution 
Imaging Interferometry', ed., J. M. Beckers \& F. Merkle, Garching bei 
M\"unchen, FRG, 141.

Karovska M., Nisenson P., Noyes R., 1986b, Ap J, 308, 260.

Karovska M., Nisenson P., Papaliolios C., Boyle R. P., 1991, Ap J, 374, 
L51.

Karovska M., Nisenson P., Stachnik R. V., 1986a, A J, 92, 4.
 
Kasle D. B., Morgan J. S., 1991, Proc. SPIE on `EUV, X-Ray and Gamma Ray
Instrumentation for Astronomy II', 1549, 52.
   
Keller C. U., Johannesson A., 1995, A \& A Suppl. Ser., 110, 565.

Keller C. U., Von der L\"uhe O., 1992a, Proc. ESO-NOAO conf. `High Resolution 
Imaging Interferometry', ed., J. M. Beckers \& F. Merkle, Garching bei 
M\"unchen, FRG, 453.

Keller C. U., Von der L\"uhe O., 1992b, A \& A, 261, 321.
 
Kinahan B. F., 1976, Ap J, 209, 282.

Kl\"uckers V. A., Edmunds G., Morris R. H., Wooder N., 1997, MNRAS, 284, 
711.

Knox K. T., 1976, J. Opt. Soc. Am., 66, 1236.

Knox K. T., Thompson B. J., 1974, Ap J, 193, L45.

Koechlin L., Lawson P. R., Mourard D., Blazit A., Bonneau D., Morand F., Stee P.,
Tallon-Bosc I., Vakili F., 1996, App. Opt., 35, 3002.

Koechlin L., Morel S., 1998, A \& A, (to appear).

Kolmogorov A., 1941a, in `Turbulence', ed., S. K. Friedlander \& 
L. Topper, 1961, Wiley-Inerscience, N. Y., 151.

Kolmogorov A., 1941b, in `Turbulence', ed., S. K. Friedlander \& 
L. Topper, 1961, Wiley-Inerscience, N. Y., 156.

Kolmogorov A., 1941c, in `Turbulence', ed., S. K. Friedlander \& 
L. Topper, 1961, Wiley-Inerscience, N. Y., 159.

Korff D., 1973, J. Opt. Soc. Am., 63, 971.

Koutchmy S., Belmahdi M., Coulter R. L., D\'emoulin P., Gaizauskas V.,
MacQueen R. M., Monnet G., Mouette J., No\"ens J. C., November I. J.,
Noyes R. W., Sime D. G., Smartt R. N., Sovka J., Vial J. C., Zimmermann J. P.,
Zirker J. B., 1994, A \& A, 281, 249.

Krishan V., Witta P. J., 1994, Ap J, 423, 172.

Kuwamura S., Baba N., Miura N., Hege E. K., 1993a, A J, 105, 665.
 
Kuwamura S., Baba N., Miura N., Noguchi M., Norimoto Y., 1993b, A J, 106,
2532.
 
Kuwamura S., Baba N., Miura N., Noguchi M., Norimoto Y., Isobe S., 1992, Proc. 
ESO-NOAO conf. `High Resolution Imaging Interferometry', ed., J. M. Beckers 
\& F. Merkle, Garching bei M\"unchen, FRG, 461.

Labeyrie A., 1970, A \& A, 6, 85.

Labeyrie A., 1974, Nouv. Rev. Optique, 5, 141.

Labeyrie A., 1975, Ap J, 196, L71.

Labeyrie A., 1978, AR A \& A, 16, 77.

Labeyrie A., 1980, Proc. K P N O conf., 786.

Labeyrie A., 1985, 15th. Advanced Course, Swiss Society of Astrophys. and
Astron., ed. A. Benz, M. Huber \& M. Mayor, 170.

Labeyrie A., 1988, Proc. NATO-ASI, `Diffraction Limited Imaging with Very Large
Telescopes', ed. D. M. Alloin \& J. -M. Mariotti, Carg\'ese, France, 327.

Labeyrie A., 1995, A \& A, 298, 544.

Labeyrie A., 1996, A \& AS, 118, 517.

Labeyrie A., 1998a, 'Exo-earth discoverer, a free-flyer interferometer for 
snapshot imaging and coronography', Proc. Conf., 'Extrasolar planets: formation,
detection and modeling', 27 April-1 May, 1998, Lisbon, Portugal,
(to appear), http://www.obs-hp.fr.

Labeyrie A., 1998b, 'Direct searches: imaging, dark speckle and coronography',
Proc. NATO-ASI, `Planets outside the solar system', 5-15 May, 1998,
Carg\'ese, France, http://www.obs-hp.fr.

Labeyrie A., Koechlin L., Bonneau D., Blazit A., Foy R., 1977, Ap J, 218, 
L75.

Labeyrie A., Schumacher G., Dugu$\acute{e}$ M., Thom C., Bourlon P., Foy
F., Bonneau D., Foy R., 1986, A \& A, 162, 359.

Lai O., Rouan D., Rigaut F., Arsenault R., Gendron E., 1998, A \& A, 334,
783.

Lai O., Rouan D., Rigaut F., Doyon F., Lacombe F., 1999, A \& A, (to appear).

Lampton M., Carlson C. W., 1979, Rev. Sci. Instrum., 50, 1093.

Lampton M., Siegmund O., Raffanti R., 1987, Rev. Sci. Instrum., 58, 2298.

Lane R. G., Bates R. H. T., 1987, J. Opt. Soc. Am. A, 4, 180.

Lawrence R. S., Ochs G. R., Clifford S. F., 1970, J. Opt. Soc. Am., 60,
826.

Lawrence T. W., Goodman D. M., Johansson E. M., Fitch J. P., 1992, App. Opt.
31, 6307.

Lawson P. R., 1994, App. Opt., 33, 1146.

Ledoux C., Th\'eodore B., Petitjean P., Bremer M. N., Lewis G. F., Ibata R. A.,
Irwin M. J., Totten E. J., 1998, astro-ph/9810140, A \& A. (to appear).

Leinert C., Haas M., 1989, A \& A, 221, 110.

Leinert C., Richichi A., Haas M., 1997, A \& A, 318, 472.

Lemonier M., Richard J. -C., Riou D., Fouassier M, 1988, Proc. SPIE.,
`Underwater Imaging', 980, 27.

L\'ena P., 1997, Experimental Astr., 7, 281.

L\'ena P., Lai O., 1999a, `Adaptive Optics in Astronomy', ed. F. Roddier, Cambridge
Univ. Press, 351.

L\'ena P., Lai O., 1999b, `Adaptive Optics in Astronomy', ed. F. Roddier, Cambridge
Univ. Press, 371.

Liang J., Williams D. R., Miller D. T., 1997, J. Opt. Soc. Am. A, 14, 
2884.

Lillard R. L., Schell J. D., 1994, Proc. SPIE, on `Adaptive Optics in Astronomy',
2201, 740.

Liu Y. C., Lohmann A. W., 1973, Opt. Comm., 8, 372.

Lohmann A. W., Weigelt G. P., Wirnitzer B., 1983, App. Opt., 22, 4028.

Lopez B., 1991, `Last Mission at La Silla, April 19 $-$ May 8, on the Measure
of the Wave-front Evolution Velocity', E S O Internal Report.

Love G. D., Major J. V., Purvis A., 1994, Opt. Lett., 19, 1170.

Love G. D., Gourlay J., 1996, Opt. Lett., 21, 1496.

Lowne C. M., 1979, MNRAS, 188, 249.

Lynds C. R., Worden S. P., Harvey J. W., 1976, Ap J, 207, 174.

Malbet F., 1995, astro-ph/9509072, A \& AS.

Malbet F., Berger J. -P., Colavita M. M., Koresko C. D., Beichman C., Boden A. 
F., Kulkarni S. R., Lane B. F., Mobley D. W., Pan X. -P., Shao M., van Belle G. 
T., Wallace J. K., 1998, astro-ph/9808326, Ap. JL. (accepted).

Mali R. K., Bifano T. G., Vandelli N., Horenstein M. N., 1997, 
Opt. Eng., 36, 542.

Marco O., Encrenaz T., Gendron E., 1997, Planet Sp. Sci., 46, 547.

Mariotti J. -M., Denise C., D\'eric F., Ferrari M., Glindemann A., Koehler B.,
Lev\`eque S., Paresce F., Sch\"oller M., Tarenghi M., Verola M., 1998, Proc.
SPIE on `Astronomical Interferometry', Hawaii, 3350, 800.

Martin C., Jelinsky P., Lampton M., Malina R. F., Anger H. O., 1981, Rev. Sci. 
Instrum., 52, 1067.

Masciadri E., Vernin J., Bougeault P., 1999, A \& AS, 137, 203.

Mason B. D., 1995, PASP, 107, 299.

Mason B. D., 1996, A J, 112, 2260.

Mason B. D., Douglass G. G., Hartkopf W. I., 1999, A J, 117, 1023.

Mason B. D., Gies D. R., Hartkopf W. I., Bagnuolo W. G., (Jr.), Brummelaar t.
T., McAlister H. L., 1998, A J, 115, 821.

Mason B. D., Martin C., Hartkopf W. I., Barry D. J., Germain M. E., Douglass G.
G., Worley C. H., Wycoff G. L., Brummelaar t. T., Franz O. G., 1999, A J, 
117, 1890.

Mason B. D., McAlister H. L., Hartkopf W. I., Bagnuolo W. G., (Jr.),  
1993, A J, 105, 220.

Mason B. D., Brummelaar t. T., Gies D. R., Hartkopf W. I., Thaller M. L., 
1997, A J, 114, 2112.

Matson C. L., 1991, J. Opt. Soc. Am. A, 8, 1905.

Max C., 1994, Private communication.

McAlister H. A., 1988, Proc. ESO-NOAO conf. `High Resolution Imaging 
Interferometry', ed. F. Merkle, Garching bei M\"unchen, FRG, 3. 

McAlister H. A., Bagnuolo W. G., Brummelaar t. T., Hartkopf W. I., Shure M. A.,
Sturmann L., Turner N. H., Ridgway S. T., 1998, Proc.
SPIE on `Astronomical Interferometry', Hawaii, 3350, 947.

McAlister H. A., Hartkopf W. I., Hutter D. J, Franz O. G., 1987, A J, 93, 
688.

McAlister H. A., Hartkopf W. I., Franz O. G., 1990, A J, 99, 965.
 
McAlister H. A., Hartkopf W. I., Sowell J. R., Dombrowski E. G., Franz O. G., 
1989, A J, 97, 510.

McAlister H. A., Hartkopf W. I., Mason B. D., Roberts L. C. (Jr.), 
Shara M. M., 1996, A J, 112, 1169.
   
McAlister H. A., Mason B. D., Hartkopf W. I., Shara M. M., 1993, A J, 106,
1639.

Meikle W. P. S., Matcher S. J., Morgan B. L., 1987, Nature, 329, 608.

Meng J. G., Aitken J. M., Hege E. K., Morgan J. S., 1990, J. Opt. Soc. Am. A,
7, 1243.

Men'shchikov A. B., Balega Y. Y., Osterbart R., Weigelt G., 1998, New Astr.,
3, 601. 

Men'shchikov A. B., Henning Th., 1997, A \& A, 318, 879.

Mertz L., 1970, `Transformations in Optics', ed. J. Wiley \& Sons, 16.

Metcalf J. L., 1975, J. Atmos. Sci., 32, 362.
 
Michelson A. A., 1920, Ap J, 51, 257.

Michelson A. A., Pease F. G., 1921, Ap J, 53, 249.

Miura N., Baba N., Isobe S., Noguchi M., Norimoto Y., 1992, J. Modern Opt., 
39, 1137.

Monnier J. D., Tuthill P. G., Lopez B., Cruzal\'ebes P., Danchi W. C., Haniff C. A.,
1999, Ap J, 512, 351.

Morel S., Koechlin L., 1998, Proc. SPIE on `Astronomical Interferometry',
Hawaii, 3350, 257.

Morris M., 1987, PASP, 99, 1115.

Mouillet D., Larwood J. D., Papaloizou J. C., Lagrange A. M., 1997, MNRAS,
292, 896.

Mourard D., Bonneau D., Koechlin L., Labeyrie A., Morand F., Stee P.,
Tallon-Bosc I., Vakili F., 1997, A \& A, 317, 789.

Mourard D., Bosc I., Labeyrie A., Koechlin L., Saha S., 1989, Nature, 342, 
520.

Mourard D., Blazit A., Bonneau D., Koechlin L., Labeyrie A., Morand F., 
Percheron I., Tallon-Bosc I., Vakili F., 1992, 
Proc. ESO-NOAO conf. `High Resolution Imaging Interferometry', 
ed., J. M. Beckers \& F. Merkle, Garching bei M\"unchen, FRG, 707.

Mourard D., Tallon-Bosc I., Blazit A., Bonneau D., Merlin G., Morand F., 
Vakili F., Labeyrie A., 1994a, A \& A, 283, 705.

Mourard D., Tallon-Bosc I., Rigal F., Vakili F., Bonneau D., Morand F., Stee P., 
1994b, A \& A, 288, 675.
 
Mozurkewich D., Johnston K. J., Simon R., Hutter D. J.,
Colavita M. M., Shao M., Pan X. -P., 1991, A J, 101, 2207.

Nakajima T., Golimowski D. A., 1995, A J, 109, 1181.

Nakajima T., Kulkarni S. R., Gorham P. W., Ghez A. M., Neugebauer G., Oke J. B.,
Prince T. A., Readhead A. C. S., 1989, A J, 97, 1510.

Neri R., Grewing M., 1988, A \& A, 196, 338.

Nightingale N. S., 1991, Experimental Astr., 1, 6.

Nightingale N. S., Buscher D. F., 1991, MNRAS, 251, 155.

Nisenson P., 1988, Proc. NATO-ASI, `Diffraction Limited Imaging with Very Large 
Telescopes', ed. D. M. Alloin \& J. -M. Mariotti, Carg\'ese, France, 157.

Nisenson P., 1992, Proc. ESO-NOAO conf. `High Resolution Imaging Interferometry', 
ed., J. M. Beckers \& F. Merkle, Garching bei M\"unchen, FRG, 299.

Nisenson P., 1997, Private communication.

Nisenson P., Papaliolios C., 1999, Ap J, 518, L29.

Nisenson P., Papaliolios C., Karovska M., Noyes R., 1987, Ap J, 320, L15.

Nisenson P., Stachnik R. V., Karovska M., Noyes R., 1985, Ap J, 297, L17.

Nisenson P., Standley C., Gay D., 1990, Proc. Space Telescope Science Institute
Workshop on HST Image Processing, Baltimore, Md.

Northcott M. J., Ayers G. R., Dainty J. C., 1988, J. Opt. Soc. Am. A, 5,
986.

Nota A., Leitherer C., Clampin M., Greenfield P., Golimowski D. A., 1992, A J,
398, 621.

Nulsen P. E. J., Wood P. R., Gilliangham P. R., Bessel M. S., Dopita M. A.,
McCowage C., 1990, Ap J, 358, 266.

O'Byrne J. W., 1988, PASP, 100, 1169.

Ochs G. R., Lawrence R. S., 1972, N O A A Tech. Rep. E R L 251, W P L, 22.

Osterbart R., Balega Y. Y., Weigelt G., Langer N., 1996, Proc. IAU symp. 180,
on `Planetary Nabulae', ed., H. J. Habing \& G. L. M. Lamers, 362.

Osterbart R., Langer N., Weigelt G., 1997, A \& A, 325, 609.
 
Pan X. -P., Shao M., Colavita M. M., 1992, IAU Colloq. 135., ASP Conf. Proc. 32,
`Complementary Approaches to Double and Multiple Star Research', ed. H. A.
McAlister and W. I. Hartkopf, 502.

Pan X. -P., Kulkarni S. R., Colavita M. M., Shao M., 1996, Bull. Am. Astron. 
Soc., 28, 1312.

Papaliolios C., Karovska M., Koechlin L., Nisenson P., Standley C., Heathcote S.,
1989, Nature, 338, 565.

Papaliolios C., Nisenson P., Ebstein, S., 1985, App. Opt., 24, 287.

Paresce F., Clampin M., Cox C., Crocker J., Rafal M., Sen A., Hiltner W. A., 
1988, Proc., `Instrumentation for Ground Based Optical Astronomy - 
Present and Future', ed., L. Robinson, Springer Verlag, New York, 542.

Pauls T. A., Mozurkewich D., Armstrong J. T., Hummel C. A., Benson J. A.,
Hajian A. R., 1998, Proc. SPIE on `Astronomical Interferometry', 3350, 467.

Paxman R. G., Schulz T. J., Fienup J. R., (1992), J. Opt. Soc. Am., 9, 1072.

Pedretti E., Labeyrie A., 1999, A \& A, (to appear).

Pehlemann E., Hofmann K. -H., Weigelt G., 1992, A \& A, 256, 701.

Peraiah A., 1999, Private communication.

Peraiah A., Rao M. S., 1998, A \& AS, 132, 45.

Perrin G., du Foresto V. C., Ridgway S. T., Mariotti J. -M., Traub W. A.,
Carleton N. P., Lacasse M. G., 1998, A \& A, 331, 619.

Petr M. G., Du Foresto V., Beckwith S. V. W., Richichi A., McCaughrean M. J.,
1998, Ap J, 500, 825.

Petrov R. G., Balega Y. Y., Blazit A., Borgnino, J., Foy R., Lagarde S., 
Martin F., Vasyuk V. V., 1992, Proc. ESO-NOAO conf. `High Resolution Imaging 
Interferometry', ed., J. M. Beckers \& F. Merkle, Garching bei M\"unchen, 
FRG, 435.

Petrov R., Roddier F., Aime C., 1986, J. Opt. Soc. Am. A, 3, 634.

Poulet F., Sicardy B., 1996, Bull. Astr. Am. Soc., 28, 1124.

Prieur J. -L., Koechlin C., Andr\'e C., Gallou G., Lucuix C., 1998, Experimental
Astr., 8, 297.

Primmerman C. A., Murphy D. V., Page D. A., Zollars B. G., Barclays H. T.,
1991, Nature, 353, 141.

Racine R., 1984, in `Very Large Telescopes, their Instrumentation and
Programs', IAU Colloq. no. 79, ESO, ed., M. -H. Ulrich \& Kj\"ar, Garching 
bei M\"unchen, FRG, 235.

Racine R., Salmon D., Cowley D., Sovka J., 1991, PASP, 103, 1020.

Readhead A. C. S., Nakajima T. S., Pearson T. J., Neugebauer G., Oke J. B.,
Sargent W. L. W., 1988, A J, 95, 1278.

Reinheimer T., Weigelt G., 1987, A \& A, 176, L17.

Rhodes W. T., Goodman J. W., 1973, J. Opt. Soc. Am., 63, 647.

Ribak E., 1986, J. Opt. Soc. Am. A, 3, 2069.

Ribak E., Lipson S. G., 1981, App. Opt., 20, 1102.

Ribak E., Leibowitz E., Hege E. K., 1985, App. Opt., 24, 3094.

Richardson L. F., 1922, `Weather Prediction by Numerical Process, Intro':
S. Chapman., Cambridge Uni. Press, N. Y.

Ricort G., Aime C., 1979, A \& A, 76, 324.

Ridgway S. T., 1988, Proc. NATO-ASI, `Diffraction Limited Imaging with Very 
Large Telescopes', ed. D. M. Alloin \& J. -M. Mariotti, Carg\'ese, 
France, 307.

Ridgway S. T., 1992, Proc. ESO-NOAO conf. `High Resolution Imaging 
Interferometry', ed., J. M. Beckers \& F. Merkle, Garching bei M\"unchen, FRG, 
653.

Riguat F., Cuby J. -G., Caes M., Monin J. -L., Vittot M., Richard J. -C.,
Rousset G., L\'ena P., 1992, A \& A, 259, L57.

Rigaut F., Salmon D., Arsenault R., Thomas J., Lai O., Rouan D., V\'eran J. P.,
Gigan P., Crampton D., Fletcher J. M., Stilburn J., Boyer C., Jagourel P.,
1998, PASP, 110, 152.

Robertson J. G., Bedding T. R., Aerts C., Waelkens C., Marson R. G., Barton 
J. R., 1999, MNRAS, 302, 245.

Roddier C., Roddier F., 1983, Ap J, 270, L23.
 
Roddier C., Roddier F., 1988, Proc. NATO-ASI, `Diffraction Limited 
Imaging with Very Large Telescopes', ed. D. M. Alloin \& J. -M. Mariotti, 
Carg\'ese, France, 221.

Roddier C., Roddier F., Northcott M. J., Graves J. E., Jim K., 1996, Ap J, 
463, 326.

Roddier F., 1981, Progress in Optics, XIX, 281.

Roddier F., 1986, Opt. Commun., 60, 145.

Roddier F., 1988a, Proc. NATO-ASI, `Diffraction Limited Imaging with Very Large 
Telescopes', ed. D. M. Alloin \& J. -M. Mariotti, Carg\'ese, France, 33.

Roddier F., 1988b, Phys. Rep., 170, 97.
 
Roddier F., 1988c, App. Opt., 27, 1223.

Roddier F., 1990, Opt. Eng., 29, 1239.

Roddier F., 1994, in `Adaptive Optics in Astronomy', ed., M. A. Esley \& F. 
Merkle, SPIE, Bellingham, WA, 1487, 123.

Roddier F. J., Graves J. E., McKenna D., Northcott M. J., 1991, SPIE Conf., 
1524, 248.
 
Roddier F., Roddier C., 1985, Ap J, 295, L21.

Roddier F., Roddier C., 1997, PASP., 109, 815.

Roddier F., Roddier C., Brahic A., Dumas C., Graves J. E., Northcott M. J., 
Owen T., 1997, Planet Sp. Sci., 45, 1031.

Roddier F., Roddier C., Graves J. E., Northcott M. J., 1995, Ap J, 443, 
249.

Rodriguez L. F., Sosa N., Rosa F., Fuensalida J. J., 1992, Proc. ESO-NOAO conf. 
`High Resolution Imaging Interferometry', ed., J. M. Beckers \& F. Merkle, 
Garching bei M\"unchen, FRG, 621.

Roggemann M. C., Welsh B. M., Fugate R. Q., 1997, Rev. Modern Phys., 69, 437.

Rogstad D. H., 1968, App. Opt., 7, 585.

Rouan D., 1996, Proc. NATO-ASI, `High Resolution in Astrophysics', Les Houches,
France, ed. A. M. Lagrange, D. Mourard \& P. L\'ena, 293.

Rouan D., Field D., Lemaire J. -L., Lai O., de Foresto G. P., Falgarone E., 
Deltorn J. -M., 1997, MNRAS, 284, 395.

Rouan D., Rigaut F., Alloin D., Doyon R., Lai O., Crampton D., Gendron E.,
Arsenault R., 1998, astro-ph/9807053, A \& A.
 
Rousset G., 1999, `Adaptive Optics in Astronomy', ed. F. Roddier, Cambridge
Univ. Press, 91.

Rousset G., Fontanella J. C., Kem P., Gigan P., Rigaut F., L$\acute{e}$na P.,
Boyer P., Jagourel P., Gaffard J. P., Merkle F., 1990, A \& A, 230, L29.

Ryan S. G., Wood P. G., 1995, Pub. ASA., 12, 89.

Saha S. K., 1999, Ind. J. Phys., 73B, 552.

Saha S. K., Chinnappan V., 1999, BASI, 27, 327.

Saha S. K., Jayarajan A. P., Rangarajan K. E., Chatterjee S., 1988, Proc.
ESO-NOAO conf. `High Resolution Imaging Interferometry', ed. F. Merkle,
Garching bei M\"unchen, FRG, 661.

Saha S. K., Jayarajan A. P., Sudheendra G., Umesh Chandra A., 1997a, 
BASI, 25, 379.

Saha S. K., Nagabhushana B. S., Ananth A. V., Venkatakrishnan P., 
1997b, Kod. Obs. Bull., 13, 91. 

Saha S. K., Rajamohan R., Vivekananda Rao P., Som Sunder G., Swaminathan
R., Lokanadham B., 1997c, BASI, 25, 563.

Saha S. K., Sankarasubramanian K., Sridharan R., 1999d, `Auto-correlation
of Binary stars' (in preparation).

Saha S. K., Sridharan R., Sankarasubramanian K., 1999b, `Speckle image 
reconstruction of binary stars', Submitted to BASI.

Saha S. K., Sridharan R., Sankarasubramanian K., 1999c, `Triple correlation
of Binary stars' (in preparation).
 
Saha S. K., Sudheendra G., Umesh Chandra A., Chinnappan V., 1999, 
Experimental Astr., 9, 39. 

Saha S. K., Venkatakrishnan P., 1997, BASI, 25, 329.

Saha S. K., Venkatakrishnan P., Jayarajan A. P., Jayavel N., 1987,
Curr. Sci., 56, 985.

Sams B. J., Schuster K., Brandl B., 1996, Ap. J., 459, 491.

Sandler D., Cuellar L., Lefebvre M., Barrett T., Arnold R., Johnson P., Rego A.,
Smith G., Taylor G., Spivey B., 1994, J. Opt. Soc. Am. A, 11, 858.
 
Saxena A. K., 1993, Private communication.
 
Saxena A. K., Jayarajan A. P., 1981, App. Opt., 20, 724.

Saxena A. K., Lancelot J. P., 1982, App. Opt., 21, 4030.

Schertl D., Hofmann K. -H., Seggewiss W., Weigelt G., 1996, A \& A, 302,
327.

Sch\"oller M., Brandner W., Lehmann T., Weigelt G., Zinnecker H., 1996, A \& A,
315, 445.

S\'echaud M., 1999, `Adaptive Optics in Astronomy', ed. F. Roddier, Cambridge
Univ. Press, 57.

Seldin J. H., Paxman R. G., 1994, Proc. SPIE., 2302, 268.

Shao M., 1988, Proc. ESO-NOAO conf. `High Resolution Imaging 
Interferometry', ed., F. Merkle, Garching bei M\"unchen, FRG, 823.

Shao M., Colavita M. M., 1987, Proc. ESO-NOAO Workshop., ed. J. W. Goad, 115.
 
Shao M., Colavita M. M., 1994, IAU Symp. 158, `Very high resolution imaging', 
ed., J. G. Robertson and W. J. Tango, 413.

Shao M., Colavita M. M., Hines B. E., Staelin D. H., Hutter D. J., Johnston
K. J., Mozurkewich D., Simon R. H., Hershey J. L., Hughes J. A., Kaplan G. H.,
1988, A \& A, 193, 357.

Shao M., Colavita M. M., Hines B. E., Hershey J. L., Hughes J. A., 
Hutter D. J., Kaplan G. H., Johnston K. J., Mozurkewich D., Simon R. H., 
Pan X. -P., 1990, A J, 100, 1701.

Shao M., Colavita M. M., Staelin D. H., 1986a, Proc. SPIE, 628, 250.

Shao M., Colavita M. M., Staelin D. H., Simon R., Johnston K. J., 1986b, Proc.
IAU symposium 109, 46.
 
Shao M., Staelin D. H., 1980, App. Phys., 19, 1519.

Shelton J. C., Baliunas S. L., 1993, Proc. SPIE., on `Active and Adaptive
Optical Components and Systems II', 1920, 371.

Shields G. A., 1999, astro-ph/9903401., PASP.

Sicardy B., Roddier F., Roddier C., Perozzi E., Graves J. E., Guyon O.,
Northcott M. J., 1999, Nature, 400, 731.

Siegmund O. H. W., Lampton M., Bixler J., Chakrabarti S., Vallaga J., Bowyer
S., Malina R. F., 1986, J. Opt. Soc. Am., 3, 2139.

Siegmund O. H. W., Stock J. M., Marsh D. R., Gummin M. A., Raffanti R., Hull J.,
Gaines G. A., Welsh B. Y., Donakowski B., Jelinsky P. N., Sasseen T., Tom J. L.,
Higgins B., Magoncelli T., Hamilton W., Battel S. J., Poland A. L., Jhabvala M. 
D., Sizemore J., Shannon J., 1994, Proc. SPIE, 2280, 89.

Simon M., Close L. M., Beck T. L., 1999, A J, 117, 1375.

Singh K., 1999, Private communication.

Sridharan R., Venkatakrishnan P., 1999, `Speckle image reconstruction of solar
features', Presented at the XIX ASI meeting, Bangalore, India, Feb. 1-4, 1999.

Stachnik R. V., Nisenson P., Ehn D. C., Hudgin R. H., Schirf V. E., 1977,
Nature, 266, 149.

Stachnik R. V., Nisenson P., Noyes R. W., 1983, Ap J, 271, L37.

Stahl S. S., Sandler D. G., 1995, Ap J, 454, L153.

Stee P., 1999, http://www.obs-nice.fr/stee/welcome.html

Stee P., de Ara\'ujo, Vakili F., Mourard D., Arnold I., Bonneau D., Morand F.,
Tallon-Bosc I., 1995, A \& A, 300, 219.

Stee P., Vakili F., Bonneau D., Mourard D., 1998, A \& A, 332, 268.

Tallon M., Foy R., 1990, A \& A, 235, 549.

Tallon M., Foy R., Blazit A., 1988, ESO Conf. on `Very Large Telescopes and
their Instrumentation', ed. M.-H. Ulrich, 743.

Tatarski V. I., 1967, `Wave Propagation in a Turbulent Medium', Dover, N. Y.

Tatarski V. I., 1968, `The Effect of the Turbulent Atmosphere on Wave 
propagation', N S F Report TT-68-50464.

Tatarski V. I., 1993, J. Opt. Soc. Am. A, 56, 1380.

Taylor G. L., 1921, in `Turbulence', ed., S. K. Friedlander \&
L. Topper, 1961, Wiley-Inerscience, N. Y., 1.

Taylor G. L., 1935, in `Turbulence', ed., S. K. Friedlander \& 
L. Topper, 1961, Wiley-Inerscience, N. Y., 18.

Tej A., Chandrasekhar T., Ashok M. N., Ragland S., Richichi A., Stecklum B.,
1999, A J, 117, 1857.

Thi\'ebaut E., 1994, A \& A, 284, 340.

Thi\'ebaut E., Balega Y., Balega I., Belkine I., Bouvier J., Foy R., Blazit A.,
Bonneau D., 1995, A \& A, 304, L17.

Thom C., Granes P., Vakili F., 1986, A \& A, 165, L13.

Thompson L. A., Gardner C. S., 1988, Nature, 328, 229.

Thureau N., Chesneau O., Berio P., Bonneau D., Mourard D., Stee P., Vakili F.,
Verinaud C., 1998, Proc. SPIE on `Astronomical Interferometry', 3350, 505.

Timothy J. G., 1986, SPIE Ultraviolet Technology, 687, 109.

Timothy J. G., 1993, Proc. SPIE., 1982, 4.

Timothy J. G., Morgan J.S., Slater D. C., Kasle D. B., Bybee R. L., Culver H. 
K., 1989, Proc., SPIE, 1158, Ultraviolet Technology III, ed. R. E. 
Huffmann, 104.

Tokovinin A., 1992, Proc. ESO-NOAO conf. `High Resolution Imaging Interferometry', 
ed., J. M. Beckers \& F. Merkle, Garching bei M\"unchen, FRG, 425.

Tokovinin A., 1997, A \& AS, 124, 75.
 
Torres G., Stefanik R. P., Latham D. W., 1997, Ap J, 485, 167.

Troxel S. E., Welsh B. M., Roggemann M. C., 1994, J. Opt. Soc. Am A, 11, 2100.

Tsvang L. R., 1969, Radio Sci., 4, 1175.

Tuthill P. G., Haniff C. A., Baldwin J. E., 1997, MNRAS, 285, 529.

Tuthill P. G., Haniff C. A., Baldwin J. E., 1999a, MNRAS, 306, 353.

Tuthill P. G., Monnier J. D., Danchi W. C., 1999b, Nature, 398, 487.

Twichell J. C., Burke B. E., Reich R. K., McGonagle W. H., Huang C. M., Bautz
M. W., Doty J. A., Ricker G. R., Mountain R. W., Dolat V. S., 1990, A. Rev. Sci. 
Instrum., 61, 2744.

Ulrich M. -H., 1988, Proc. ESO-NOAO conf. `High Resolution Imaging 
Interferometry', ed. F. Merkle, Garching bei M\"unchen, FRG, 33.

Vakili F., Berio P., Bonneau D., Chesneau O., Mourard D., Stee P., Thureau N.,
1998a, Proc., `Be stars', ed. A. M. Hubert and C. Jaschek, 173.

Vakili F., Mourard D., Bonneau D., Morand F., Stee P., 1997, A \& A, 323,
183.

Vakili F., Mourard D., Stee P., Bonneau D., Berio P., Chesneau O., Thureau N.,
Morand F., Labeyrie A., Tallon-Bosc I., 1998b, A \& A, 335, 261.

van Belle G. T., Boden A. F., Colavita M. M., Shao M., Vasisht G., Wallace J. 
K., 1998, Proc. SPIE on `Astronomical Interferometry', 3350, 362.

van Belle G. T., Dyck H. M., Benson J. A., Lacasse M. G., 1996, A J, 112,
2147.

van Belle G. T., Dyck H. M., Thompson R. R., Benson J. A., Kannappan S. J., 
1997, A J, 114, 2150.

Vasisht G., Boden A. F., Colavita M. M., Crawford S. L., Shao M., Swanson P.,
van Belle G. T., Wallace J. K., 1998, Proc. SPIE on `Astronomical Interferometry', 
3350, 354.

Venkatakrishnan P., 1987, MNRAS, 229, 379.

Venkatakrishnan P., Chatterjee S., 1987, MNRAS, 224, 265.

Venkatakrishnan P., Saha S. K., Shevgaonkar R. K., 1989, Proc., `Image
Processing in Astronomy', ed. T. Velusamy, 57.

Vernin J., Weigelt G., Caccia J. L., M\"uller M., 1991, A \& A, 243, 553.

Von der L\"uhe O., 1984, J. Opt. Soc. Am. A, 1, 510. 

Von der L\"uhe O., 1989, in High Spatial Resolution Solar Observation, Proc. of the
Tenth Sermento Peak Summer Workshop, ed., O. Von der L\"uhe, Sunspot, New Mexico.
 
Von der L\"uhe O., Dunn R. B., 1987, A \& A, 177, 265. 

Von der L\"uhe O., Pehlemann E., 1988, Proc. ESO-NOAO conf. `High Resolution Imaging 
Interferometry', ed., F. Merkle, Garching bei M\"unchen, FRG, 159.

Von der L\"uhe O., Zirker J. B., 1988, Proc. ESO-NOAO conf. `High Resolution Imaging 
Interferometry', ed., F. Merkle, Garching bei M\"unchen, FRG, 77.

Walker G., Chapman S., Mandushev G., Racine R., Nadeau D., Doyon R., V\'eran 
J. P., 1998, astro-ph 9810443.

Walters D. L., Kunkel K. E., 1981, J. Opt. Soc. Am., 71, 397.

Weigelt G., 1977, Opt. Communication, 21, 55.

Weigelt G., 1978, App. Opt., 17, 2660.

Weigelt G., 1988, Proc. NATO-ASI, `Diffraction Limited Imaging with Very Large
Telescopes', ed. D. M. Alloin \& J. -M. Mariotti, Carg\'ese, France, 191.

Weigelt G., Bair G., 1985, A \& A, 150, L18.

Weigelt G., Balega Y., Bl\"ocker T., Fleischer A. J., Osterbart R., Winters J. M.,
1998, A \& A, 333, L51.

Weigelt G., Balega Y., Hofmann K. -H., Scholz M., 1996, A \& A, 316, L21.

Weigelt G., Balega Y., Preibisch T., Schertl D., Sch\"oller M., Zinnecker H., 
1999, astro-ph/9906233 (to appear in A \& A).

Weigelt G., Ebersberger J., 1986, A \& A, 163, L5.

Weigelt G., Grieger F., Hofmann K. -H., Pausenberger R., 1992, Proc. 
ESO-NOAO conf. `High Resolution Imaging Interferometry', ed., J. M. Beckers 
\& F. Merkle, Garching bei M\"unchen, FRG, 471.

Weinberger A. J., Neugebauer G., Mathews K., 1996, MNRAS, 196, 1005.

Weitzel N., Haas M., Leinert Ch., 1992, Proc. ESO-NOAO conf. `High Resolution 
Imaging Interferometry', ed., J. M. Beckers \& F. Merkle, Garching bei 
M\"unchen, FRG, 511.

Wilken V., de Boer C. R., Denkar C., Kneer F., 1997, A \& A, 325, 819.

Wilson R. W., Baldwin J. E., Busher D. F., Warner P. J., 1992, MNRAS, 257,
369.

Wilson R. W., Dhillon V. S., Haniff C. A., 1997, MNRAS, 291, 819.

Wirnitzer B., 1985, J. Opt. Soc. Am. A, 2, 14.

Wittkowski M., Balega Y., Beckert T., Duschi W. J., Hofmann K. -H., Weigelt G., 
1998b, A \& A, 329, L45.

Wittkowski M., Langer N., Weigelt G., 1998a, A \& A, 340, L39.

Wizinovitch P. L., Nelson J. E., Mast T. S., Glecker A. D., 1994, Proc. SPIE.,
on `Adaptive Optics in Astronomy', 2201, 22.

Wood P. R., 1985, Proc., A S A, 6, 120.                                                                                                                                                                                      

Wood P. R., Bessel M. S., Dopita M. A., 1986, Ap J, 311, 632.

Wood P. R., Meatheringham S. J., Dopita M. A., Morgan, D. M., 1987, Ap 
J, 320, 178.

Wood P. R., Nulsen P. E. J., Gillingham P. R., Bessel M. S., Dopita M. A., 
McCowage C., 1989, Ap J, 339, 1073.

Woolf N. J., Cheng A., 1988, ESO Conf. on `Very Large Telescopes and
their Instrumentation', ed. M.-H. Ulrich, 845.

Worden S. P., Lynds C. R., Harvey J. W., 1976, J. Opt. Soc. Am., 66, 1243.

Wyngaard J. C., Izumi Y., Collins S. A., 1971, J. Opt. Soc. Am., 60, 1495.

Wyant J. C., 1974, App. Opt., 13, 200.

Wyant J. C., Kaliopoulos C. L., 1981, in Agard Conf. Proc., 300.

Young J. S., Baldwin J. E., Beckett G., Boysen R. C., Haniff C. A., Lawson P. 
R., Mackay C. D., Rogers J., St-Jacques D., Warner P. J., Wilson D. M. A., 1998, 
Proc. SPIE., on `Astronomical Interferometry', 3350, 746.

Young S., Packham C., Hough J. H., Efstathiou A., 1996, MNRAS, 283, L1.

Zago L., 1995, http://www.eso.org/gen-fac/pubs/astclim/lz-thesis/node4-html.

Zappa F., Lacaita A. L., Cova S. D., Lovati P., 1996, Opt. Eng., 35, 938.

Zeidler P., Appenzeller I., Hofmann K. -H., Mauder W., Wagner S., Weigelt G.,
1992, Proc. ESO-NOAO conf. `High Resolution Imaging Interferometry', 
ed., J. M. Beckers \& F. Merkle, Garching bei M\"unchen, FRG, 67.

Zienkiewicz O. C., 1967, `The Finite Element Methods in Structural and
Continuum Mechanics', McGrawhill Publication.

}                                         
\vskip 20pt
\centerline {\bf Appendix I}
\bigskip
\centerline {\bf Table of symbols}
\settabs\+ OOOOOOOOO &00000000000000000000000000  \cr
\+                    & \cr
\+ $V{\bf (r}, t)$ &  the monochromatic optical wave \cr
\+ ${\bf r} = (x, y, z)$ & position vector of a point in space \cr
\+ $t$ & time \cr
\+ $\nu_\circ$ & frequency of the wave \cr 
\+ ${\it a}({\bf r})$ & complex amplitude of the wave \cr 
\+ ${\cal U}{\bf (r}, t)$ & complex representation of the analytical signal \cr
\+ $\Psi({\bf r})$  & complex vector functions of position ${\bf r}$\cr 
\+ $\Psi({\bf x})$ & complex amplitude in the image plane \cr
\+ ${\cal I}({\bf x})$ & intensity of light \cr
\+ $< \ >$ & ensemble average \cr
\+ $^\ast$ & complex operator \cr
\+ $\lambda$ & wavelength \cr
\+ ${\bf x} = (x, y)$ & 2-dimensional space vector \cr
\+ $\Psi_\circ({\bf x_\circ})$ & complex amplitude in the telescope pupil plane \cr
\+ ${\cal T}({\bf x, x_\circ})$ & impulse response of the optical system \cr
\+ ${\cal P}({\bf x})$ & the pupil transmission function \cr
\+ $\ast$ & convolution operator \cr
\+ $\widehat{\cal P}({\bf u})$ & transfer function \cr
\+ ${\cal S}({\bf x})$ & point spread function (PSF) \cr 
\+ $\widehat {\cal S}({\bf u})$ & optical transfer function (OTF) \cr
\+ $\mid\widehat{\cal S}({\bf u})\mid^2$ & modulus transfer function (MTF) \cr
\+ ${\cal R}$ & resolving power of an optical system \cr
\+ $\omega$ & angular frequency \cr
\+ ${\tt T}$ & period \cr
\+ ${\bf V}_j$ & monochromatic wave vector \cr
\+ $j$ & = 1, 2, 3 \cr
\+ ${\cal J}_{12}$ & interference term \cr 
\+ $\delta$ & phase difference \cr
\+ $\Delta\varphi$ & optical path difference (OPD) \cr  
\+ $\lambda_\circ$ & wavelength in vacuum \cr
\+ ${\bf K}_j$ & constants \cr
\+ $r_j$ & pin-holes in the wave field \cr
\+ $s_j$ & distances of a meeting point of the two beams from the two pin-holes \cr
\+ $\tau_j$ &  travel time from the respective pin-holes to the meeting point \cr
\+ $c$ & velocity of light \cr
\+ $\pmb{\gamma}_{12}(\tau)$ & complex degree of (mutual) coherence \cr
\+ $\pmb{\Gamma}_{12}(\tau)$ & mutual coherence \cr
\+ $\pmb{\Gamma}_{11}(\tau)$ & self coherence \cr
\+ $\tau$ & delay \cr
\+ $\tau_c$ & temporal width or coherence time \cr 
\+ $\Delta\nu$ & spectral width \cr 
\+ $l_c$ & coherence length \cr 
\+ $\pmb{\gamma}_{12}(0)$ & spatial coherence \cr 
\+ $\psi_{12}$ & argument of $\pmb{\gamma}_{12}(\tau)$ \cr
\+ $x$ & distance from the origin \cr
\+ $z$ & distance from the aperture \cr
\+ $d(x)$ & OPD corresponding to $x$ \cr 
\+ $b$ & distance between the two apertures \cr
\+ ${\cal V}$ & contrast of the fringes \cr 
\+ ${\bf B}$ & baseline vector\cr
\+ $D$ &  diameter of the sub-apertures \cr
\+ $f$ & focal length \cr
\+ $\Psi(D\cdot x)$ & envelop of the image of each sub-apertures (Airy disk) \cr
\+ $\psi_B$ & phase of $\gamma_B(0)$ \cr
\+ $d$ &  extra optical path in front of one aperture \cr
\+ $v_a$ &  average velocity of a viscous fluid \cr
\+ $l$ & characteristic size of viscous fluid \cr
\+ $R_e$ & Reynolds number \cr
\+ $\nu_v$ & kinematic viscosity of fluid \cr
\+ $L_\circ$ & outer scale length \cr 
\+ $k_{L_{\circ}}$ & spatial frequency of outer scale \cr 
\+ $n({\bf r, t})$ & refractive index of the atmosphere \cr
\+ $n_\circ$ & mean refractive index of air \cr
\+ $P$ & pressure \cr 
\+ $T$ & temperature \cr
\+ $\Phi_n({\bf k})$ & power spectral density \cr 
\+ $k_\circ$ & critical wave number \cr
\+ $l_\circ$ & inner scale length \cr
\+ $k_{l_{\circ}}$ & spatial frequency of inner scale \cr
\+ ${\cal C}_n^2$ & refractive index structure constant \cr
\+ ${\cal D}_n({\bf r})$ & refractive index structure function \cr
\+ ${\cal B}_n({\bf r})$ & covariance function \cr
\+ ${\cal D}_v ({\bf r})$ & velocity structure function \cr
\+ ${\cal C}_v^2$ & velocity structure constant \cr
\+ ${\cal D}_T ({\bf r})$ & temperature structure function \cr
\+ ${\cal C}_T^2$ & temperature structure constant \cr
\+ $h$ & height \cr
\+ $({\bf x}, h)$ & co-ordinate \cr
\+ $\Psi_h({\bf x})$ & complex amplitude at co-ordinate, $({\bf x}, h)$ \cr
\+ $<\psi_h({\bf x})>$ & average value of the phase at $h$ \cr 
\+ $\delta h_j$ & thickness of the turbulence layer \cr
\+ ${\cal D}_{\psi_j}\left(\pmb{\xi}\right)$ & phase structure function \cr
\+ ${\cal D}_n\left(\pmb{\xi}, \zeta\right)$ & refractive index structure function \cr
\+ ${\cal B}_{h_j}\left(\pmb{\xi}\right)$ & covariance of the phase \cr
\+ ${\cal B}\left(\pmb{\xi}\right)$ & coherence function \cr
\+ $\gamma$ & distance from the zenith \cr 
\+ $r_\circ$ & Fried's parameter \cr
\+ ${\cal O}({\bf x})$ & object illumination \cr
\+ $<\widehat{{\cal S}}({\bf u})>$ & transfer function for long-exposure images \cr
\+ ${\bf u}$ & spatial frequency vector with magnitude $u$ \cr
\+ $\widehat{\cal I}({\bf u})$ & image spectrum \cr
\+ $\widehat{\cal O}({\bf u})$ & object spectrum \cr
\+ ${\cal B}({\bf u})$ &  atmosphere transfer function \cr
\+ ${\cal T}({\bf u})$ &  telescope transfer function \cr
\+ $\psi({\bf u})$ & Fourier phase at ${\bf u}$ \cr
\+ $arg\mid\mid$ & the phase of ` \ \ ' \cr  
\+ ${\cal P}$ & aperture of large telescope \cr
\+ $p_j$ & sub-apertures \cr
\+ $\sigma$ & coherence area \cr
\+ ${\cal N}_j({\bf x})$ & noise contamination \cr
\+ ${\cal W}({\bf x})$ & intensity distribution of objective prism speckle spectrogram \cr
\+ ${\cal G}_m({\bf x})$ & spectrum of the object pixel \cr
\+ ${\cal D}_{\cal I}({\bf x})$ & differential image \cr
\+ $\beta_{123}$ & closure phase \cr  
\+ $\theta_i, \theta_j$ & error terms introduced by errors at the individual antennae \cr
\+ ${\cal A}\delta({\bf x})$ & Dirac impulse of a point source \cr
\+ ${\cal C}_{{\cal O}_1}{(\bf x)}$ & autocorrelation of the object \cr 
\+ ${\bf x}_k$ & position of the brightest pixel \cr
\+ ${\cal I}_{sa}({\bf x})$ & shift-and-add image \cr
\+ ${\it w}[{\cal I}_k({\bf x}_k)]$ & weighting of the brightest pixel \cr
\+ $mask_m({\bf x})$ &  m$^{th}$ speckle mask \cr
\+ $m{\cal I}_m({\bf x})$ & m$^{th}$ masked speckled image \cr
\+ $\otimes$ & correlation \cr
\+ ${\bf x^\prime}_k$ & constant position vectors \cr
\+ $d_k$ & positive constant \cr
\+ $\widehat{\cal N}({\bf u})$ & noise spectrum \cr
\+ $<\mid\widehat{\cal I}({\bf u})\mid^{2}>$ & image energy spectrum \cr
\+ ${\bf \Delta u}$ & offset spatial frequency \cr
\+ $\theta^{TC}_{\cal O}$ & phase of the bispectrum \cr
\+ $\theta_j$ & apertures \cr

\end